\documentclass[aps,prb,twocolumn,superscriptaddress,10pt,article,nofootinbib,longbibliography]{revtex4-2}
\usepackage{graphicx,amsfonts,amssymb,amsmath,hyperref,hypcap,enumerate,bm,color,mathrsfs,latexsym,psfrag,tabularx}

\newcommand{\bra}[1]{\left\langle#1\right|}
\newcommand{\ket}[1]{\left|#1\right\rangle}

\newcommand{\beq}{\begin{equation}}
\newcommand{\eneq}{\end{equation}}

\def\qq{\mathbf{q}}
\def\kk{\mathbf{k}}
\def\KK{\mathbf{K}}

\def\pp{\mathbf{p}}
\def\PP{\mathbf{P}}
\def\RR{\mathbf{R}}

\def\rr{\mathbf{r}}
\def\GG{\mathbf{G}}
\def\QQ{\mathbf{Q}}
\def\aa{\mathbf{a}}
\def\bb{\mathbf{b}}
\def\uu{\mathbf{u}}

\def\KK{\mathbf{K}}
\def\qq{\mathbf{q}}
\def\pp{\mathbf{p}}
\def\pp{\mathbf{p}}
\def\GG{\mathbf{G}}
\def\QQ{\mathbf{Q}}
\def\RR{\mathbf{R}}

\def\dd{\mathbf{d}}
\def\aa{\mathbf{a}}
\def\bb{\mathbf{b}}
\def\ee{\epsilon}

\def\mS{{\mathcal{S}}}

\def\hH{{ \hat{H} }}

\def\mF{{\mathcal{F}}}

\def\pr{\prime}

\def\mS{{\mathcal{S}}}

\def\ie{{\it i.e.},\ }

\begin{document}

\title{Electron-$K$-Phonon Interaction In Twisted Bilayer Graphene}

\author{Chao-Xing Liu$^*$}
\affiliation{Department of Physics, The Pennsylvania State University, University Park,  Pennsylvania 16802, USA}
\affiliation{Department of Physics, Princeton University, Princeton, NJ 08544} 
\author{Yulin Chen}
\affiliation{Department of Physics, University of Oxford, Oxford, OX1 3PU, United Kingdom}
\author{Ali Yazdani}
\affiliation{Department of Physics, Princeton University, Princeton, NJ 08544} 
\author{B. Andrei Bernevig$^*$} 
\affiliation{Department of Physics, Princeton University, Princeton, NJ 08544} 
\affiliation{Donostia International Physics Center, P. Manuel de Lardizabal 4, 20018 Donostia-San Sebastian, Spain}
\affiliation{IKERBASQUE, Basque Foundation for Science, Bilbao, Spain}

\begin{abstract} 
We develop an analytic theory to describe the interaction between electrons and $K$-phonons and study its influence on superconductivity in the \emph{bare bands} of twisted bilayer graphene (TBG). We find that, due to symmetry and the two-center approximation, only one optical $K$  phonon ($\sim 160meV$) of graphene is responsible for inter-valley electron-phonon interaction. This phonon has recently been found in angular-resolved photo-emission spectroscopy to be responsible for replicas of the TBG flat bands. By projecting the interaction to the TBG flat bands, we perform the full symmetry analysis of phonon-mediated attractive interaction and pairing channels in the Chern basis, and show that several channels are guaranteed to have gapless order parameters. From the linearized gap equations, we find that the highest $T_c$ pairing induced by this phonon is a singlet gapped s-wave inter-Chern-band order parameter, followed closely by a gapless nematic d-wave intra-Chern-band order parameter. We justify these results analytically, using the topological heavy fermion mapping of TBG which has allowed us to obtain an analytic form of phonon-mediated attractive interaction and to analytically solve the linearized and $T=0$ gap equations. For the intra-Chern-band channel, the nematic state with nodes is shown to be stabilized in the chiral flat band limit.
While the flat band Coulomb interaction can be screened sufficiently enough - around Van-Hove singularities - to allow for electron-phonon based superconductivity, it is unlikely that this effect can be maintained in the lower density of states excitation bands around the correlated insulator states. 
\end{abstract}

\date{\today}


\maketitle

{\it Introduction - } Superconductivity in twisted bilayer graphene (TBG) appears within its phase diagram around the correlated insulator states \cite{cao2018correlated,cao2018unconventional,lu2019superconductors,codecido2019correlated,yankowitz2019tuning,cao2021nematicity,oh2021evidence,liu2021tuning,saito2020independent,arora2020superconductivity,codecido2019correlated,andrei2020graphene,liu2021orbital,kwan2021kekule,nuckolls2023quantum,cualuguaru2022spectroscopy,kang2021cascades,kang2018symmetry,kang2019strong,vafek2020renormalization}. Amongst the mechanisms suggested for superconductivity are the phonons, spin fluctuations, skyrmions, and others \cite{khalaf2021charged,khalaf2022symmetry,po2018origin,you2019superconductivity,lian2019twisted,wu2018theory,wu2019topological,wu2019phonon,wu2019identification,chichinadze2020nematic,fernandes2021charge,wang2021topological,yu2022euler2,sharma2020superconductivity,isobe2018unconventional,roy2019unconventional,peltonen2018mean,chichinadze2020nematic,kennes2018strong,gonzalez2019kohn,cea2021coulomb,shavit2021theory,liu2018chiral,fidrysiak2018unconventional,sharma2020superconductivity,chou2019superconductor,kozii2019nematic,you2019superconductivity,wang2021topological}. 
Based on a recent experiment that suggest a strong coupling between the graphene $K$-phonon and the flat bands in TBG \cite{chen2023strong}, we perform a comprehensive analysis of the electron-$K$-phonon (e-K-ph) interaction and the resulting phonon-mediated superconductivity on the bare flat bands of TBG. We develop an exhaustive numerical, analytical, and symmetry based description of the e-K-ph interaction in TBG and the symmetry classifications of the order parameter, and find the competing singlet gapped inter-Chern-band channel and nematic gapless intra-Chern-band channel. Armed with the heavy-fermion description of TBG\cite{song2022magic,cualuguaru2023tbg,hu2023symmetric,hu2023kondo,lau2023topological,datta2023heavy,yu2023magic,chou2022kondo,zhou2023kondo}, the form factors of the $K$-phonon induced attractive interaction can be analytically computed and matched well to a full numerical calculations. An analysis of the Coulomb screening shows that, due to the high density of states (DOS) of flat bands, the Coulomb interaction might be strongly renormalized down near the Van Hove singularities. However, it remains unclear if the Hartree-Fock bands of the correlated insulator, with the lower DOS, can provide a similar result. 
 
\begin{figure}[hbt!]
   \centering
    \includegraphics[width=3.4in]{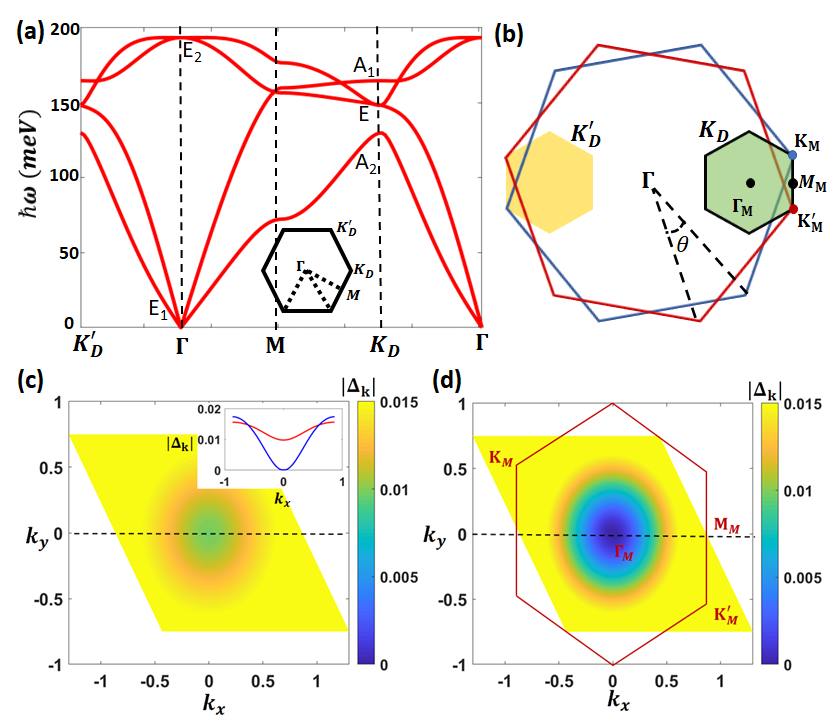} 
    \caption{ (a) Phonon dispersion of graphene. The irreps for phonon modes at $\Gamma$ and $\KK_D$ are labelled. Inset: BZ of graphene. (b) MBZ of TBG. 
    (c) and (d) shows the momentum dependence of the normalized gap function $|\Delta_\kk|$ for the inter-Chern-band $A_1$ singlet (or the $A_2$ triplet) channel and intra-Chern-band pairing 2D $E_2$ singlet channel, respectively. 
    The inset in (c) shows the $|\Delta_\kk|$ along the dashed line $k_y=0$ for both the inter-Chern-band (red) and intra-Chern-band (blue) channels.  
    The momenta ${\bf \Gamma}_M, \KK_M, {\bf M}_M$ are labelled in MBZ in (b) and (d). 
  }
    \label{fig:gapinterintra}
\end{figure}

{\it Model Hamiltonian for electron-phonon interaction in TBG -}
We consider the deformation potential type of theory, described by a tight-binding (TB) model for the electron Hamiltonian with the hopping parameters 
depending on the atom positions $\tilde{\bf R}_\alpha^l  ={\bf R}^l + {\bf \tau}_{\alpha}^l + {\bf u}^l({\bf R}^l_\alpha)$ with a displacement field 
${\bf u}^l({\bf R}^l_\alpha={\bf R}^l + {\bf \tau}_{\alpha}^l)$, where ${\bf R}^l$ and ${\bf \tau}_{\alpha}^l$ label the lattice vector and 
the sublattice atom position ($\alpha=A,B$) at the layer $l$, respectively. By treating ${\bf u}$ as a perturbation, we expand the intra-layer Hamiltonian up to the linear order in $\uu$ (Supplementary Material (SM) Sec. II \cite{sm2022}). We only keep $\uu$-independent terms for the inter-layer Hamiltonian for TBG, thus focusing on intra-layer electron-phonon (e-ph) interaction in this work. 
As only the Dirac bands appear around the Fermi energy close to $\pm \KK_D=\pm \frac{4\pi}{3a_0}(1,0)$ in Brillouin zone (BZ) with lattice constant $a_0$ in graphene, we also expand the Hamiltonian 
around $\eta \KK_D$ ($\eta=\pm$ labelling two valleys) and focus on Dirac electrons around two valleys. Our full Hamiltonian consists of three parts 
\begin{eqnarray}
H=H_{el}+H_{ph}+H_{eph}. 
\end{eqnarray}
Here $H_{el}$ describes the Dirac electrons located at valley $\eta=\pm$ momentum $\eta \KK_D$ that are coupled through inter-layer tunnelings and is given by the Bistritzer-MacDonald (BM) model \cite{bistritzer2011moire} (SM Sec. V \cite{sm2022}), 
\begin{equation}\label{eq:BM_Hamiltonian1}
\hH_{\rm el} = \sum_{\eta s} \sum_{\kk \in {\rm MBZ}} \sum_{\alpha \alpha'} \sum_{\QQ,\QQ'} h_{\QQ\alpha,\QQ'\alpha'}^{(\eta)}(\kk) c_{\kk,\QQ,\alpha,\eta,s}^\dagger c_{\kk,\QQ',\alpha',\eta,s}
\end{equation}
where $c_{\kk,\QQ,\alpha,\eta,s}$ is the fermion annihilation operator  $\kk$ is a momentum in the Moir\'e Brillouin zone (MBZ) (Fig. \ref{fig:gapinterintra}b), $\alpha$ is the sublattice index, and $s$ is spin. The vector $\QQ$ belongs to the lattice set 
$\mathcal{Q}_{l \eta }=\{l \eta \qq_2 + n_1 \bb_{M1} + n_2 \bb_{M2}\ |\ n_{1,2}\in \mathbb{Z} \}$, where $l$ is the layer index, 
$\qq_2=k_{\theta}(\frac{\sqrt{3}}{2},\frac{1}{2})$, $\bb_{M1}=k_{\theta}(\frac{\sqrt{3}}{2},\frac{3}{2})$,  $\bb_{M2}=k_{\theta}(-\frac{\sqrt{3}}{2},\frac{3}{2})$
and $k_\theta=2|\KK_D|\sin\frac{\theta}{2}$ with $\theta$ the twist angle. $h_{\QQ\alpha,\QQ'\alpha'}^{(\eta)}(\kk)$ is given in SM Sec. V.A \cite{sm2022}.  
 $\hH_{\rm el}$ exhibits $C_{6v}$ and time-reversal symmetries, generated by valley-switching $\pi/6$-rotation along z axis ($\hat{C}_{6z}$), time reversal ($\hat{T}$), and $\pi$-rotation along y axis ($C_{2y}$), and valley-preserving $\pi$-rotation along x axis ($\hat{C}_{2x}$), and the composite anti-unitary $C_{2z}T$.  In addition,  $\hH_{\rm el}$ has a unitary particle-hole ($\hat{P}$) symmetry, 
as well as a chiral symmetry $\hat{C}$ in the limit with vanishing $AA$ region hopping ($w_0=0$) \cite{vafek2020renormalization}. A full discussion of symmetry of BM model \cite{bernevig2021twisted,song2021twisted} is found in SM Sec.V.B \cite{sm2022}. 

$H_{ph}$ describes the intra-layer in-plane phonon modes. Out-of-plane phonon modes are decoupled from Dirac electrons for intra-layer e-ph interaction. 
The dynamical matrix for a single layer graphene is derived in SM Sec.III \cite{sm2022} based both on symmetry considerations and the microscopic model, up to the next nearest neighbor interaction. The resulting in-plane phonon dispersion in Fig. \ref{fig:gapinterintra}a reproduces that in literature \cite{neto2009electronic,sahoo2012phonon,thingstad2020phonon,maultzsch2004phonon,mohr2007phonon} (SM Sec.III and IV.B \cite{sm2022}). 
The phonon modes at $\Gamma$ and $\eta \KK_D$ can induce intra-valley and inter-valley e-ph interactions, respectively. In this paper we focus on the $\eta \KK_D$ phonons; the $\Gamma$ phonons will be derived in \cite{liu2022future1}.
At $\eta \KK_D$, we have one $A_1$ ($\sim 160 meV$), one $A_2$ ($\sim 140 meV$) and one 2D $E$ mode ($\sim 150 meV$) of $C_{3v}$ group. Based on the deformation potential theory, we derive the e-ph interaction $H_{eph}$ by expanding the TB Hamiltonian treating both the momentum
and phonon displacement field $\uu$ as perturbations. For e-ph interaction, we only keep the dominant zeroth-order in momentum for the $\eta \KK_D$ phonons. 
We find, due to both symmetry and the two-center approximation (SM Sec.II.E \cite{sm2022}), that only the $A_1$ phonons at $\KK_D$ can scatter an electron from $\KK_D$
to $-\KK_D$ \cite{mohr2007phonon}. The corresponding Hamiltonian reads
\begin{widetext}
\begin{eqnarray}\label{eq:Hepintervall} 
&  H_{inter-vall}^{op,A_1}   \approx \frac{\gamma_3}{ \sqrt{2 N_G M \omega_{A1}}} \sum_{\tilde{\kk},\tilde{\kk}',\eta,\alpha \beta } (b_{-\eta \KK_D+\tilde{\kk}-\tilde{\kk}',A_1} 
+b_{\eta \KK_D-\tilde{\kk}+\tilde{\kk}', A_1 }^\dagger)
c^\dagger_{\tilde{\kk}+ \eta \KK_D, \alpha } (\sigma_x)_{\alpha\beta} c_{\tilde{\kk}' - \eta \KK_D, \beta},
\end{eqnarray}
\end{widetext}
where $\tilde{\kk}$ is the electron momentum away from $\eta\KK_D$, $N_G$ is the number of atomic unit cells, $M$ is the atomic mass, $\omega_{A1}$ is the $A_1$ phonon frequency, and $b$ and $c$ are phonon and electron annihilation operators. The material dependent parameter 
$\gamma_3$ can be derived from the hopping potential as $\gamma_3= 2 i \sum_{\GG} e^{i(\tau_A- \tau_B)\cdot \GG } (\GG+ \KK_D)_y t(\GG+ \KK_D,0)\approx 17 \text{eV/ \AA}$, where $\GG$ is the reciprocal lattice vector and $t(\qq)$ is the Fourier transform of the $\pi$-bond hopping function
between two carbon $p_z$ orbitals in graphene (Eq. 6 in SM Sec.I \cite{sm2022}). Our next step is to re-write the electron momentum $\tilde{k}$ into the MBZ by 
$\tilde{\kk}=\kk-\QQ_{l\eta}$ with $\kk\in \text{MBZ}$, so that $c_{\kk, \QQ_{l\eta}, \alpha, \eta, s}= c_{\eta \KK_{D}^l + \tilde{\kk}, \alpha, l, s} $
and $\sum_{\tilde{\kk}}\rightarrow \sum_{\kk \in \text{MBZ}} \sum_{\QQ_{l\eta}}$, where we have added the spin index $s$ and layer index $l$. 
Finally, we project the e-ph interaction $H_{inter-vall}^{op,A_1}$ into the flat bands of the BM Hamiltonian as
\begin{eqnarray}\label{eq:Hinter-vall_opA1}
&  H_{inter-vall}^{op,A_1} \approx\frac{1}{ \sqrt{N_G}} \sum G_{\kk,\kk', \QQ_{-l\eta}}^{\eta n n' l} \gamma_{\kk,n,\eta, s}^\dagger \gamma_{\kk',n',-\eta, s} \nonumber\\
& (b_{-\eta \KK_D+\kk-\kk'- \QQ_{-l\eta} ,l, A_1} +b_{\eta \KK_D-\kk+\kk'+ \QQ_{-l\eta} ,l,  A_1 }^\dagger)
 \end{eqnarray} 
where the summation includes $\kk,\kk',n,n',\eta,s,l, \QQ_{-l\eta}$, $\gamma_{\mathbf{k},n,\eta, s}^{\dagger}=\sum_{\mathbf{Q}\alpha}u_{\mathbf{Q}\alpha;n\eta}\left(\mathbf{k}\right)c_{\mathbf{k},\mathbf{Q},\eta,\alpha s}^{\dagger}$
with $u^n_{\kk, \QQ'_{l\eta}, \alpha, \eta}$ the eigenstates of $h_{\QQ\alpha,\QQ'\alpha'}^{(\eta)}(\kk)$.
The matrix element
\begin{eqnarray}\label{eq:eph_Gfunction1}
&   G_{\kk,\kk', \QQ_{-l\eta}}^{\eta n n' l}= \frac{\gamma_3}{ \sqrt{2 M\omega_{A_1} }} 
 \sum_{\QQ'_{l\eta},\alpha \beta} \nonumber \\& 
 u^{n\star}_{\kk, \QQ'_{l\eta}, \alpha, \eta} \sigma^x_{\alpha\beta} u^{n'}_{\kk', \QQ'_{l\eta}-\QQ_{-l\eta}, \beta, -\eta}  
\end{eqnarray}
characterizes the e-ph interaction strength for TBG and can be evaluated numerically (and later analytically), as shown in SM Sec. VI.F \cite{sm2022}. 
We focus on two flat bands (per valley per spin) of TBG, labelled by $n=\pm$. Instead of the eigen-state basis, we work on the so-called "Chern-band" basis, defined by 
\begin{equation}
u^{e_Y}_{\kk,\mathbf{Q},\alpha,\eta}=\frac{1}{\sqrt{2}}(u^{n=+}_{\kk,\mathbf{Q},\alpha,\eta}+ie_Y u^{n=-}_{\kk,\mathbf{Q},\alpha,\eta})
\label{eq:Chern-band-u}
\end{equation}
with $e_Y=\pm 1$. $u^{e_Y}_{\kk,\mathbf{Q},\alpha,\eta}$ carries the Chern number $\pm 1$. 
On the Chern-band basis, the expressions for e-ph interaction can be obtained by replacing the $n,n'$ indices in Eqs. (\ref{eq:Hinter-vall_opA1}) and (\ref{eq:eph_Gfunction1}) with 
$e_Y, e'_Y$ indices and $u^{n}_{\kk, \QQ, \alpha, \eta}$ in Eq. (\ref{eq:eph_Gfunction1}) with $u^{e_Y}_{\kk, \QQ, \alpha, \eta}$. 
Discrete symmetries can constrain the form of the function 
$G_{\kk,\kk', \QQ_{-l\eta}}^{\eta e_Y e_Y' l}$, as discussed in SM Sec.VI.D \cite{sm2022}. 
In particular, in the chiral limit $w_0=0$ one can show that $G_{\kk,\kk', \QQ_{-l\eta}}^{\eta e_Y e_Y' l}=\delta_{e_Y, e_Y'} G_{\kk,\kk', \QQ_{-l\eta}}^{\eta e_Y e_Y l}$ has diagonal form on the Chern-band basis. 

{\it Phonon-mediated Electron-electron Interaction and Symmetry Classification of Superconducting Pairing Channels -}
We next apply the Schrieffer-Wolff transformation \cite{schrieffer1966relation} to integrate out the phonon modes and obtain the phonon-mediated electron-electron (el-el) interaction \cite{wu2018theory,yu2022euler2}. 
We focus on the Cooper pair channel of the attractive interaction, which takes the form 
\begin{eqnarray}\label{eq:elelinteraction_phononmediate}
& H_{ee}=-\frac{1}{N_M} \sum_{\kk,\kk',s, s_1, e_Y, e_Y'} V^{\eta, e_Y, e_Y'}_{\kk, \kk'} \nonumber\\
& \gamma_{\kk e_Y\eta s}^\dagger  \gamma_{-\kk e'_{Y}, - \eta s_1}^\dagger \gamma_{-\kk' e'_{Y},\eta s_1}  \gamma_{\kk' e_Y,-\eta s}, 
\end{eqnarray}
where 
\begin{equation}\label{eq:Ffunction_def}
V^{\eta, e_Y, e_Y'}_{\kk, \kk'}=\frac{1}{N_0\omega_{A_1}} \sum_{\GG_M, l} G_{\kk,\kk', -l\eta \qq_2+ \GG_M}^{\eta, e_Y, l}  
G_{-\kk,-\kk', l\eta \qq_2 - \GG_M}^{-\eta,  e_{Y}',   l} \nonumber
\end{equation}
with $\GG_M$ the Moir\'e reciprocal lattice vectors, $N_M$ the number of Moir\'e unit cells and $N_0$ the number of atomic unit cells in one Moir\'e unit cell ($N_G=N_0\times N_M$). Discrete symmetries  constrain the form of the interaction parameter $V^{\eta, e_Y, e_Y'}_{\kk, \kk'}$. 
The ones leaving the momentum $(\kk,\kk')$ unchanged are: (1) 
$\hat{C}_{2z}\hat{P}$:  $V^{\eta, e_Y, e_Y'}_{\kk, \kk'}=V^{-\eta, e_Y, e_Y'}_{\kk, \kk'} $; (2) $\hat{C}_{2z}\hat{T}$: 
$V^{\eta, e_Y, e_Y'}_{\kk, \kk'}= V^{\eta, -e_Y, -e_Y'\star}_{\kk, \kk'}$; and (3) the combination of index reshuffling and $\hat{P}$ symmetry:
$V^{\eta, e_Y e_Y'}_{\kk,\kk'}= V^{-\eta, e_Y' e_Y}_{\kk, \kk'}$. These three symmetry operations reduce the number of the independent components
of the $V$-function for a fixed $(\kk,\kk')$ from 8 complex parameters to 1 real ($V^{+,+-}_{\kk,\kk'}$) and 1 complex parameter ($V^{+,++}_{\kk,\kk'}$). Other discrete symmetries, including $\hat{P}$, reshuffling, hermicity, $\hat{C}_{3z}$ and $\hat{C}_{2z}$, relate the $V$-function at different $(\kk,\kk')$. In particular, $\hat{C}_{3z}$ guarantees $V^{\eta, e_Y, e_Y}_{\KK_M,0}=0$ for the intra-Chern-band channels.  The projected Coulomb interaction into the flat bands of the BM model possesses a large 
$U(4)\times U(4)$ spin-valley continuous symmetry \cite{bernevig2021twisted,kang2019strong,bultinck2020ground}. the el-el interaction (\ref{eq:elelinteraction_phononmediate}) breaks this symmetry down to the $U(2)_{e_Y=+}\times U(2)_{e_Y=-}$ in the chiral limit and further to a total spin $SU(2)$ together with a valley charge $U(1) \otimes U(1)$  (SM Sec.VI.E \cite{sm2022}).   

At the mean field level, the attractive interaction (\ref{eq:elelinteraction_phononmediate}) is decomposed into the fermion bilinear form
$H_{\Delta}=\hat{\Delta}+\hat{\Delta}^\dagger$ with
\beq\label{eq:GapFun1}
 \hat{\Delta} = \sum \gamma_{\kk, e_{Y_1}, \eta , s_1 }^\dagger \Delta^{\eta}_{\kk; e_{Y_1} s_1, e_{Y_2} s_2}  \gamma_{-\kk, e_{Y_2}, -\eta , s_2}^\dagger,  \eneq
where the summation above includes the indices $\kk, e_{Y_1}, e_{Y_2}, s_1, s_2, \eta $ and the gap function 
\beq\label{eq:gap-eqn1}
\Delta^{\eta}_{\kk; e_{Y_1} s_1, e_{Y_2} s_2}=-\frac{1}{N_M}\sum_{\kk'} V_{\kk\kk'}^{\eta e_{Y_1} e_{Y_2}} \langle \gamma_{-\kk'  e_{Y_2} \eta s_2} \gamma_{\kk' e_{Y_1} -\eta s_1}\rangle. \eneq 
Since the  interaction $V$-function  does not involve spin, we can decompose $\Delta^{\eta}_{\kk; e_{Y_1} s_1, e_{Y_2} s_2}=\sum_{S,M} \Delta^{\eta,SM}_{\kk; e_{Y_1} e_{Y_2}} \mS^{SM}_{s_1s_2} $, where $S=0$ for spin singlet and $S=1$ ($M=-S,...,S$) for spin triplet (SM Sec.VI.G.1 \cite{sm2022}). 

The gap function can be classified according to the discrete symmetries. The $C_{6v}$ group includes four 1D irreducible representations (irreps), e.g. $A_{1,2}$ and $B_{1,2}$, and two 2D irreps, $E_{1,2}$. 1D irreps $A_{1,2}$ and $B_{1,2}$ channels differ by their $\hat{C}_{2z}$ eigen-values, $\lambda_{C_{2z}}=+1$ for $A_{1,2}$ and $\lambda_{C_{2z}}=-1$ for $B_{1,2}$. Combining $\hat{C}_{2z}$ and reshuffling symmetries leads to $\Delta^\eta_{\kk; e_{Y_1}, e_{Y_2}} = \lambda_{C_{2z}}\Delta^\eta_{\kk; e_{Y_2}, e_{Y_1}}$ for spin singlet and $\Delta^\eta_{\kk; e_{Y_1}, e_{Y_2}} = -\lambda_{C_{2z}}\Delta^\eta_{\kk; e_{Y_2}, e_{Y_1}}$ for spin triplet. Thus, for intra-Chern-band pairing ($e_{Y_1}=e_{Y_2}$), the $A_{1,2}$ channel must be spin singlet while the $B_{1,2}$ channel must be spin triplet. Furthermore, the rotation $\hat{C}_{3z}$ ensures the existence of nodes at $\KK_M$ for the gap function of any 1D irrep intra-Chern-band channel ($\Delta^\eta_{\KK_M; e_{Y}, e_{Y}}=0$), while the inter-Chern-band channel does not have such constraint. The 2D irreps $E_1$ and $E_2$ have different $\hat{C}_{2z}$ eigen-values, $\lambda_{C_{2z}}=+1$ for $E_2$ and $\lambda_{C_{2z}}=-1$ for $E_1$, similarly to the 1D irrep case. Consequently, the $E_2$ channel must be spin singlet while the $E_1$ channel must be spin triplet for intra-Chern-band pairings. $\hat{C}_{3z}$ guarantees nodes at $\Gamma_M$ for both intra- and inter-Chern-band channels, and it requires additional nodes at $\KK_M$ for the inter-Chern-band channels for both 2D $E_{1,2}$ pairings. Besides discrete
symmetries, the continuous $U(2)_{e_Y=1}\times U(2)_{e_Y=-1}$ spin symmetry in the chiral limit guarantees the singlet and triplet pairings of inter-Chern-band channel to be degenerate in the chiral flat band limit. The full symmetry analysis of the gap functions can be found in SM Sec.VI.G \cite{sm2022}.
 
{\it Gap Equations and Self-consistent Solution of Pairing Channels -}
The linearized gap equation (LGE) for the attractive interaction (\ref{eq:elelinteraction_phononmediate}) can be derived by evaluating $\langle \gamma_{-\kk'  e_{Y_2} \eta s_2} \gamma_{\kk' e_{Y_1} -\eta s_1}\rangle$ in Eq. (\ref{eq:gap-eqn1}) and expanding it to 
linear order of the gap function. In the chiral flat band limit, e.g. the band width is much smaller than the critical temperature $T_c$,
the LGE is derived as
\beq\label{eq:LGE_gapequation1}
2k_B T\Delta^{\eta,SM}_{\kk;e_{Y_1} e_{Y_2}}=\frac{1}{N_M} \sum_{\kk'}V^{\eta e_{Y_1} e_{Y_2}}_{\kk,\kk'} \Delta^{-\eta,SM}_{\kk';e_{Y_1} e_{Y_2}}. 
\eneq
This is eigen-equation problem for the matrix $V^{\eta e_{Y_1} e_{Y_2}}_{\kk,\kk'}$: the $T_c$ is determined by the largest
eigen-value and the symmetry of the gap function is determined by that of its eigenvector. As mentioned, the only two independent components of the $V$-function (complex $V^{+++}_{\kk,\kk'}$ and real $V^{++-}_{\kk,\kk'}$) leads to two independent LGEs for the intra- and inter-Chern-band channels, respectively. 
The form of the LGE suggests that all the gap functions are doubly degenerate at $T_c$ in the flat-band limit. They belong either to two degenerate 1D irreps or one 2D irrep. We first numerically solve these two LGEs from Eq. (\ref{eq:LGE_gapequation1}), and find the forms of the gap functions with the largest eigen-values, as shown in Fig. \ref{fig:gapinterintra}.  
Our numerical calculations show $k_B T_c \sim 0.21 meV$ for the inter-Chern-band channel, slightly larger than $k_B T_c \sim 0.16 meV$ for the intra-Chern-band channel. For the inter-Chern-band channels, the gap function is almost a constant in Fig. \ref{fig:gapinterintra}a, featuring a fully gapped s-wave pairing with even $\hat{C}_{2z}$-parity ($A_1$ or $A_2$ irrep). 
In the chiral flat-band limit, spin singlet and triplet pairings are degenerate, as required by the continuous $U(2)\times U(2)$ spin symmetry
(SM Sec.VI.E \cite{sm2022}). Including kinetic energy splits this degeneracy and makes the spin singlet $A_1$ irrep channel to have the highest $T_c$. 
For the intra-Chern-band channel, one can see nodes appear at the $\Gamma_M$ in Fig. \ref{fig:gapinterintra}b. As our previous symmetry analysis shows that the gap function should have nodes at $\KK_M$ for the 1D irrep ($A_{1,2}, B_{1,2}$) and $\Gamma_M$ for the 2D irrep ($E_{1,2}$), numerical results should correspond to a 2D irrep. Numerically analyzing the symmetry property of the gap function suggests that the intra-Chern-band channel belongs to the 2D $E_2$ irrep with spin singlet. Full numerical results are discussed in SM Sec.VI.H.2 and 3 \cite{sm2022}. 

\begin{figure}[hbt!]
   \centering
    \includegraphics[width=3.3in]{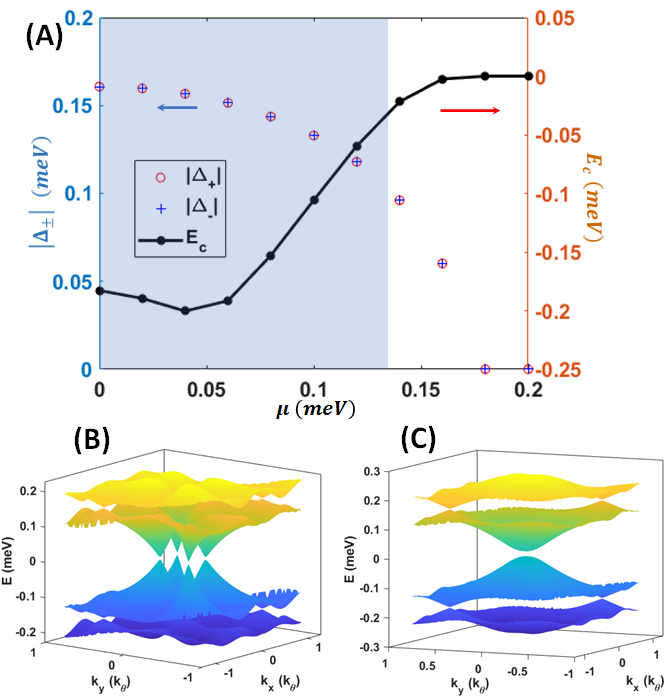} 
    \caption{ (A) The superconductor order parameter amplitudes $|\Delta_{\pm}|$ (red circles and blue crosses) and the ground state energy (black dots) as a function of $\mu$. The superconducting phase has nodes in the shadowed regime. (B) and (C) show the BdG spectrum with and without nodes at $\mu=0.04 meV$ and $0.14 meV$, respectively. The single-particle bandwidth is set around $0.3 meV$. 
  }
    \label{fig:gapBdG}
\end{figure}

Our results for the intra-Chern-band channels reveal a d-wave character of the gap. Using the heavy fermion formalism of TBG \cite{song2022magic,liu2022future1}, we analytically obtain $V^{\eta,e_Y,e_Y}_{\kk,\kk'}$: 
\begin{eqnarray}\label{eq:Interactionform_intra_Chern1}
V^{\eta,e_Y,e_Y}_{\kk,\kk'}=U^*_{e_Y,\kk} U_{e_Y,\kk'};\;\; U_{e_Y,\kk}=\frac{\sqrt{V_0}}{k^2+b^2} k_{e_Y}^2, 
\end{eqnarray}
with $k_{e_Y}=k_x+i e_Y k_y$ ($e_Y=\pm$). This interaction allows us to solve the LGE analytically to obtain the $T_c$,
\beq
k_BT_c=\frac{\tilde{V}_0}{2},\;\;  \tilde{V}_0=\frac{1}{N_M} \sum_\kk V_0\frac{k^4}{(k^2+b^2)^2},
\eneq
where $V_0$ and $b$ are material dependent parameters. The corresponding self-consistent gap function takes the d-wave form 
\begin{eqnarray}
\left(\begin{array}{c}
\Delta^{+,00}_{\kk,e_Y e_Y}\\
\Delta^{-,00}_{\kk,e_Y e_Y}
\end{array}\right)
=\Delta_{e_Y} \frac{ k_{-e_Y}^2}{k^2+b^2} \left(\begin{array}{c}
1\\1\end{array}\right) 
\end{eqnarray}
with $e_Y=\pm$ and $\Delta_{e_Y}$ a parameter to be determined. Time reversal, if exists, requires 
$\left(\Delta^{-,00}_{\kk; --}, \Delta^{+,00}_{\kk; --}\right) =\left( \Delta^{+,00}_{\kk; ++}, \Delta^{-,00}_{\kk; ++}\right)^* $.
The d-wave nature of the gap function suggests the possibility of the nodal superconductivity. However, one should note that the single-particle
Hamiltonian is {\it not} diagonal in the Chern-band basis. The Bogoliubov-de Gennes (BdG) spectrum must be checked with kinetic energy added. The BdG Hamiltonian for the intra-Chern-band pairing is block diagonal and one block $\mathcal{H}_{BdG}^{+,+}$ on the basis
$(\gamma^\dagger_{\kk, e_Y = \pm ,+,s =\uparrow },\gamma_{-\kk, e_Y=\pm,-,s=\downarrow})$ reads
\begin{eqnarray}\label{eq:HBdG_intra}
H_{BdG}^{+,+}(\kk) =\begin{pmatrix}
h_+(\kk) & \Delta^+_{\kk} \\
(\Delta^+_{\kk})^\dagger & - h_{-}^\star(-\kk)
\end{pmatrix}
\end{eqnarray}
with $h_\eta(\kk)= (d_{0,\eta}(\kk)-\mu)\zeta^0+d_{x,\eta}(\kk)\zeta^x$ and $\Delta^+_{\kk}=Diag[\Delta^{+,00}_{\kk,++},\Delta^{+,00}_{\kk,--}]$. 
Here $d_{0,\eta}(\kk)=(\epsilon_{+,\eta}(\kk)+\epsilon_{-,\eta}(\kk))/2$ and $d_{x,\eta}(\kk)=(\epsilon_{+,\eta}(\kk)-\epsilon_{-,\eta}(\kk))/2$,
where $\epsilon_{\pm, \eta}(\kk)$ are the eigen-energies for the two low-energy flat bands (per valley per spin) of the BM model $\hH_{\rm el}$ (\ref{eq:BM_Hamiltonian1}). 
The corresponding energy spectrum can possess nodes when the pairing amplitudes of two Chern-band channels are equal, $|\Delta_{e_Y=+}|=|\Delta_{e_Y=-}|=\Delta_{0}$, which corresponds to the Euler pairing discussed in Ref.\cite{yu2022euler1,yu2022euler2}. 
Point nodes appear at the location defined by two conditions (1) $\cos((\Phi_{\kk,-}-\Phi_{\kk,+})/2)=0$,
where $\Phi_{\kk,e_Y}=\varphi_{e_Y}-2 e_Y \theta_{\kk}$ with $\Delta_{e_Y}=\Delta_0 e^{i \varphi_{e_Y}}$ and $k_{e_Y}=k e^{i e_Y \theta_{\kk}}$;
and (2) $d_{x,\kk}^2=(d_{0,\kk}-\mu)^2+\Delta_{0,\kk}^2$ with $\Delta_{0,\kk}=\Delta_0\frac{k^2}{k^2+b^2}$, as discussed in SM Sec.VI.H.5 \cite{sm2022}. The first condition determines the momentum angle for the nodes while the second gives the momentum amplitude, thus together fixing the location of point nodes in the 2D momentum space. 
We next solve the self-consistent gap equation at zero temperature for the interaction form 
(\ref{eq:Interactionform_intra_Chern1}). With the gap function ansatz $\Delta_{\kk;e_Y} = \Delta_{e_Y} \frac{k^2_{-e_Y}}{k^2+b^2}$, we find a self-consistent gap equation 
\beq \Delta_{e_Y}= \frac{V_0 }{N_M} \sum_{\kk', e_{Y_1} } \frac{{k'}^2_{e_Y}}{k'^2+b^2}  u_{-\kk',e_Y e_{Y_1}} w^*_{-\kk',e_Y e_{Y_1}}, \eneq
where $\psi_{\kk,e_{Y_1}}=(u_{\kk,\pm, e_{Y_1}},w_{\kk,\pm e_{Y_1}})$ ($e_{Y_1}=\pm$) are the eigen-wave functions with the positive eigen-energies of the BdG Hamiltonian $H_{BdG}^{+,+}(\kk)$ (\ref{eq:HBdG_intra}). Fig. \ref{fig:gapBdG}A shows the chemical potential dependence of the gap functions and the condensation energy. The Euler pairing $|\Delta_{+}|=|\Delta_-|$ is always energetically favored for a non-flat band width $\sim 0.3 meV$, quite different from chiral d-wave pairing in doped graphene \cite{black2014chiral,nandkishore2012chiral,wu2019topological}. For the chemical potential $\mu$ below $0.1 meV$, a nodal superconductor phase with four point nodes (Fig. \ref{fig:gapBdG}B) located at the positions determined by two conditions discussed above \cite{yu2022euler2}. With increasing $\mu$, four nodes move towards $\Gamma_M$ and eventually a gapped superconductor phase (Fig. \ref{fig:gapBdG}C) appears for $\mu > 0.1 meV$.  

The energy scale of the Coulomb interaction in TBG is $\sim 24 meV$ \cite{bernevig2021twisted}, much larger than the estimated energy scale of e-K-ph mediated attractive interaction $\sim 0.3 meV$ \cite{note1}. Near the Van-Hove singularities of flat bands, the screening can significantly reduce the Coulomb interaction to a similar order as e-K-ph mediated interaction due to the large DOS (SM Sec.VI.H.7 \cite{sm2022}), thus making superconductivity from this mechanism possible. If, however, the DOS is that of the Hartree-Fock bands of correlated insulators, the screening might not be enough to reduce the Coulomb interaction. Hence, superconductivity from this $K$-phonon flat bare band mechanism could appear only when the correlated insulator states are suppressed. 

{\it Conclusion - }
In conclusion, we develop a theory for the projected e-K-ph interaction of the flat bands and the resulting superconductor pairing channels in TBG. We find the inter-Chern-band s-wave singlet pairing and the intra-Chern-band d-wave nematic singlet pairing have highest $T_c$, and the $T_c$ of inter-Chern-band channel is slightly higher than the intra-Chen-band channel. The intra-Chen-band channel can have nodes in a large parameter regime. From the estimate of the screened Coulomb interaction, we argue that this mechanism requires the correlated insulators to be suppressed.

{\it Acknowledgement} --
We would like to acknowledge Biao Lian, Xi Dai and Zhida Song for the helpful discussion. B.A.B.’s heavy fermion in twisted bilayer research was supported by the DOE Grant No. DE-SC0016239. BAB’s sabbatical support also comes from the Simons Investigator Grant No. 404513, the Gordon and Betty Moore Foundation through Grant No.GBMF8685 towards the Princeton theory program, the Gordon and Betty Moore Foundation’s EPiQS Initiative (Grant No. GBMF11070), Office of Naval Research (ONR Grant No. N00014-20-1-2303), and the European Research Council (ERC) under the European Union’s Horizon 2020 research and innovation program (grant agreement no. 101020833). CXL also acknowledges the support through the Penn State MRSEC–Center for Nanoscale Science via NSF award DMR-2011839. AY acknowledges support from the Gordon and Betty Moore Foundation’s EPiQS initiative grant GBMF9469, DOE- BES grant DE-FG02-07ER46419, NSF-DMR-1904442, ARO MURI (W911NF-21-2-0147), and ONR N00012- 21-1-2592. CXL and AY also acknowledges the support from the NSF-MERSEC (Grant No. MER- SEC DMR 2011750). YLC acknowledges the support from the Oxford-ShanghaiTech collaboration project and the Shanghai Municipal Science and Technology Major Project (grant 2018SHZDZX02).

\bibliographystyle{apsrev4-2}
\bibliography{ref_epTBG}

\begin{thebibliography}{76}
\bibitem [{not()}]{note1}%
  \BibitemOpen
  \href@noop {} {}\bibinfo {note} {Note that our calculation is performed in
  the chiral limit, so we may underestimate the electron-phonon coupling
  strength.}\BibitemShut {Stop}%
\end{thebibliography}

\newpage

\onecolumngrid
\begin{center}
{\Large Supplementary Material for "Electron-$K$-Phonon Interaction In Twisted Bilayer Graphene"}\\
\end{center}

\tableofcontents
\section{Notation}
In this section, we will first describe our notations for the whole paper. 
We denote ${\bf a}_1^l , {\bf a}_2^l$ as the un-strained, microscopic lattice vectors of the graphene layer $l$  ($l=\pm$ for the top and bottom layers), 
while ${\bf G}_{1,2}^l$ are the corresponding reciprocal lattice vectors. ${\bf G}_a^l\cdot {\bf a}_b^l = 2 \pi \delta_{ab}$. 
In the layer $l$, there are two sub-lattice positions ${\bf \tau}_A^l, {\bf \tau}_B^l$. We assume the graphene layer $l$ is rotated by an in-plane angle 
$l\theta/2$ followed by a shift $l \frac{l_0 }{2} \hat{z}$ along the z direction, 
so any vector ${\bf v}^l$ is related to the un-rotated vector ${\bf v}$ by ${\bf v}^l=\hat{R}^l_{\theta/2} {\bf v} +l \frac{l_0 }{2} \hat{z}=l \frac{\theta}{2}\hat{z}\times {\bf v} +l \frac{l_0 }{2} \hat{z}$, where $\hat{R}^l_{\theta/2}$ labels the rotation operator, 
$\hat{z}$ is the out-of-plane unit vector and $l_0$ is the equilibrium distance between two graphene layers. 
Thus, we only need to give the values of all the un-rotated vectors below. 
Two primitive lattice vectors are ${\bf a}_1 = a_0 (\frac{-1}{2}, \frac{\sqrt{3}}{2},0)$ and ${\bf a}_2 = - a_0 (\frac{1}{2},\frac{\sqrt{3}}{2},0)$, 
where $a_0 = 2.46\AA$ 
is the lattice constant. The two primitive reciprocal lattice vectors spanning the Brillouin zone (BZ) are 
\beq
{\bf G}_1= \frac{4\pi}{\sqrt{3} a_0} (-\frac{\sqrt{3}}{2}, \frac{1}{2});\;\; {\bf G}_2= \frac{4\pi}{\sqrt{3} a_0} (-\frac{\sqrt{3}}{2}, -\frac{1}{2}). 
\eneq
Since the Fermi surface in graphene is around two $K$-valley, we also label them as $\eta \KK_D$ with $\KK_D=\frac{4\pi}{3a_0}(1,0)$ and $\eta=\pm$. 
The atoms are at positions 
\beq
{\bf \tau}_A= \frac{a_0}{2}(1, -\frac{1}{\sqrt{3}}),\;\;  {\bf \tau}_B= \frac{a_0}{2}(1, \frac{1}{\sqrt{3}}). 
\eneq
For two layers $l=\pm$, the atoms on position $\alpha= A,B$ in the top/bottom later reside at
\beq
{\bf R}^l_\alpha = \hat{R}_{\theta/2}^l {\bf R}+ l \frac{l_0 }{2} \hat{z} + {\bf \tau}_\alpha^l  = {\bf R}^l + {\bf \tau}_{\alpha}^l
\eneq where ${\bf R} = m {\bf a}_1 + n {\bf a}_2$ is a unit cell vector. 

The hopping between two atomic centers at a distance $r$ is 
\begin{eqnarray}
& t({\bf r}_\parallel , z)= V_{pp\pi} (1- (\frac{{\bf r} \cdot \hat{z}}{r} )^2 ) e^{- (r- \frac{a}{\sqrt{3}})/r_0 } + V_{pp\sigma} (\frac{{\bf r} 
\cdot \hat{z}}{r})^2 e^{-\frac{r-l_0}{r_0}} \label{hoppingtwocenterapproximation1}
\end{eqnarray}
where ${\bf r}_\parallel=(x,y)$, $r_0$ characterizes the decaying length of atomic orbitals of carbon atoms, 
$a= a_0/\sqrt{3}$ is the distance between the $A$ and $B$ atoms, and $V_{pp\pi}$ and $V_{pp\sigma}$ are the hybridization parameters of $\pi$ 
and $\sigma$ bonds. We performed the Fourier transform to the in-plane coordinate of $t({\bf r}_\parallel , z)$, 
\beq
t_{{\bf q},z} = \frac{1}{S_0}\int d^2 {\bf r}_\parallel e^{-i {\bf q}\cdot {\bf r}_\parallel} t({\bf r}_\parallel, z)
\eneq where $S_0$ is the unit cell area. 

\begin{figure}[hbt!]
    \centering
    \includegraphics[width=3.5in]{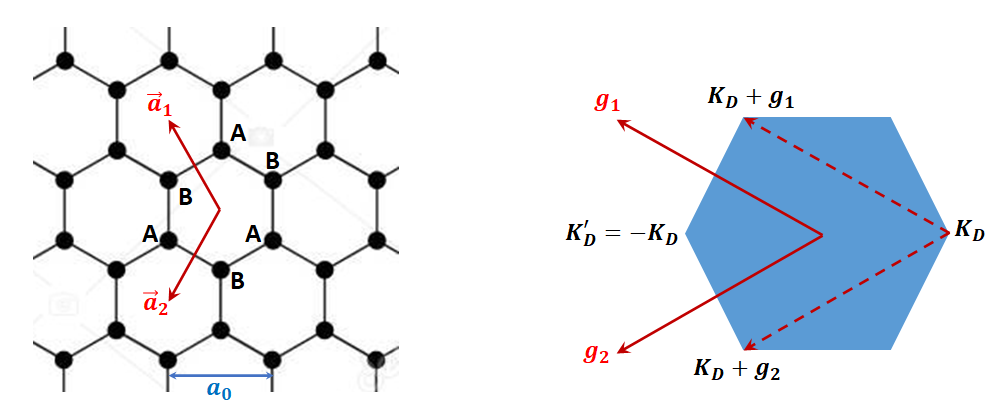} 
    \caption{Lattice and Brillouin zone for graphene. }
    \label{fig1:lattice}
\end{figure}


In the discussion below, we mainly focus on the in-plane hopping term, and direct calculation gives 
\beq
  t_{{\bf q},0} =\frac{V_{pp\pi} e^{ \frac{a_0}{\sqrt{3}r_0} }}{S_0} \frac{ 2\pi r_0^2}{(1+(q r_0)^{2})^{3/2}}
\eneq

In the calculation we have used \beq
r_0= 0.184 a_0;\; V_{pp\pi}= -2.7eV;\;  S_0=\frac{\sqrt{3}a_0^2}{2};  
\; |\KK_D|=1.7028 \AA^{-1}
\eneq

In the presence of a phonon field, the atom position should include the displacement field ${\bf u}^l({\bf R}^l_\alpha)$ as
\beq
{\bf R}_\alpha^l ={\bf R}^l + {\bf \tau}_{\alpha}^l \rightarrow \tilde{\bf R}_\alpha^l  ={\bf R}^l + {\bf \tau}_{\alpha}^l + {\bf u}^l({\bf R}^l_\alpha)
\eneq 
where 
\beq 
{\bf u}^l({\bf R}^l+\tau_\alpha^l) = \frac{1}{\sqrt{N_{G}}}\sum_{\bf p} {\bf u}_{\bf p}^{l,\alpha} e^{i {\bf p}\cdot ({\bf R}^l+\tau_\alpha^l) }
\eneq where the sum over ${\bf p}$ extends of the graphene BZ and the $N_{G}$ are the total number of atomic unit cells in graphene. 
Another important property of the phonon field is its periodic gauge. With the form of inversion Fourier transform
\beq
\uu_\qq^{l,\alpha} = \frac{1}{\sqrt{N_G}} \sum_\RR \uu^l(\RR^l + \tau_{\alpha}^l) e^{- i \qq\cdot(\RR^l + \tau_{\alpha}^l)}, 
\eneq
we can see
\beq
\uu^\alpha_{\qq+\GG} = e^{- i \GG \cdot \tau_\alpha } \uu^{l\alpha}_\qq
\eneq
which is called the "periodic gauge". 

Here we only introduce our notation for the single layer graphene, and the notation for the Moir\'e lattice for twisted bilayer graphene (TBG)
will be introduced in Sec. \ref{sec:BM_model}.

\section{Electron-phonon interaction in a single layer graphene}
On the basis states
\beq
\ket{\psi_{{\bf p},\alpha}^l} = \frac{1}{\sqrt{N_{G}}}\sum_{{\bf R}^l}  e^{i {\bf p}\cdot ({\bf R}^l + {\bf \tau}_\alpha^l ) }  \ket{\psi({\bf R}_\alpha^l + {\bf u}^l({\bf R}^l + {\bf \tau}_\alpha^l)},
\eneq
the matrix Hamiltonian between the momentum ${\bf p}, {\bf p}'$ in layers $a,b=1,2$ and orbitals $\alpha, \beta = A,B$ is given by 
\begin{eqnarray}
& H_{l l'}^{\alpha \alpha'}({\bf p p}') = \frac{1}{N_{G}} \sum_{{\bf R}^l, {\bf R}^{l'}} t(\tilde{\bf R}_\alpha^l - \tilde{\bf R}_{\alpha'}^{l'})e^{- i ({\bf R}_\alpha^l \cdot {\bf p} - {\bf R}_{\alpha'}^{l'} \cdot {\bf p}')  }
\end{eqnarray}
or 

\beq
H_{l l'}^{\alpha \alpha'}({\bf p p}') = \frac{S_0}{(2\pi)^2 N_G}  \int d^2{\bf q} \left( \sum_{{\bf R}^l, {\bf R}^{l'}} e^{- i ({\bf R}_\alpha^l \cdot ({\bf p -q})- {\bf R}_{\alpha'}^{l'} \cdot ({\bf p}'-{\bf q}))}e^{i{\bf q}\cdot ( {\bf u}({\bf R}_\alpha^l) - {\bf u}({\bf R}_{\alpha'}^{l'}))} \right) t({\bf q},z_{ll'})
\eneq
where $z_{ll'}$ is the height difference between the layers $l$ and $l'$. 
In this section, we first discuss the electron-phonon (e-ph) interaction in the same layer $l=l'$, and may drop the layer index $l$ in some expressions for 
simplicity. 

\subsection{Zero Displacement}
We have
\beq
\frac{1}{N_G} \sum_{{\bf R}^l }e^{-i {\bf R}^l \cdot ({\bf p-q})} = \sum_{{\bf G}^l} \delta_{{\bf p-q}, {\bf G}^l}
\eneq where ${\bf G}^l$ is a reciprocal lattice vector in the layer $l$. 

For the layer $l=l'$, we have (since $z_{ll}=0$)

\begin{eqnarray}
&H_{0,l l}^{\alpha \alpha'}({\bf p p}') = \frac{S_0 N_G}{(2\pi)^2 }  \int d^2{\bf q} \sum_{{\bf G}_a^l, {\bf G}_b^l} \left( \delta_{{\bf p-q}, - {\bf G}_a^l }\delta_{{\bf p'- q}, - {\bf G}_b^l } e^{-i {\bf G}_b^l \cdot {\bf \tau}_{\alpha'}^l + i {\bf G}_a^l \cdot {\bf \tau}_{\alpha}^l } \right) t({\bf q},0)   \nonumber \\ & = \sum_{{\bf G}_a^l, {\bf G}_b^l} \delta_{{\bf p} + {\bf G}_a^l ,{\bf p}'+ {\bf G}_b^l } e^{-i {\bf G}_b^l \cdot {\bf \tau}_{\alpha'}^l + i {\bf G}_a^l
\cdot {\bf \tau}_{\alpha}^l } t({\bf p+  G}_a^l,0) \label{samelayerhamiltonian1}
\end{eqnarray}
where ${\bf G}_a^l$, ${\bf G}_b^l$ are two sets of reciprocal lattice vectors in the BZ of the layer $l$. 
The summation in Eq. (\ref{samelayerhamiltonian1}) respects the symmetries of the problem. When we restrict ${\bf p, p}'$ to the same (first) BZ of graphene, 
we have $\delta_{{\bf p + G}_a^l ,{\bf p'+ G}_b^l } = \delta_{\bf p,p'}\delta_{{\bf G}_a^l, {\bf G}_b^l}$ and hence 
\beq
H_{0l l}^{\alpha \alpha'}({\bf p p}')=\delta_{\bf p,p'} \sum_{{\bf G}_a^l}  e^{-i {\bf G}_a^l \cdot ({\bf \tau}_{\alpha'}^l - {\bf \tau}_{\alpha}^l )} 
t({\bf p+  G}_a^l,0)\label{eq:H0gra}
\eneq 
We then have the tight-binding Hamiltonian for the single layer graphene
\beq
t_{\bf p}^{\alpha \alpha'}= \sum_{{\bf G}_a^l}  e^{-i {\bf G}_a^l \cdot ({\bf \tau}_{\alpha'}^l - {\bf \tau}_{\alpha}^l )} t({\bf p+  G}_a^l,0),
\eneq 
which has a node at the ${\bf K}_D$ point.

\subsection{Finite Out of-Plane Displacement}

For the out-of-plane displacement, we can use  hopping $t({\bf q}, z=u^z({\bf R}_\alpha^l) - u^z({\bf R}_\alpha'^l))$. 
With the small displacement approximation, we can linearize the expression in small $u$, and sum over graphene lattice, 
leading to the intra-layer Hamiltonian
\begin{eqnarray}
&H_{\perp l l}^{\alpha \alpha'}({\bf p p}') = \frac{S_0}{(2\pi)^2 N_{G}}  \int d^2{\bf q} \left( \sum_{{\bf R}_1^l, {\bf R}_2^l} 
e^{- i ({\bf R}_{1\alpha}^l ({\bf p -q})- {\bf R}_{2\alpha'}^{l} ({\bf p'-q}))}   
( u^z({\bf R}_\alpha^l) - u^z({\bf R}_\alpha'^l) ) \partial_z t({\bf q},z)|_{z=0}\right)  \nonumber  \\
& =\frac{1}{N^{1/2}_{G} }  \sum_{{\bf p}_1 \in BZ} \sum_{{\bf G}_a^l, {\bf G}_b^l} 
e^{ i (  \tau_\alpha^l \cdot {\bf G}_a^l - \tau_{\alpha'}^l\cdot {\bf G}_b^l)} \delta_{{\bf p'+p_1 + G}_b^l, {\bf p+G}_a^l} \nonumber \\ 
& \left(    u_{{\bf p}_1}^{z l,\alpha}   \partial_z t({\bf G}_a^l+ {\bf p-p_1},z)|_{z=0}      - u_{\bf p_1}^{z l,\alpha'} \partial_z t( {\bf G}_b^l + 
{\bf p' +p_1},z)|_{z=0}  \right). 
\end{eqnarray}
One can show that
\beq
\frac{\partial t({\bf r},t)}{\partial z }|_{z=0} =0
\eneq
and hence there is no out-of-plane phonon coupling to the graphene electrons (this will also be derived by symmetry), and thus
\beq
H_{\perp l l}^{\alpha \alpha'}({\bf p p}')=0
\eneq
Thus, we only focus on in-plane phonon coupling below.

\subsection{Finite In-Plane Displacement}
\label{Sec:inplane_displacement}

For the in-plane displacement, we can use the Fourier transform form of the hopping term, $t({\bf q}, z=0)$. 
With the small displacement approximation, we can linearize ${\bf u}$, and sum over graphene lattice, leading to the intra-layer Hamiltonian 

\begin{eqnarray}
&H_{\parallel l l}^{\alpha \alpha'}({\bf p p}') = \frac{S_0}{(2\pi)^2 N_{G}}  \int d^2{\bf q} \left( \sum_{{\bf R}_1^l, {\bf R}_2^l} e^{- i ({\bf R}_{1\alpha}^l 
\cdot ({\bf p -q})- {\bf R}_{2\alpha'}^{l} ({\bf p'-q}))} i{\bf q}\cdot ( {\bf u}({\bf R}_{1\alpha}^l) - {\bf u}({\bf R}_{2 \alpha'}^{l})) \right) t({\bf q},0)
 \nonumber \\ 
&= \frac{S_0 N^{1/2}_{G}}{(2\pi)^2 }  \int d^2{\bf q} \times t({\bf q},0) i{\bf q}\cdot  \nonumber \\ & \sum_{{\bf p}_1 \in BZ} \sum_{{\bf G}_a^l, {\bf G}_b^l} \left(\delta_{ {\bf -p_1 +p -q}, - {\bf G}_a^l } \delta_{{\bf p'-q},-{\bf G}_b^l} e^{ i (  {\bf \tau}_\alpha^l\cdot {\bf G}_a^l - {\bf \tau}_{\alpha'}^l\cdot {\bf G}_b^l))} {\bf u}_{{\bf p}_1}^{l,\alpha}         -\delta_{{\bf p_1+ p'-q},-{\bf G}_b^l}\delta_{{\bf p -q}, -{\bf G}_a^l   }  e^{i ( {\bf \tau}_\alpha^l \cdot 
{\bf G}_a^l-  {\bf \tau}_{\alpha'}^l \cdot {\bf G}_b^l)}  {\bf u}_{{\bf p}_1}^{l,\alpha'}   \right)  
\end{eqnarray}
where ${\bf G}_a^l, {\bf G}_b^l$ are two sets of reciprocal lattice vectors in the the BZ of layer 
$l$. Transforming the $\delta$ functions
\begin{eqnarray}
&\delta_{ {\bf -p_1 +p -q}, - {\bf G}_a^l } \delta_{{\bf p'-q},-{\bf G}_b^l} = \delta_{{\bf q}, \frac{{\bf G}_a^l+{\bf G}_b^l+ {\bf p+p'-p}_1}{2} } 
\delta_{{\bf p'+p}_1+ {\bf G}_b^l, {\bf p+G}_a^l} \nonumber \\ & \delta_{{\bf p_1+ p'-q},-{\bf G}_b^l}\delta_{{\bf p -q}, -{\bf G}_a^l   } 
= \delta_{{\bf q}, \frac{{\bf G}_a^l + {\bf G}_b^l + {\bf p + p' +p_1}}{2}}\delta_{{\bf p'+p_1 + G}_b^l, {\bf p+G}_a^l},
\end{eqnarray} 
we find the matrix elements of the intra-layer Hamiltonian, 
\begin{eqnarray}\label{eq:eph_Ham_inplane}
&H_{\parallel l}^{\alpha \alpha'}({\bf p p}') = \frac{1 }{N^{1/2}_{G} }  \sum_{{\bf p}_1 \in BZ} \sum_{{\bf G}_a^l, {\bf G}_b^l} e^{ i (  {\bf \tau}_\alpha^l \cdot {\bf G}_a^l - {\bf \tau}_{\alpha'}^l \cdot {\bf G}_b^l)} \delta_{{\bf p'+p_1 + G}_b^l, {\bf p+G}_a^l}  \nonumber \\ & \left( i \frac{{\bf G}_a^l+{\bf G}_b^l+ {\bf p+p'-p_1}}{2} \cdot   {\bf u}_{{\bf p}_1}^{l,\alpha}   t(\frac{{\bf G}_a^l+{\bf G}_b^l+ {\bf p+p'-p_1}}{2},0)      -i\frac{{\bf G}_a^l + {\bf G}_b^l + {\bf p + p' +p_1}}{2} \cdot {\bf u}_{{\bf p}_1}^{l,\alpha'} t(\frac{{\bf G}_a^l + {\bf G}_b^l + {\bf p + p' +p_1}}{2},0)  \right) = \nonumber \\ & = \frac{1 }{N^{1/2}_{G} }  \sum_{{\bf p}_1 \in BZ} \sum_{{\bf G}_a^l, {\bf G}_b^l} e^{ i (  {\bf \tau}_\alpha^l \cdot {\bf G}_a^l - {\bf \tau}_{\alpha'}^l \cdot {\bf G}_b^l)} \delta_{{\bf p'+p_1 + G}_b^l, {\bf p+G}_a^l}  \nonumber \\ & \left( i ({\bf G}_a^l+{\bf  p-p_1}) \cdot   {\bf u}_{{\bf p}_1}^{l,\alpha}   t({\bf G}_a^l+ {\bf p-p_1},0)      
-i( {\bf G}_b^l + {\bf p' +p_1}) \cdot {\bf u}_{{\bf p}_1}^{l,\alpha'} t( {\bf G}_b^l + {\bf p' +p_1},0)  \right) \label{eq:ephintra}
\end{eqnarray}

It should be noted that ${\bf p}_1$ summation in the above expression is over one BZ. For convenience, we can choose it within the first BZ. From the $\delta$-function, we have ${\bf p_1}={\bf p+G}_a^l-{\bf p'-G}_b^l$, so if we keep the summation over ${\bf G}_b^l$, only one ${\bf G}_a^l$ is allowed to make 
${\bf p}_1$ in the first BZ. 
Actually the choice of ${\bf G}_a^l$ will not influence the results. To see that, let's assume the choice of ${\bf G}_{a0}^l$ can make ${\bf p}_{10}={\bf p+G}_{a0}^l-{\bf p'-G}_b^l$ within the first BZ, and now we consider another choice ${\bf G}_a^l={\bf G}_{a0}^l+\Delta {\bf G}^l$, so ${\bf p_1=p+G}_a^l-{\bf p'-G}_{b}^l={\bf p}_{10}+\Delta {\bf G}^l$ is outside the first BZ. For the first term in Eq. (\ref{eq:ephintra}), we have

\begin{eqnarray}
&e^{ i (\tau_\alpha^l \cdot {\bf G}_a^l - \tau_{\alpha'}^l \cdot {\bf G}_b^l)}  i ({\bf G}_a^l+ {\bf p-p_1}) \cdot   {\bf u}_{{\bf p}_1}^{l,\alpha}   t({\bf G}_a^l+ {\bf p-p_1},0) \nonumber \\
&= e^{ i (\tau_\alpha^l \cdot ({\bf G}_{a0}^l+\Delta {\bf G}^l) - \tau_{\alpha'}^l \cdot {\bf G}_b^l)}  i ({\bf G}_{a0}^l+ {\bf p-p_{10}}) \cdot   
{\bf u}_{{\bf p}_{10}+\Delta {\bf G}^l}^{l,\alpha}   t({\bf G}_{a0}^l+ {\bf p-p_{10}},0)\nonumber\\
&= e^{ i (\tau_\alpha^l \cdot {\bf G}_{a0}^l - \tau_{\alpha'}^l \cdot {\bf G}_b^l)}  i ({\bf G}_{a0}^l+ {\bf p-p}_{10}) \cdot  {\bf u}_{{\bf p}_{10}}^{l,\alpha}   t({\bf G}_{a0}^l+ {\bf p-p}_{10},0)
\end{eqnarray}

where we have used the periodic gauge $\uu^{l,\alpha}_{\pp_{10}+\Delta \GG^l} =  e^{-i \tau_\alpha^l \cdot \Delta {\bf G}^l}  {\bf u}_{\pp_{10}}^{l,\alpha}$. 
From the above expression, one can see that the first term in Eq. (\ref{eq:ephintra}) remains the same if one changes from the summation over ${\bf p}_{10}$ to that over ${\bf p}_1$. This means one just needs to pick one ${\bf G}_a^l$ (e.g. one possible choice ${\bf G}_a^l={\bf G}_b^l$), while which choice will not influence the results. Similar analysis can also be applied to the second term in Eq. (\ref{eq:ephintra}).  
 

Both in graphene and TBG, the electronic states with the energies around the Fermi energy is only close to the $\eta {\bf K}_D$ point where $\eta=\pm$ is the valley. Thus, one needs to expand the electron-phonon Hamiltonian around $\eta {\bf K}_D$. The electron momenta $\pp$ and $\pp'$ in Eq. (\ref{eq:eph_Ham_inplane}) can be expanded around the same valley ($\eta {\bf K}_D$ for a fixed $\eta$) or opposite valleys ($\eta {\bf K}_D$ and $-\eta {\bf K}_D$), corresponding to intra-valley and inter-valley e-ph interaction, respectively. In this work, we focus on inter-valley e-ph interaction.

\subsection{Inter-valley Electron-Phonon Interaction with In-plane Phonons}
In this section, we choose $\pp= \kk+ \eta \KK_D$, $\pp'= \kk'- \eta \KK_D$ where $k,k' \ll |\KK_D|$,
and correspondingly, we also need to decompose the phonon momentum into a large value (the difference between the two valleys) 
and a small value around that,
\beq
\pp_1 = \pp+ \PP_1,;\;\; p\ll P_1
\eneq 
where $\PP_1$ determines the major part of phonon momentum. The momentum $\delta$-function in $H_{\parallel l l}^{\alpha \alpha'}(\pp \pp')$ becomes
\beq
\delta_{\kk+ \eta \KK_D + \GG_a^l, \kk' - \eta \KK_D+ \pp_1+ \GG_b^l} = \delta_{\kk, \kk'+ \pp}\delta_{2 \eta K_D + \GG_a^l-  \GG_b^l,\PP_1}. 
\eneq 
For the momentum $\pp$ in the first BZ, there are $3$ possible values for $\PP_1$ momenta  
\beq
\PP_1 = - \eta \KK_D; - \eta (\KK_D+ \GG_1) ; - \eta (\KK_D+ \GG_2). 
\eneq 
However, as discussed in Sec. \ref{Sec:inplane_displacement}, the summation over $\pp_1$ in the Hamiltonian (\ref{eq:ephintra}) is within one BZ
and the choice of different BZs will not influence the result due to the periodic gauge of the phonon modes. 
Thus, we make the choice of $\PP_1 = - \eta \KK_D$ below, which corresponds to the K-phonon, and the Hamiltonian reads
\begin{eqnarray}
&H_{\parallel l}^{\eta\alpha, -\eta\alpha'}(\kk \kk') = \frac{i}{N^{1/2}_{G} }  \sum_{\GG_a^l} e^{ i (  \tau_\alpha^l  - \tau_{\alpha'}^l )\cdot \GG_a^l}
e^{i\tau_{\alpha'}^l \cdot ( \PP_1 - 2 \eta  \KK_D) } \nonumber \\ &   \left(  (\GG_a^l+\eta \KK_D + \kk'-\PP_1) \cdot   \uu_{\PP_1+ \kk-\kk'}^{l,\alpha}   t(\GG_a^l+\eta \KK_D +\kk' -\PP_1,0)-(\GG_a^l + \eta \KK_D +\kk) \cdot \uu_{\PP_1+ \kk-\kk'}^{l,\alpha'} t( \GG_a^l +\eta \KK_D  + \kk,0)  \right) \nonumber\\
&=\frac{i }{N^{1/2}_{G} }   \sum_{\GG_a^l} e^{ i (  \tau_\alpha^l  - \tau_{\alpha'}^l )\cdot \GG_a^l} \left(  (\GG_a^l+2 \eta \KK_D + \kk') \cdot   \uu_{-\eta \KK_D+ \kk-\kk'}^{l,\alpha}   t(\GG_a^l+ 2 \eta \KK_D +\kk' ,0) \right. \nonumber \\ &  \left. -( \GG_a^l + \eta \KK_D +\kk) \cdot \uu_{-\eta \KK_D+ \kk-\kk'}^{l,\alpha'} 
t( \GG_a^l +\eta \KK_D  + \kk,0)  \right),  
\end{eqnarray}
where we have used
\begin{eqnarray}
& 3\KK_D= \GG_1+ \GG_2 \nonumber \\ & e^{i \tau_A \cdot (\GG_1+ \GG_2) }=e^{i \tau_B \cdot (\GG_1+ \GG_2) }=1. 
\end{eqnarray}

We next parameterize the above expression and define
\begin{eqnarray}
    &\gamma_0= \sum_{\GG_a} t(\GG_a+ \eta \KK_D, 0),  \nonumber \\ 
    &\gamma_1=\sum_{\GG_a} (\widehat{\GG_a+ \eta \KK_D})_x  (\GG_a+ \eta \KK_D)_x \frac{ \partial {t(\qq,0)} }{\partial q}|_{q= |\KK_D+ \eta \GG_a|  },
\end{eqnarray}
\beq\gamma_2 = \sum_{\GG_a} e^{i(\tau_A- \tau_B)\cdot \GG_a }(\widehat{\GG_a+ \KK_D})_x  (\GG_a+ \KK_D)_x 
\frac{ \partial {t(\qq,0)} }{\partial q}|_{q= |\KK_D+ \GG_a|  }\eneq
and
\beq
\gamma_3= 2 i \sum_{\GG_a} e^{i(\tau_A- \tau_B)\cdot \GG_a } (\GG_a+ \KK_D)_y t(\GG_a+ \KK_D,0). 
\eneq

We also define the acoustic and optical basis for the displacement field as
\beq
\uu_\pp^{l,\alpha}=\uu_\pp^{l,ac} + (-1)^\alpha \uu_\pp^{l,op}
\eneq  
where we define $(-1)^{\alpha=A(B)}=-1(+1)$. 


With the above definitions, we can simplify the inter-valley e-ph Hamiltonian as
\begin{eqnarray}\label{eq:Ham_parallel1}
&  H_{\parallel l}^{\eta , -\eta }(\kk \kk') =  - \frac{i}{\sqrt{N_G}} (\gamma_0 + \gamma_1) ( \kk-\kk') \cdot  
(\uu_{-\eta \KK_D+\kk-\kk'}^{l,ac}\sigma_0 - \uu_{-\eta \KK_D+\kk-\kk'}^{l,op}\sigma_z)  \nonumber \\ &+\frac{1}{\sqrt{N_G}}\gamma_3\sigma_x (- u^{l,op}_{-\eta \KK_D+\kk-\kk',y} -i\eta  u^{l,ac}_{-\eta \KK_D+\kk-\kk',x})  \nonumber \\ &  + \frac{i}{\sqrt{N_G}} \gamma_2 [  ( -(k_x-k'_x)\sigma_x + \eta (k_y + k'_y) \sigma_y )   u^{l,ac}_{-\eta \KK_D+\kk-\kk',x}  +  ( (k_y-k'_y)\sigma_x + \eta (k_x + k'_x) \sigma_y )  u^{l,ac}_{-\eta \KK_D+\kk-\kk',y} ]  \nonumber \\&  + \frac{1}{\sqrt{N_G}} \gamma_2 [   ((k_x+k'_x)  \sigma_y  +\eta (k_y-k_y')   \sigma_x) u^{l,op}_{-\eta \KK_D+\kk-\kk',x}      +   ( -(k_y+k'_y)\sigma_y + \eta (k_x- k_x')\sigma_x ) u^{l,op}_{-\eta \KK_D+\kk-\kk',y}    ] ,
\end{eqnarray}
where the Pauli matrix $\sigma$ is for the sublattice basis ($(\sigma_i)_{\alpha, \alpha'}$ with $\alpha,\alpha'=A,B$). 

In addition, we also define 
\beq
    v_F=\left|\sum_{\GG_a}e^{i \GG_a \cdot (\tau_A- \tau_B)}\widehat{(\KK_D+ \GG_a)}_x \frac{ \partial {t(\qq,0)} }{\partial q}|_{q= |\KK_D+ \GG_a|  }\right| 
\eneq
for the Fermi velocity of graphene. 

All these parameters can be numerically evaluated and their values are listed in Table (\ref{Table1}) for both the $3$ $\GG_a$ and for 2000 (infinite) $\GG_a$'s.
One can see a large difference for the obtained values for all the parameters when keeping only $3$ $\GG_a$, as compared to keeping 2000 (infinite) $\GG_a$'s. 
For example, when keeping of order 2000 $G_a$'s, we find $\gamma_1= 63.506$ and $\gamma_0= -61.4306$, which almost cancel each other, and hence 
$\gamma=\gamma_0+\gamma_1= 2.07565$. For the first $3$ $\GG_a$'s approximation, $\gamma_0 = -22.7898$ and $\gamma_1= 12.7393$, and hence $\gamma=-10.0505$, 
which is larger and of opposite sign as compared to keeping all $\GG_a$. 
Here $\gamma_0, \gamma_1$ and $\gamma_2$ is in the unit of eV while $\gamma_3$ is in the unit of $eV/\AA$. 


\begin{table}[h!]
\centering
\begin{tabular}{||c c c c||} 
 \hline
  & Infinite & 3 Shells  & 3/Infinite  \\ 
 \hline\hline
 $v_f$ &  $5.229$  & $7.3206$  & $1.4$ \\ 
 \hline
 $\gamma_0$ & $-61.4306$ & $-22.7898$ & $0.371$ \\
 \hline
 $\gamma_1$ & $63.5062$ & $12.7393$ & $0.2006$ \\
 \hline
 $\gamma=\gamma_0+\gamma_1$ & $2.07565$ & $-10.0505$ & $-4.8421$ \\
 \hline
 $\gamma_2$ & $6.7281$ & $6.3917$ & $0.95$ \\
 \hline
 $\gamma_3$ & $-17.088$ & $-37.5936$ & $2.2$ \\
 \hline 
\end{tabular}
\caption{Table for all the parameters in the Hamiltonian (\ref{eq:Ham_parallel1}
). Here $\gamma_0, \gamma_1$ and $\gamma_2$ is in the unit of eV while $\gamma_3$ is in the unit of $eV/\AA$.  }
\label{Table1}
\end{table}

In the above Hamiltonian (\ref{eq:Ham_parallel1}), the term in the second line ($\gamma_3$ term) is zeroth order in $\kk$ while all the other terms linearly depend on $\kk$. 
Thus, we below only keep the zeroth-order term for e-ph interaction for K-phonons. 



\subsubsection{Physical Meaning of the phonon Modes at $\KK_D$}


We now focus on the leading term of the inter-valley electron-phonon Hamiltonian
\begin{eqnarray}
&  H = \frac{\gamma_3}{\sqrt{N_G}}\sum_{\kk,\kk',\eta,\alpha \beta } (- u^{l,op}_{-\eta \KK_D+\kk-\kk',y} -i\eta  u^{l,ac}_{-\eta \KK_D+\kk-\kk',x})  
c^\dagger_{\kk+ \eta \KK_D, \alpha } (\sigma_x)_{\alpha\beta} c_{\kk' - \eta \KK_D, \beta}  
\end{eqnarray}
where the Hamiltonian is hermitian once the summation over $\eta$ is performed. We take the Fourier transform form of the phonon field
\begin{eqnarray}
& \uu_\alpha(\kk-\kk'- \eta \KK_D) =  \frac{1}{\sqrt{N_G}} \sum_{\RR} e^{i \eta \KK_D\cdot (\RR+ \tau_\alpha)}
e^{-i (\kk-\kk')\cdot (\RR+ \tau_\alpha)} \uu_\alpha(\RR+\tau_\alpha). 
\end{eqnarray} 
We can see that the term $ \eta \KK_D\cdot \RR$ is not periodic in $\RR$, and this problem can be fixed by enlarging the unit cell. 
By choosing $\RR=n \aa_1+ m \aa_2$, we have
\beq
\KK_D \cdot \RR = -\frac{2\pi}{3} (m+n). 
\eneq
By choosing 
\beq
m+n= 3 n_1+ s\;\;\;
\eneq  with $s=0,1,2$ and 
\beq
\sum_{\RR}= \sum_{m,n} = \sum_{n_1,m, s},\eneq
We have
\beq
\RR'= 3 n_1 \aa_1 + m (\aa_2-\aa_1),\;\; \RR= \RR' + s \aa_1, 
\eneq 
so 
\begin{eqnarray}
& \uu_\alpha(\kk-\kk'-\eta \KK_D) =  \frac{1}{\sqrt{N_G}} \sum_{s,\RR'} e^{i \eta \KK_D\cdot (s \aa_1+\tau_\alpha)} 
e^{-i (\kk-\kk')\cdot(\RR'+ s \aa_1+ \tau_\alpha)} \uu(\RR' + s \aa_1+\tau_\alpha)  \nonumber \\ & = \frac{1}{\sqrt{3} }\sum_{s} e^{i \frac{2\pi}{3} \eta (1-s)} \uu_{\alpha s}(\kk-\kk')
\end{eqnarray} 
where we have defined the $\uu$ modes in the larger unit cell as
\beq
\uu_{\alpha s}(\qq) = \sqrt{\frac{3}{N_G}}\sum_{\RR'} e^{-i \qq\cdot (\RR'+ s \aa_1+ \tau_\alpha)} \uu(\RR' + s \aa_1+\tau_\alpha). 
\eneq
For our choices of vectors, we note that 
\beq
\tau_A \cdot \KK_D =\tau_B \cdot \KK_D= \frac{2\pi}{3}
\eneq The fact that they have the same value allows us to separate them once again into optical and acoustic combinations
\beq
u_{\alpha s}= u_{ac,s} + (-1)^\alpha u_{op,s}
\eneq
and to have 
\begin{eqnarray}
& \uu_{ac/op} (\kk-\kk'- \eta \KK_D) =  \frac{1}{\sqrt{3}}   \sum_{s} e^{i \frac{2\pi}{3} \eta (1-s)} \uu_{ac/op, s}(\kk-\kk')
\end{eqnarray}
And the Hamiltonian becomes
\begin{eqnarray}
&  H = \frac{\gamma_3}{\sqrt{3N_G}}\sum_{\kk,\kk',\eta,\alpha \beta } \sum_{s=1,2,3} e^{i \frac{2\pi}{3} \eta (1- s)}  (- u_{op, s,y}(\kk-\kk')
 -i\eta u_{ac, s,x}(\kk-\kk'))  c^\dagger_{\kk+ \eta \KK_D, \alpha } (\sigma_x)_{\alpha\beta} c_{\kk' - \eta \KK_D, \beta}  \end{eqnarray}
We may separate the valley index from electron momentum and introduce the field operator
\beq
\hat{\Psi}_{\kk,\alpha}=(c_{\kk+\KK_D,\alpha},c_{\kk-\KK_D,\alpha})^T
\eneq
and the Pauli matrix $\tau$ for the valley index. The Hamiltonian is then written as
\begin{eqnarray}
&  H = \frac{\gamma_3}{2\sqrt{3N_G}}\sum_{\kk,\kk'} \hat{\Psi}^\dagger_{\kk}\sigma_x\left(
\begin{array}{cc}
0 &  \sum_{\alpha, s} (\dd^r_{\alpha s}\cdot\uu_{\alpha, s}-i \dd^i_{\alpha s}\cdot \uu_{\alpha, s})\\
\sum_{\alpha, s} (\dd^r_{\alpha s}\cdot\uu_{\alpha s}+i \dd^i_{\alpha s}\cdot \uu_{\alpha, s}) & 0 
\end{array}\right)\hat{\Psi}_{\kk'}\nonumber\\
&=\frac{\gamma_3}{2\sqrt{3N_G}}\sum_{\kk,\kk',\alpha,s} \hat{\Psi}^\dagger_{\kk}\sigma_x\left( (\dd^r_{\alpha s}\cdot\uu_{\alpha, s})\tau_x+
(\dd^i_{\alpha s}\cdot \uu_{\alpha, s})\tau_y\right) \hat{\Psi}_{\kk'}, 
\end{eqnarray}
where we have used 
\begin{eqnarray}
&  \sum_{s} e^{i \eta\theta_{s} }  (- u_{op, s,y}-i\eta u_{ac, s,x}))  = \frac{1}{2} \sum_{s} e^{i\eta \theta_{s}}  (- (u_{B, s,y}-u_{A, s,y})
 -i\eta (u_{A, s,x}+u_{B, s,x}))\nonumber\\
& = \frac{1}{2} \sum_{\alpha, s} (\dd^r_{\alpha s}\cdot\uu_{\alpha s}+i \eta \dd^i_{\alpha s}\cdot \uu_{\alpha, s})
\end{eqnarray}
with 
\begin{eqnarray}
& \dd^r_{A, s}=\cos{\theta_{s}} \hat{e}_y+\sin{\theta_{ s}}\hat{e}_x, \;\; \dd^r_{B, s}=-\cos{\theta_{s}} \hat{e}_y+\sin{\theta_{s}}\hat{e}_x, \nonumber\\
& \dd^i_{A, s}=\sin{\theta_{s}} \hat{e}_y-\cos{\theta_{s}}\hat{e}_x, \;\; \dd^i_{B, s}=-\sin{\theta_{ s}} \hat{e}_y- \cos{\theta_{ s}}\hat{e}_x, \label{eq:dvector}
\end{eqnarray}
and $\theta_{ s}=\frac{2\pi}{3} (1-s)$. Here $\uu_{\alpha, s}$ is a function of $\kk-\kk'$. The above e-ph Hamiltonian
takes the same form as in Ref. \cite{wu2018theory}. 

\begin{figure}[hbt!]
    \centering
    \includegraphics[width=3.5in]{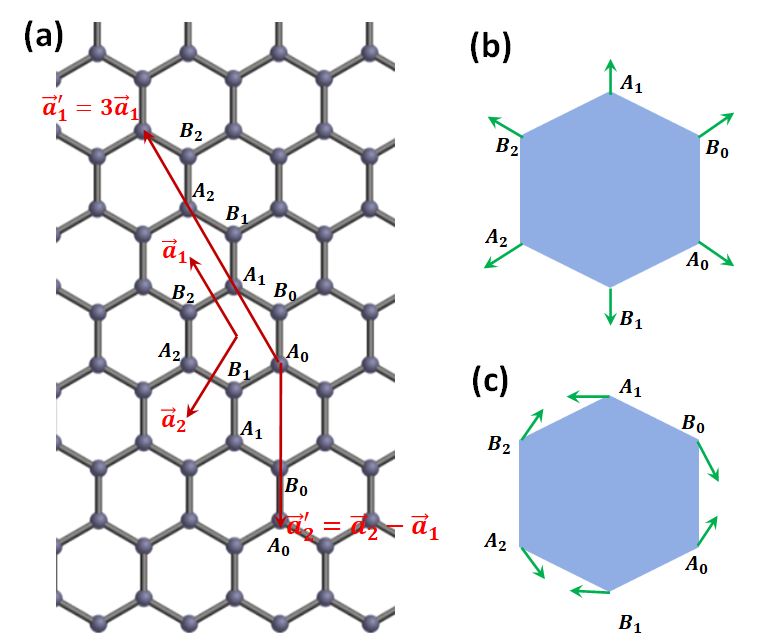} 
    \caption{ (a) Enlarged unit cell for phonon modes at $\KK_D$. (b) and (c) are for the $A_1$ and $A_2$ phonon modes at $\KK_D$. }
    \label{fig2:Kphonon}
\end{figure}

Now let's focus on the vibration mode at $\KK_D$ ($\kk\sim \kk'\sim 0$), so $\uu_{\alpha, s}(0)=\sqrt{3/N_G}\sum_{\RR'}\uu(\RR'+s\aa_1+\tau_\alpha)$. 
We can view $\dd^{r(i)}_{\alpha s}$ as the vector, along which the $(\alpha,s)$-atom vibrates. Explicitly, the $\dd^r$-vectors are given by 
$\dd^r_{A,0}=-\frac{1}{2} \hat{e}_y+\frac{\sqrt{3}}{2}\hat{e}_x$, $\dd^r_{B,0}=\frac{1}{2} \hat{e}_y+\frac{\sqrt{3}}{2}\hat{e}_x$,
$\dd^r_{A,1}=\hat{e}_y$, $\dd^r_{B,1}=-\hat{e}_y$, $\dd^r_{A,2}=-\frac{1}{2} \hat{e}_y-\frac{\sqrt{3}}{2}\hat{e}_x$
and $\dd^r_{B,2}=\frac{1}{2} \hat{e}_y-\frac{\sqrt{3}}{2}\hat{e}_x$, which can be depicted by the green arrows in Fig. \ref{fig2:Kphonon}b
and correspond to the $A_1$ phonon mode. The vibration configuration of each atoms for the $\dd^i$ phonon mode is shown 
in Fig. \ref{fig2:Kphonon}c and corresponds to the $A_2$ mode (Here $A_1$ and $A_2$ corresponds to the irreducible representations of the $C_{3v}$ group). 


\subsection{Symmetry properties of the Electron-Phonon Interaction}
In this section, we will present a construction of the electron-phonon Hamiltonian from the symmetry principle. 
We now write all the symmetry operators on the electron and phonon modes (we consider the phonon as a field and expand it around $\eta \KK_D$):
\begin{eqnarray}
&\hat{C}_3 \hat{c}^\dagger _{\kk+ \eta \KK_D, \alpha } \hat{C}_3^{-1}= e^{i \eta \tau_\alpha \cdot \GG_1 } \hat{c}^\dagger_{\hat{C}_3 \kk+ \eta \KK_D, \alpha }\nonumber\\ 
& \hat{M}_y\hat{c}^\dagger_{\kk+ \eta \KK_D, \alpha } \hat{M}_y^{-1} = \sum_\beta (\sigma_x)_{\beta \alpha} \hat{c}^\dagger_{\hat{M}_y \kk + \eta \KK_D, \beta}\nonumber \\
& \hat{M}_x\hat{c}^\dagger_{\kk+ \eta \KK_D, \alpha} \hat{M}_x^{-1} = \hat{c}^\dagger_{\hat{M}_x\kk- \eta \KK_D, \alpha} \nonumber \\ 
& \hat{T} \hat{c}^\dagger_{\kk+ \eta \KK_D, \alpha} \hat{T}^{-1}= \hat{c}^\dagger_{-\kk- \eta \KK_D, \alpha}   \nonumber \\ 
& \hat{c}^\dagger_{\kk+ \GG,\alpha}= e^{i \tau_\alpha\cdot \GG} \hat{c}^\dagger_{\kk,\alpha}\nonumber \\ 
& u_{\alpha i }(\pp+\GG) = e^{- i\tau_\alpha\cdot \GG} u_{\alpha i }(\pp)  \nonumber \\ 
& \hat{C}_3 u_{\alpha i}(\kk + \eta \KK_D)\hat{C}_3^{-1}=e^{-i \eta \tau_\alpha\cdot \GG_1} \sum_{i'}  u_{\alpha i'}(\hat{C}_3 \kk + \eta \KK_D)C_{3i'i}\nonumber \\ 
& \hat{M}_y u_{\alpha i}(\kk + \eta \KK_D) \hat{M}_y^{-1}= \sum_{i'\beta } (\sigma_x)_{\beta \alpha} u_{\beta i'}(\hat{M}_y \kk + \eta \KK_D)M_{yi'i}\nonumber \\ 
& \hat{M}_x u_{\alpha i}(\kk + \eta \KK_D) \hat{M}_x^{-1}= \sum_{i' }  u_{\alpha i'}(\hat{M}_x \kk - \eta \KK_D)M_{xi'i}\nonumber \\ 
& \hat{T} u_{\alpha i}(\kk + \eta \KK_D) \hat{T}^{-1}= u_{\alpha i}(-\kk - \eta \KK_D) 
\end{eqnarray} 


In the above, the matrix representations are defined as
\begin{eqnarray}
C_3=\left(\begin{array}{cc}-1/2&-\sqrt{3}/2\\\sqrt{3}/2&-1/2\end{array}\right)\nonumber\\
M_x=\left(\begin{array}{cc}-1&0\\0&1\end{array}\right)\nonumber\\
M_y=\left(\begin{array}{cc}1&0\\0&-1\end{array}\right). 
\end{eqnarray}
We now use the above symmetry transformation to construct the inter-valley electron-phonon Hamiltonian.  

The generic form of the intra-valley electron-phonon Hamiltonian reads
\begin{eqnarray}
&H_2=\frac{1}{\sqrt{N_G}}\sum_{\kk,\kk', \eta; \alpha \beta \theta; j} F_{\alpha \beta \theta; j}^\eta(\kk, \kk') 
c^\dagger_{\kk+ \eta \KK_D,\alpha}u_{\theta j}(\kk-\kk' - \eta \KK_D) c_{\kk' - \eta \KK_D, \beta}
\end{eqnarray}
where $F_{\alpha \beta \theta; j}^\eta(\kk, \kk')$ is the generic inter-valley electron-phonon interaction parameter. 
We consider the case of $F_{\alpha \beta \theta; j}^\eta(\kk, \kk')$ independent of $\kk,\kk'$ in order to obtain the leading term. 
However, we first obtain the symmetry conditions considering the entire functional dependence. 
Under the generic $\hat{g}$-symmetry transformations, we have
\begin{eqnarray} & \hat{g} c^\dagger_{\kk+\eta \KK_D,\alpha}\hat{g}^{-1}= \sum_{\eta', \alpha' } c^\dagger_{\hat{g}\kk+\eta' \KK_D,\alpha'} D_c(\hat{g})_{\eta'\alpha'; \eta \alpha }\nonumber \\ 
& \hat{g} u_{\alpha i}(\pp+ \eta \KK_D) \hat{g}^{-1}= \sum_{\alpha' i' \eta'} u_{\alpha' i'}(\hat{g}\pp + \eta' \KK_D) 
S_{i'i}^{\alpha'\eta'; \alpha\eta}(\hat{g}) \end{eqnarray} 
where $S_{i'\eta';i\eta}^{\alpha' \alpha}$ is a combination of the space rotation and sublattice and valley rotation. The equation $\hat{g} H \hat{g}^{-1}=H$ gives
\begin{eqnarray}& \sum_{\eta \alpha \beta\theta i}
D_c(\hat{g})_{\eta'\alpha'; \eta \alpha }S_{i', i}^{\theta'-\eta'; \theta-\eta}D^\star_c(\hat{g})_{-\eta'\beta';- \eta \beta }
F_{\alpha \beta \theta;i}^\eta(\kk, \kk')=  F_{\alpha' \beta' \theta';i'}^{\eta'}(\hat{g}\kk, \hat{g}\kk')
\end{eqnarray}
We now write the explicit form of the matrices and the constraints
\begin{eqnarray}
&\hat{M}_y: D_{c\eta'\alpha'; \eta \alpha } = \delta_{\eta'\eta} \sigma^x_{\alpha'\alpha},\;\; S_{i'i}^{\theta'\eta'; \theta\eta} = \sigma^x_{\theta'\theta} (-1)^{i+1}\delta_{i'i} \delta_{\eta, \eta'}\nonumber\\ &\implies F_{\bar{\alpha}\bar{\beta}\bar{\theta}; i}^\eta(\hat{M}_y\kk , \hat{M}_y \kk') = (-1)^{i+1} 
F_{\alpha \beta \theta;i}^\eta(\kk, \kk')\nonumber \\& \hat{M}_x:D_{c\eta'\alpha'; \eta \alpha } = \delta_{\eta'\bar{\eta}} \delta_{\alpha'\alpha},\;\; S_{i'i}^{\theta'\eta'; \theta\eta} = \delta_{\theta'\theta} (-1)^{i}\delta_{i'i} \delta_{\eta'\bar{\eta}}\nonumber\\ &\implies F_{\alpha \beta \theta;i}^{\bar{\eta}}(\hat{M}_x\kk , \hat{M}_x \kk') = (-1)^{i} F_{\alpha \beta \theta;i}^\eta(\kk, \kk')\nonumber \\& \hat{C}_3: D_{c\eta'\alpha'; \eta \alpha } = \delta_{\eta'\eta} \delta_{\alpha'\alpha} e^{i\eta\tau_\alpha \cdot \GG_1};\; \; S_{i'i}^{\theta'\eta'; \theta\eta} =e^{-i\eta \tau_\theta \cdot \GG_1}\delta_{\eta'\eta}  \delta_{\theta'\theta} C_{3 i' i} \nonumber \\ &\implies  F_{\alpha \beta \theta;i' }^\eta(\hat{C}_3\kk,\hat{C}_3 \kk')= 
e^{i \eta (\tau_\alpha + \tau_\beta+\tau_\theta)\cdot \GG_1 }\times \sum_i F_{\alpha \beta \theta;i}^\eta(\kk, \kk') C_{3i'i} \nonumber \\ 
& \hat{T}: D_{c\eta'\alpha'; \eta \alpha } = \delta_{\eta'\bar{\eta}} \delta_{\alpha'\alpha},\;\; S_{i'i}^{\theta'\eta' ;\theta\eta} = \delta_{\theta'\theta}\delta_{i'i}\delta_{\eta',\bar{\eta}}\nonumber \\ &\implies F_{\alpha \beta \theta;i}^{\bar{\eta}}(-\kk, -\kk')  = 
F_{\alpha \beta \theta;i}^{\eta\star}(\kk, \kk')
\end{eqnarray}
The Hermiticity of the $H_2$ Hamiltonian requires
\beq
\text{Hermiticity:}\;\;\; F_{\alpha \beta \theta;i}^{\eta}(\kk, \kk')  = F_{ \beta\alpha \theta;i}^{\bar{\eta} \star }(\kk', \kk) 
\eneq  We are aided by the following identity
\beq
e^{i \eta (\tau_\alpha + \tau_\beta+\tau_\theta)\cdot \GG_1 }= e^{i (-1)^\eta ((-1)^{\alpha} + (-1)^{\beta}+(-1)^{\theta})\frac{2\pi}{3} }
\eneq
This is a simple system of linear equations to solve, giving rise to the \emph{two} possible solutions. The first solution is given by
\begin{eqnarray}
&F^\eta_{\alpha\beta \theta x} = -\frac{\gamma_3}{2} \eta\sigma^x_{\alpha\beta }, \;\; F^\eta_{\alpha\beta \theta y} = -\frac{\gamma_3}{2} \sigma^x_{\alpha\beta }(-1)^{\theta}
\end{eqnarray} 
which corresponds to the inter-valley phonon contribution that we have obtained, $-\gamma_3 (i \eta u^{ac}_x + u^{op}_y)\sigma_x$. 

The second solution is the missing term whose only nonzero components are
\beq
F_{\alpha\alpha\bar{\alpha};x}^\eta= i \eta ;\;\;  F_{\alpha\alpha\bar{\alpha};y}^\eta=(-1)^\alpha
\eneq  which would correspond to a new term in the Hamiltonian
\begin{eqnarray}
& \sum_{\kk\kk'\eta} c^\dagger_{\kk+\eta \KK_D, \alpha} c_{\kk'-\eta \KK_D \alpha }\times (i \eta u_{\bar{\alpha} x}(\kk-\kk'-\eta \KK_D) 
+ (-1)^\alpha u_{\bar{\alpha} y}(\kk-\kk'-\eta \KK_D) )\nonumber
\end{eqnarray}
from which we can see two crucial properties of the extra terms

\begin{itemize}
    \item First, it is diagonal in the electron sublattice index, hence it can only come from second nearest neighbor, or larger. 
    
    \item Crucially, the sublattice index of the electron $\alpha$ and that of the phonon $\bar{\alpha}$ are opposite. It means that the phonon on one sublattice couples to the electron annihilation and creation operators on the other sublattice. Hence it is now clear why this term does not appear in our Hamiltonian or \emph{in any other two-center approximation}. The two center approximation reads
    \beq
    t(R_\alpha +u_\alpha - R_\beta-u_\beta) c_{R_\alpha}^\dagger c_{R_\beta}
    \eneq and hence the phonon sublattice index has to be either that of $c^\dagger$ or that of $c$. It cannot be neither that  of $c^\dagger$ nor that of $c$. We hence see that the absence of this term is due to the two-center approximation of our Hamiltonian (and of most tight-binding Hamiltonians).

\end{itemize}

\section{Phonons in a Single Layer Graphene}
In this section, we will discuss the phonon dynamical matrix and phonon eigen-modes. In real-space, in any dimensions, the dynamica matrix $D$ reads
\beq
(D_{\RR\alpha, \RR', \alpha'})_{ij}
\eneq
where $\RR,\RR'$ are the unit cells, the $\alpha, \alpha'$ are the atoms, and the $i, j \in x,y,z$ are the components in 3D. The presence of stability (the free energy is invariant to adding any constant vector $\vec{a}$ to the displacements: $\uu_{r\alpha} \rightarrow \uu_{r\alpha} + \aa$ ) gives
\beq
\sum_{\RR', \alpha'} (D_{\RR\alpha, \RR', \alpha'})_{ij} =0
\eneq Which gives, in a spring model, 
\beq
\sum_{\beta} D_{\alpha \beta}(\qq=0)=0, \forall \alpha;\; \sum_{\alpha} D_{\alpha \beta}(\qq=0)=0, \forall \beta
\eneq Which guarantees the presence of an acoustic phonon mode. 

\subsection{Dynamical Matrix}
The dynamical matrix can be obtained from the potential $V$ of the system
\beq
(D_{\RR\alpha, \RR' \alpha'})_{ii'} = \frac{\partial^2 V}{\partial u_{\RR\alpha i }\partial u_{\RR'\alpha ' i' }}
\eneq As such, if there exists a symmetry $\hat{g}$ of the system with representation matrix $G_{ij}$, we have the relation
\beq
D_{\hat{g} \RR, \hat{g}\alpha; \hat{g}\RR',\hat{g} \alpha'} = G D_{\RR\alpha; \RR' \alpha'} G^{-1}
\eneq

Next let's consider the constraint on the dynamical matrix from the symmetry operators, $\hat{C}_3, \hat{M}_x$ $\hat{M}_y$ and $\hat{T}$, with their 
representation matrices given in the above. Due to translation symmetry, $(D_{\RR\alpha, \RR' \alpha'})_{ii'}=(D_{\alpha\alpha'}(\RR_\alpha-\RR'_{\alpha'}))_{ii'}$ 
only depends $\RR_\alpha-\RR'_{\alpha'}$. We only consider the nearest neighbor (nn) and next nearest neighbor (nnn) interaction. 
Three nn vectors are labelled as $\delta^{nn}_{AB,\mu=1,2,3}=(a_0/2,a_0/2\sqrt{3}),(-a_0/2,a_0/2\sqrt{3}),(0,-a_0/\sqrt{3})$ that satisfies $\hat{C}_3\delta^{nn}_i=\delta^{nn}_{i+1}$, 
and six nnn vectors are $\delta^{nnn}_{\alpha\alpha,\mu=1,2,3}=-\aa_1-\aa_2,\aa_1,\aa_2$ and $-\delta^{nnn}_{\alpha\alpha,\mu=1,2,3}$ with $\hat{C}_3\delta^{nnn}_i=\delta^{nnn}_{i+1}$. 
In the above expressions, we use the convention $\delta^{nn}_{4}=\delta^{nn}_1$ and $\delta^{nnn}_{4}=\delta^{nnn}_1$. 

For the nn interaction that couples the A and B atoms, $\delta^{nn}_{AB,3}=(0,-a_0/\sqrt{3})$ remains the same under $\hat{M}_x$ and thus 
\beq M_x D_{AB}(\delta^{nn}_3)M_x^{-1}=D_{AB}(\delta^{nn}_3), \eneq 
which requires the diagonal form of $D_{AB}(\delta^{nn}_3)$, given by
\begin{equation}
D_{AB}(\delta^{nn}_3)=-\left(\begin{array}{cc}K_{0t}&0\\0&K_{0r}
\end{array}\right). 
\end{equation}
Other nn interaction terms can be related to $D_{AB}(\delta^{nn}_3)$ through $C_3$ rotation, 
\beq
D_{AB}(\delta^{nn}_{i+1})=C_3D_{AB}(\delta^{nn}_i)C_3^{-1}. \eneq 
$\hat{M}_y$ reverses the sign of $\delta^{nn}_{AB,\mu}$ and interchanges A and B sublattice, and thus
\beq 
D_{BA}(-\delta^{nn}_i)=M_y D_{AB}(\delta^{nn}_i) M_y^{-1}. 
\eneq

For the nnn interaction, we first consider the matrix $D_{AA}(\delta^{nnn}_1)$. Mirror $\hat{M}_x$ requires 
\beq D_{AA}(-\delta^{nnn}_1)
=M_x D_{AA}(\delta^{nnn}_{1})M_x^{-1}\eneq. 
Furthermore, the symmetric condition for the dynamical matrix gives rise to 
\beq D_{AA}(\delta^{nnn}_1)=(D_{AA}(-\delta^{nnn}_1))^T. \eneq Thus, we have the condition 
\beq (D_{AA}(\delta^{nnn}_1))^T
=M_x D_{AA}(\delta^{nnn}_{1})M_x^{-1}, \eneq which leads to the form
\begin{equation}
D_{AA}(\delta^{nnn}_1)=-\left(\begin{array}{cc}K_{r1}&K_2\\-K_2&K_{t1}
\end{array}\right). 
\end{equation}
$D_{AB}(\delta^{nnn}_{2,3})$ can be related to $D_{AB}(\delta^{nnn}_1)$ from $C_3$ rotation, 
\beq D_{AB}(\delta^{nnn}_{i+1})=C_3D_{AB}(\delta^{nnn}_{i})C_3^{-1}, \eneq 
while $D_{AB}(-\delta^{nnn}_i)$ can be related to $D_{AB}(\delta^{nnn}_i)$ from mirror $\hat{M}_x$, 
\beq D_{AB}(-\delta^{nnn}_i)=M_xD_{AB}(\delta^{nnn}_i)M_x^{-1}. \eneq 
Finally, time reversal $\hat{T}$ requires all the five parameters 
$K_{t0}, K_{r0}, K_{t1}, K_{r1}, K_2$ to be real. 
With all the components in the real space, we can transform the dynamical matrix to the momentum space, which is given by
\begin{equation}
D_{\alpha\alpha'}(\kk)=\sum_{\RR}e^{-i\kk\cdot(\RR+\tau_\alpha-\tau_{\alpha'})} D_{\alpha\alpha'}(\RR+\tau_\alpha-\tau_{\alpha'})
\end{equation}


The explicit form of the dynamical matrix under the basis $(u_{Bx}, u_{By}, u_{Ax}, u_{Ay})^T$ is given by 
\begin{eqnarray}
& D(\kk) = d_{ij}(k_{x}, k_y) \sigma_i \otimes \xi_j \nonumber \\ 
&  d_{00}(k_{x}, k_y) =\left(K_{\text{r1}}+K_{\text{t1}}\right) \left(-2 \cos \left(\frac{k_x}{2}\right) \cos \left(\frac{\sqrt{3} k_y}{2}\right)-\cos \left(k_x\right)+3\right)+\frac{1}{2} \left(3 K_{\text{r0}}+3 K_{\text{t0}}\right)\nonumber \\ 
& d_{01}(k_{x}, k_y)  =\sqrt{3} \sin \left(\frac{k_x}{2}\right) \sin \left(\frac{\sqrt{3} k_y}{2}\right) \left(K_{\text{r1}}-K_{\text{t1}}\right)\nonumber \\ 
&  d_{03}(k_{x}, k_y) =-\left(K_{\text{r1}}-K_{\text{t1}}\right) \left(\cos \left(k_x\right)-\cos \left(\frac{k_x}{2}\right) \cos \left(\frac{\sqrt{3} k_y}{2}\right)\right)
\nonumber \\ 
&  d_{11}(k_{x}, k_y) =\frac{1}{2} \sqrt{3} \left(K_{\text{r0}}-K_{\text{t0}}\right) \sin \left(\frac{k_x}{2}\right) \sin \left(\frac{k_y}{2 \sqrt{3}}\right)
\nonumber \\ 
&  d_{21}(k_{x}, k_y) =\frac{1}{2} \sqrt{3} \left(K_{\text{r0}}-K_{\text{t0}}\right) \sin \left(\frac{k_x}{2}\right) \cos \left(\frac{k_y}{2 \sqrt{3}}\right)
 \nonumber \\ 
&  d_{10}(k_{x}, k_y) =-\frac{1}{2} \left(K_{\text{r0}}+K_{\text{t0}}\right)\left( 2 \cos \left(\frac{k_x}{2}\right) \cos \left(\frac{k_y}{2 \sqrt{3}}\right)+\cos \left(\frac{k_y}{\sqrt{3}}\right)\right)
\nonumber \\ 
&  d_{20}(k_{x}, k_y) =\frac{1}{2} \left(K_{\text{r0}}+K_{\text{t0}}\right)\left( 2 \cos \left(\frac{k_x}{2}\right) \sin \left(\frac{k_y}{2 \sqrt{3}}\right)-\sin \left(\frac{k_y}{\sqrt{3}}\right)\right)
\nonumber \\ 
&  d_{13}(k_{x}, k_y) =-\frac{1}{2} \left(K_{\text{r0}}-K_{\text{t0}}\right)\left( \cos \left(\frac{k_x}{2}\right) \cos \left(\frac{k_y}{2 \sqrt{3}}\right)-\cos \left(\frac{k_y}{\sqrt{3}}\right)\right)
\nonumber \\ 
&  d_{23}(k_{x}, k_y) =\frac{1}{2} \left(K_{\text{r0}}-K_{\text{t0}}\right)\sin \left(\frac{k_y}{2 \sqrt{3}}\right) \left(\cos \left(\frac{k_x}{2}\right)+2 \cos \left(\frac{k_y}{2 \sqrt{3}}\right)\right)
 \nonumber \\ 
&  d_{32}(k_{x}, k_y) =4 K_2 \sin \left(\frac{k_x}{2}\right) \left(\cos \left(\frac{k_x}{2}\right)-\cos \left(\frac{\sqrt{3} k_y}{2}\right)\right)
\end{eqnarray}
where $\sigma$ operates in the $A,B$ sub-lattices while $\xi$ operates in the $x,y$ components. 
The eigen-equation of dynamical matrix is given by 
\beq D(\kk) \epsilon_{\kk} = M \omega^2 \epsilon_{\kk}. \eneq
Below we focus on the dynamical matrix and the corresponding phonon modes at $\KK_D$.

\subsection{Dynamical Matrix and Phonon Modes at $\KK_D$}
Close to the $K_D$ point, the dynamical matrix is not solvable, so we first consider the eigen-states at the $\KK_D$ point
\begin{eqnarray}
&D(\eta \KK_D)= \frac{3}{2} \left(K_{\text{r0}}+3 K_{\text{r1}}+K_{\text{t0}}+3 K_{\text{t1}}\right) \sigma_0\otimes \sigma_0 +\nonumber \\ &+ \frac{3}{4} \left(K_{\text{r0}}-K_{\text{t0}}\right) ( \sigma_1\otimes \sigma_3+ \eta \sigma_2\otimes \sigma_1)- \nonumber \\ &-\eta3 \sqrt{3} K_2\sigma_3\otimes \sigma_3
\end{eqnarray}
Let's define
\begin{eqnarray}
&\alpha_1=\frac{3}{2} \left(K_{\text{r0}}+3 K_{\text{r1}}+K_{\text{t0}}+3 K_{\text{t1}}\right)\nonumber \\ &\alpha_2 =\frac{3}{2} \left(K_{\text{r0}}-K_{\text{t0}}\right)\nonumber\\ &\alpha_3= 3 \sqrt{3} K_2
\end{eqnarray}
and the eigenvalues and eigenstates are
\begin{eqnarray}
&E_{E_-}= \alpha_1 -\alpha_3; \;  E_{E_+}=\alpha_1 -\alpha_3 \\ & E_{A_1}=\alpha_1+ \alpha_2+ \alpha_3 \nonumber \\& E_{A_2}=\alpha_1-\alpha_2 +\alpha_3 \nonumber \\ & \epsilon_{\eta K_D}^{E_-}=\frac{1}{\sqrt{2}} (0,0,\eta i,1)^T,\;\; \epsilon_{\eta K_D}^{E_+}=\frac{1}{\sqrt{2}} (-\eta i,1,0,0)^T,\nonumber \\ & \epsilon_{\eta K_D}^{A_1}=\frac{1}{2}(-\eta i,-1,-\eta i,1)^T,\;\;\epsilon_{\eta K_D}^{A_2} =\frac{1}{2}(\eta i,1,-\eta i,1)^T\nonumber 
\end{eqnarray}
and the eigen-frequencies are related to eigenvalues by 
\beq \omega_r = \sqrt{E_{r}/M}, \; \; r=E_{\pm}, A_2, A_1. \eneq
The projected Dynamical matrix into the basis of the doublet $E_\pm$ is 
\begin{eqnarray}
&\bra{\epsilon_{\eta \KK_D}^{E_\mp} }D(\eta \KK_D) \ket{ \epsilon_{\eta \KK_D}^{E_\mp} }=\nonumber \\ 
&=  E_{E_-} \sigma_0 +\frac{\alpha_2}{2\sqrt{3}}(k_y\sigma_y - \eta k_x \sigma_x) 
\end{eqnarray}
The symmetry properties of these states are:
\begin{eqnarray}
& \hat{M}_y \epsilon_{\eta \KK_D}^{A_1} = \epsilon_{\eta \KK_D}^{A_1};\;\;\hat{M}_y \epsilon_{\eta \KK_D}^{A_2} =- \epsilon_{\eta \KK_D}^{A_2}\nonumber \\ 
& \hat{M}_y \epsilon_{\eta \KK_D}^{E_-} =- \epsilon_{\eta \KK_D}^{E_+},\;\;\;  \hat{M}_y \epsilon_{\eta \KK_D}^{E_+} =- \epsilon_{\eta \KK_D}^{E_-}
\end{eqnarray} 
The energies of the phonon bands away from $\eta \KK_D $ are
\begin{eqnarray}
&\bra{\epsilon_{\eta \KK_D}^{A_{1,2}} }D(\eta \KK_D) \ket{ \epsilon_{\eta \KK_D}^{A_{1,2}} }= \nonumber \\ 
& =E_{A_{1,2}} + \frac{k^2}{8} (-6\sqrt{3} K_2-K_{r0}-3 K_{r1}+K_{t0}-3K_{t1})
\end{eqnarray} 
Hence the effective masses of phonons away from $\eta \KK_D$ are the same for the $A_{1,2}$ modes.


\section{Electron and Phonon-Band Projected Electron-Phonon Hamiltonian}
Since the inter-valley phonons always have a finite frequency, we keep only the zeroth-order terms of momenta and neglect the rest for the inter-valley electron-phonon Hamiltonian
\begin{eqnarray}
&  H_{\parallel l}^{\eta, -\eta}(\kk, \kk') =  \frac{1}{\sqrt{N_G}}\gamma_3\sigma_x (- u^{l,op}_{-\eta \KK_D+\kk-\kk',y} 
-i\eta  u^{l,ac}_{-\eta \KK_D+\kk-\kk',x}) 
\end{eqnarray}


For the electron Dirac Hamiltonian $v_f(\eta k_x \sigma_x + k_y \sigma_y) $, we have the energies $E_\pm = \pm v_f k$ and the corresponding eigen-states
\begin{eqnarray}
\psi^\pm_{\kk,\eta}= \frac{1}{\sqrt{2}} \left[
    \begin{array}{c}
          \pm \eta e^{-i \eta \phi} \\
          1
    \end{array}
  \right],\;\;\; e^{i\phi}=  \frac{k_x+ i k_y}{k}
\end{eqnarray}
The electron operator can also be expanded on the eigen-states with
\beq
c_{\kk+\eta \KK_D, \alpha} = \sum_{m=1,2} \psi^m_{\kk,\eta \alpha} \gamma_{\kk,\eta}^m. 
\eneq

For the phonon part, with the eigenstates of the dynamical matrix $\epsilon^n_{\qq\alpha}$, 
we have that the displacements $b_{\qq n}$ become decomposed in phonon band basis 
\beq
u_{\qq t} =\sum_{n=1,2,3,4} \epsilon_{\qq t}^n \frac{b_{\qq n} + b_{-\qq n}^\dagger}{\sqrt{2M \omega_{\qq n}}} \label{eq:phononeigenexpansion}
\eneq where $t=1,2,3,4$ corresponds to $u_{Bx}, u_{By}, u_{Ax}, u_{Ay}$, respectively, and $n$ labels the energy eigenstates (from lower energy
to higher energy). 

Below we will project the electron-phonon Hamiltonian into the eigen-state basis of both electrons and phonons.

\subsection{Projected Inter-Valley Electron-Phonon Interaction}\label{sec:IntervalleyPorjected}
We next consider the inter-valley e-ph interaction and focus on the $k$-independent terms. 

\begin{eqnarray}
&  H = \frac{1}{\sqrt{N_G}}\gamma_3\sum_{\kk,\kk',\eta,\alpha \beta } (- u^{l,op}_{-\eta \KK_D+\kk-\kk',y} -i\eta  u^{l,ac}_{-\eta \KK_D+\kk-\kk',x})  
c^\dagger_{\kk+ \eta \KK_D, \alpha} (\sigma_x)_{\alpha\beta} c_{\kk' - \eta \KK_D, \beta}  \nonumber \\ 
& =   -\frac{1}{2 \sqrt{N_G}}\gamma_3\sum_{m,n=\pm} \sum_{r=1}^4 \sum_{\kk,\kk',\eta,\alpha \beta } \frac{b_{-\eta \KK_D+\kk-\kk',r} 
+b_{\eta \KK_D-\kk+\kk', r }^\dagger}{\sqrt{2M\omega_{ -\eta \KK_D+\kk-\kk',r } }}
\gamma_{\kk,\eta}^{n\dagger} \gamma_{\kk',-\eta}^m \\ 
&(( \epsilon^r_{-\eta \KK_D+\kk-\kk',2}- \epsilon^r_{-\eta \KK_D+\kk-\kk',4}) + i \eta (\epsilon^r_{-\eta \KK_D+\kk-\kk',1}+ 
\epsilon^r_{-\eta \KK_D+\kk-\kk',3}))  \psi_{\kk,\eta, \alpha}^{n\star}  (\sigma_x)_{\alpha\beta} \psi_{\kk',-\eta, \beta}^m \nonumber 
\end{eqnarray}
Using the phonon eigen-states obtained at $\KK_D$, we can obtain the approximate form of the couplings for $\kk=\kk'$ as 
\begin{eqnarray}\label{eq:InterValley_r}
&r= E_-: \;\; (( \epsilon^r_{-\eta \KK_D,2}- \epsilon^r_{-\eta \KK_D,4}) + i \eta (\epsilon^r_{-\eta \KK_D,1}+ \epsilon^r_{-\eta \KK_D,3})) = 0\nonumber \\ 
& r= E_+: \;\; (( \epsilon^r_{-\eta \KK_D,2}- \epsilon^r_{-\eta \KK_D,4}) + i \eta (\epsilon^r_{-\eta \KK_D,1}+ \epsilon^r_{-\eta \KK_D,3})) = 0\nonumber \\ 
&r= A_2: \;\; (( \epsilon^r_{-\eta \KK_D,2}- \epsilon^r_{-\eta \KK_D,4}) + i \eta (\epsilon^r_{-\eta \KK_D,1}+ \epsilon^r_{-\eta \KK_D,3})) = 0\nonumber \\ 
&r= A_1: \;\; (( \epsilon^r_{-\eta \KK_D,2}- \epsilon^r_{-\eta \KK_D,4}) + i \eta (\epsilon^r_{-\eta \KK_D,1}+ \epsilon^r_{-\eta \KK_D,3})) = -2 
\end{eqnarray}
Hence the electrons at the $\KK_D$ point only couple with the $A_1$ phonons, while the couplings to all other 3 phonon modes are proportional to 
the momentum $\kk-\kk'$, being zero when $\kk=\kk'$. With 
\beq
\sum_{\alpha\beta}\psi^{n\star}_{\kk\eta \alpha } \sigma^x_{\alpha \beta} \psi^m_{\kk'-\eta \beta} = \frac{\eta}{2}(ne^{i \eta \phi} -me^{i \eta \phi'}  )
\eneq
We then have the inter-valley electron-phonon coupling 
\begin{eqnarray}\label{eq:elph_Hop_intervalley}
&  H^{op}_{inter-vall} =    \frac{1}{ \sqrt{N_G}}\gamma_3\sum_{m,n=\pm}  \sum_{\kk,\kk',\eta,\alpha \beta } \frac{b_{-\eta \KK_D+\kk-\kk',A_1} 
+b_{\eta \KK_D-\kk+\kk', A_1 }^\dagger}{\sqrt{2M\omega_{ -\eta \KK_D+\kk-\kk',A_1 } }}
\gamma_{\kk,\eta}^{n\dagger} \gamma_{\kk',-\eta}^m    \psi_{\kk,\eta, \alpha}^{n\star}  (\sigma_x)_{\alpha\beta} \psi_{\kk',-\eta, \beta}^m  \nonumber \\  
&\approx \frac{1}{2\sqrt{N_G}}\gamma_3\sum_{m,n=\pm}  \sum_{\kk,\kk',\eta} \frac{b_{-\eta \KK_D+\kk-\kk',A_1} +b_{\eta \KK_D-\kk+\kk', 
A_1 }^\dagger}{\sqrt{2\sqrt{M(\alpha_1+ \alpha_2 + \alpha_3)} }}\gamma_{\kk,\eta}^{n\dagger} \gamma_{\kk',-\eta}^m   \eta(ne^{i\eta\phi} -me^{i\eta\phi'}) 
  \end{eqnarray}
where $\phi, \phi'$ are the angles of the momenta $\kk, \kk'$. 

\begin{itemize}
    \item Intra-band  ($m=n$) electrons have the matrix element
\beq
m=n: \eta n(e^{i\eta\phi} -e^{i\eta\phi'}) 
\eneq this is zero for $\phi=\phi'$, i.e. for $\kk, \kk'$ being parallel, and is maximum for  $\phi=\pi+ \phi'$, i.e. for $\kk,\kk'$ being anti-parallel.

 \item Inter-band  ($m\neq n$) electrons have the matrix element
\beq
m=-n: \eta n(e^{i\eta\phi} + e^{i\eta\phi'}) 
\eneq this is zero for $\phi=\pi+ \phi'$, i.e. for $\kk,\kk'$ antiparallel and is maximum for $\phi=\phi'$, i.e. for $\kk,\kk'$ parallel.

\end{itemize}

For further reference, we will also need the electron-phonon interaction on the sublattice basis, \emph{not} projected into the graphene bands - 
as we will need to project it into the TBG bands, which are different from the graphene bands.  
This form of the inter-valley electron-phonon interaction reads
\begin{eqnarray}\label{eq:Hepintervall} 
&  H_{inter-vall}^{op}   \approx \frac{1}{ \sqrt{N_G}}\gamma_3 \sum_{\kk,\kk',\eta,\alpha \beta } \frac{b_{-\eta \KK_D+\kk-\kk',A_1} 
+b_{\eta \KK_D-\kk+\kk', A_1 }^\dagger}{\sqrt{2 \sqrt{ M(\alpha_1+ \alpha_2 + \alpha_3)} }}
c^\dagger_{\kk+ \eta \KK_D, \alpha } (\sigma_x)_{\alpha\beta} c_{\kk' - \eta \KK_D, \beta}. 
\end{eqnarray}

\subsection{Numerical Calculations of Electron-Phonon Interaction in a Single-Layer Graphene}\label{sec:numerical}
\begin{figure}[hbt!]
    \centering
    \includegraphics[width=7in]{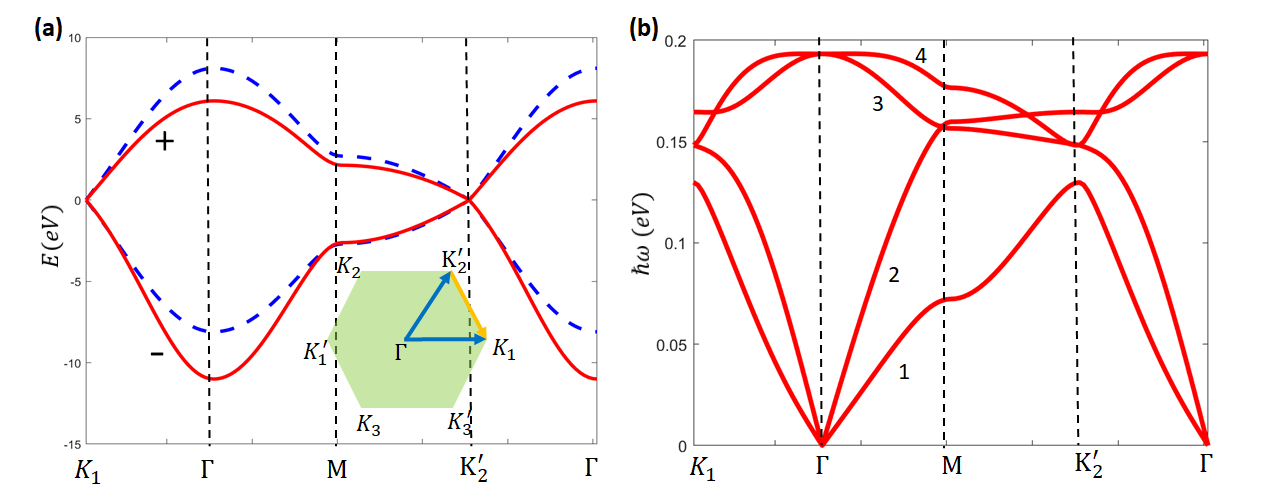} 
    \caption{(a) Energy dispersion for electrons in single-layer graphene. Here the blue dashed lines are for the nearest neighbor approximation, while
    the red solid lines are for the full long-range hopping. The inset shows the Brillouin zone of graphene. Here $K_1$ is just $K_D$ defined
    in the main text. (b) Phonon dispersion in single-layer graphene.  }
    \label{fig:dispersion}
\end{figure}

In this section, we will describe our numerical calculations of energy spectrum for both electrons and phonons in a single layer graphene. 
The energy spectrum for electrons and phonons are shown in Fig. \ref{fig:dispersion}(a) and (b), respectively,
which are consistent with the literature \cite{neto2009electronic,sahoo2012phonon,thingstad2020phonon,maultzsch2004phonon,mohr2007phonon}. 
The parameters for the calculation of energy spectrum of electrons and phonons are summarized in table \ref{tab:parameters}. 
For the phonon dispersion, we label four in-plane phonon modes as the modes 1, 2, 3 and 4, with their irreducible representations (irreps) at high symmetry momenta shown in table \ref{tab:phononsymmetry}. 
Here we only consider the $C_{6v}$ symmetry group for the in-plane modes of graphene, as the z-direction symmetry will be lost when considering TBG. 
The phonon modes 1 and 2 are degenerate have zero energy at ${\bf \Gamma}$, so they describe the acoustic phonon modes with $E_1$ irrep of $C_{6v}$, while
the phonon modes 3 and 4 are degenerate at ${\bf \Gamma}$ and describe two optical phonon modes with $E_2$ irrep. 
These degeneracies are split away from $\Gamma$. At $\KK_1$ (the same as $\KK_D$), the modes 2 and 3 become degenerate, forming the $E$ irrep, while the mode 1 belongs to $A_2$ irrep and the mode 4 to $A_1$ irrep, as shown in Fig. \ref{fig:dispersion} (b). 

\begin{table}[h!]
\centering
\begin{tabular}{|| c | c  || }
\hline
 Quantity & Value  \\ 
 \hline
 $a_0$ & $2.46 $\AA  \\  
 \hline
 $r_0$ & $0.148 a_0$  \\
 \hline
 $V_{pp\pi}$ & $-2.7 eV$  \\
 \hline
 $K_{t0}$ & $2.94\times 10^5 cm^{-2}$ \\
 \hline
 $K_{r0}$ & $5.15\times 10^5 cm^{-2}$ \\
 \hline
 $K_{t1}$ & $-8.58\times 10^4 cm^{-2}$ \\
 \hline
 $K_{r1}$ & $1.33\times 10^5 cm^{-2}$  \\
 \hline
 $K_2$  &  0   \\
 \hline
\end{tabular}
\caption{Table for parameters used for calculating energy spectrum of electrons and phonons in single-layer graphene. }
\label{tab:parameters}
\end{table}

\begin{table}[h!]
\centering
\begin{tabular}{|| c | c | c | c | c || }
\hline
 Phonon modes & 1 & 2 & 3 & 4 \\ 
 \hline
$\Gamma$ ($C_{6v}$)  & $E_1$ & $E_1$ & $E_2$ & $E_2$   \\  
 \hline
 $K$ ($C_{3v}$) & $A_2$ & $E$ & $E$ & $A_1$  \\
  \hline
\end{tabular}
\caption{Symmetry representations for the in-plane phonons at high symmetry momenta $\Gamma$ and $\KK$ in single-layer graphene. }
\label{tab:phononsymmetry}
\end{table}

For the numerical evaluation of electron-phonon interaction, we introduce the interaction parameter $g$ as 
(Since we only consider a single-layer graphene here, we have dropped the layer index) 
\begin{eqnarray}
&H_{ep}=\frac{1}{\sqrt{N_G}}\sum_{\pp\pp',\alpha\beta r} g^r_{\alpha\beta}(\pp,\pp') c^{\dagger}_{\pp,\alpha}c_{\pp',\beta}(b_{\pp-\pp',r}+b^{\dagger}_{\pp'-\pp,r}) \nonumber \\
&=\frac{1}{\sqrt{N_G}}\sum_{\pp\pp',nm,r} g^s_{nm}(\pp,\pp') \gamma^{n\dagger}_{\pp}\gamma^{m}_{\pp'}(b_{\pp-\pp',r}+b^{\dagger}_{\pp'-\pp,r})
\end{eqnarray}
where $N_G$ is the number of unit cells, $\alpha,\beta=A,B$, $n,m$ label the eigen-states of electrons and $s$ labels the eigen-state of phonon modes. 
Here $g^r_{\alpha\beta}(\pp,\pp')$ is in the sub-lattice basis for electrons while $g^r_{nm}(\pp,\pp')$ is in the eigen-state basis. 
Explicitly, $g^r_{\alpha\beta}(\pp,\pp')$ is given by 
\begin{eqnarray}
g^r_{\alpha\beta}(\pp,\pp')=i\sqrt{\frac{\hbar}{2M\omega_{\pp-\pp',r}}}\sum_{G} e^{i \GG\cdot(\tau_{\alpha}-\tau_{\beta})} 
\left(t(\pp'+\GG) (\pp'+\GG)\cdot{\bf \epsilon}^r_{\pp-\pp',\alpha}-t(\pp+\GG) (\pp+\GG)\cdot{\bf \epsilon}^r_{\pp-\pp',\beta}\right)
\end{eqnarray}
which can be directly obtained from substituting the phonon eigen-mode expansion Eq. (\ref{eq:phononeigenexpansion}) into Eq. (\ref{eq:ephintra}). 
Here we have recovered the constant $\hbar$ for the correct unit. 
\begin{eqnarray}
g^r_{nm}(\pp,\pp')=\sum_{\alpha \beta}\psi^{n*}_{\alpha,\pp} g^r_{\alpha\beta}(\pp,\pp') \psi^{m}_{\beta,\pp'}. 
\end{eqnarray}
With the expansion of the momentum around $\eta \KK_D$, one can directly relate the function $g$ to the parameters $\gamma_{1,2,3}$ defined in the 
early sections. We can give an estimate of the magnitude of $g$ from the value of $\gamma_3$ as $|\gamma_3|\sqrt{\frac{\hbar}{2M\omega_{op,K}}}= 
|\gamma_3|a\sqrt{\frac{\hbar^2}{2m_e a_B^2}\frac{1}{\hbar\omega_{op,K}}\frac{m_e}{M} \frac{a_B^2}{a^2}}$, where $\frac{\hbar^2}{2m_e a_B^2}$ is Rydberg
energy, $\hbar\omega_{op,K}$ is optical phonon energy at $K$,
$m_e/M$ is the mass ration between electron and carbon atom, 
$a_B$ is the Bohr radius and $a=a_0/\sqrt{3}$ is the NN distance between two carbon atoms in graphene. With $|\gamma_3|\sim 17 eV/$\AA, 
$a\sim 1.42 \AA$ and the dimensionless quantity $\sqrt{\frac{\hbar^2}{2m_e a_B^2}\frac{1}{\hbar\omega_{op,\Gamma}}\frac{m_e}{M} \frac{a_B^2}{a^2}}
\sim 0.02$, we find the magnitude of $g$ is around $0.56$ eV. 
To evaluate $g$ numerically, it is more efficient to perform the calculation in the real space instead of the momentum space since the hopping function $t$ 
decays exponentially in the real space, but only in power-law in the momentum space. To get a real space expression, we can use the identity
\begin{eqnarray}
\sum_{\GG}e^{i \GG\cdot(\tau_{\alpha}-\tau_{\beta})} i (\pp+\GG) t(\pp+\GG)=\sum_{\RR} e^{-i \pp\cdot(\RR+\tau_{\alpha}-\tau_{\beta})}\nabla t(\rr)|_{\rr=\RR+\tau_{\alpha}-\tau_{\beta}},
\end{eqnarray}
and obtain
\begin{eqnarray}
&g^r_{\alpha\beta}(\pp,\pp')=\sqrt{\frac{\hbar}{2M\omega_{r,\pp-\pp'}}}\sum_{\RR} 
\left( e^{-i \pp'\cdot\delta_{\alpha\beta}}{\bf \epsilon}^r_{\alpha,\pp-\pp'} - e^{-i \pp\cdot\delta_{\alpha\beta}} {\bf \epsilon}^r_{\beta,\pp-\pp'}\right)
\cdot \left.\nabla t(\rr)\right|_{\rr=\delta_{\alpha\beta}}\nonumber\\
&=-\sqrt{\frac{\hbar}{2M\omega_{r,\pp-\pp'}}}\sum_{\RR} \frac{t(\delta_{\alpha\beta})}{r_0}
\left( e^{-i \pp'\cdot(\RR+\tau_{\alpha}-\tau_{\beta})}{\bf \epsilon}^r_{\alpha,\pp-\pp'} - 
e^{-i \pp\cdot(\RR+\tau_{\alpha}-\tau_{\beta})} {\bf \epsilon}^r_{\beta,\pp-\pp'}\right)
\cdot \frac{\delta_{\alpha\beta}}{|\delta_{\alpha\beta}|},\label{eq:ephinteractionparameterfull}
\end{eqnarray}
where $\delta_{\alpha\beta}=\RR+\tau_{\alpha}-\tau_{\beta}$. This is the expression for our numerical simulations. 

\begin{figure}[hbt!]
    \centering
    \includegraphics[width=7in]{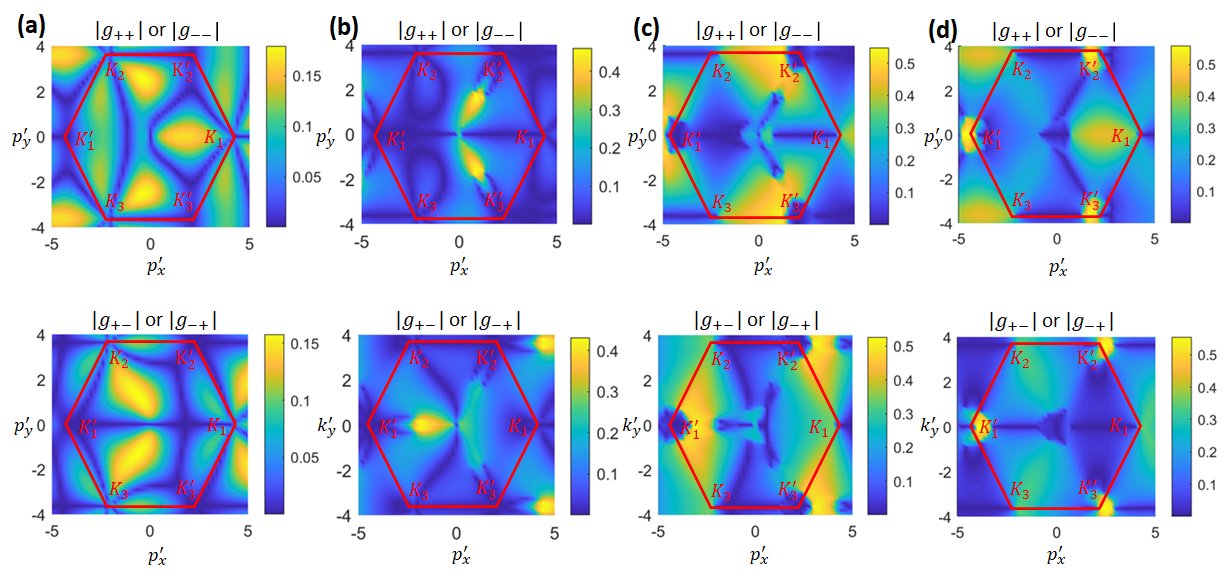} 
    \caption{ The electron-phonon interaction parameter $g^r_{nm}(\pp,\pp')$ as a function of $p'$ for a fixed $\pp=\KK_1+\delta \pp$, 
    where $\delta \pp=\frac{1}{a_0}(0.001,0)$. (a), (b), (c)
    and (d) are for four phonon modes $r=1,2,3,4$, respectively, and $n,m=\pm$. The red hexagon shows the Brillouin zone.
    Here the momenta $\pp$ and $\pp'$ are in the unit of $1/a_0$ with $a_0=2.46$\AA. }
    \label{fig:ephpara1}
\end{figure}

\begin{figure}[hbt!]
    \centering
    \includegraphics[width=7in]{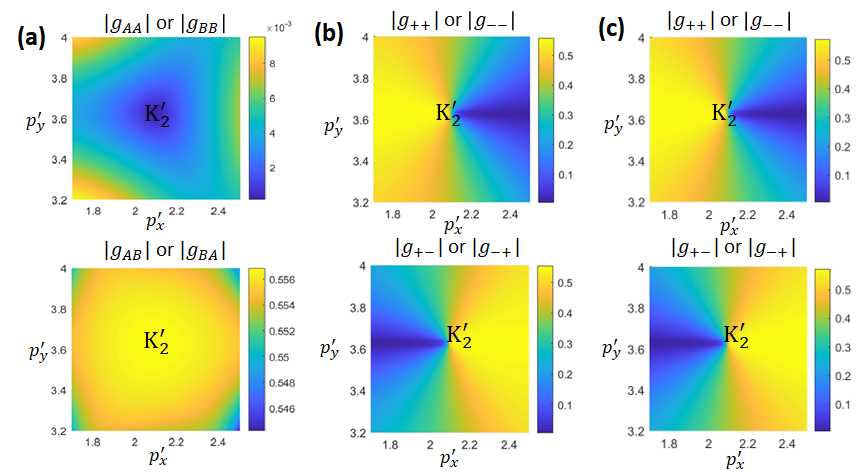} 
    \caption{ Zoom in for the electron-phonon parameter $g(\pp,\pp')$ as a function of $\pp'$ with a fixed $\pp=\KK_1+\delta \pp$ ($\delta \pp=\frac{1}{a_0}(0.001,0)$)
    for the phonon mode $r=4$ (inter-valley scattering).
    (a) and (b) show the zoom in of $g^{r=4}_{\alpha\beta}(\pp,\pp')$ ($\alpha,\beta=A,B$) and $g^{r=4}_{nm}(\pp,\pp')$ ($n,m=\pm$) for $\pp'$ around 
    $\KK_2'$. (c) shows the analytical solutions for the electron-phonon parameters $g^{r=4}_{nm}(\pp,\pp')$ ($n,m=\pm$) for $\pp'$ around $\KK_2'$ for the phonon mode 4 from Eq. (\ref{eq:elph_Hop_intervalley}). Here the momenta $\pp$ and $\pp'$ are in the unit of $1/a_0$ with $a_0=2.46$\AA. }
    \label{fig:ephpara2_inter}
\end{figure}

Next let's discuss our numerical results in Figs. \ref{fig:ephpara1} and \ref{fig:ephpara2_inter} based on the full expression Eq. (\ref{eq:ephinteractionparameterfull}). 
Fig. \ref{fig:ephpara1} shows $g^r_{nm}(\pp=\KK_1+\delta \pp,\pp')$ ($\KK_1=\KK_D$) as a function of $\pp'$ in 
the whole Brillouin zone for different phonon modes (a small $\delta \pp=\frac{1}{a_0}(0.001,0)$ is chosen to get rid of the degeneracy of electron states at
the Dirac point). 
Since electron Fermi surface is only around $\pm \KK_D$, we can focus on the corners of the Brillouin zone (BZ), which is shown by the red hexagon in Fig. \ref{fig:ephpara1}. 
Furthermore, we focus on the inter-valley e-ph interaction, which means $\pp'$ around $\KK_1',\KK_2',\KK_3'$, as we fix $\pp=\KK_1+\delta \pp$ and the phonon momentum is $\qq=\pp-\pp'$. 
We first notice that the e-ph interaction for the phonon modes 1, 2 and 3 are approaching zero when the momentum $\pp'$ is close to $\KK_1',\KK_2',\KK_3'$, which is consistent with our analytical solution (See Eq. \ref{eq:InterValley_r}) in Sec. \ref{sec:IntervalleyPorjected}. Thus, only phonon modes 4 couple to electron for the $K$-phonons. 
The phonon mode 4 belongs to $E_2$ irrep of $C_{6v}$ at ${\bf \Gamma}$ and $A_1$ irrep of $C_{3v}$ at $\KK_D$, and we show the zoom-in for $\pp'$ around $\KK_2'$ (inter-valley scattering) in Fig. \ref{fig:ephpara2_inter}(a) and (b). Here (a) and (b) correspond to the parameter $g$ in the sublattice basis and the eigen-state basis, respectively, for the electron part. 
On the sublattice basis, we find $g$ is non-zero for the inter-sublattice components, while its value approaches zero for the intra-valley components, consistent with the analytical results shown in Eq. (\ref{eq:Hepintervall}). On the eigen-state basis, we find $g$ shows a singular behavior at $\KK_2'$ for the inter-valley scattering. In Fig. \ref{fig:ephpara2_inter}(c), we also show the analytical results of inter-valley scattering from Eq. (\ref{eq:elph_Hop_intervalley}), which is consistent with the numerical results in Fig. \ref{fig:ephpara2_inter} (b) on the eigen-state basis. 



\section{Review of Bistritzer-MacDonald model for Twisted Bilayer Graphene}  \label{sec:BM_model}
Here we first briefly review the Bistritzer-MacDonald (BM) model \cite{bistritzer2011moire} of magic-angle twisted bilayer graphene (MATBG)
by introducing our notations. Readers may refer to the supplementary materials of Ref.~\cite{song2019all} for more details. 

\subsection{Basis and Single-particle Hamiltonian } \label{sec:BM_basis}
\begin{figure}
\centering
\includegraphics[width=0.6\linewidth]{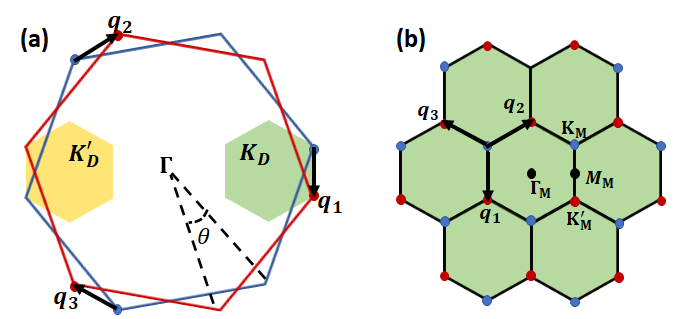}
\caption{The moir\'e BZ and the $\QQ$ lattice. (a) The blue and red hexagons represent the BZ's of the top and bottom graphene layers, respectively. The blue and red dots represent the Diract points ($\KK_D$ momentum) from the two layers. The vectors $\qq_1,\qq_2,\qq_3$ connecting the Dirac points of the two layers span the moir\'e BZ in the valley $\KK_D$ (shaded with light green). Similarly, the Dirac points at the $\KK_D^\prime$ momentum form another Moir\'e BZ in the valley $\KK_D^\prime$ (shaded with light yellow). (b) We denote the blue and red $\QQ$ lattices as $\mathcal{Q}_+$ and $\mathcal{Q}_{-}$ respectively. In the BM model, the states in the valley $\KK_D$ ($\KK_D^\prime$) of the top layer and in the valley $\KK_D^\prime$ ($\KK_D$) of the bottom layer contribute to the lattice $\mathcal{Q}_+$ ($\mathcal{Q}_-$).}
\label{fig:QQ}
\end{figure}

As shown in Fig. \ref{fig:QQ}(a), the low-energy states around the Dirac points at $\KK_D$ of the two graphene layers form the Moir\'e Brillouin zone (MBZ) in the valley $\KK_D$. Similarly, those at the $\KK_D^\prime$ momenta of the two layers form another MBZ in the valley $\KK_D^\prime$.
Thus, in the BM model there are two independent sets of basis from two valleys.
We use the index $\eta$ ($=+$ for $\KK_D$ and $-$ for $\KK_D^\prime$) to label the two graphene valleys.
We denote the basis as $c_{\kk,\QQ,\alpha,\eta,s}$, where $\kk$ is a momentum in the MBZ, $\QQ$ takes values in the lattice shown in Fig. \ref{fig:QQ}(b) (illustrated below), $\alpha=1,2$ represents the graphene sublattice, $\eta=\pm$ represents the graphene valley, and $s=\uparrow,\downarrow$ 
is the spin index. To construct the MBZ and the $\QQ$ lattices in Fig. \ref{fig:QQ}(b), we define 
${\bf q}_1= {\bf K}_D^-- {\bf K}_D^+= k_\theta(0,-1) , {\bf q}_2 =\hat{C}_{3z} {\bf q}_1= {\bf K}_D^-+ {\bf G}_1^2 - 
{\bf K}_D^+- {\bf G}_1^1 = k_\theta (\frac{\sqrt{3}}{2}, \frac{1}{2}   ), {\bf q}_3 =C_{3z}^2 {\bf q}_1 = {\bf K}_D^-+ {\bf G}_2^2 - {\bf K}_D^+ - {\bf G}_2^1 = k_\theta (-\frac{\sqrt{3}}{2}, \frac{1}{2}   )$, where $k_\theta=2|{\bf K}_D|\sin\frac{\theta}{2}$ and $\theta$ is the twist angle between two layers. Here $\KK_D^+$ and $\KK_D^-$ are the $\KK_D$ momentum of the top and bottom layers 
There are two types of $\QQ$ lattices in Fig. \ref{fig:QQ}(b): the blue lattice $\mathcal{Q}_+=\{ \qq_2 + n_1 \bb_{M1} + n_2 \bb_{M2}\ |\ n_{1,2}\in \mathbb{Z} \}$ and the red lattice $\mathcal{Q}_- = \{- \qq_2 + n_1 \bb_{M1} + n_2 \bb_{M2}\ |\ n_{1,2}\in \mathbb{Z} \}$, where $\bb_{M1},\bb_{M2}$ are Moir\'e reciprocal lattice basis
\begin{equation}
    \bb_{M1} = \qq_2 - \qq_1 =k_\theta (\frac{\sqrt{3}}{2},\frac{3}{2}) ,\qquad 
    \bb_{M2} = \qq_3 - \qq_1 = k_\theta (-\frac{\sqrt{3}}{2},\frac{3}{2}) . 
\end{equation}
For $\QQ \in \mathcal{Q}_+$, the basis is defined as 
\begin{equation} \label{eq:BM-basis1}
(\QQ\in \mathcal{Q}_+)\qquad
c_{\kk,\QQ,\alpha,\eta,s}^\dagger = 
\begin{cases}\displaystyle
\frac{1}{\sqrt{N_{\rm G}}} \sum_{\RR \in \mathrm{top}} e^{i(\KK_D^+ + \kk-\QQ) \cdot (\RR + \tau_\alpha) }  c_{\RR,\alpha,s}^\dagger \qquad & \text{if}\; \eta=+ \\
\displaystyle
\frac{1}{\sqrt{N_{\rm G}}} \sum_{\RR' \in \mathrm{bottom}} e^{i(-\KK_D^- + \kk-\QQ) \cdot (\RR' + \tau_\alpha') }  c_{\RR',\alpha,s}^\dagger \qquad & \text{if}\; \eta=-
\end{cases},
\end{equation}
and for $\QQ \in \mathcal{Q}_-$, the basis is defined as 
\begin{equation}\label{eq:BM-basis2}
(\QQ\in \mathcal{Q}_-)\qquad
c_{\kk,\QQ,\alpha,\eta,s}^\dagger = 
\begin{cases}\displaystyle
\frac{1}{\sqrt{N_{\rm G}}} \sum_{\RR' \in \mathrm{bottom}} e^{i(\KK_D^- + \kk-\QQ) \cdot (\RR'+ \tau_\alpha') }  c_{\RR',\alpha,s}^\dagger \qquad & \text{if}\; \eta=+ \\
\displaystyle
\frac{1}{\sqrt{N_{\rm G}}} \sum_{\RR \in \mathrm{top}} e^{i(-\KK_D^+ + \kk-\QQ) \cdot (\RR + \tau_\alpha) }  c_{\RR,\alpha,s}^\dagger \qquad & \text{if}\; \eta=-
\end{cases}\ . 
\end{equation}
Here $N_{\rm G}$ is the number of graphene unit cells in each layer, $\RR$ and $\RR'$ index graphene lattices in the the top and bottom layers, respectively, $\tau_\alpha$ and $\tau_\alpha'$ are the sublattice vectors of the two layers, respectively, $\KK_D^+$  and $\KK_D^-$ are the $\KK_D$ momentum of the top and bottom layers, respectively, and $c_{\RR,\alpha,s}$ ($c_{\RR',\alpha,s}$) is the fermion operator with spin $s$ at the atom site $\RR+\tau_\alpha$ ($\RR'+\tau_\alpha'$).
The $\mathcal{Q}_{+}$ ($\mathcal{Q}_{-}$) lattice is defined in such a way that $\eta \KK_D^+ +\kk - \QQ$ ($\eta \KK_D^- +\kk - \QQ$) with $\QQ\in \mathcal{Q}_+$ ($\QQ\in \mathcal{Q}_-$) is the Dirac point position $\eta \KK_D^+$ ($\eta \KK_D^-$) when $\kk$ equals to the high symmetry point $\eta \KK_M$ ($\eta \KK_M'$) of the MBZ, as shown in Fig. \ref{fig:QQ}(b).

\begin{figure}
\centering
\includegraphics[width=0.9\linewidth]{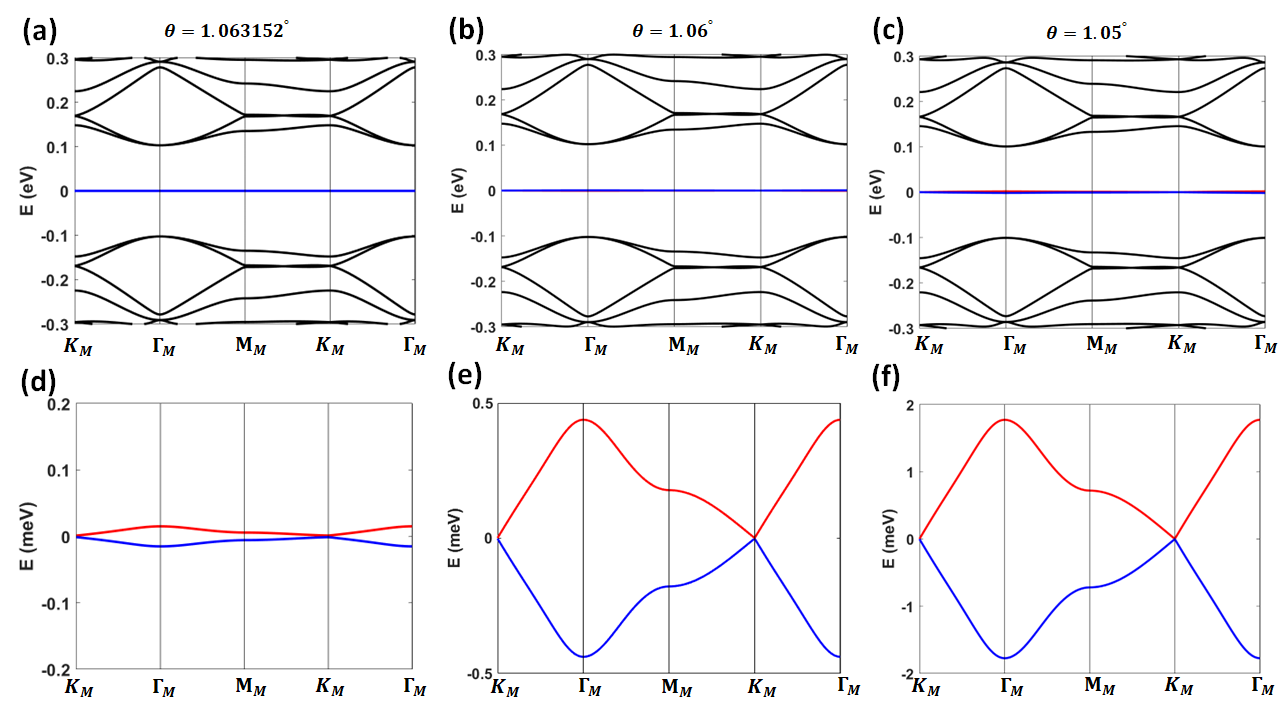}
\caption{ The band structure in the valley $\eta=+$ of TBG near the magic angle in the chiral limit ($w_0=0$). (a), (b) and (c) are for three twist angles $\theta=1.063152^\circ$, $\theta=1.06^\circ$ and  $\theta=1.05^\circ$. (d), (e) and (f) show the zoom in for the dispersion of the flat bands for these three angles. One can see the profiles of the dispersion look similar for these three angles, while the band width changes $0.02$ meV to 2 meV. The other parameters of the BM model are given by  $v_F = 5.944\mathrm{eV}\cdot\mathring{A}$, $|\KK|=1.703\mathrm{\mathring{A}^{-1}}$, $w_1=110\mathrm{meV}$. 
}
\label{fig:bands_free}
\end{figure}

The BM model is given by 
\begin{equation} \label{eq:H0}
\hH_{\rm 0} = \sum_{\eta s} \sum_{\kk \in {\rm MBZ}} \sum_{\alpha \alpha'} \sum_{\QQ,\QQ'} h_{\QQ\alpha,\QQ'\alpha'}^{(\eta)}(\kk) c_{\kk,\QQ,\alpha,\eta,s}^\dagger c_{\kk,\QQ',\alpha',\eta,s}
\end{equation}
where the single particle Hamiltonian reads 
\begin{equation}\label{eq:BM-Hamiltonian}
h_{\QQ\alpha,\QQ'\alpha'}^{(+)}(\kk) = v_F (\kk-\QQ)\cdot \boldsymbol{\sigma} \delta_{\QQ,\QQ'} 
+ \sum_{j=1}^3 [T_j]_{\alpha\alpha'} \delta_{\QQ,\QQ'\pm \qq_j},\qquad 
h_{\QQ\alpha,\QQ'\alpha'}^{(-)}(\kk) = h_{-\QQ\alpha,-\QQ'\alpha'}^{(+)*}(-\kk), 
\end{equation}
\begin{equation}
T_j = w_0 \sigma_0 + w_1 \sigma_x \cos \frac{2\pi(j-1)}3 + w_1 \sigma_y \sin \frac{2\pi(j-1)}3 \ .
\end{equation}
Here $w_0$ and $w_1$ are the interlayer couplings in the AA-stacking and AB-stacking regions, respectively. 
In this work, we adopt the parameters $v_F = 5.944\mathrm{eV\cdot\mathring{A}}$, $k_\theta=0.0312 \mathrm{\mathring{A}^{-1}}$, $w_1=110\mathrm{meV}$. 
The relation $h_{\QQ\alpha,\QQ'\alpha'}^{(-)}(\kk) = h_{-\QQ\alpha,-\QQ'\alpha'}^{(+)*}(-\kk)$ is due to the time-reversal symmetry that transform the two valleys to each other. 
The single-particle Hamiltonian (upon to a unitary transformation \cite{song2019all}) is periodic with respect to the reciprocal lattice vectors $\bb_{M1}$, $\bb_{M2}$. The MBZ and high symmetry momenta are defined in Fig. \ref{fig:QQ}(b). 
The corresponding real space unit cell and maximal Wyckoff positions are shown in Fig. \ref{fig:QQ}(c). 
The $1a$ and $2c$ positions correspond to the AA-stacking and AB-stacking regions, respectively. 
We denote the real space lattice basis as $\aa_{M1}$, $\aa_{M2}$, they satisfy $\aa_{M i}\cdot\bb_{Mj} = 2\pi \delta_{ij}$.
We plot the band structures in the valley $\eta=+$ for different twisted angles around the magic angle in the chiral limit $w_0=0$ in Fig. \ref{fig:bands_free} . 

The fact that the band structure is labeled by $\kk$ in the MBZ implies that $\kk$ labels the eigenvalues of translation operators of the Moir\'e lattice.  
We hence {\it define} the translation operator $T_\RR$ as 
\begin{equation} \label{eq:moire-translation-momentum-basis}
    T_{\RR} c_{\kk,\QQ,\alpha,\eta,s}^\dagger T_{\RR}^{-1}
= e^{-i\kk\cdot\RR} c_{\kk,\QQ,\alpha,\eta,s}^\dagger\ ,
\end{equation}
where $\RR = n_1 \aa_{M1} + n_2 \aa_{M2}$, with $n_{1,2}\in \mathbb{Z}$, is a Moir\'e lattice vector. 
One should not confuse $T_\RR$ with the time-reversal symmetry ($\hat{T}$) defined in the early section. 
We now verify Eq. (\ref{eq:moire-translation-momentum-basis}) at the commensurate twist angles, 
where $\aa_{M1}$ and $\aa_{M2}$ are integer linear combinations of the microscopic graphene lattice vectors $\aa_1,\aa_2$. 
$T_{\RR}$ can be defined as translation acting on the atomic orbitals: $\RR'\to \RR'+\RR$, leading to
\begin{align}
T_{\RR} c_{\kk,\QQ,\alpha,\eta,s}^\dagger T_{\RR}^{-1} = \frac1{\sqrt{N_{\rm G}}} 
    \sum_{\RR' \in l} e^{i(\eta \KK_D^l + \kk -\QQ)\cdot (\RR'+\tau_\alpha)} c_{\RR'+\RR,\alpha,s}^\dagger 
= \frac1{\sqrt{N_{\rm G}}} 
\sum_{\RR' \in l} e^{-i(\eta\KK_D^{l}+\kk-\QQ)\cdot\RR} e^{i(\eta \KK_D^l + \kk -\QQ)\cdot (\RR'+\tau_\alpha)} c_{\RR',\alpha,s}^\dagger \ ,
\end{align}
where $l=\eta$ ($-\eta$) for $\QQ \in \mathcal{Q}_+$ ($\mathcal{Q}_-$). 
For commensurate angle, $\eta\KK_D^{l}-\QQ$ is a Moir\'e reciprocal lattice and hence $e^{-i(\eta\KK_D^{l}+\kk-\QQ)\cdot\RR} = e^{-i\kk\cdot\RR}$. 
Then Eq. (\ref{eq:moire-translation-momentum-basis}) is justified at commensurate angles.
We emphasize that Eq. (\ref{eq:moire-translation-momentum-basis}) is also well-defined even at non-commensurate angles.


\subsection{Symmetries of the BM model} \label{sec:BM-sym}
The single-particle Hamiltonian in each valley has the symmetry of the magnetic space group $P6'2'2$ (\# 177.151 in the BNS setting), 
which is generated by $\hat{C}_{2z}\hat{T}$, $\hat{C}_{3z}$, $\hat{C}_{2x}$ and translation symmetries. 
Since the two valleys are related by time-reversal symmetry $\hat{T}$, the total system also has the $\hat{C}_{2z}$ symmetry 
(product of $\hat{C}_{2z}T$ and $\hat{T}$). The full crystalline symmetries of the two valleys form the space group $P622$ (\#177), 
which is generated by $\hat{C}_{6z}$, $\hat{C}_{2x}$, and translation symmetries. Thus, these symmetry operators satisfy
\begin{equation}
[\hat{C}_{3z},H_{0}]=[\hat{C}_{2z},H_{0}]=[\hat{C}_{2x},H_{0}]=[\hat{T},H_{0}]=0,
\end{equation}
and we write the symmetry action on the fermion operators as 
\begin{equation}
\hat{g} c_{\kk,\QQ,\eta,\alpha,s}^\dagger \hat{g}^{-1} = \sum_{\QQ' \alpha'\eta'} c_{\hat{g}\kk,\QQ',\eta',\alpha',s}^\dagger 
[D(\hat{g})]_{\QQ'\eta'\alpha',\QQ\eta\alpha}  \ . 
\end{equation}
The $D$ matrices for the above discrete symmetries are given by 
\beq
[D(C_{3z})]_{\QQ^\pr \eta^\pr \beta,\QQ \eta \alpha} = \delta_{\QQ^\pr, C_{3z} \QQ} \delta_{\eta^\pr, \eta} (e^{i\eta  \frac{2\pi}{3}\sigma_{z}})_{\beta\alpha} \;\;\;
[D (C_{2x})]_{\QQ^\pr \eta^\pr \beta, \QQ \eta \alpha} = \delta_{\QQ^\pr, C_{2x}\QQ} \delta_{\eta^\pr, \eta} (\sigma_x)_{\beta\alpha}, \label{eq:C2x}
\eneq
\beq
[D(C_{2z})]_{\QQ^\pr \eta^\pr \beta, \QQ\eta \alpha} = \delta_{\QQ^\pr,- \QQ} \delta_{\eta^\pr,-\eta} (\sigma_x)_{\beta\alpha}, \;\;\;
[D(T)]_{\QQ^\pr \eta^\pr \beta, \QQ \eta \alpha}= \delta_{\QQ^\pr,-\QQ} \delta_{\eta^\pr,-\eta} \delta_{\beta,\alpha, \label{eq:TRS}
}\eneq
\beq
[D(C_{2z} T)]_{\QQ^\pr \eta^\pr \beta, \QQ \eta \alpha}= [D(C_{2z}) D(T)]_{\QQ^\pr \eta^\pr \beta, \QQ \eta \alpha} = \delta_{\QQ^\pr,\QQ} \delta_{\eta^\pr,\eta} (\sigma_x)_{\beta,\alpha}. \label{eq:C2T}
\eneq
Here and below we will use $\sigma_{x,y,z}$ ($\sigma_0$) and $\tau_{x,y,z}$ ($\tau_0$) for the Pauli (identity) matrices in the sub-lattice and valley spaces.
For the sublattice, we choose $\alpha=A,B$ to be $1,2$ for $\sigma$ matrix. 


Besides the crystalline symmetries, the BM model also has a unitary particle-hole (PH) symmetry $\hat{P}$ that anti-commutes with the single-particle Hamiltonian, 
$\{\hat{P},H_{0}\}=0$. $\hat{P}$ transforms $\kk$ to $-\kk$ and the corresponding $D$ matrix reads
\begin{equation}
P c^\dagger_{\kk,\QQ,\eta,\alpha,s} P^{-1} = \sum_{\QQ^\pr \eta^\pr \beta} 
[D(P)]_{\QQ^\pr \eta^\pr \beta, \QQ \eta \alpha} c^\dagger_{-\kk,\QQ^\pr,\eta^\pr,\beta,s},\;\;\;
\quad[D(P)]_{\QQ^\pr \eta^\pr \beta, \QQ \eta \alpha} = \delta_{\QQ^\pr,-\QQ} \delta_{\eta^\pr, \eta} \delta_{\beta,\alpha} \zeta_{\QQ}\ , \label{eq:P}
\eneq
where $\zeta_{\QQ} = \pm 1$ for $\QQ\in\mathcal{Q}_\pm$, respectively. 
Note that $P$ transforms creation operators to creation operators (rather than annihilation operators), and maps sites $\QQ\in \mathcal{Q}_\pm$ into $-\QQ \in\mathcal{Q}_\mp$. The PH transformation $P$ satisfies
\beq\label{seq:P-com}
P^2=-1,\qquad
[P,C_{3z}] = 0,\qquad
\{P,C_{2x}\} = 0,\qquad
\{P,C_{2z}\} = 0,\qquad
\{P,T\} = 0,\qquad
[P,C_{2z}T]=0.
\eneq
The particle-hole symmetry will be broken if the $\theta$-dependence of the single-layer Hamiltonian or quadratic terms in $\kk$ of the single-layer Hamiltonian are taken into account. 


Besides the above discrete symmetry, the BM model also possesses the continuous U(2)$\times$U(2) spin-charge rotation symmetry,
and the corresponding symmetry operators are given by 
\begin{equation}\label{seq:U2U2-gene}
\hat{S}^{ a b}=\sum_{\mathbf{k}} (\tau^a)_{\eta\eta'}(s^b)_{ss'}c_{\mathbf{k},\mathbf{Q},\eta,\alpha,s}^{\dagger} c_{\mathbf{k},\mathbf{Q},\eta',\alpha,s'}\ , \qquad (a=0,z,\quad b=0,x,y,z)\ ,
\end{equation}
where $\tau$ and $s$ are Pauli matrices for valley and spin degree of freedom. 

If, furthermore, $w_0=0$, the model acquires an effective chiral symmetry $\hat{C}$ that anti-commute with the single-particle Hamiltonian \cite{tarnopolsky2019origin}. $\hat{C}$ is referred to as the first chiral symmetry in Refs.~\cite{song2021twisted,bernevig2021twisted}. 
$\hat{C}$ leaves $\kk$ invariant and its $D$ matrix reads 
\begin{equation}
[D(\hat{C})]_{\QQ'\alpha'\eta',\QQ\alpha\eta}  = \delta_{\QQ',\QQ} [\sigma_z]_{\alpha'\alpha} [\tau_0]_{\eta'\eta}\ .
\end{equation}
In the first-quantized formalism the algebra between $C$ and other symmetries are given by
\begin{equation} \label{eq:C-first-algebra}
C^2 =1,\qquad 
[C,T]=0,\qquad    [C,C_{3z}]=0,\qquad \{C,C_{2x}\}=0,\qquad \{C,C_{2z}T\}=0,\qquad 
[C,P]=0 \ .  
\end{equation}

\section{Intra-Layer Inter-Valley Electron-phonon Interaction in Twisted Bilayer Graphene}

\subsection{Projected Intra-Layer Inter-Valley Electron-Phonon Interaction for the Flat Bands}
We define the short-hand notation for the $\QQ$ lattice in the layer $l =\pm $ (top/bottom) and valley $\eta=\pm $ as
\beq \mathcal{Q}_{l \eta }=\{l \eta \qq_2 + n_1 \bb_{M1} + n_2 \bb_{M2}\ |\ n_{1,2}\in \mathbb{Z} \}, \eneq
which is consistent with the definition in the previous Sec. \ref{sec:BM_basis}. 
We have the equivalence between the operators $c_{\kk,\QQ,\alpha,\eta,s}$ in Eq. (\ref{eq:BM-basis1}) in the MBZ 
and the ones $c_{\kk,\alpha,l, s}$ in a single layer graphene,
\beq c_{\kk, \QQ_{l\eta}, \alpha, \eta, s}= c_{\eta \KK_{D}^l + \kk - \QQ_{l\eta}, \alpha, l, s},\;\;\; k\in MBZ
\eneq 
where $\KK_D^l$ is the $K$ momentum in the layer $l$. 

With the electron-phonon Hamiltonian now summed over the two layers $l=\pm$, and with $\KK_{D}^l$ being the graphene Dirac momentum in the layer $l$, 
we have 
\begin{eqnarray}
&  H_{inter-vall}^{op}     \approx \frac{\gamma_3}{ \sqrt{2N_G {\sqrt{M (\alpha_1+ \alpha_2 + \alpha_3)} }}} \sum_{\kk,\kk',\eta,\alpha \beta, l ,s} 
(b_{-\eta \KK_D+\kk-\kk',l, A_1} +b_{\eta \KK_D-\kk+\kk',l,  A_1 }^\dagger)
c^\dagger_{\kk+ \eta \KK_D,l, \alpha, s} (\sigma_x)_{\alpha\beta} c_{\kk' - \eta \KK_D,l, \beta, s} \nonumber\\
 \end{eqnarray} 
 where we have added the layer $l=\pm$ index as well as the spin index in the electrons. 
 In this summation, $\kk$ is summed over the vicinity of the Dirac node, over a radius much larger than the first MBZ 
 (but  much smaller smaller than $\KK_D$). We hence break the sum over the momentum into
 \beq
 \sum_{\kk}\rightarrow \sum_{\kk \in MBZ} \sum_{\QQ_{l\eta}}
\eneq 
where the lattice $\QQ_{l, \eta}$ depends on $\eta, l$ for electrons around the $\eta \KK_D$ point of the $l$ layer,
and find
\begin{eqnarray}
&  H_{inter-vall}^{op}     \approx \frac{\gamma_3}{ \sqrt{2N_G {\sqrt{M (\alpha_1+ \alpha_2 + \alpha_3)} }}} \sum_{\kk,\kk' \in MBZ} 
\sum_{l,\eta} \sum_{\QQ_{l\eta},\QQ'_{-l\eta}} \sum_{\alpha \beta,s} \nonumber \\ &(b_{-\eta \KK_D+\kk-\kk'- \QQ_{l\eta} + \QQ'_{-l\eta},l, A_1} 
+b_{\eta \KK_D-\kk+\kk'+ \QQ_{l\eta} -\QQ'_{-l\eta},l,  A_1 }^\dagger)
c^\dagger_{ \eta \KK_D+\kk-\QQ_{l\eta} ,l, \alpha s} (\sigma_x)_{\alpha\beta} c_{- \eta \KK_D+\kk'- \QQ'_{-l\eta} ,l, \beta,s} \nonumber \\ 
&=\frac{\gamma_3}{ \sqrt{2 N_G {\sqrt{M (\alpha_1+ \alpha_2 + \alpha_3)} }}} \sum_{\kk,\kk' \in MBZ} \sum_{l,\eta} \sum_{\QQ_{l\eta},\QQ'_{-l\eta}} 
\sum_{\alpha \beta,s} \nonumber \\ &(b_{-\eta \KK_D+\kk-\kk'- \QQ_{l\eta} + \QQ'_{-l\eta},l, A_1} +b_{\eta \KK_D-\kk+\kk'+ \QQ_{l\eta} -\QQ'_{-l\eta},l,  
A_1 }^\dagger) c^\dagger_{ \kk, \QQ_{l\eta}, \alpha,\eta, s} (\sigma_x)_{\alpha\beta} c_{\kk', \QQ'_{-l\eta} , \beta,-\eta, s}  
 \end{eqnarray}
 
We next project the above Hamiltonian into the flat bands of the BM model with the expansion
\beq
c_{ \kk, \QQ_{l\eta}, \alpha,\eta, s} =\sum_{n=1,2} u^n_{\kk, \QQ_{l\eta}, \alpha, \eta} \gamma_{\kk n\eta s}
\eneq where $u_{\kk, \QQ_{l\eta}, \alpha, \eta}$ is the eigenstate of the BM model. This gives
\begin{eqnarray}
&  H_{inter-vall}^{op} 
\approx\frac{\gamma_3}{ \sqrt{2N_G {\sqrt{M (\alpha_1+ \alpha_2 + \alpha_3)} }}} \sum_{\kk,\kk' \in MBZ}\sum_{n,n'=1,2}\sum_s \gamma_{\kk,n,\eta, s}^\dagger \gamma_{\kk',n',-\eta, s} \nonumber \\ 
&\sum_{l,\eta} \sum_{\QQ_{l\eta}} \sum_{\QQ'_{-l\eta}} (b_{-\eta \KK_D+\kk-\kk'- \QQ_{l\eta} + \QQ'_{-l\eta},l, A_1} +b_{\eta \KK_D-\kk+\kk'+ \QQ_{l\eta} -\QQ'_{-l\eta},l,  A_1 }^\dagger) \sum_{\alpha \beta} u^{n\star}_{\kk, \QQ_{l\eta}, \alpha, \eta} \sigma^x_{\alpha\beta} u^{n'}_{\kk', \QQ'_{-l\eta}, \beta, -\eta}. 
 \end{eqnarray}
 
By defining
\beq
\QQ_{l\eta} - \QQ'_{-l\eta}= \QQ''_{-l\eta}, 
\eneq 
the Inter-valley Hamiltonian becomes
\begin{eqnarray}
&  H_{inter-vall}^{op} 
\approx\frac{\gamma_3}{ \sqrt{2N_G {\sqrt{M (\alpha_1+ \alpha_2 + \alpha_3)} }}} \sum_{\kk,\kk' \in MBZ}\sum_{n,n'=1,2}\sum_{\eta, s }
\gamma_{\kk,n,\eta, s}^\dagger \gamma_{\kk',n',-\eta, s} \nonumber \\ 
&\sum_{l}  \sum_{\QQ''_{-l\eta}} (b_{-\eta \KK_D+\kk-\kk'- \QQ''_{-l\eta} ,l, A_1} +b_{\eta \KK_D-\kk+\kk'+ \QQ''_{-l\eta} ,l,  A_1 }^\dagger)
 \sum_{\QQ_{l\eta}}\sum_{\alpha \beta} u^{n\star}_{\kk, \QQ_{l\eta}, \alpha, \eta} \sigma^x_{\alpha\beta} u^{n'}_{\kk', \QQ_{l\eta}-\QQ''_{-l\eta}, \beta, -\eta}. 
\end{eqnarray}

\subsection{Phonon-Mediated Electron-Electron Interaction for the Flat Bands}
We rewrite 
\begin{eqnarray}
&  H_{inter-vall}^{op} \approx\frac{1}{ \sqrt{N_G}} \sum_{\kk,\kk' \in MBZ}\sum_{n,n'=1,2}\sum_{\eta, s }\sum_{l=\pm} \sum_{\QQ_{-l\eta}} \times \nonumber \\ 
& \times G_{\kk,\kk', \QQ_{-l\eta}}^{\eta n n' l} \gamma_{\kk,n,\eta, s}^\dagger \gamma_{\kk',n',-\eta, s} 
 (b_{-\eta \KK_D+\kk-\kk'- \QQ_{-l\eta} ,l, A_1} +b_{\eta \KK_D-\kk+\kk'+ \QQ_{-l\eta} ,l,  A_1 }^\dagger)\nonumber \\ 
 &  G_{\kk,\kk', \QQ_{-l\eta}}^{\eta n n' l}= \frac{\gamma_3}{ \sqrt{2M\omega_{A_1} }} 
 \sum_{\QQ'_{l\eta}}\sum_{\alpha \beta} u^{n\star}_{\kk, \QQ'_{l\eta}, \alpha, \eta} \sigma^x_{\alpha\beta} u^{n'}_{\kk', \QQ'_{l\eta}-\QQ_{-l\eta}, \beta, -\eta} \nonumber \\
 & \omega_{A1}=  \sqrt{ (\alpha_1+ \alpha_2 + \alpha_3)/M}
 \end{eqnarray} 
 In the above expression, $\QQ_{\pm l\eta} = \pm l \eta \qq_2+ \GG_M $ and the summation is made over the $\GG_M$. 
 
 With the electron Hamiltonian
 \beq
 H_{el} = \sum_{m, \kk\in MBZ, \eta, s} E_{m\kk} \gamma_{\kk m\eta s}^\dagger \gamma_{\kk m\eta s}
 \eneq 
 and phonon Hamiltonian for the $A_1$ mode
 \beq
 H_{ph} = \sum_{\qq,l} \omega_{A1} b_{\qq l, A_1 }^\dagger b_{\qq l, A_1 }
\eneq
 we apply the transformation $S$
 \beq
 - H_{inter-vall}^{op}  = [H_{el} + H_{ph}, S]
 \eneq 
 and obtain
 \begin{eqnarray}
 &S=\frac{1}{ \sqrt{N_G}} \sum_{\kk,\kk' \in MBZ}\sum_{n,n'=1,2}\sum_{\eta, s }\sum_{l=\pm} \sum_{\QQ_{-l\eta}} \times \nonumber \\ 
 &  \gamma_{\kk,n,\eta, s}^\dagger \gamma_{\kk',n',-\eta, s} (A_{\kk,\kk', \QQ_{-l\eta}}^{\eta n n' l}  b_{-\eta \KK_D+\kk-\kk'- \QQ_{-l\eta} ,l, A_1} 
 - A_{\kk',\kk, -\QQ_{-l\eta}}^{-\eta n' n l\star}  b_{\eta \KK_D-\kk+\kk'+ \QQ_{-l\eta} ,l,  A_1 }^\dagger) \nonumber \\ 
 & A_{\kk,\kk', \QQ_{-l\eta}}^{\eta n n' l} = -\frac{\GG_{\kk,\kk', \QQ_{-l\eta}}^{\eta n n' l} }{E_{n\kk}- E_{n'\kk'}- \omega_{A_1} }
 \end{eqnarray}
 One should note that it is important to keep $-\QQ_{-l\eta}$ in $A_{k',k, -\QQ_{-l\eta}}^{-\eta n' n l\star}$ and not rewrite it as $\QQ_{l\eta}$ since 
 $-\QQ_{-l\eta} = l\eta \qq_2 - \GG_M$ whereas $\QQ_{l\eta} = l\eta \qq_2+ \GG_M$. They hence differ by a Moir\'e reciprocal lattice vector, 
 but since this vector is what we sum over, it is important to keep the correct form. 
 
We then obtain the phonon-mediated electron-electron interaction (we now expand explicitly the $\QQ_{l\eta}$ and perform the summation over $\GG_M$)
\begin{eqnarray}
 &H_{el-el}=\frac{1}{N_G}\sum_{\kk\kk',\kk_1\kk_1', \GG_M,l, s, s_1} \frac{G_{\kk,\kk', -l\eta \qq_2+ \GG_M}^{\eta n n' l}  
 G_{\kk_1,\kk_1', l\eta \qq_2 - \GG_M}^{-\eta n_1 n_1' l}  \omega_{A_1}}{(E_{n_1\kk_1}- E_{n_1' \kk_1'})^2  - \omega_{A_1}^2}  
 \delta_{\kk-\kk', -\kk_1+ \kk_1'} \gamma_{\kk, n,\eta, s}^\dagger \gamma_{\kk_1, n_1, -\eta, s_1}^\dagger \gamma_{\kk_1', n_1',\eta, s_1} 
 \gamma_{\kk',n',-\eta, s}\nonumber\\
\end{eqnarray}
Since the optical phonon frequency $\omega_{A1}$ is much larger than the bandwidth of the flat bands in TBG, we neglect the latter. 
In the Cooper channel $\kk_1=- \kk, \kk_1'=-\kk'$, we have
\begin{eqnarray}
&H_{el-el}=-\frac{1}{N_G}\frac{1}{\omega_{A_1}}\sum_{\kk,\kk', \GG_M,l, s, s_1} G_{\kk,\kk', -l\eta \qq_2+ \GG_M}^{\eta n n' l}  
G_{-\kk,-\kk', l\eta \qq_2 - \GG_M}^{-\eta n_1 n_1' l} \gamma_{\kk, n,\eta, s}^\dagger \gamma_{-\kk, n_1,- \eta, s_1}^\dagger \gamma_{-\kk', n_1',\eta, s_1} \gamma_{\kk',n',-\eta, s}. 
\end{eqnarray}
If one wants to include more bands in the system, one can extend the indices $n,n'$, although due to the larger dispersion of the remote bands, one might need to keep the kinetic energies. 

It should be noted as the summation over $\kk,\kk'$ is within the MBZ, so we should split the factor $N_G$ (which represents the total number of atomic unit cells) into two parts, $N_G=N_M\times N_0$, with $N_M$ the total number of Moir\'e unit cell and $N_0$ the number of atomic unit cells in one Moir\'e unit cell. 
Consequently, we can rewrite the above Hamiltonian as
\begin{eqnarray}\label{eq:Helel1}
&H_{el-el}=-\frac{1}{N_M}\sum_{\kk,\kk', s, s_1} V^{\eta nn',n_1n_1'}_{\kk, \kk'; -\kk, -\kk' } \gamma_{\kk,n,\eta, s}^\dagger \gamma_{-\kk, n_1, - \eta, s_1}^\dagger \gamma_{-\kk', n_1',\eta, s_1} \gamma_{\kk',n',-\eta, s}
\end{eqnarray}
where 
\beq
V^{\eta nn',n_1n_1'}_{\kk, \kk'; \kk_1, \kk_1' }=\frac{1}{N_0} \frac{1}{\omega_{A_1}} \sum_{\GG_M}\sum_l G_{\kk,\kk', -l\eta \qq_2+ \GG_M}^{\eta n n',l} 
G_{\kk_1,\kk_1', l\eta \qq_2 - \GG_M}^{-\eta  n_1 n_1', l}
\eneq
represents the effective attractive electron-electron interaction mediated by phonons in TBG.

\subsection{Gauge Fixing for Eigen-state Basis and Chern Band Basis}

\subsubsection{Eigen-state Basis}

The eigen-energy basis of the bands in the TBG is given by
\begin{equation}
\gamma_{\mathbf{k},n,\eta, s}^{\dagger}=\sum_{\mathbf{Q}\alpha}u_{\mathbf{Q}\alpha;n\eta}\left(\mathbf{k}\right)c_{\mathbf{k},\mathbf{Q},\eta,\alpha s}^{\dagger}\ ,\label{eq:solution}
\end{equation}
where $u_{\mathbf{Q}\alpha;n\eta}(\kk)$ is the eigenstate  of energy band $n$ of the first quantized single-particle Hamiltonian $h_{\mathbf{Q},\mathbf{Q}^{\prime}}^{\left(\eta\right)}\left(\mathbf{k}\right)$ in valley $\eta$ 
\begin{equation}
\sum_{\QQ',\beta}[h_{\mathbf{Q},\mathbf{Q}^{\prime}}^{\left(\eta\right)}\left(\mathbf{k}\right)]_{\alpha\beta}u_{\mathbf{Q}'\beta;n\eta}(\kk)=\epsilon_{n, \eta}(\kk) u_{\mathbf{Q}\alpha;n\eta}(\kk)\ ,
\end{equation}
where $\epsilon_{n, \eta}(\kk)$ is the single-particle energy of the eigen-state $u_{\mathbf{Q}\alpha;n\eta}(\kk)$.  In each valley and spin, we  use integers $n>0$ to label the $n$-th conduction band, and use integer $n<0$ to label the $|n|$-th valence band (thus $n\neq0$). The lowest conduction and valence bands in each valley-spin flavor is thus labeled by $n=\pm1$.

The BM Hamiltonian becomes
\beq
\hat{H}_{0} = \sum_{\kk} \sum_{n \eta s} \epsilon_{n, \eta}(\kk) \gamma_{\kk n \eta s}^\dagger \gamma_{\kk n\eta s}. 
\eneq
Bloch periodicity and periodic gauge require
 \beq
 c_{\mathbf{k}+\mathbf{b}_{Mi},\mathbf{Q},\eta\alpha s}^{\dagger}=c_{\mathbf{k},\mathbf{Q}-\mathbf{b}_{Mi},\eta\alpha s}^{\dagger} \implies u_{\mathbf{Q}\alpha; n\eta}\left(\mathbf{k}+\mathbf{b}_{Mi}\right)=u_{\mathbf{Q}-\mathbf{b}_{Mi},\alpha;n\eta}\left(\mathbf{k}\right)\  \implies \gamma_{\mathbf{k}+\mathbf{b}_{Mi},n\eta s}^{\dagger}=\gamma_{\mathbf{k}n\eta s}^{\dagger} \label{eq:U-embedding}
\end{equation} 
and the $\hat{C}_{2z}$ and $\hat{P}$ symmetries require
\beq\label{seq:energy-relation}
\hat{C}_{2z}: \;\; \epsilon_{n,\eta}(\kk) = \epsilon_{n,-\eta}(-\kk)\ ,\qquad
\hat{P}:\;\; \epsilon_{n,\eta}(\kk) =-\epsilon_{-n,\eta}(-\kk)\ .
\eneq

\subsubsection{Gauge Fixing for Eigen-state Basis}\label{app:gaugefixing}
We fix the gauge for the eigen-state basis of energy bands $\gamma_{\mathbf{k}n\eta s}^{\dagger}$ in Eq.~(\ref{eq:solution}) for the symmetry operators $\hat{C}_{2z},\hat{T}$ and $\hat{P}$. 
We denote the wave function $u_{\mathbf{Q}\alpha;n\eta}(\kk)$ as a column vector $u_{n\eta}(\kk)$ in the space of indices $\{\QQ,\alpha\}$. 
A representation matrix $D(\hat{g})$ of an operation $\hat{g}$ acts on a wave function $u_{n\eta'}(\kk)$, we denote the resulting wave function in the valley $\eta$ for short as $[D(g)]_{\eta\eta'} u_{n\eta'}(\kk)$, the components of which are given by $\sum_{\mathbf{Q}'\beta\eta'}[D(g)]_{\mathbf{Q}\alpha\eta,\mathbf{Q}'\beta\eta'}u_{\mathbf{Q}'\beta;n\eta'}(\kk)$. 
For the symmetries $\hat{C}_{2z},\hat{T}$ and $\hat{P}$, the sewing matrices $B^{\hat{g}}(\kk)$ in the band and valley space are defined by
\begin{equation}
[D(\hat{C}_{2z})]_{\eta\eta'} u_{n\eta'}(\kk)= \sum_{m}[B^{C_{2z}}(\kk)]_{m\eta,n\eta'}u_{m\eta}(-\kk)\ ,\;\;\;
[D(\hat{T})]_{\eta\eta'} u_{n\eta'}^*(\kk)= \sum_{m}[B^{T}(\kk)]_{m\eta,n\eta'}u_{m\eta}(-\kk)\ ,
\end{equation}
\begin{equation}
[D(\hat{P})]_{\eta\eta'} u_{n\eta'}(\kk)=\sum_{m}[B^{P}(\kk)]_{m\eta,n\eta'}u_{m\eta}(-\kk)\ .
\end{equation}
For non-degenerate wave function $u_{n\eta'}(\kk)$ in the valley $\eta'$, since $\hat{C}_{2z}$ and $\hat{T}$ commute with the $\hat{H}_{0}$ and flips the valley $\eta$, while $\hat{P}$ anti-commutes with $\hat{H}_{0}$ and preserves the valley $\eta$, we generically have
\begin{equation}\label{seq:sewing}
\begin{split}
&[B^{C_{2z}}(\kk)]_{m\eta,n\eta'}=\delta_{\eta,-\eta'}\delta_{m,n} e^{i\varphi^{C_{2z}}_{n,\eta'}(\kk)}\ , \quad [B^{T}(\kk)]_{m\eta,n\eta'}=\delta_{\eta,-\eta'}\delta_{m,n} e^{i\varphi^{T}_{n,\eta'}(\kk)}\ ,\\
& [B^{P}(\kk)]_{m\eta,n\eta'}=\delta_{\eta,\eta'}\delta_{-m,n} e^{i\varphi^{P}_{n,\eta'}(\kk)}\ .
\end{split}
\end{equation}
Accordingly the energy band fermion operators transform as
\begin{equation}
\hat{g} \gamma_{\mathbf{k},n,\eta' ,s}^{\dagger} \hat{g}^{-1}= \sum_{m\eta} [B^{g}(\kk)]_{m\eta,n\eta'} \gamma_{\hat{g}\mathbf{k},m,\eta, s}^{\dagger}\ .
\end{equation}
Since these three symmetries satisfy the relations
\begin{equation}
C_{2z}^2=1\ ,\ T^2=1\ ,\ P^2=-1\ ,\ \{P,C_{2z}\}=0\ ,\ \{P,T\}=0\ ,\ [C_{2z},T]=0\ ,
\end{equation}
With the above notations, the symmetries $\hat{C}_{2z},\hat{T}$ and $\hat{P}$ allows us to define
\begin{equation}
\begin{split}
&B^{C_{2z}}(-\kk)B^{C_{2z}}(\kk)=B^{T}(-\kk)B^{T*}(\kk)=-B^{P}(-\kk)B^{P}(\kk)=I\ ,\qquad B^{P}(-\kk)B^{C_{2z}}(\kk)=-B^{C_{2z}}(-\kk)B^{P}(\kk)\ , \\
&\qquad\qquad B^{P}(-\kk)B^{T}(\kk)=-B^{T}(-\kk)B^{P*}(\kk)\ ,\qquad 
B^{T}(-\kk)B^{C_{2z}*}(\kk)=B^{C_{2z}}(-\kk)B^{T}(\kk)\ ,
\end{split}
\end{equation}
where $B^{g*}(\kk)$ stands for the complex conjugation of matrix $B^g(\kk)$, and $I$ is the identity matrix in the $n,\eta$ space. 
We have two independent symmetry operations $\hat{C}_{2z}\hat{T}$ (anti-unitary) and $\hat{C}_{2z}\hat{P}$ (unitary), which do not change $\kk$. 
Their sewing matrices are defined by
\begin{equation}\label{seq:B-C2T}
[D(\hat{C}_{2z})D(\hat{T})]_{\eta\eta'} u_{n\eta'}^*(\kk)=\sum_{m}[B^{C_{2z}T}(\kk)]_{m\eta,n\eta'}u_{m\eta}(\kk)\ ,
\end{equation}
\begin{equation}\label{seq:B-C2P}
[D(\hat{P})D(\hat{C}_{2z})]_{\eta\eta'} u_{n\eta'}(\kk)= \sum_{m}[B^{C_{2z}P}(\kk)]_{m\eta,n\eta'}u_{m\eta}(\kk)\ .
\end{equation}
For non-degenerate eigenstates at momentum $\kk$ (non-degenerate within one valley), we have
\begin{equation}\label{seq:sewing-C2T-C2P}
[B^{C_{2z}T}(\kk)]_{m\eta,n\eta'}=\delta_{\eta,\eta'}\delta_{m,n} e^{i\varphi^{C_{2z}T}_{n,\eta'}(\kk)}\ , \qquad
[B^{C_{2z}P}(\kk)]_{m\eta,n\eta'}=\delta_{-\eta,\eta'}\delta_{-m,n} e^{i\varphi^{C_{2z}P}_{n,\eta'}(\kk)}\ ,
\end{equation}
where by definition we have $\varphi^{C_{2z}T}_{n,\eta'}(\kk)=\varphi^{T}_{n,\eta'}(\kk)+\varphi^{C_{2z}}_{n,-\eta'}(-\kk)$, and $\varphi^{C_{2z}P}_{n,\eta'}(\kk)=\varphi^{C_{2z}}_{n,\eta'}(\kk)+\varphi^{P}_{n,-\eta'}(-\kk)$. The sewing matrices of $C_{2z}T$ and $C_{2z}P$ are subject to the constraint that
\begin{equation}
(C_{2z}T)^2=(C_{2z}P)^2=1\ ,\qquad [C_{2z}T,C_{2z}P]=0\ ,
\end{equation}
and thus they satisfy
\begin{equation}
B^{C_{2z}T}(\kk)B^{C_{2z}T*}(\kk)=[B^{C_{2z}P}(\kk)]^2=I\ , \qquad B^{C_{2z}P}(\kk)B^{C_{2z}T}(\kk)=B^{C_{2z}T}(\kk)B^{C_{2z}P*}(\kk)\ .\label{seq:sewing-C2T-C2P-relation}
\end{equation}


We will now fixe the gauge of the wave functions and sewing matrices of the $\kk$ preserving symmetry operations $C_{2z}T$ and $C_{2z}P$. By Eqs. (\ref{seq:sewing-C2T-C2P}) and (\ref{seq:sewing-C2T-C2P-relation}), we make the $\kk$-independent choices for the sewing matrices 
\begin{equation}\label{eq:gauge-n}
[B^{C_{2z}T}(\kk)]_{m\eta,n\eta'}=\delta_{\eta,\eta'}\delta_{m,n} \ , \qquad
[B^{C_{2z}P}(\kk)]_{m\eta,n\eta'}=-\text{sgn}(n)\eta' \delta_{-\eta,\eta'}\delta_{-m,n}\ .
\end{equation}
Accordingly, the symmetry actions on the band basis fermion operators are given by
\begin{equation}
(C_{2z}T) \gamma_{\mathbf{k},n,\eta ,s}^{\dagger} (C_{2z}T)^{-1}=\gamma_{\mathbf{k},n,\eta ,s}^{\dagger}\ ,\qquad (C_{2z}P) \gamma_{\mathbf{k},n,\eta ,s}^{\dagger} (C_{2z}P)^{-1}=-\text{sgn}(n)\eta \gamma_{\mathbf{k},-n,-\eta ,s}^{\dagger}\ .
\end{equation}
This does not yet fix the entire phases of the energy basis at momentum $\kk$, since the sewing matrices in Eq.~(\ref{eq:gauge-n}) are invariant under the unitary transformation of wave functions $u_{n\eta}(\kk)\rightarrow \text{sgn}(n)\eta u_{n\eta}(\kk)$ at each individual $\kk$. To further fix this gauge freedom for different $\kk\in\text{MBZ}$, we start by choosing a momentum $\kk=\kk_0$ where the eigenstates within one valley are non-degenerate, and choose a gauge fixing of the eigen-state basis at $\kk_0$ satisfying Eq.~(\ref{eq:gauge-n}). We then fix the band basis of bands $\pm n$ at other $\kk\neq\kk_0$ by requiring
\begin{equation}
f_{n,\eta}(\kk+\qq,\kk)= \left|u^\dag_{n,\eta}(\kk+\qq) u_{n,\eta}(\kk)- u^\dag_{-n,\eta}(\kk+\qq) u_{-n,\eta}(\kk)\right|
\end{equation} 
to be a continuous function of $\kk$ and $\qq$, and satisfy
\begin{equation}\label{seq:c-continuous}
\lim_{\qq\rightarrow \mathbf{0}}f_{n,\eta}(\kk+\qq,\kk)=0
\end{equation} 
for all $\kk$. This fixes the relative sign between wave functions $u_{n,\eta}(\kk)$ and $u_{-n,\eta}(\kk)$ in a way that is continuous in $\kk$. We do not require the wave function $u_{n,\eta}(\kk)$ itself to be globally continuous in $\kk$ of the entire MBZ; locally $u_{n,\eta}(\kk)$ can always be chosen to be continuous in $\kk$, provided $u_{n,\eta}(\kk)$ is non-degenerate at momentum $\kk$. 

 Within the space of each pair of PH symmetric bands with band indices $n=\pm n_\text{B}$, if we use $\zeta^a$ and $\tau^a$ ($a=0,x,y,z$) to denote the identity and Pauli matrices in the energy band $n=\pm n_\text{B}$ space and the valley space:
\begin{equation}\label{eq:gauge-0}
B^{C_{2z}T}(\kk)=\zeta^0\tau^0\ ,\qquad B^{C_{2z}P}(\kk)=-\zeta^y\tau^y\ .
\end{equation}
For $n_\text{B}=1$ (i.e., within the lowest conduction and valence bands $n=\pm1$ per spin per valley) when $\kk$ is at $\KK_M$ or $\KK_M'$ point of the MBZ,
the bands $n=+1$ and $n=-1$ are degenerate. In this case, we can still choose the eigenstate basis at $\KK_M$ or $\KK_M'$ point such that Eqs. (\ref{eq:gauge-0}) and (\ref{seq:c-continuous}) are satisfied.

We  further fix the relative gauge between wave functions at momenta $\kk$ and $-\kk$ by fixing the sewing matrices of $\hat{C}_{2z}$ and $\hat{P}$. 
For $\kk$ not at the $P$-invariant momenta, which are $\Gamma_M$ and the three equivalent $M_M$ in TBG, we have
\begin{equation}\label{eq:gauge-1}
B^{C_{2z}}(\kk)=\zeta^0\tau^x\ ,\qquad B^{T}(\kk)=\zeta^0\tau^x\ , \qquad B^{P}(\kk)=-i\zeta^y\tau^z\ .
\end{equation}
which are consistent with Eq.~(\ref{eq:gauge-0}). 
With the continous gauge condition, the sewing matrix $B^P(\kk)$ must have additional minus signs, \ie $B^P(\kk)=i\zeta^y\tau^z$, at an odd (even) number of the four $P$-invariant momenta if the the two bands $n=\pm n_B$ have an odd (even) topological winding number protected by $C_{2z}T$; and at the other odd (even) $P$-invariant momenta $B^P(\kk)$ are $-i\zeta^y\tau^z$, same as those at generic momenta. Accordingly, the sewing matrices $B^{C_{2z}}(\kk)$ and $B^{T}(\kk)$ also have the additional minus at momenta where $B^P(\kk)$ has the minus sign.
In this work, we choose $B^P(\kk_{\Gamma_M})=-i\zeta^y \tau^z$ and $B^P(\kk_{M_M})=i\zeta^y \tau^z$.

\subsubsection{Gauge Fixing for Chern band basis}\label{app:chern}
After the gauge fixing in Eqs.~(\ref{eq:gauge-0}) and~(\ref{seq:c-continuous}), we define a new irrep basis $\gamma^{(n_\text{B})\dag}_{\mathbf{k},e_Y,\eta, s}$ 
\begin{equation}\label{eq-irrepbasis}
\gamma^{(n_\text{B})\dag}_{\mathbf{k},e_Y,\eta, s}=\frac{\gamma^\dag_{\mathbf{k},n_\text{B},\eta, s}+ie_Y \gamma^\dag_{\mathbf{k},-n_\text{B},\eta, s}}{\sqrt{2}}\ , \quad (e_Y=\pm1).
\end{equation}
For $n_\text{B}=1$, we call them the Chern band basis within the lowest two bands  $\gamma^{(1)\dag}_{\mathbf{k},e_Y,\eta, s}=\gamma^\dagger_{\kk,e_Y,\eta,s}$, as given in Eq. (\ref{eq-irrepbasis}), where $e_Y=\pm1$. 

The basis $\gamma^{(n_\text{B})\dag}_{\mathbf{k},e_Y,\eta, s}$ defines a
band with well-defined Berry curvature; for fixed $e_Y,\eta,s$, Eq.(\ref{eq-irrepbasis}) gives a band with Chern number
\begin{equation}\label{seq:C-chern-band}
C^{n_\text{B}}_{e_Y,\eta,s} =e_Ye_{2,n_\text{B}}\ ,
\end{equation}
where $e_{2,n_\text{B}}\in\mathbb{Z}$ is the Wilson loop winding number of the two bands $n=\pm n_\text{B}$, provided the pair of bands $n=\pm n_\text{B}$ are disconnected with other bands.  The wave functions of the Chern basis in Eq.~(\ref{eq-irrepbasis}) are given by
\begin{equation}
u_{e_Y,n_\text{B},\eta}'(\kk)=\frac{u_{+n_\text{B},\eta}(\kk)+ie_Yu_{-n_\text{B},\eta}(\kk)}{\sqrt{2}}\ . \label{szdeq:Chern-band-u}
\end{equation}
Due to the continuous gauge condition, we see that $\lim_{\qq\rightarrow \mathbf{0}} u^\dag_{+n_\text{B},\eta}(\kk+\qq) u_{+n_\text{B},\eta}(\kk)=\lim_{\qq\rightarrow \mathbf{0}} u^\dag_{-n_\text{B},\eta}(\kk+\qq) u_{-n_\text{B},\eta}(\kk)$. Therefore, the Chern band wave functions satisfy the continuous condition
\begin{equation}
\lim_{\qq\to 0} |u_{e_Y,n_\text{B},\eta}'^\dag(\kk+\qq)u_{e_Y',n_\text{B},\eta}'(\kk)| = \frac{1}{2}\lim_{\qq\to 0} |u^\dag_{+n_\text{B},\eta}(\kk+\qq) u_{+n_\text{B},\eta}(\kk)+e_Ye_Y'u^\dag_{-n_\text{B},\eta}(\kk+\qq) u_{-n_\text{B},\eta}(\kk)| = \delta_{e_Y,e_Y'}, \label{eq:Chern-band-gauge}
\end{equation} This allows us to define a continuous Berry curvature for the Chern band wave function $u_{e_Y,n_\text{B},\eta}'(\kk)$.

We first focus in the valley $\eta=+$. The sewing matrix for $C_{2z}T$ restricted in the valley $\eta=+$ is given by $B^{C_{2z}T}(\kk)=\zeta^0$ (see Eq.~(\ref{eq:gauge-0})). Under this gauge, according to \cite{ahn2019failure}, the non-abelian Berry's connection 
$[\mathbf{A}(\kk)]_{mn}=i u^\dag_{m,+}(\kk)\partial_\kk u_{n,+}(\kk)$ will take the form 
\begin{equation}\label{seq:berry0}
\mathbf{A}(\kk) = \begin{pmatrix}
0 & i\aa(\kk) \\
-i\aa(\kk) & 0
\end{pmatrix}. 
\end{equation}
The sign of wave functions $u_{n,+}(\kk)$ is fixed in such a way that $\aa(\kk)$ is globally continuous in the MBZ excluding the Dirac nodes between the two bands $\pm n_\text{B}$ (recall that we assume the bands $\pm n_\text{B}$ are disconnected from other bands, and thus there can be Dirac nodes between them only if $n_B=1$), which is always possible \cite{ahn2019failure}. In particular, this way of sign fixing is consistent with Eq. (\ref{seq:c-continuous}), since the vanishing of the diagonal Berry's connection requires $\lim_{\qq\to 0} |u^\dag_{m,\eta}(\kk+\qq) u_{n,\eta}(\kk)| =\delta_{m,n}$. 

It is known that the Wilson loop winding number of two bands isolated from other bands is given by the Euler class \cite{ahn2019failure}:
\begin{equation}\label{seq:e2}
e_{2,n_\text{B}} = \frac{1}{2\pi} \sum_i \oint_{\partial D_i} d\kk \cdot \aa(\kk)
= \frac1{2\pi} \int_{\mathrm{MBZ}-\sum_i D_i } d^2\kk\ \Omega(\kk)\ ,
\end{equation}
where $D_i$ is a sufficiently small region containing the $i$th Dirac point in the BZ, and $\Omega(\kk) = \nabla_\kk \times \aa(\kk)$. 

With Eq.~(\ref{seq:berry0}), we can derive the Berry connection of the irrep band basis $\gamma^\dag_{\kk,e_Y,+,s}$ at $\kk$ away from Dirac points as
\begin{align}
&\mathbf{A}^{\pr}_{e_Y}(\kk) =i u'^\dag_{e_Y,n_\text{B},+}(\kk)\partial_\kk u_{e_Y,n_\text{B},+}'(\kk) \nonumber \\ 
&=\frac{i}{2} [u^\dag_{+n_\text{B},+}(\kk)\partial_\kk u_{+n_\text{B},+}(\kk) +i e_Y u^\dag_{+n_\text{B},+}(\kk)\partial_\kk u_{-n_\text{B},+}(\kk)-ie_Y u^\dag_{-n_\text{B},+}(\kk)\partial_\kk u_{+n_\text{B},+}(\kk)+u^\dagger_{-n_\text{B},+}(\kk)\partial_\kk u_{-n_\text{B},+}(\kk)] \nonumber \\
&=e_Y \aa(\kk)\ .
\end{align}
Furthermore, the Berry curvature can be shown to be non-divergent at the Dirac points between the two bands $n=\pm n_\text{B}$ (see proof in \cite{song2021twisted}). 
If $n_\text{B}>1$, there are no Dirac points between bands $n=\pm n_\text{B}$). Therefore, by Eq.~(\ref{seq:e2}), we find the irrep band basis $\gamma^{(n_\text{B})\dagger}_{\kk,e_Y,+,s}$ carries a Chern number given by Eq. (\ref{seq:C-chern-band}). We also note that the $\hat{C}_{2z}$ symmetry maps the irrep band basis $\gamma^{(n_\text{B})\dagger}_{\kk,e_Y,+,s}$ into $\gamma^{(n_\text{B})\dagger}_{-\kk,e_Y,-,s}$ (see Eq.~(\ref{eq:gauge-1})). Since $\hat{C}_{2z}$ does not change the Chern number, we conclude that the Chern number of the Chern band basis $\gamma^{(n_\text{B})\dagger}_{\kk,e_Y,\eta,s}$ in the MBZ is simply given by Eq. (\ref{seq:C-chern-band}).

In particular, for the lowest two bands $n_\text{B}=1$, the bands are topological and carry a winding number $e_2=1$ \cite{song2019all,po2019faithful,ahn2019failure}. Therefore, for the Chern band basis (the irrep basis with $n_\text{B}=1$) $\gamma^{\dagger}_{-\kk,e_Y,-,s}$, we have Chern number
\begin{equation}
C_{e_Y,\eta,s} =e_Y\ ,
\end{equation}
thus the name ``Chern band basis" within the lowest two bands (see \cite{song2021twisted} for a more careful treatment at the Dirac points at CNP, which does not change the conclusion).

Now we show that if $e_{2,n_B}$ is odd, then the sign  of the sewing matrix $B^P(\kk)$ \emph{must} be $\kk$-dependent: for $\eta=+$, $B^P(\kk)$ can be chosen as $-i\zeta^y$ at all the momenta except one or three of the $P$-invariant momenta, where $B^P(\kk)$ must be $i\zeta^y$.
To see this, we assume $B^P(\kk)=-i\chi(\kk) \zeta^y$, where $\chi(\kk)=\pm1$, and transform it into the Chern band basis (\ref{szdeq:Chern-band-u}).
We obtain 
\begin{equation}
B^{P\prime}_{e_Y,e_Y'}(\kk) = u^{\prime\dagger}_{e_Y,n_B,+}(-\kk) D(P) u^\prime_{e_Y,n_B,+}(\kk) = -i \chi(\kk) e_Y \delta_{e_Y,e_Y'}.
\end{equation}
Therefore, $\hat{P}$ leaves each branch of the Chern band basis, which has the Chern numbers $e_{2,n_B}e_Y$, invariant.
$i\hat{P}$ can be equivalently thought as an inversion symmetry for each Chern band since it squares to 1 and changes $\kk$ to $-\kk$. 
The ``inversion'' eigenvalues of the Chern band $e_Y$ are given by $\chi(\kk) e_Y$ for $\kk$ being the $P$-invariant momentum. 
Due to the relation between Chern number and inversion eigenvalues, we have
\begin{equation}
    (-1)^{C^{n_\text{B}}_{e_Y,\eta,s}} = \prod_{K} \chi(K), 
\end{equation}
where $K$ indexes the four $P$-invariant momenta. 
Therefore, the right hand side must be -1 (1) if $C^{n_\text{B}}_{e_Y,\eta,s}$ is odd (even), implying $\chi(K)=-1$ at one or three (zero, two, or four) of the four $P$-invariant momenta. The sign of $B^{P}(\kk)$ in the other valley $\eta=-$ can be obtained from the constraint of $B^{C_{2z}P}(\kk)$ and $B^{P}(\kk)$.

\subsubsection{The chiral symmetry at $w_0=0$}

At $w_0=0$, the single-particle Hamiltonian of TBG acquires an additional unitary chiral symmetry $\hat{C}$, 
which satisfies the anti-commutation relation with the full single-particle Hamiltonian $\hat{H}_{0}$ in Eq.~(\ref{eq:H0})
\begin{equation}
\{\hat{C},\hat{H}_0\}=0; \;\;\;\;
\hat{C} c^\dagger_{\kk,\QQ,\eta,\alpha,s} \hat{C}^{-1} = \sum_{\QQ^\pr \eta^\pr \beta} 
[D(\hat{C})]_{\QQ^\pr \eta^\pr \beta, \QQ \eta \alpha} c^\dagger_{\kk,\QQ^\pr,\eta^\pr,\beta,s}\ ,
\eneq
with the representation matrix
\beq
\quad[D(\hat{C})]_{\QQ^\pr \eta^\pr \beta, \QQ \eta \alpha} = \delta_{\QQ^\pr,\QQ} \delta_{\eta^\pr, \eta} (\sigma_z)_{\beta,\alpha}\ . \label{eq:C}
\eneq
Note that $\hat{C}$ preserves the electron momentum $\kk$. Since $\hat{C}$ flips the single-particle Hamiltonian $\hat{H}_0$, 
it is not a commuting symmetry of TBG, but only reflects a relation between the positive and negative energy spectra. The transformation $\hat{C}$ satisfies
\beq
\hat{C}^2=1,\qquad
\{\hat{C},\hat{C}_{2z}\} = 0,\qquad
[\hat{C},\hat{T}] = 0,\qquad
[\hat{C},\hat{P}]=0, \qquad
\{\hat{C},\hat{C}_{2z}T\}=0,\qquad
\{\hat{C},\hat{C}_{2z}\hat{P}\}=0.
\eneq

When transformed into the energy band basis, the chiral symmetry $\hat{C}$ implies
\begin{equation}\label{seq:C-H0energy}
\ee_{n,\eta}(\kk)=-\ee_{-n,\eta}(\kk), \;\;\;
[D(\hat{C})]_{\eta\eta'} u_{n\eta'}(\kk)= \sum_{m}[B^{C}(\kk)]_{m\eta,n\eta'}u_{m\eta}(\kk)\ ,
\end{equation}
where
\begin{equation}\label{seq:sewing-C0}
[B^{C}(\kk)]_{m\eta,n\eta'}=\delta_{\eta,\eta'}\delta_{-m,n} e^{i\varphi^{C}_{n,\eta'}(\kk)};\;\;\;\hat{C} \gamma_{\mathbf{k},n,\eta' ,s}^{\dagger} \hat{C}^{-1}= \sum_{m\eta} [B^C(\kk)]_{m\eta,n\eta'} \gamma_{\mathbf{k},m,\eta, s}^{\dagger}\ .
\end{equation}
By the relations $\{\hat{C},\hat{C}_{2z}T\}=\{\hat{C},\hat{C}_{2z}P\}=0$, the sewing matrix of $C$ satisfies
\beq\label{seq:sewing-C1}
B^C(\kk) B^{C_{2z}T}(\kk) = - B^{C_{2z}T}(\kk) B^{C*}(\kk)\ ,\qquad B^C(\kk) B^{C_{2z}P}(\kk) = - B^{C_{2z}P}(\kk) B^{C}(\kk)\ .
\eneq
Under the gauge fixing of Eq.~(\ref{eq:gauge-0}), we have $B^{C_{2z}T}(\kk)=\zeta^0\tau^0$, and $B^{C_{2z}P}(\kk)=-\zeta^y\tau^y$. The only $\kk$-independent gauge for sewing matrix of $C$ in consistency with Eqs. (\ref{seq:sewing-C0}) and (\ref{seq:sewing-C1}) within each pair of bands $n=\pm n_\text{B}$ is then (up to a global minus sign)
\begin{equation}\label{seq:sewing-C}
B^C(\kk)=\zeta^y\tau^z\ .
\end{equation}
In particular, this $\kk$-independent gauge fixing (\ref{seq:sewing-C}) of $C$ automatically ensures the continuous gauge fixing condition (\ref{seq:c-continuous}), which is crucial for defining the irrep band basis in Eq.~(\ref{eq-irrepbasis}). To see this, note that Eq.~(\ref{seq:sewing-C}) tells us that $u_{-n,\eta}(\kk)=i\text{sgn}(n)\eta u_{n,\eta}(\kk)$ for band $n=\pm n_\text{B}$, and thus we have
\begin{equation}
f_{n,\eta}(\kk+\qq,\kk)= \left|u^\dag_{n,\eta}(\kk+\qq) u_{n,\eta}(\kk)- u^\dag_{-n,\eta}(\kk+\qq) u_{-n,\eta}(\kk)\right|=\left|u^\dag_{n,\eta}(\kk+\qq) u_{n,\eta}(\kk)[1-\text{sgn}(n)^2\eta^2]\right|=0
\end{equation}
for any $\kk$ and $\qq$, satisfying Eq.~(\ref{seq:c-continuous}).

We also note that this gauge fixing of $C$ is consistent with the gauge fixings of both $C_{2z}$ and $P$ separately in Eq.~(\ref{eq:gauge-1}). Basically, the relations $\{\hat{C},\hat{C}_{2z}\}=0$ and $[\hat{C},\hat{P}]=0$ requires
\begin{equation}
B^C(-\kk) B^{C_{2z}}(\kk) = - B^{C_{2z}}(\kk) B^{C}(\kk)\ , \qquad B^C(-\kk) B^{P}(\kk) = B^{P}(\kk) B^{C}(\kk)\ ,
\end{equation}
which is satisfied by Eq.~(\ref{seq:sewing-C})

\subsection{Discrete Symmetry Analysis of the Intra-Layer Inter-valley Electron-Phonon Interaction and the phonon-induced electron-electron interaction}
 
We transform the phonon-mediated electron-electron interaction $H_{el-el}$ into the Chern band basis and the Hamiltonian is written as
 \begin{eqnarray}
 &H_{el-el}=-\frac{1}{N_G\omega_{A_1}} \sum_{\kk,\kk',\kk_1,\kk_1', \GG_M,l, s, s_1} G_{\kk,\kk', -l\eta \qq_2+ \GG_M}^{\eta e_Y  e_Y' l} 
 G_{\kk_1,\kk_1', l\eta \qq_2 - \GG_M}^{-\eta  e_{Y_1}  e_{Y_1}' l} \delta_{\kk-\kk', -\kk_1+\kk_1'} \gamma_{\kk e_Y\eta s}^\dagger \gamma_{\kk_1 e_{Y_1} - \eta s_1}^\dagger \gamma_{\kk_1', e_{Y_1}',\eta, s_1} \gamma_{\kk', e_Y',-\eta, s}\nonumber\\
\end{eqnarray}
with
\beq
  G_{\kk,\kk', \QQ_{-l\eta}}^{\eta e_Y e_Y' l}= \frac{\gamma_3}{ \sqrt{2M\omega_{A_1} }}
 \sum_{\QQ'_{l\eta}}\sum_{\alpha \beta} u^{e_Y\star}_{\kk, \QQ'_{l\eta}, \alpha, \eta} \sigma^x_{\alpha\beta} u^{e_Y'}_{\kk', \QQ'_{l\eta}-\QQ_{-l\eta}, \beta, -\eta}. 
 \eneq
 
Here we consider the chiral limit with $w_0=0$ and will discuss the constraint on the form of $G_{\kk,\kk', \QQ_{l\eta}}^{\eta e_Y e_Y' l}$, 
as well as $V^{\eta e_Y e_Y',e_{Y1} e_{Y1}'}_{\kk, \kk'; \kk_1, \kk_1' }$, from different discrete symmetries.   
 
\begin{itemize}
 
     \item Chiral Symmetry $\hat{C}$:
     
     \begin{eqnarray} 
     & [D(\hat{C})]_{\eta\eta'} u_{n\eta'}(\kk)= \sum_{m}[B^{C}(\kk)]_{m\eta,n\eta'}u_{m\eta}(\kk),\;\; B^C(\kk)=\zeta^y\tau^z,\nonumber \\ 
     &  [B^{C}(\kk)]_{m\eta,n\eta'} = -i m \delta_{m, -n} \eta \delta_{\eta \eta'},\;\; \quad[D(\hat{C})]_{\QQ^\prime \eta^\prime \beta, \QQ \eta \alpha} = \delta_{\QQ^\prime,\QQ} \delta_{\eta^\pr, \eta} (\sigma_z)_{\beta,\alpha}=\delta_{\QQ^\pr,\QQ} \delta_{\eta^\prime, \eta} (-1)^{\alpha+1} \delta_{\beta,\alpha}
\end{eqnarray} 
which leads to
\beq
(-1)^{\alpha+1} u_{\kk, \QQ, \alpha, n,\eta} =i n \eta u_{\kk, \QQ, \alpha, -n,\eta} \implies  \boxed{ u_{\kk, \QQ, \alpha, n,\eta} =- i n \eta (-1)^{\alpha} u_{\kk, \QQ, \alpha, -n,\eta}}
\eneq in the eigen-state basis. 
On the Chern band basis, we have 
\begin{eqnarray}
& u_{\kk, \QQ, \alpha, e_Y,\eta}= \frac{ u_{\kk, \QQ, \alpha,+,\eta} + i e_Y  u_{\kk, \QQ, \alpha,-,\eta}     }{\sqrt{2}} = \frac{   - i  \eta (-1)^{\alpha} u_{\kk, \QQ, \alpha, -,\eta  } + i e_Y (- i) (-1)  \eta (-1)^{\alpha} u_{\kk, \QQ, \alpha, +,\eta}       }{\sqrt{2}} \nonumber \\ 
& = \frac{   - i  \eta (-1)^{\alpha} u_{\kk, \QQ, \alpha, -,\eta  } - e_Y   \eta (-1)^{\alpha} u_{\kk, \QQ, \alpha, +,\eta}       }{\sqrt{2}} =  - e_Y \eta (-1)^{\alpha}  \frac{    i e_Y  u_{\kk, \QQ, \alpha, -,\eta  }  +    u_{\kk, \QQ, \alpha, +,\eta}       }{\sqrt{2}} 
=  - e_Y \eta (-1)^{\alpha}  u_{\kk, \QQ, \alpha, e_Y,\eta}. 
\end{eqnarray} 
This shows that, depending on the valley $\eta$ and Chern number $e_Y$, the Chern band basis is fully sub-lattice polarized (in the chiral limit) and that the Chern number for $u_{\kk \QQ \alpha e_Y \eta}$ is $e_Y= \eta (-1)^{\alpha+1} $. 
The sewing matrix in the Chern band basis can be obtained from
\beq
B^g_{e_{Y_1} \eta; e_Y \eta'} = \sum_{\QQ,\alpha} u^\star_{\hat{g}\kk, \QQ, \alpha, e_{Y_1}, \eta } \sum_{\QQ'\alpha'}  [D(\hat{g})]_{\QQ\eta \alpha; \QQ' \eta' \alpha'} u_{\kk \QQ' \alpha' e_Y \eta' }
\eneq which leads to
\beq\label{eq:sew_chiral_Chern}
B^C_{e_{Y_1} \eta; e_Y \eta'} = e_Y\eta \delta_{e_{Y_1}, e_Y} \delta_{\eta \eta'}. 
\eneq

From the above constraint equation for the wave function, the constraint equation for $G_{\kk,\kk', \QQ_{l\eta}}^{\eta e_Y e_Y' l}$ is given by 
\begin{eqnarray} \label{eq:eph_G_chiral_symmetryconstraint}
 & G_{\kk,\kk', \QQ_{-l\eta}}^{\eta e_Y e_Y' l}
 =e_Y   e_Y' G_{\kk,\kk', \QQ_{-l\eta}}^{\eta e_Y e_Y' l}\nonumber \\ & \implies G_{\kk,\kk',\QQ_{-l\eta}}^{\eta e_Y e_Y' l} = \delta_{e_Y, e_Y'} G_{\kk,\kk', \QQ_{-l\eta}}^{\eta e_Y e_Y l} =: \delta_{e_Y, e_Y'} G_{\kk,\kk', \QQ_{-l\eta}}^{\eta e_Y l}
\end{eqnarray}
We only keep one $e_Y$ index in the last equality. This greatly simplifies the form of the electron-electron interaction, which becomes
 \begin{eqnarray}
 &H_{el-el}=-\frac{1}{N_G\omega_{A_1}} \sum_{\kk,\kk',\kk_1,\kk_1', \GG_M,l, s, s_1, e_Y, e_Y'} G_{\kk,\kk', -l\eta \qq_2+ \GG_M}^{\eta e_Y   l}  
 G_{\kk_1,\kk_1', l\eta \qq_2 - \GG_M}^{-\eta  e_{Y1}   l} \gamma_{\kk,e_Y,\eta, s}^\dagger \gamma_{\kk_1, e_{Y_1}, - \eta s_1}^\dagger 
 \gamma_{\kk_1', e_{Y_1},\eta, s_1} \gamma_{\kk',e_Y,-\eta, s}
\end{eqnarray}
 with
\beq
  G_{\kk,\kk', \QQ_{-l\eta}}^{\eta e_Y  l}= \frac{\gamma_3}{ \sqrt{2M\omega_{A_1} }}
 \sum_{\QQ'_{l\eta}}\sum_{\alpha \beta} u^{e_Y\star}_{\kk, \QQ'_{l\eta}, \alpha, \eta} \sigma^x_{\alpha\beta} u^{e_Y}_{\kk', \QQ'_{l\eta}-\QQ_{-l\eta}, \beta, -\eta}.
 \eneq
 
 \item $\hat{C}_{2z} \hat{P}$: 
 

\begin{eqnarray}
& [D(C_{2z} P)]_{\eta\eta'} u_{n\eta'}(\kk)=\sum_{m}[B^{C_{2z} P}(\kk)]_{m\eta,n\eta'}u_{m\eta}(\kk),\;\;\; B^{C_{2z}P}(\kk)=-\zeta^y\tau^y \nonumber \\ 
&  [B^{C_{2z} P}(\kk)]_{m\eta,n\eta'} 
= m \eta \delta_{m, -n} \delta_{\eta, - \eta'}, \;
[D(\hat{C}_{2z}\hat{P})]_{\QQ\alpha \eta; \QQ' \beta  \eta'  }=  
\delta_{\QQ, \QQ'}\zeta_{\QQ'} \sigma^x_{\alpha\beta} \delta_{\eta, - \eta'}\nonumber \\ 
&\implies  \delta_{\eta,- \eta'} \zeta_\QQ \sigma^x_{\alpha \beta} u_{\kk, \QQ, \beta,n  \eta'} =  m \eta \delta_{m, -n} \delta_{\eta, - \eta'} 
u_{\kk \QQ \alpha  m \eta } \implies \nonumber \\ &   \boxed{u_{\kk, \QQ, \alpha,m  \eta} =  m \eta  \zeta_Q u_{\kk, \QQ, \bar{\alpha}, - m, -\eta }}
\end{eqnarray}

On the Chern band basis, this gives
\begin{eqnarray}
& u_{\kk, \QQ, \alpha, e_Y,\eta}=  - i e_Y \eta \zeta_Q  u_{\kk, \QQ, \bar{\alpha}, e_Y,-{\eta}}
\end{eqnarray}
and thus
\beq
B^{C_{2z}P}_{e_{Y_1} \eta; e_Y \eta'} = -i e_Y\eta'  \delta_{e_{Y_1}, e_Y} \delta_{\eta, - \eta'}. 
\eneq

$\hat{C}_{2z} \hat{P}$ can lead to the constraint 
\begin{eqnarray}
& G_{\kk,\kk', \QQ_{-l\eta}}^{\eta e_Y  l} = G_{\kk,\kk', \QQ_{-l\eta}}^{- \eta e_Y  -l}
\end{eqnarray} 
where we have used $\zeta_{ \QQ'_{l\eta} } = l\eta; \;\;  \zeta_{ \QQ'_{l\eta}-\QQ_{-l\eta}} =  \zeta_{ \QQ''_{-l\eta} }= - l \eta$. 


\item $\hat{C}_{2z}\hat{T} $: 
\beq u_{\kk,\QQ,\beta, n,\eta} = \sigma^x_{\beta\alpha }u_{\kk,\QQ,\alpha, n,\eta}^\star  \implies  u_{\kk,\QQ,\alpha, e_Y, \eta}  = u_{\kk,\QQ, \bar{\alpha}, - e_Y \eta}^\star \eneq
and the sewing matrix is
\beq
B^{C_{2z}T}_{e_{Y_1} \eta; e_Y \eta'} = \delta_{e_{Y_1},-  e_Y} \delta_{\eta \eta'}
\eneq

This now gives the following transformation property
\beq
G_{\kk,\kk', -l\eta \qq_2+ \GG_M}^{\eta e_Y l} = (G_{\kk,\kk', -l\eta \qq_2+ \GG_M}^{\eta -e_Y  l})^\star. 
\eneq

\item Particle-hole symmetry $\hat{P}$: 
we next deal with the particle-hole symmetry, of which the gauge fixing cannot be done with a constant matrix over the entire MBZ as discussed in 
Sec. \ref{app:gaugefixing} and \ref{app:chern}. 
We write the sewing matrix for $\hat{P}$ as
\beq
B^P(\kk) =e^{i\phi_\kk} i \zeta^y \tau^z, 
\eneq 
where $e^{i\phi_{\Gamma_M}}=-1$, $e^{i\phi_{M_M}}=1$ and $e^{i\phi_{\kk}}=-1$ for a generic $\kk$. 
With
\beq
\sum_{\QQ'\alpha' } [D(\hat{P})]_{\QQ\eta\alpha; \QQ'\eta'\alpha' } u_{\kk \QQ'\alpha' n \eta'} = \sum_{m} B^P_{m\eta; n \eta' }u_{-\kk,\QQ,\alpha, m, \eta} 
\eneq
and $ B^P_{m\eta; n \eta' } =e^{i\phi_\kk} m\eta \delta_{m, -n}\delta_{\eta, \eta'}$, we obtain
\beq  
\zeta_\QQ u_{\kk,\QQ,\alpha n \eta' } =  e^{i\phi_\kk} m\eta \delta_{m, -n}\delta_{\eta, \eta'} u_{-\kk, -\QQ,\alpha m \eta} \implies u_{\kk,\QQ,\alpha -m \eta  }= e^{i\phi_\kk} \zeta_\QQ m\eta u_{- \kk, -\QQ,\alpha m \eta}.  
\eneq on the eigen-state basis and
\begin{eqnarray}
& u_{\kk, \QQ, \alpha, e_Y,\eta}= 
i e_Y e^{i\phi_\kk} \zeta_\QQ \eta   u_{-\kk, -\QQ, {\alpha}, e_Y,{\eta}}
\end{eqnarray} on the Chern band basis. The corresponding sewing matrix on the Chern band basis reads  
\beq
B^P_{e_{Y_1} \eta; e_Y \eta'} = i e_Y\eta e^{i \phi_\kk} \delta_{e_{Y_1}, e_Y} \delta_{\eta \eta'},
\eneq
which leads to the constraint 
\begin{eqnarray}
& G_{\kk,\kk', \QQ_{-l\eta} = - l \eta\qq_2+ \GG}^{\eta e_Y  l} = e^{i (\phi_{\kk'}- \phi_\kk)}  G_{-\kk,-\kk', \QQ_{l\eta} =  l \eta\qq_2- \GG}^{\eta e_Y  -l}. 
\end{eqnarray}


\item $\hat{C}_{3z}$: 
The sewing matrix $B^{C_{3z}}(\kk)$ satisfies the following conditions (we write the $\kk$ dependence on the sewing matrices of $\hat{C}_{2z}\hat{T}$ and $\hat{C}_{2z} P$ for completeness, although we know these matrices have been chosen $\kk$-independent):
\begin{eqnarray}
&Unitarity: \;\;\; B^{C_{3z}}(\kk) B^{C_{3z}\dagger}(\kk) = B^{C_{3z}\dagger }(\kk) B^{C_{3z}}(\kk)  = \zeta_0\tau_0\nonumber \\ &
\hat{C}_{3z}^3=1:\;\;\; B^{C_{3z}}(\hat{C}_3^2\kk)B^{C_{3z}}(\hat{C}_3 \kk)B^{C_{3z}}(\kk)= \zeta_0\tau_0\nonumber \\ 
& \hat{C}_{2z}\hat{T}:\;\;\; B^{C_{3z}}(\kk) B^{C_{2z}T}(\kk)= B^{C_{2z}T}(\hat{C}_{3z} \kk) B^{C_{3z}\star}(\kk) \nonumber \\ 
& \hat{C}_{2z}\hat{P}:\;\;\; B^{C_{3z}}(\kk) B^{C_{2z}P }(\kk) = B^{C_{2z}P}(\hat{C}_{3z} \kk) B^{C_{3z}}(\kk)
\end{eqnarray}
It is easiest to solve the constraint from $\hat{C}_{2z}\hat{T}$ first, since its sewing matrix is identity. Hence, the matrix elements of $B^{C_{3z}}(\kk)$ are all real. We also know $\hat{C}_{3z}$ does not change the valley and hence 
\beq 
B^{C_{3z}}(\kk)_{m\eta; n\eta'}= A_{mn}^\eta \delta_{\eta, \eta'}\eneq 
with the real matrix $A_{mn}^\eta$ to be determined. The unitarity condition gives $AA^T = 1$ and hence $A$ is a rotation matrix,
\beq
[B^{C_{3z}}(\kk)]_{m\eta; n\eta'}  = \delta_{\eta \eta'} \begin{pmatrix}
\cos \theta^\eta(\kk)  & - \sin  \theta^\eta(\kk) \\
\sin  \theta^\eta(\kk)  & \cos  \theta^\eta(\kk) 
\end{pmatrix}
\eneq
Since $[B^{C_{2z}P}(\kk)]_{m\eta; n\eta'} =m \eta \delta_{m,-n}\delta_{\eta, - \eta'} $, the condition becomes
\begin{eqnarray}
&A^{-\eta}_{r,-n}(-n) = rA^{\eta}_{-r,n} \implies A^{-\eta}_{1,1} = A^{\eta}_{2,2},\;\; A^{-\eta}_{1,2} =- A^{\eta}_{2,1}
\end{eqnarray} where $n,r=\pm$  translates into $r=1,2$ in the matrix. This condition implies the $C_{3z}$ sewing matrix is the same at either of the valleys,
namely
\beq
B^{C_{3z}}(\kk)_{m\eta; n\eta'}  = \delta_{\eta \eta'} \begin{pmatrix}
\cos \theta(\kk)  & - \sin  \theta(\kk) \\
\sin  \theta(\kk)  & \cos  \theta(\kk) 
\end{pmatrix}
\eneq 
Finally $\hat{C}_{3z}^3= 1$ implies
\beq\label{eq:C3z_SW_theta_equality1}
\theta(\kk) + \theta(\hat{C}_{3z}\kk) + \theta(\hat{C}_{3z}^2\kk)=2 \pi n,\;\;\; n\in Z
\eneq where $n$ is an integer. On the eigen-state basis, this reads
\beq
e^{i \eta \frac{2\pi}{3}  (-1)^{\alpha+1} } u_{\kk, \QQ, \alpha n\eta}=\sum_{m=1,2} A_{mn}u_{\hat{C}_3 \kk, \hat{C}_3 \QQ, \alpha m \eta}
\eneq
while on the Chern band basis, this then gives
\beq
u_{\hat{C}_3 \kk, \hat{C}_3 \QQ, \alpha e_Y \eta}= e^{i e_Y \theta(\kk) }e^{i \eta \frac{2\pi}{3}  (-1)^{\alpha+1} }  u_{\kk,\QQ,\alpha e_Y\eta},
\eneq  
from which we find the sewing matrix of $\hat{C}_{3z}$ on the Chern basis is diagonal
\beq
B^{C_{3z}}_{e_{Y_1}\eta, e_Y\eta'}(\kk)
= e^{- i e_Y \theta(\kk) }\delta_{e_Y, e_{Y_1}} \delta_{\eta\eta'}. 
\eneq
This can be understood from the fact that in the chiral limit, there exists a relation between the Chern number and the sublattice/valley polarization, 
$e_Y= \eta (-1)^{\alpha+1} $. 

We can see that $\theta(\kk)$ cannot be constant over the whole MBZ. At the $\hat{C}_{3z} $-invariant points, i.e. $\kk=\Gamma_M$, $\kk =\KK_M$, or $\kk= \KK_M'$, $e^{- i e_Y \theta(\kk) }$ is the $\hat{C}_{3z}$-eigenvalue, and there exists a relation between the Chern number and $\hat{C}_{3z}$-eigenvalues, 
\beq
e^{ i \frac{2\pi}{3}C}= B(\Gamma_M) B(\KK_M)B(\KK_M')= e^{- i e_Y (\theta(\Gamma_M) + \theta(\KK_M) + \theta(\KK_M')) },
\eneq 
in which $C$ is the Chern number and it should take $C=e_Y$ for the Chern band basis. The $\theta$ value at $\Gamma_M$ can be obtained from 
the relation (\ref{eq:C3z_SW_theta_equality1}),
\beq
\theta(\Gamma_M) + \theta(\Gamma_M) + \theta(\Gamma_M)=2 \pi n,\;\;\; n\in Z \implies \theta(\Gamma_M) = 2 \pi/3 \mod 2 \pi.
\eneq 
So if we assume $\theta$ is a constant over the whole MBZ, we will find $e^{- i e_Y (\theta(\Gamma_M) + \theta(\KK_M) + \theta(\KK_M')) }=1$,
which would mean zero Chern number (mod 3) and not $e_Y=1$ as it is. From the TBG, we know the $\hat{C}_{3z}$-eigenvalues of the flat bands is $1$ at $\Gamma$,
and $\exp(\pm 2\pi i/3)$ at $\KK_M$, $\KK_M'$ (for the degenerate point). When we separate the bands into Chern band basis, the $\hat{C}_{3z}$-eigenvalue at $\KK_M$ of the Chern band and that at $\KK_M'$ have to be the same (if they were opposite, and with eigenvalue $1$ at $\Gamma_M$, we would get zero Chern number). We hence deduce
\beq
\theta(\Gamma_M) = 0 \mod 2 \pi n;\;\;\;\theta(\KK_M) = \frac{2\pi}{3} \mod 2 \pi ;\;\;\;\theta(\KK_M') = \frac{2\pi}{3} \mod 2 \pi 
\eneq 
which gives
\beq e^{ i \frac{2\pi}{3}C}= B(\Gamma_M) B(K_M)B(K_M')= e^{- i e_Y (\theta(\Gamma_M) + \theta(K_M) + \theta(K_M')) } =  e^{ i \frac{2\pi}{3}e_Y},
\eneq 
leading to $C=e_Y$ as advertised. 

The $\hat{C}_{3z}$ symmetry gives a constraint on the $G$ function,
\beq
G_{\kk,\kk', \QQ_{-l\eta} }^{\eta e_Y  l}=e^{ie_Y(\theta(\kk) - \theta(\kk')) } G_{\hat{C}_{3z}\kk,\hat{C}_{3z} \kk', \hat{C}_{3z}\QQ_{-l\eta} }^{\eta e_Y  l}
\eneq
 

\item $\hat{C}_{2x}$: 
The sewing matrix for $\hat{C}_{2x}$ takes the form
\beq
  B^{C_{2x}}_{m\eta; n \eta'}(\kk) = A^\eta_{mn} \delta_{\eta \eta'} 
\eneq 
and we need to again determine the matrix $A^\eta_{mn}$. For $\hat{C}_{2z}\hat{T}$, 
\beq  \hat{C}_{2z}\hat{T}:\;\;\; B^{C_{2x}}(\kk) B^{C_{2z}T}(\kk)= B^{C_{2z}T}(\hat{C}_{2x} \kk) B^{C_{2x}\star}(\kk) 
\implies A = A^\star
\eneq
and the unitarity condition requires $AA^T=1$. Furthermore, for the chiral symmetry $\hat{C}$, 
 \begin{eqnarray}
&  \{C,C_{2x} \}=0: \;\;  B^{C_{2x}}(\kk)B^C(\kk)=-B^C(\hat{C}_{2x}\kk)B^{C_{2x}}(\kk) \implies A^\eta(\kk) = \cos[\alpha^\eta(\kk)]\zeta_z + \sin[\alpha^\eta(\kk)] \zeta_x
  \end{eqnarray}  
 with $\zeta$ being in $n=1,2$ band space and $\alpha^\eta$ being an angle (not to be confused for the orbital $\alpha$). For $\hat{C}_{2z} \hat{P}$, 
\begin{eqnarray}
& \{\hat{C}_{2z} \hat{P}, \hat{C}_{2x}\} =0: \;\; B^{C_{2x}}(\kk) B^{C_{2z}P}(\kk)= -B^{C_{2z}P}(\hat{C}_{2x} \kk) B^{C_{2x}}(\kk), \nonumber\\
& A^{-\eta}_{-m,-n} = -mn A^{\eta}_{m,n}\implies \alpha^{\eta}(\kk)=\alpha^{-\eta}(\kk) = \alpha(\kk). 
\end{eqnarray} 
For the $\hat{C}_{3z}$ rotation, 
\begin{eqnarray}
&B^{C_{2x}}(\hat{C}_{3z}^2\kk)B^{C_{3z}}(\hat{C}_{3z}\kk)B^{C_{3z}}(\kk) = B^{C_{3z}}(\hat{C}_{2x}\kk)B^{C_{2x}}(\kk) \implies \nonumber \\ 
& \alpha_{\hat{C}_{3z}^2\kk }- \alpha_\kk - \theta_{\hat{C}_{3z}\kk} - \theta_\kk - \theta_{\hat{C}_{2x}\kk} =0 \mod 2 \pi
\end{eqnarray}
Putting all the above together, one can find the transformation for the Chern band basis
\beq
u_{\kk,\QQ,\alpha e_Y\eta} = e^{i e_Y \alpha(\kk) } u_{\hat{C}_{2x}\kk, \hat{C}_{2x}\QQ, \bar{\alpha},-e_Y, \eta}
\eneq from which we see that
\beq
\alpha(\kk) - \alpha(\hat{C}_{2x}\kk) =0 \mod 2\pi. 
\eneq 
The sewing matrix is then
\beq
B^{C_{2x}}_{e_{Y_1} \eta; e_Y \eta'} = \delta_{e_{Y_1}, -e_Y} e^{i e_Y \alpha(\kk) } \delta_{\eta \eta'}.
\eneq
$\hat{C}_{2x}$ imposes the following constraint
\begin{eqnarray}
& G_{\kk,\kk', Q_{-l\eta} = - l \eta \qq_2+ \GG}^{\eta, e_Y, l}=e^{ie_Y( \alpha(\kk')-\alpha(\kk))} G_{\hat{C}_{2x}\kk,\hat{C}_{2x} \kk',\hat{C}_{2x} \QQ_{-l\eta} =\hat{C}_{2x}( - l \eta \qq_2+ \GG)}^{\eta,- e_Y, -l}
\end{eqnarray}

 
\item Hermiticity: From the definition of $ G_{\kk,\kk', -l\eta \qq_2+ \GG_M}^{\eta n n' l} $, one can obtain the identity
\beq
 G_{\kk',\kk, -l\eta \qq_2+ \GG_M}^{\eta, n' n l\star} =  G_{\kk,\kk', l\eta \qq_2- \GG_M}^{- \eta, n n' l} 
\eneq 
which is required for the hermiticity condition $H_{el-el}^\dagger= H_{el-el}$.

\end{itemize}

We now define the $F$-function
\beq\label{eq:Ffunction_def}
F^{\eta, e_Y, e_Y'}_{\kk, \kk'; \kk_1, \kk_1' }=\frac{1}{N_0\omega_{A_1}} \sum_{\GG_M, l} G_{\kk,\kk', -l\eta \qq_2+ \GG_M}^{\eta, e_Y, l}  
G_{\kk_1,\kk_1', l\eta \qq_2 - \GG_M}^{-\eta,  e_{Y}',   l}
\eneq
so that the phonon-mediated electron-electron interaction takes the form 
\begin{equation}
 H_{el-el}=-\frac{1}{N_M}\sum_{\kk,\kk',\kk_1,\kk_1',s, s_1, e_Y, e_Y'} F^{\eta, e_Y, e_Y'}_{\kk, \kk'; \kk_1, \kk_1' } \gamma_{\kk,e_Y,\eta, s}^\dagger 
 \gamma_{\kk_1, e_{Y_1}, -\eta, s_1}^\dagger \gamma_{\kk_1', e_{Y_1},\eta, s_1} \gamma_{\kk',e_Y,-\eta, s}
\end{equation} 
We need to keep in mind that this term actually came from neglecting the one-body term in the phonon-mediated electron-electron interaction, which is 
\beq
 H_{el-el}=-\frac{1}{N_M} \sum_{\kk,\kk',\kk_1,\kk_1',s, s_1, e_Y, e_Y'} F^{\eta, e_Y, e_Y'}_{\kk, \kk'; \kk_1, \kk_1' } 
 \gamma_{\kk,e_Y,\eta, s}^\dagger \gamma_{\kk',e_Y,-\eta, s} \gamma_{\kk_1, e_{Y_1}, - \eta, s_1}^\dagger \gamma_{\kk_1', e_{Y_1},\eta, s_1} 
\eneq

We can apply all the symmetry operators to derive the constraint equations for the $F$-function, which are summarized as follows. 
\begin{eqnarray}
&\hat{C}_{2z} \hat{P}: \boxed{ F^{\eta, e_Y, e_Y'}_{\kk, \kk'; \kk_1, \kk_1' }=F^{-\eta, e_Y, e_Y'}_{\kk, \kk'; \kk_1, \kk_1' } }\nonumber\\ 
&  \hat{C}_{2z} \hat{T}: \boxed{ F^{\eta e_Y, e_Y'}_{\kk, \kk'; \kk_1, \kk_1' }= F^{\eta, - e_Y, -e_Y' \star}_{\kk, \kk'; \kk_1, \kk_1' }} \nonumber \\ 
 & Hermiticiy: \boxed{ F^{\eta, e_Y, e_Y'}_{\kk, \kk'; \kk_1, \kk_1' } = F^{\eta, e_Y', e_Y \star}_{\kk_1', \kk_1; \kk', \kk }}\nonumber \\ 
 & Reshuffling: \boxed{ F^{\eta, e_Y, e_Y'}_{\kk, \kk'; \kk_1, \kk_1' }= F^{-\eta, e_Y', e_Y}_{\kk_1, \kk_1'; \kk, \kk' } }\nonumber \\ 
 &\hat{P}: \boxed{ F^{\eta, e_Y, e_Y'}_{\kk, \kk'; \kk_1, \kk_1' } =e^{i (\phi_{\kk'}- \phi_\kk + \phi_{\kk_1'} - \phi_{\kk_1})}   F^{\eta, e_Y, e_Y'}_{-\kk, -\kk'; -\kk_1, -\kk_1' } } \nonumber \\ & \hat{C}_{3z}: \;\; \boxed{ F^{\eta, e_Y, e_Y'}_{\kk, \kk'; \kk_1, \kk_1' } =e^{ie_Y(\theta(\kk) - \theta(\kk')) + i e_Y'(\theta(\kk_1) - \theta(\kk_1'))} F^{\eta, e_Y, e_Y'}_{\hat{C}_{3z}\kk, \hat{C}_{3z}\kk'; \hat{C}_{3z}\kk_1, \hat{C}_{3z}\kk_1' }   } \nonumber \\ 
 & \hat{C}_{2x}: \;\; \boxed{ F^{\eta, e_Y, e_Y'}_{\kk, \kk'; \kk_1, \kk_1' } =e^{-ie_Y(\alpha(\kk) - \alpha(\kk')) - i e_Y'(\alpha(\kk_1) - \alpha(\kk_1'))} 
 F^{\eta, -e_Y, -e_Y'}_{\hat{C}_{2x}\kk, \hat{C}_{2x}\kk'; \hat{C}_{2x}\kk_1, \hat{C}_{2x}\kk_1' }   } 
\end{eqnarray}

Since we have already derived all the symmetry constraint for the $G$-function, all the symmetry constraint on the $F$-function can be derived from
the definition of $F$-function (Eq. \ref{eq:Ffunction_def}). Here we only show an example of reshuffling symmetry. 
\begin{eqnarray}
&F^{\eta e_Y, e_Y'}_{\kk, \kk'; \kk_1, \kk_1' }= \sum_{\GG_M}\sum_l G_{\kk,\kk', -l\eta \qq_2+ \GG_M}^{\eta e_Y l}  G_{\kk_1,\kk_1', l\eta \qq_2 - \GG_M}^{-\eta  e_{Y}' l}  \nonumber \\ & F^{- \eta e_Y', e_Y}_{\kk_1 \kk_1'; \kk, \kk' }= \sum_{\GG_M}\sum_l G_{\kk_1,\kk_1', l\eta \qq_2+ \GG_M}^{-\eta e_Y' l}  G_{\kk,\kk',- l\eta \qq_2 - \GG_M}^{\eta  e_{Y}   l} = F^{\eta e_Y, e_Y'}_{\kk, \kk'; \kk_1, \kk_1' }. 
\end{eqnarray}

The above symmetry relations relate different components of the $F$-function and let's next count the number of independent parameters. Out of $2\times 16=32$ complex numbers for each set of $\kk,\kk';\kk_1, \kk_1'$ momenta, we are left with $2$ complex numbers $F^{+ ++ }_{\kk, \kk'; \kk_1, \kk_1' }$, $F^{+ + -}_{\kk, \kk'; \kk_1, \kk_1' }$ out of which we can obtain the entire set of values. 

Next we consider the phonon-mediated electron-electron interaction in the Cooper channel (similar to Eq. \ref{eq:Helel1} but in the Chern band basis), 
\begin{eqnarray}
&H_{el-el}=-\frac{1}{N_M} \sum_{\kk,\kk',s, s_1, e_Y, e_Y'} V^{\eta, e_Y, e_Y'}_{\kk, \kk'}  \gamma_{\kk e_Y\eta s}^\dagger  \gamma_{-\kk e_{Y_1} - \eta s_1}^\dagger \gamma_{-\kk', e_{Y_1},\eta, s_1}  \gamma_{\kk' e_Y,-\eta, s}
\end{eqnarray}
with 
\begin{eqnarray}
 V^{\eta, e_Y, e_Y'}_{\kk, \kk'} = F^{\eta e_Y, e_Y'}_{\kk, \kk'; -\kk, -\kk' } .  
\end{eqnarray}

From the above constraint equations for the $F$-function, we can easily extract the corresponding relations for $V^{\eta, e_Y, e_Y'}_{\kk, \kk'} $,
which are listed as follows. 
\begin{eqnarray}\label{eq:Vfun_constraint1}
 &\hat{C}_{2z} \hat{P}: \boxed{V^{\eta, e_Y, e_Y'}_{\kk, \kk'}=V^{-\eta, e_Y, e_Y'}_{\kk, \kk'} }\nonumber\\ & 
 \hat{C}_{2z} \hat{T}: \boxed{V^{\eta, e_Y, e_Y'}_{\kk, \kk'}= V^{\eta, -e_Y, -e_Y'\star}_{\kk, \kk'}} \nonumber \\ 
 & Hermiticiy: \boxed{ V^{\eta, e_Y, e_Y'}_{\kk, \kk'}  = V^{\eta, e_Y', e_Y\star}_{-\kk', -\kk}} 
\nonumber \\ & Reshuffling: \boxed{ V^{\eta, e_Y, e_Y'}_{\kk, \kk'}= V^{-\eta, e_Y', e_Y}_{-\kk,-\kk'} } 
\nonumber \\ &\hat{P}: \boxed{ V^{\eta, e_Y, e_Y'}_{\kk, \kk'} =V^{\eta, e_Y, e_Y'}_{-\kk,- \kk'} } \nonumber \\ 
& \hat{C}_{3z}: \;\; \boxed{ V^{\eta, e_Y, e_Y'}_{\kk, \kk'} =e^{ie_Y(\theta(\kk) - \theta(\kk')) + i e_Y'(\theta(-\kk) - \theta(-\kk'))} V^{\eta, e_Y, e_Y'}_{\hat{C}_{3z}\kk, \hat{C}_{3z}\kk'}  } \nonumber \\ & \hat{C}_{2x}: \;\; \boxed{ V^{\eta, e_Y, e_Y'}_{\kk, \kk'} =e^{-ie_Y(\alpha(\kk) - \alpha(\kk')) 
- i e_Y'(\alpha(-\kk) - \alpha(-\kk'))} V^{\eta, -e_Y, -e_Y'}_{\hat{C}_{2x}\kk, \hat{C}_{2x}\kk'}  } 
\end{eqnarray}


We now have the constraint equations for $\hat{C}_{2z} \hat{P}$ and $\hat{C}_{2z} \hat{T}$ for the $V$ potential at the same momentum. In addition, 
the combination of reshuffling and particle-hole symmetry also give one more constraint equation that does not change the momentum, 
\begin{eqnarray}\label{eq:Vfun_constraint2}
\text{Reshuffling and } \hat{P}:  \boxed{  V^{\eta, e_Y e_Y'}_{\kk,\kk'}= V^{-\eta, e_Y' e_Y}_{\kk, \kk'}}. 
\end{eqnarray}
Putting the above results together, we can explicitly show the following constraints for different components of the $V$ potential at the same momentum
\begin{eqnarray}\label{eq:Vfun_constraint3}
& V^{+++}_{\kk\kk'} = V^{-++}_{\kk\kk'} = V^{+--\star}_{\kk\kk'}= V^{---\star}_{\kk\kk'}\nonumber \\ & V^{++-}_{\kk\kk'} = V^{--+}_{\kk\kk'} = V^{+-+}_{\kk\kk'}= V^{-+-}_{\kk\kk'} =V^{+-+\star}_{\kk\kk'} = V^{-+-\star}_{\kk\kk'} = V^{++-\star}_{\kk\kk'}= V^{--+\star}_{\kk\kk'}    
\end{eqnarray} 

We notice that there are more constraints for the case of $e_Y\ne e'_Y$, provided by the combination of Reshuffling and $P$. This extra constraint is trivial
for the case of $e_Y=e'_Y$ since the constraint from Reshuffling and $P$ is the identical to that from $C_{2z}P$. 
Hence fundamentally, in the chiral limit, the phonon-mediated electron-electron interaction at every $\kk,\kk'$ depends only on one complex parameter $V^{+ ++}_{\kk\kk'}$ and one \emph{real} parameter $V^{+ + -}_{\kk\kk'}$. 

For the above symmetry constraint results, we have two more comments, as discussed below. 
\begin{itemize}
\item While we see the change of $e^{i \phi_\kk}$ in the sewing matrix of $\hat{P}$ between $k=\Gamma_M$ and $k=M_M$, this does not help to determine the sign changes of the $V$ potential. The reason is that, for the Cooper channel ($\kk_1=-\kk, \kk_1'=-\kk'$), we have
 \beq
 F^{\eta e_Y, e_Y'}_{\kk, \kk'; -\kk, -\kk' } =e^{i (\phi_{\kk'}- \phi_\kk + \phi_{-\kk'} - \phi_{-\kk})}   F^{\eta e_Y, e_Y'}_{-\kk, -\kk'; \kk, \kk' }
 \eneq
 We can also find a relation between $\phi_\kk$ and $\phi_{-\kk}$
 \begin{eqnarray}
& u_{\kk, \QQ, \alpha, e_Y,\eta}=   i e_Y e^{i\phi_\kk} \zeta_\QQ \eta   u_{-\kk, -\QQ, \bar{\alpha}, e_Y,{\eta}} \implies - i e_Y e^{-i\phi_\kk} \eta = \sum_{\QQ, \alpha}   u_{\kk, \QQ, \alpha, e_Y,\eta}^\star \zeta_\QQ    u_{-\kk, -\QQ, \bar{\alpha}, e_Y,{\eta}}\implies \nonumber \\ 
& - i e_Y e^{-i\phi_{-\kk}} \eta = 
- i e_Y e^{i\phi_\kk} \eta \implies  
e^{i(\phi_\kk + \phi_{-\kk} )}=1
\end{eqnarray} 
Thus, we have
\beq  F^{\eta e_Y, e_Y'}_{\kk, \kk'; -\kk, -\kk' } = F^{\eta e_Y, e_Y'}_{-\kk, -\kk'; \kk, \kk' },\eneq
which corresponds to the constraint equation for $V$ potential from $\hat{P}$. Hence, the derivation here show that no sign change occurs because of the $\hat{P}$ sewing matrix. 

 \item We now analyze the constraint from $\hat{C}_{3z}$. First we put $\kk'=0$ and we have
 \beq
 V^{\eta, e_Y, e_Y'}_{\kk,0} =e^{ie_Y(\theta(\kk) - 2 \theta(0)) + i e_Y'\theta(-\kk)} V^{\eta, e_Y, e_Y'}_{\hat{C}_3\kk, 0}  
\eneq 
Now notice that if we put $\kk=\KK_M$ then $-\kk= \KK_M'$ but $\theta(\KK_M) = \theta(\KK_M') = 2\pi/3 \mod 2\pi $, 
and we have ($\theta_0 = 0 \mod 2 \pi$)
\beq
V^{\eta, e_Y, e_Y'}_{\KK_M,0} =e^{i(e_Y+ e_Y' ) \frac{2\pi}{3}} V^{\eta, e_Y, e_Y'}_{\KK_M, 0} 
\eneq
(one notes that $V_{\kk+ \GG, \kk'}= V_{\kk, \kk'+\GG}= V_{\kk,\kk'}$). 
Hence, there are nodes for the channel with $e_Y= e_Y'$, but not for $e_Y= -e_Y'$, as shown below. 
\beq\label{eq:zeroVatKMGamma}
e_Y= e_Y': \;\; V^{\eta, e_Y, e_Y}_{\KK_M,0} =e^{ie_Y \frac{4\pi}{3}} V^{\eta, e_Y, e_Y}_{\KK_M, 0} \implies  V^{\eta, e_Y, e_Y}_{\KK_M,0}=0. 
\eneq

\end{itemize}

\subsection{Continuous Symmetry Analysis of the Intra-layer and Inter-valley Electron-Phonon Interaction and the phonon-induced electron-electron interaction}
 
 It is known that the projected Coulomb interaction Hamiltonian into the flat bands, in the chiral limit, enjoys a large $U(4) \times U(4)$ symmetry with generators
 \beq
 S^{ab}_{e_Y}= \sum_{\kk \eta s \eta' s'} \gamma^\dagger_{\kk e_Y \eta s  } \tau^{a}_{\eta \eta'}s^b_{ss'}\gamma_{\kk e_Y \eta' s'},\;\;\;\ e_Y= \pm 1,\; a, b= 0,1,2,3
 \eneq
 However, this enlarged symmetry will be broken by the electron-phonon coupling. To see that, we calculate $[H_{el-el}, S^{ab}_{e_{Y_2}}]$ and with
 some tedious calculation, we find 
  \begin{eqnarray}
 &[H_{el-el}, S^{ab}_{e_{Y_2}}] =
2 F^{\eta e_Y, e_{Y_2}}_{\kk, \kk'; \kk_1, \kk_1' }  \gamma_{\kk e_Y\eta s}^\dagger \gamma_{\kk'e_Y-\eta s}(  \gamma_{\kk_1 e_{Y_2} - \eta s_2}^\dagger  \tau^{a}_{\eta, \eta_2}s^b_{s_2s_1}\gamma_{\kk_1' e_{Y_2} \eta_2 s_1}  - \gamma_{\kk_1 e_{Y_2} \eta_2 s_2}^\dagger  \tau^{a}_{\eta_2, -\eta}s^b_{s_2s_1} 
\gamma_{\kk_1' e_{Y_2}\eta s_1}) +\nonumber \\ & +  \text{one-body terms},
\end{eqnarray}
from which one can see that the two body terms only vanish for 
\beq
a= 0, \;\;\forall b \implies U_s(2)_{e_Y=1} \times U_s(2)_{e_Y=-1 }
\eneq 
where $s$ means spin. The electron-phonon coupling breaks the $U(4)\times U(4)$ symmetry in the chiral flat-band limit to a $U_s(2)_{e_Y=1} \times U_s(2)_{e_Y=-1 }$ spin $U(2)$ symmetry in each Chern sector, with the generators
\beq
 S^{b}_{e_Y}= \sum_{\kk \eta s} \gamma^\dagger_{\kk e_Y \eta s } s^b_{ss'}\gamma_{\kk e_Y \eta s'},\;\;\;\ e_Y= \pm 1,\;  b= 0,1,2,3
\eneq 
This symmetry is broken away from the chiral flat-band limit, leaving behind just the total $U(2)$ spin symmetry.

We also have $a=3, b=0$ (Valley $U(1)$ charge \emph{per Chern sector} ) for which $\tau^{a}_{\eta\eta'} = \eta \delta_{\eta \eta'}$. 
In this case, one can show 
\begin{eqnarray}
 &[H_{el-el}, S^{30}_{e_{Y_2}}] = 
-\frac{2}{N_M}\sum_{\kk,\kk',\kk_1,\kk_1',s, s_1, e_Y}  \eta  \nonumber \\ &  [    F^{\eta e_Y, e_{Y_2}}_{\kk, \kk'; \kk_1, \kk_1' } \gamma_{\kk,e_Y,\eta s}^\dagger \gamma_{\kk',e_Y,-\eta s} \gamma_{\kk_1, e_{Y_2}, - \eta, s_1}^\dagger  \gamma_{\kk_1', e_{Y_2}, \eta, s_1} -    
F^{\eta e_{Y_2}, e_{Y} }_{\kk, \kk'; \kk_1, \kk_1' } \gamma_{\kk,e_{Y_2}, \eta s}^\dagger  \gamma_{\kk',e_{Y_2},-\eta, s } \gamma_{\kk_1, e_{Y}, -\eta, s_1}^\dagger   \gamma_{\kk_1', e_{Y},\eta, s_1}.  
\end{eqnarray}
And we can see that Valley $U(1)$ charge per Chern band sector is \emph{not} conserved. However, the \emph{total (summed over both chern sectors)} $U(1)$  Valley  charge 
  \beq
  Q_v= \sum_{\kk e_Y \eta s} \gamma^\dagger_{\kk e_Y \eta s } \tau^z_{\eta \eta'}\gamma_{\kk e_Y \eta s}
 \eneq
  is a good quantum number; it remains so away from the chiral flat-band limit. 
 
  



\subsection{Numerical results for the phonon-induced electron-electron interaction }
In this section, we will discuss our numerical results for the electron-phonon interaction and 
the phonon-induced electron-electron interaction in TBG and compare the results
with the symmetry analysis of $V^{\eta e_1 e_2}_{\kk \kk'}$ in the previous section.

Our numerical calculation is based on the expression (\ref{eq:ephinteractionparameterfull}) for electron-phonon coupling $g^s_{\alpha\beta}(\pp,\pp')$ 
in one single layer and then project it into the flat band basis in TBG as
\begin{eqnarray}
G_{\kk,\kk', Q_{-l\eta}}^{\eta n n' ls}= \sum_{\QQ'_{l\eta},\alpha \beta} u^{n\star}_{\kk, \QQ'_{l\eta}, \alpha, \eta} g^{s,l}_{\alpha\beta}(\pp,\pp') 
u^{n'}_{\kk', \QQ'_{l\eta}-\QQ_{-l\eta}, \beta, -\eta},
\end{eqnarray} 
with $\pp=\kk+\eta \KK_D-\QQ'_{l\eta}$ and $\pp'=\kk'-\eta \KK_D-\QQ'_{l\eta}+\QQ_{-l\eta}$ and $\kk,\kk'$ is chosen within the Moir\'e Brillouin zone. 
The corresponding inter-valley electron-phonon interaction is then given by
\begin{eqnarray}
&H_{inter-vall} =\frac{1}{ \sqrt{N_G}} \sum_{\kk,\kk' \in MBZ}\sum_{n,n'=1,2}\sum_{\eta, s }\sum_{l=\pm} \sum_{Q_{-l\eta}} 
G_{\kk,\kk', \QQ_{-l\eta}}^{\eta n n' ls} \gamma_{\kk,n,\eta, s}^\dagger \gamma_{\kk',n',-\eta, s} \nonumber\\
& (b_{-\eta \KK_D+\kk-\kk'- \QQ_{-l\eta} ,l, A_1} +b_{\eta \KK_D-\kk+\kk'+ \QQ_{-l\eta} ,l,  A_1 }^\dagger)
 \end{eqnarray} 

Since the Moir\'e Hamiltonian is expanded around $\KK_D$, namely $|\KK_D|\gg k,k',|\QQ'_{l\eta}|$ and $g^s_{\alpha\beta}(\pp,\pp')$ is a smooth
function of $\pp$ and $\pp'$, we can approximate $g^s_{\alpha\beta}(\pp,\pp')$ by $g^s_{\alpha\beta}(\eta \KK_D,-\eta \KK_D)$ in our numerical calculations. 
We test this approximation by directly evaluating $g^s_{\alpha\beta}(\pp,\pp')$ and find that the derivation of $g^s_{\alpha\beta}(\pp,\pp')$ from 
$g^s_{\alpha\beta}(\eta \KK_D,-\eta \KK_D)$ is only within 5\% at the relevant momentum scale of Moir\'e reciprocal lattice vectors. 
For $g^s_{\alpha\beta}(\eta \KK_D,-\eta \KK_D)$, only the phonon mode $s=A1$ give non-zero contribution, so we only keep $s=A1$ below and drop the
s-index in calculating coupling strength 
\begin{eqnarray}
G_{\kk,\kk', \QQ_{-l\eta}}^{\eta n n' l}= \sum_{\QQ'_{l\eta},\alpha \beta} u^{n\star}_{\kk, \QQ'_{l\eta}, \alpha, \eta} 
g^{l}_{\alpha\beta}(\eta \KK_D,-\eta \KK_D) u^{n'}_{\kk', \QQ'_{l\eta}-\QQ_{-l\eta}, \beta, -\eta}. 
\end{eqnarray} 
A similar expression of electron-phonon coupling parameter for the Chern band basis is given by
\begin{eqnarray}
G_{\kk,\kk', \QQ_{-l\eta}}^{\eta e_Y e_Y' l}= \sum_{\QQ'_{l\eta},\alpha \beta} u^{\star}_{\kk, \QQ'_{l\eta}, \alpha, e_Y, \eta} 
g^{l}_{\alpha\beta}(\eta \KK_D,-\eta \KK_D) u_{\kk', \QQ'_{l\eta}-\QQ_{-l\eta}, \beta, e_Y', -\eta}. 
\end{eqnarray} 
The corresponding phonon-induced electron-electron interaction in the Cooper pair channels is given by
\beq \label{eq:Elel_interaction_ChernBasis}
H_{el-el}=-\frac{1}{N_M} \sum_{\kk,\kk',s_1, s_2, e_Y, e_Y'} V^{\eta, e_Y, e_Y'}_{\kk, \kk'} 
\gamma_{\kk,e_Y,\eta, s_1}^\dagger \gamma_{-\kk, e_Y', - \eta, s_2}^\dagger \gamma_{-\kk', e_Y', \eta, s_2} \gamma_{\kk', e_Y,-\eta, s_1} 
\eneq
where
\beq
V^{\eta, e_Y, e_Y'}_{\kk, \kk'}=\frac{1}{N_0}\frac{1}{\omega_{A_1}}
\sum_{\GG_M}\sum_l G_{\kk,\kk', -l\eta \qq_2+ \GG_M}^{\eta e_Y   l}  G_{-\kk,-\kk', l\eta \qq_2 - \GG_M}^{-\eta  e_{Y}' l}. 
\eneq
Here we have already added the required factors $N_0$ and $N_M$ and divided the expression by $\omega_{A_1}$ so 
$V^{\eta, e_Y, e_Y'}_{\kk, \kk'}$ has the correct energy unit. 

In the above expression, the phonon momentum is $\qq=-\eta \KK_D+\kk-\kk'- \QQ_{-l\eta}$ and the e-ph coupling parameter 
$G_{\kk,\kk', \QQ_{-l\eta}}^{\eta e_Y e_Y' l}$ depends on the Moir\'e momentum points $\QQ_{-l\eta}$. In Fig. \ref{fig:ephcoupling_Q_TBG}
and \ref{fig:ephcoupling_Q_TBG_chiral}, we plot $|G|$ as a function of different values of $|\QQ_{-l\eta}|$ for the non-chiral and chiral limit, separately, on the eigen-state basis.
In Fig. \ref{fig:ephcoupling_Q_TBG_chiral_Chern}, we plot $|G|$ as a function of different values of $|\QQ_{-l\eta}|$ in the chiral limit on the Chern-band basis. 
The corresponding lattice of $\QQ_{-l\eta}$ and different shells are shown in Fig. \ref{fig:dispersion_Moire}(b). 
One can see that while $|G|$ generally decays away for large $|G|$'s in both cases, the decay rate is faster for the chiral limit. 
Furthermore, we notice that in the chiral limit on the Chern-band basis (Fig.\ref{fig:ephcoupling_Q_TBG_chiral_Chern}), $|G^{e_Y e_Y'}|$ is zero if $e_Y\neq e_Y'$, which has been derived in our analytical results, see Eq. \ref{eq:eph_G_chiral_symmetryconstraint}. 
The detailed form of $G$ function is quite complicated and thus, we, instead, show $V^{\eta, e_Y, e_Y'}_{\kk, \kk'}$ as a function of $\kk'$ in
Fig. \ref{fig:Vee_Gamma} for $\kk=\Gamma_M, \KK_M, M_M$. Several features are noticed: (1) Intra-Chern-band channel
($e_Y=e_Y'=\pm$) shows nodes at $\kk=\Gamma_M, \kk'=\KK_M$, which is consistent with Eq. (\ref{eq:zeroVatKMGamma}) from the symmetry analysis. 
(2) We notice that the interaction for the inter-Chern-band channels keeps its value over the whole MBZ while that for the intra-Chern-band channels drops
quickly near the MBZ boundary due to the presence of the nodes. 

\begin{figure}[hbt!]
    \centering
    \includegraphics[width=7in]{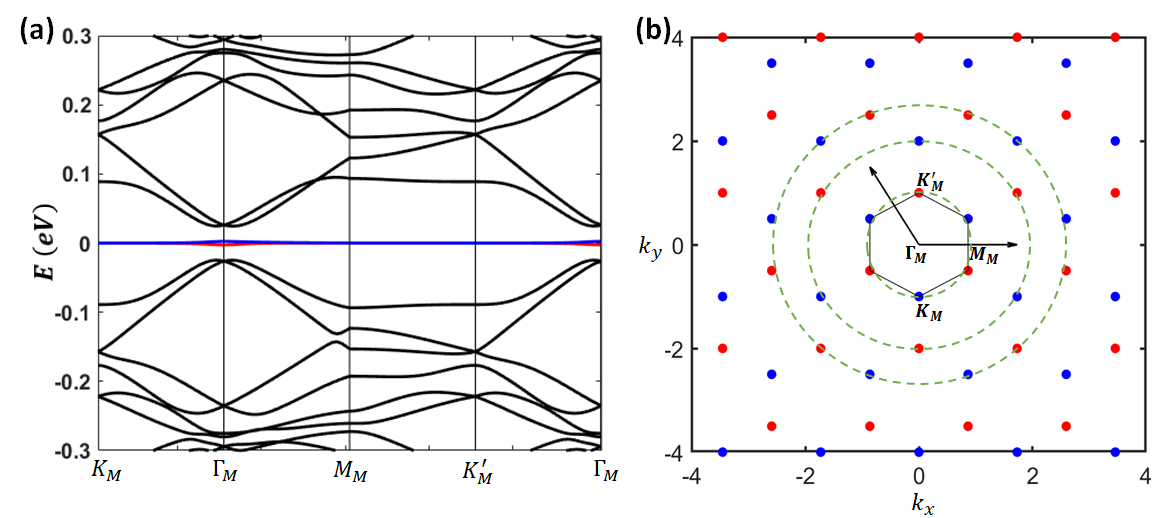} 
    \caption{ (a) The band dispersion at $\theta=1.06^\circ$ with $w_0/w_1=0.8$. Other parameters are the same as in Fig. \ref{fig:bands_free}. (b) The Moir\'e BZ and the $\mathcal{Q}_\pm$ lattice with the first three shells (dashed lines). The momentum $k_x$ and $k_y$ is in the unit of $k_\theta=2 |K_D| \sin\left(\frac{\theta}{2}\right) $ with $\theta=1.06^\circ$. 
  }
    \label{fig:dispersion_Moire}
\end{figure}

\begin{figure}[hbt!]
    \centering
    \includegraphics[width=7in]{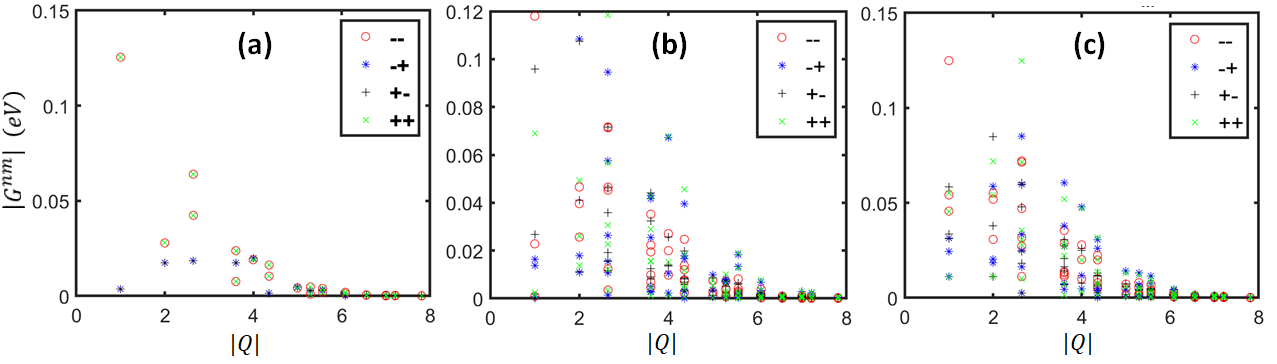} 
    \caption{ The parameter $G_{\kk,\kk', Q}^{\eta nm l}$ as a function of $|Q|$ for the projection of the electron-phonon interaction for TBG with $w_0/w_1=0.8$. 
    $|Q|$ in the unit of $k_\theta=2 |K_D| \sin\left(\frac{\theta}{2}\right) $. We choose  $\kk={\bf \Gamma}_M$ and $\kk'={\bf \Gamma}_M, \KK_M, {\bf M}_M $ for (a), (b) and (c), respectively. 
    The plot is on the eigen-state basis with $(n,m)=(-,-),(-,+),(+,-),(+,+)$ for red circles, blue stars, black plus signs, and green crosses, respectively.     
  }
    \label{fig:ephcoupling_Q_TBG}
\end{figure}

\begin{figure}[hbt!]
    \centering
    \includegraphics[width=7in]{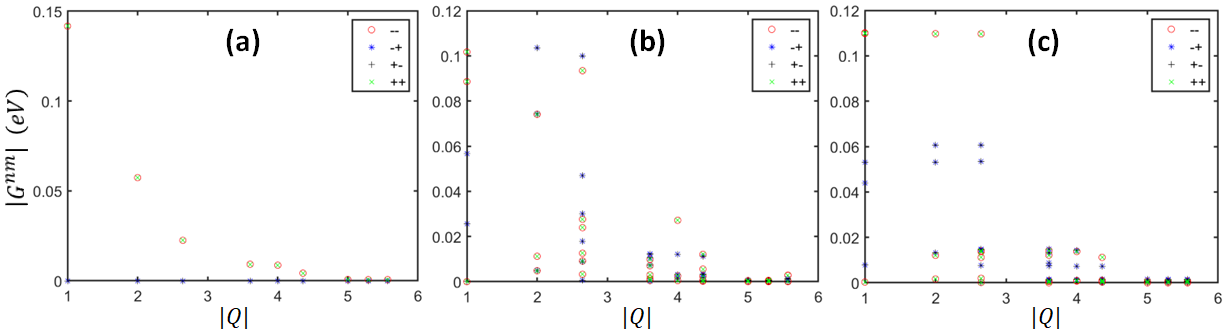} 
    \caption{ The parameter $G_{\kk,\kk', Q}^{\eta nm l}$ as a function of $|Q|$ for the projection of the electron-phonon interaction for TBG in the chiral limit ($w_0=0$). $|Q|$ in the unit of $k_\theta=2 |K_D| \sin\left(\frac{\theta}{2}\right) $. We choose  $\kk={\bf \Gamma}_M$ and $\kk'={\bf \Gamma}_M, \KK_M, {\bf M}_M $ for (a), (b) and (c), respectively. 
    The plot is on the eigen-state basis with $(n,m)=(-,-),(-,+),(+,-),(+,+)$ for red circles, blue stars, black plus signs, and green crosses, respectively.   
  }
    \label{fig:ephcoupling_Q_TBG_chiral}
\end{figure}

\begin{figure}[hbt!]
    \centering
    \includegraphics[width=7in]{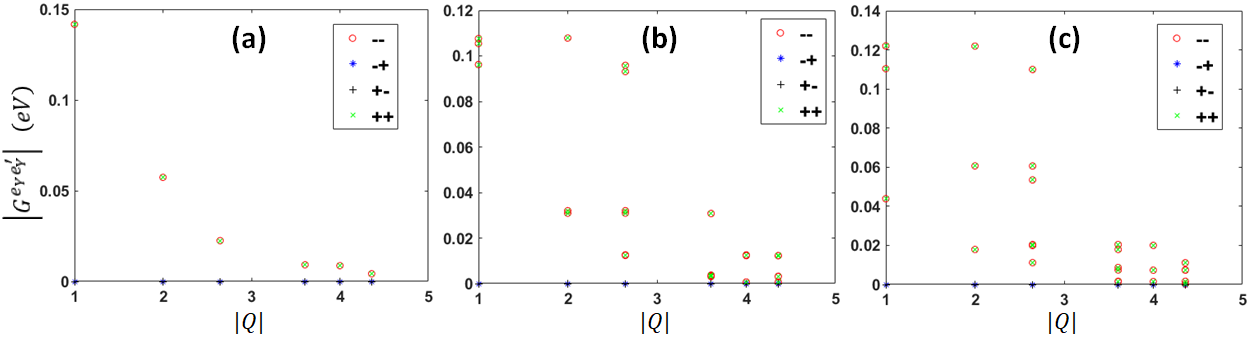} 
    \caption{ The parameter $G_{\kk,\kk', Q}^{\eta e_Y e_Y' l}$ as a function of $|Q|$ for the projection of the electron-phonon interaction for TBG on the Chern-band basis in the chiral limit ($w_0=0$). $|Q|$ in the unit of $k_\theta=2 |K_D| \sin\left(\frac{\theta}{2}\right) $. We choose  $\kk={\bf \Gamma}_M$ and $\kk'={\bf \Gamma}_M, \KK_M, {\bf M}_M $ for (a), (b) and (c), respectively. 
    The plot is on the Chern-band basis with $(e_Y,e_Y')=(-,-),(-,+),(+,-),(+,+)$ for red circles, blue stars, black plus signs, and green crosses, respectively.   
  }
    \label{fig:ephcoupling_Q_TBG_chiral_Chern}
\end{figure}

\begin{figure}[hbt!]
   \centering
    \includegraphics[width=7in]{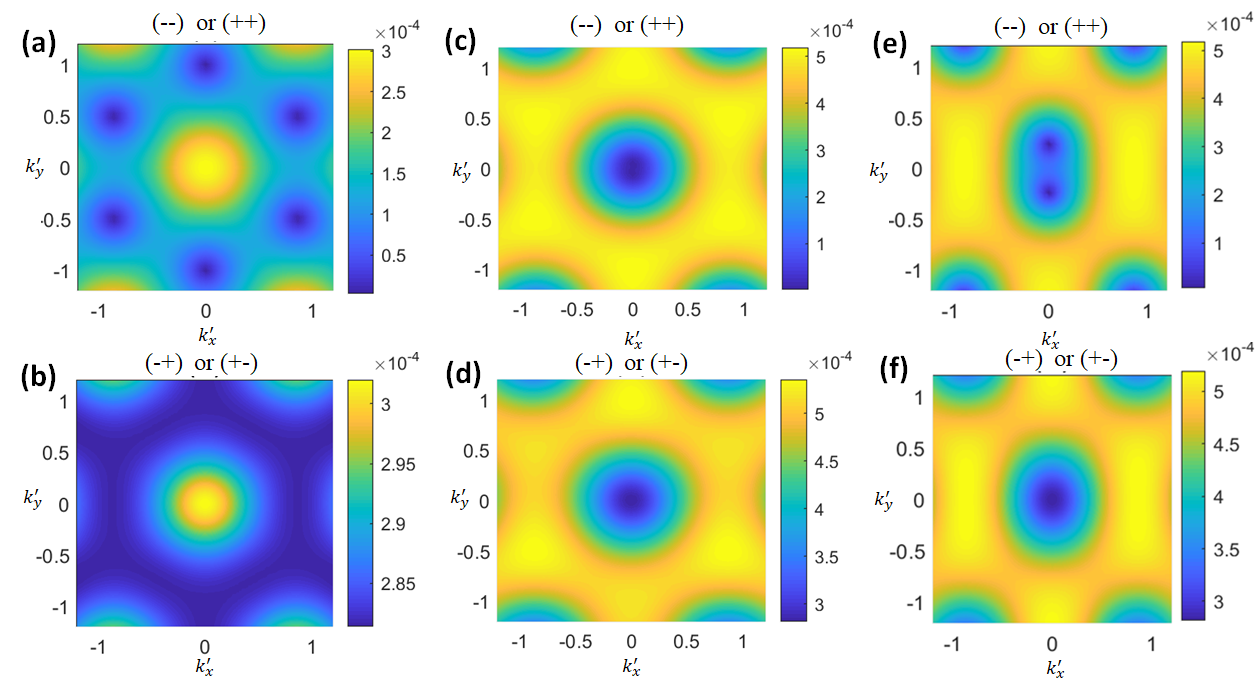} 
    \caption{ $V_{\kk,\kk'}^{+ e_Y, e'_Y}$ as a function of $\kk'$ for a fixed $\kk$.  Here $V_{\kk,\kk'}^{+ e_Y, e'_Y}$ in unit of eV and $\kk'$ in the unit of $k_\theta=2 |K_D| \sin\left(\frac{\theta}{2}\right) $. (a)
    and (b) are for $\kk={\bf \Gamma}_M$; (c) and (d) are for $\kk={\bf M}_M$; (e) and (f) are for $\kk=\KK_M$. 
  }
    \label{fig:Vee_Gamma}
\end{figure}


%

%

\subsection{Symmetry Classification of Irreducible Pairing Channels}
Since the phonon energy and the electron-phonon coupling are strong, the eige-state basis might not be the best basis to work in here. 
We hence work in the Chern band basis, for which the symmetry transformation properties are summarized as follows. 
\begin{eqnarray}\label{eq:symtrans_Chern1}
&\hat{g} \gamma^\dagger_{\kk e_Y \eta' s} \hat{g}^{-1}= \sum_{e_{Y_1}\eta }  B^{g}_{\eta e_{Y_1};e_Y \eta'}\gamma^\dagger_{\hat{g}\kk, e_{Y_1} \eta s}\nonumber \\ 
& \hat{C} \gamma_{\kk,e_Y,\eta, s}^\dagger \hat{C}^{-1} = e_Y \eta \gamma_{\kk,e_Y, \eta, s}^\dagger \nonumber\\ 
& \hat{C}_{2z}\hat{P} \gamma_{\kk,e_Y,\eta, s}^\dagger (\hat{C}_{2z}\hat{P})^{-1} = i e_Y \eta \gamma_{\kk,e_Y, -\eta, s}^\dagger \nonumber\\ & \hat{C}_{2z}\hat{T} \gamma_{\kk,e_Y,\eta, s}^\dagger (\hat{C}_{2z}\hat{T})^{-1} = \gamma_{\kk,- e_Y, \eta, s}^\dagger \nonumber\\ 
&  \hat{P} \gamma_{\kk,e_Y,\eta, s}^\dagger \hat{P}^{-1} = i e_Y \eta  e^{i \phi_\kk} \gamma_{-\kk,e_Y, \eta, s}^\dagger \nonumber\\ 
&  \hat{C}_{3z} \gamma_{\kk,e_Y,\eta, s}^\dagger \hat{C}_{3z}^{-1} = e^{- i e_Y \theta(\kk)} \gamma_{\hat{C}_{3z}\kk, e_Y, \eta, s}^\dagger \nonumber\\ 
& \hat{C}_{2x} \gamma_{\kk,e_Y,\eta, s}^\dagger \hat{C}_{2x}^{-1} = e^{ i e_Y \alpha(\kk)} \gamma_{\hat{C}_{2x}\kk, -e_Y, \eta, s}^\dagger 
\end{eqnarray}

\subsubsection{Discrete Symmetry analysis of the Gap Function} \label{Sec:DiscreteSymmetry_GapFunction}
The mean-field Bogoliubov-de Gennes (BdG) Hamiltonian for the inter-valley pairing can be written as
\beq\label{eq:GapFun1}
 H_{\Delta} =\sum_{\kk, e_{Y_1}, e_{Y_2}, s_1, s_2, \eta } \gamma_{\kk, e_{Y_1}, \eta , s_1 }^\dagger \Delta^{\eta}_{\kk; e_{Y_1} s_1, e_{Y_2} s_2}  \gamma_{-\kk, e_{Y_2}, -\eta , s_2}^\dagger +  h.c. \eneq
 
First we notice the reshuffling symmetry of the gap function, 
\beq\label{eq:GapFun_reshuffling}
\Delta^{\eta}_{\kk; e_{Y_1} s_1, e_{Y_2} s_2}= -\Delta^{-\eta}_{-\kk; e_{Y_2} s_2, e_{Y_1} s_1}, \eneq
which can be obtained by interchanging two $\gamma^\dagger$ operators and then relabelling the indices. 

Because of the $SU(2)$ spin symmetry of the system, We can define
 \beq \Delta^\eta_{\kk; e_{Y_1} s_1, e_{Y_2} s_2} = \Delta^\eta_{\kk; e_{Y_1} , e_{Y_2}} \mS_{s_1,s_2}. 
 \eneq
 Where $\mS$ is the spin matrix, which takes the form
 \beq
\text{Singlet: } \mS_{s_1s_2} = (i s_y)_{s_1s_2};\;\;\;\text{Triplet: }  \mS_{s_1s_2} = ((\dd \cdot {\bf s}) i s_y)_{s_1s_2};
 \eneq
 where $\dd$ is the three component d-vector for triplet pairing.  We note that 
 \beq
\text{Singlet: } \mS_{s_1s_2} =- \mS_{s_2s_1} ;\;\;\;\text{Triplet: }   \mS_{s_1s_2} = \mS_{s_2s_1};
 \eneq

\begin{table}[h!]
\centering
 \begin{tabularx}{0.8\textwidth} { 
  | >{\raggedright\arraybackslash}X 
  | >{\centering\arraybackslash}X 
   | >{\centering\arraybackslash}X 
    | >{\centering\arraybackslash}X 
     | >{\centering\arraybackslash}X 
  | >{\raggedleft\arraybackslash}X | }
 \hline
 Rep Symbol  & Basis Functions & Dim    & $C_{2z}$ & $C_{2x}$ & $C_{3z}$\\
\hline
 $A_1$  & 1, or $x^2+y^2$, or $z^2$   & 1 &1 &1 &1\\
 \hline
 $A_2$ & z  & 1  &1 &-1 &1 \\
\hline  $B_1$ & $y(3x^2-y^2)$   & 1 &-1 &-1 & 1\\
\hline
$B_2$  & $x(x^2- 3 y^2)$  & 1& -1&1 &  1\\
\hline  $E_1$ & $(x,y)$    & 2 &$-\sigma_0$ &$\sigma_z$ &$\exp(-i 2 \pi \sigma_y/3)$\\
\hline
$E_2$   & ($\frac{x^2-y^2}{2}, xy$)    & 2& $\sigma_0$ &$\sigma_z$ &$\exp(i 2 \pi \sigma_y/3)$ \\
\hline
\end{tabularx}
\caption{ Irreducible pairing channels. }
\label{tab:IrreduciblePairing}
\end{table}

We now write down the basis function of the irreducible representation (irrep) of the $C_{6v}$ group, 
which is the symmetry group of the TBG Hamiltonian with two valleys,
in the table \ref{tab:IrreduciblePairing}. The pairing Hamiltonian $H_{\Delta} $ can transform as a representation of the $C_{6v}$ group.
Thus, based on the table \ref{tab:IrreduciblePairing}, we next consider the symmetry classification of the irreducible pairing channels \cite{savary2017superconductivity,sigrist1991phenomenological}. 

Let's first consider (spinless) $\hat{C}_{2z}$ symmetry, which generally plays the role of inversion and allows us to identify the parity of different pairing channels. For $\hat{C}_{2z}$, the electron operator is transformed as
\beq  \hat{C}_{2z} \gamma^\dagger_{\kk, e_Y, \eta s} \hat{C}_{2z}^{-1} = e^{i\phi_\kk} \gamma^\dagger_{-\kk, e_Y, -\eta s}\eneq
From the table \ref{tab:IrreduciblePairing}, We expect the pairing Hamiltonian $H_{\Delta}$ should transform as certain representations under $C_{2z}$,
with the eigenvalue $\lambda_{C_{2z}}= \pm 1$,
\beq
\hat{C}_{2z} H_{\Delta} \hat{C}_{2z}^{-1} =\lambda_{C_{2z}} H_{\Delta}. 
\eneq 
In the case of $\lambda_{C_{2z}}=1$, the Hamiltonian is invariant under $\hat{C}_{2z}$. For the case of $\lambda_{C_{2z}}=-1$, the Hamiltonian is not invariant under $\hat{C}_{2z}$, as the gap term changes sign. However, it is invariant under $\hat{C}_{2z}$ followed by a gauge transformation $c\rightarrow i c$, i.e. a projective representation. Hence, the spectrum is still $\hat{C}_{2z}$ symmetric. 
With $\exp(i \phi_\kk+ i \phi_{-\kk}) =1$, the gap function then follows the relation
\begin{eqnarray}
&\Delta^\eta_{\kk; e_1 s_1, e_2 s_2}=\lambda_{C_{2z}} \Delta^{-\eta}_{-\kk; e_1 s_1, e_2 s_2}.
\end{eqnarray}
Together with the reshuffling symmetry (\ref{eq:GapFun_reshuffling}), we have
\begin{eqnarray}
&\Delta^\eta_{\kk; e_1, s_1, e_2, s_2}=-\lambda_{C_{2z}} \Delta^\eta_{\kk; e_2, s_2, e_1, s_1},
\end{eqnarray} 
which leads to
\beq
\text{Singlet: } \Delta^\eta_{\kk; e_1, e_2} = \lambda_{C_{2z}}\Delta^\eta_{\kk; e_2, e_1};\;\;\; \text{Triplet: } \Delta^\eta_{\kk; e_1, e_2} = -\lambda_{C_{2z}}\Delta^\eta_{\kk; e_2, e_1}. 
\eneq 

Next let's consider the (spinless) $\hat{C}_{3z}$ symmetry. Since the $1$-D irrep of $C_{6v}$ have only eigenvalue $1$ for $\hat{C}_{3z}$, 
the transformation of the pairing Hamiltonian is given by
\beq
\hat{C}_{3z} H_{\Delta} \hat{C}_{3z}^{-1} =H_{\Delta},
\eneq 
which leads to
\beq
\Delta^{\eta}_{\kk; e_1, e_2} = e^{- i (e_1 \theta_\kk + e_2 \theta_{-\kk})} \Delta^\eta_{\hat{C}_{3z}^{-1}\kk; e_1, e_2}
\eneq irrespective of the triplet and singlet paring channels. 
At the $\KK_M$ or $\KK_M'$ points, this gap then becomes
\beq
\Delta^\eta_{\KK_M; e_1, e_2} = e^{- i (e_1 \theta_{\KK_M} + e_2 \theta_{-\KK_M})} \Delta^\eta_{\KK_M; e_1, e_2} 
= e^{- i (e_1+ e_2) \theta_{\KK_M} } \Delta^\eta_{\KK_M, e_1, e_2}
\eneq 
where we have used $\theta_{\KK_M}= \theta_{-\KK_M}= 2\pi/3$ for the TBG flat bands, as we previously showed. 
This means that any $1D$ irrep will have, for the pairing between the bands with equal Chern number (called "intra-Chern-band pairing" below), 
\beq
\Delta_{\KK_M; e_Y, e_Y} = 0
\eneq 
No such condition exists for pairing between opposite Chern number bands $e_{Y_1} = - e_{Y_2}$ (called "inter-Chern-band pairing" below). 
Hence we have the following theorem: \emph{ The gap function of the 1D irrep pairing channel between two bands with the equal Chern number $\pm 1$ (intra-Chern-band pairing) of the $\hat{C}_{6v}$ symmetry group has nodes at $\KK_M$, $\KK_M'$.} 

Based on the above discussion, we can summarize the 1D irrep pairing channels as the following. 
\begin{itemize}
   \item $A_1, A_2$ irrep channels: 
\beq
\text{Singlet: } \Delta^\eta_{\kk; e_1, e_2} =  \Delta^\eta_{\kk; e_2, e_1};  \;\;\; \text{Triplet: } \Delta^\eta_{\kk; e_1, e_2} =  - \Delta^\eta_{\kk; e_2, e_1};\;\;\;  \Delta^\eta_{\kk; e_1, e_2} = e^{- i (e_1 \theta_\kk + e_2 \theta_{-\kk})} \Delta^\eta_{\hat{C}_{3z}^{-1}k, e_1, e_2}. 
\eneq
For these irrep channels, the singlet pairing between the bands with the same Chern number has a node at $\KK_M$, while the triplet pairing between the bands 
with the same Chern number vanishes identically as $\lambda_{C_{2z}}=1$ for the $A_1,A_2$ pairing channels. 

\item $B_1, B_2$ irrep channels:
\beq
\text{Singlet: } \Delta^\eta_{\kk; e_1, e_2} =  -\Delta^\eta_{\kk; e_2, e_1};  \;\;\; \text{Triplet: } \Delta^\eta_{\kk; e_1, e_2} =  \Delta^\eta_{\kk; e_2, e_1};\;\;\;  \Delta^\eta_{\kk; e_1, e_2} = e^{- i (e_1 \theta_\kk + e_2 \theta_{-\kk})} \Delta_{\hat{C}_{3z}^{-1}\kk; e_1, e_2}
\eneq For these representations, the triplet pairing between same Chern number bands has a zero at $K_M$, while the singlet pairing between the same Chern numbers vanishes identically as $\lambda_{C_{2z}}=-1$ for the $B_1,B_2$ pairing channels. 

\end{itemize}

We now move to the $2D$ irrep pairing channels, which necessarily require two pairing terms $(H_{\Delta_1}, H_{\Delta_2}) $ and two different gaps $\Delta_{1; \kk; e_1, e_2}, \Delta_{2; \kk; e_1, e_2}$. The pairing Hamiltonian is written as
\begin{eqnarray}\label{eq:GapFun2D}
&H_{\Delta} = H_{\Delta_1} + H_{\Delta_2} +  H^\dagger_{\Delta_1} + H^\dagger_{\Delta_2} \\
&H_{\Delta_1}=\sum_{\kk, e_{Y_1}, e_{Y_2}, \eta } \gamma_{\kk, e_{Y_1}, \eta }^\dagger \Delta^{\eta}_{1; \kk; e_{Y_1}, e_{Y_2} } \mS  \gamma_{-\kk, e_{Y_2}, -\eta }^\dagger \nonumber\\
&H_{\Delta_2}=\sum_{\kk, e_{Y_1}, e_{Y_2}, \eta } \gamma_{\kk, e_{Y_1}, \eta }^\dagger \Delta^{\eta}_{2; \kk; e_{Y_1}, e_{Y_2} } \mS  \gamma_{-\kk, e_{Y_2}, -\eta }^\dagger \nonumber. 
\end{eqnarray}
Here we have suppressed the spin index and all the spin information is included in the $\mS$ function (singlet or triplet).
According to the table \ref{tab:IrreduciblePairing}, the $\hat{C}_{2z}$ symmetry should give the constraints
\beq
\hat{C}_{2z} H_{\Delta_1} \hat{C}_{2z}^{-1} =\lambda_{C_{2z}} H_{\Delta_1},\;\; \hat{C}_{2z} H_{\Delta_2} \hat{C}_{2z}^{-1} =\lambda_{C_{2z}} H_{\Delta_2};\;\;\; \Delta^\eta_{1; \kk; e_1, e_2} = \pm\lambda_{C_{2z}}\Delta^\eta_{1; \kk; e_2, e_1};\;\;\Delta^\eta_{2; \kk; e_1, e_2} = \pm\lambda_{C_{2z}}\Delta^\eta_{2; \kk; e_2, e_1}
\eneq 
with $+$ for singlet pairing and $-$ for triplet pairing (and $\lambda_{C_{2z}} = - (+)$ for $E_1(E_2)$ irrep channels, respectively). 
One notices that $(H_{\Delta_1}$ and $H_{\Delta_2})$ have to be both singlet or both triplet. The actual order parameter can be a combination of $a H_{\Delta_1}+ b H_{\Delta_2}$, where $a$ and $b$ can be complex numbers. This breaks $\hat{C}_{3z}$ symmetry and is hence nematic. Another possibility would be a time-reversal breaking order parameter, which can respect $\hat{C}_{3z}$.  We now treat each of the two $2D$ irrep channels separately as follows. 
\begin{itemize}
   \item $E_2$ irrep channels: 
   We have
   \beq
   \hat{C}_{3z}H_{\Delta_{\alpha}} \hat{C}_{3z}^{-1} = H_{\Delta_\beta}(e^{-i \frac{2\pi}{3} \sigma_y})_{\beta\alpha} \implies e^{-i (e_1\theta_\kk + e_2\theta_{-\kk})} 
   \Delta^\eta_{\alpha; \hat{C}_{3z}^{-1}\kk; e_1, e_2} = \Delta^\eta_{\beta; \kk; e_1, e_2} (e^{-i \frac{2\pi}{3} \sigma_y})_{\beta\alpha},
    \eneq
    where $\alpha,\beta=1,2$ indices represent the different pairing channels and $\sigma$ is the corresponding Pauli matrix. For $\kk=\Gamma_M$ with $\theta_{\Gamma_M}=0$, the equation above becomes 
    \beq\label{eq:gapfun_E2_node}
    \Delta^\eta_{2; \Gamma_M; e_1, e_2} =\sqrt{3}  \Delta^\eta_{1;\Gamma_M; e_1, e_2};\;\;     
    \Delta^\eta_{1;\Gamma_M; e_1, e_2} =-\sqrt{3}  \Delta^\eta_{2;\Gamma_M; e_1, e_2} \implies  \Delta^\eta_{1;\Gamma_M; e_1, e_2}= \Delta^\eta_{2;\Gamma_M; e_1, e_2}=0\eneq 
    Hence the nematic $2D$ irrep $E_2$ channel has a zero in the gap function at the $\Gamma_M$ point for both the same and different Chern number pairing. 
    This is different from the $1D$ irrep pairing channels. At $\KK_M$, we have
    \beq \Delta^\eta_{1;\KK_M; e_1, -e_1}= \Delta^\eta_{2;\KK_M; e_1, -e_1}=0;\;\;\;\; i e_1 \Delta^\eta_{1;\KK_M; e_1, e_1}= \Delta^\eta_{2;\KK_M; e_1, e_1}
    \eneq Hence the pairing between the bands with opposite Chern number is zero at $\KK_M$, while the pairing between the bands with the same Chern number is nonzero and satisfies the constraint above. 
    From $\hat{C}_{2z}$, the $\lambda_{C_{2z}} =1$ for the $E_2$ representation and we have
    \beq 
   \Delta^\eta_{1;\kk e_1, e_2} = \pm \Delta_{1;\kk e_2, e_1};\;\;\Delta_{2;\kk; e_1, e_2} = \pm\Delta_{2;\kk; e_2, e_1},
   \eneq 
   where, again, we have $\pm$ for singlet/triplet. We can then see that for same chern number pairing, there cannot be a triplet component. 
   
\item  $E_1$ irrep channels: 
  We have
   \beq
   \hat{C}_{3z}H_{\Delta_{\alpha}} \hat{C}_{3z}^{-1} = H_{\Delta_\beta}(e^{i \frac{2\pi}{3} \sigma_y})_{\beta\alpha} \implies e^{-i (e_1\theta_\kk + e_2\theta_{-\kk})} \Delta^\eta_{\alpha; \hat{C}_{3z}^{-1} \kk; e_1, e_2} = \Delta^\eta_{\beta; \kk; e_1, e_2} (e^{i \frac{2\pi}{3} \sigma_y})_{\beta\alpha}.  
    \eneq
    For $\kk=\Gamma_M$ and $\theta_{\Gamma_M}=0$, the equation above becomes 
    \beq
    \Delta^\eta_{2;\Gamma_M; e_1, e_2} =-\sqrt{3}  \Delta^\eta_{1;\Gamma_M; e_1, e_2};\;\;     
    \Delta^\eta_{1;\Gamma_M; e_1, e_2} =\sqrt{3}  \Delta^\eta_{2;\Gamma_M; e_1, e_2} \implies  
    \Delta^\eta_{1;\Gamma_M; e_1, e_2}= \Delta^\eta_{2;\Gamma_M; e_1, e_2}=0. 
    \eneq 
    Hence, similar as the $E_2$ irrep case, the nematic $2D$ irrep $E_1$ channel also has a zero in the gap function at the $\Gamma_M$ point for both the same and different Chern number pairing. At $K_M$ we have
    \beq
    \Delta^\eta_{1;\KK_M; e_1, -e_1}= \Delta^\eta_{2;\KK_M; e_1, -e_1}=0;\;\;\;\; i e_1 \Delta_{1;\KK_M; e_1, e_1}=- \Delta_{2;\KK_M; e_1, e_1}
    \eneq 
    Hence, the pairing between the bands with different Chern numbers is zero at $\KK_M$ while the pairing between the bands with the same Chern number is nonzero and satisfies the constraint above. From $\hat{C}_{2z}$, the $\lambda_{C_{2z}} =-1$ for the $E_2$ representation and we have
  \beq 
   \Delta^\eta_{1;\kk; e_1, e_2} = \mp \Delta^\eta_{1;\kk; e_2, e_1};\;\;\Delta^\eta_{2; \kk; e_1, e_2} = \mp \Delta^\eta_{2;\kk; e_2, e_1}
   \eneq 
   with $\mp$ for singlet/triplet. We can then see that for the pairing between the bands with the same Chern number, there cannot be a singlet component. 
   
\end{itemize}

Finally, let's discuss the constraint due to the presence of time reversal (TR) symmetry. It should be noted that TR may be spontaneously broken by the gap function, 
so the constraint derived below can only be applied to the TR invariant pairing. 
We can combine the spinless $\hat{C}_{2z}$ with the \emph{spinful} TR $\hat{T}$ (both are independent symmetries of the TBG system), which gives
\beq
\hat{C}_{2z}T \gamma^\dagger_{\kk, e_Y, \eta, s} (\hat{C}_{2z}\hat{T})^{-1} =\sum_{s'}\gamma^\dagger_{\kk, -e_Y, \eta, s'} (i \mS_y)_{s's}.
\eneq
Applying this symmetry to the pairing Hamiltonian (\ref{eq:GapFun1}) implies that
\beq     \epsilon_{s_1' s_1} (\Delta^\eta_{\kk; -e_{Y_1} s_1, -e_{Y_2} s_2})^\star  \epsilon_{s_2's_2} = \Delta_{k; e_{Y_1} s_1', e_{Y_2} s_2'} ,
\eneq where $\epsilon_{s_1' s_1}$ is the Levi-Civita symbol. Hence, for spin singlet and triplet, we have two transformations for the gap function, 
\begin{eqnarray}\label{eq:C2zT_Gapfunction}
    &\text{Singlet: } \epsilon_{s_1' s_1} (\Delta^\eta_{\kk; -e_{Y_1}, -e_{Y_2} })^\star\epsilon_{s_1,s_2}  \epsilon_{s_2's_2} = \Delta^\eta_{\kk; e_{Y_1} , e_{Y_2}}   \epsilon_{s_1',s_2'} \implies   \boxed{ (\Delta^\eta_{\kk; -e_{Y_1}, -e_{Y_2} })^\star =  \Delta_{\kk; e_{Y_1}, e_{Y_2} } } \nonumber \\ 
    & \text{Triplet: } (\Delta^\eta_{\kk; -e_{Y_1}, -e_{Y_2} })^\star(\dd\cdot \ss)^\star_{s_1 s_2}= \Delta^\eta_{\kk; e_{Y_1}, e_{Y_2} }(\dd\cdot \ss)_{\bar{s}_1 \bar{s}_2} (-1)^{s_1+s_2} \implies  \boxed{ (\Delta^\eta_{\kk; -e_{Y_1}, -e_{Y_2} })^\star =  - \Delta^\eta_{\kk; e_{Y_1}, e_{Y_2} } } 
     \end{eqnarray}
where $s=1,2$ are $\uparrow, \downarrow$ while $\bar{1}=2$ and vice versa.


\subsubsection{Continuous Symmetry Analysis of Irreducible Pairing Channels} \label{Sec:constinuous_symmetry_pairing_Channels}
Now let's consider the constraint from the continuous symmetry in the chiral limit on the phonon-mediated electron-electron interaction in Eq. (\ref{eq:Elel_interaction_ChernBasis}) when taking the mean field decomposition for different pairing channels.



In the chiral limit, we have an $SU(2)$ for each chern number sector and we now separate the two $SU(2)$ operators per chern number sector into 
\beq
 S^{+}_{e_Y}= \sum_{\kk \eta } \gamma^\dagger_{\kk, e_Y, \eta, \uparrow } \gamma_{\kk, e_Y, \eta, \downarrow},\;\;\;\; 
 S^{-}_{e_Y}= \sum_{\kk \eta } \gamma^\dagger_{\kk, e_Y, \eta, \downarrow } \gamma_{\kk, e_Y, \eta, \uparrow},
 \eneq which can be further linearly combined to form the total (tot) spin and relative (rel) spin raising and lowering operators,
 \begin{eqnarray}
 & S^+_{tot}= \frac{1}{\sqrt{2}} (S^+_{+}+S^+_{-}) ,\;\;\;  S^-_{tot}= \frac{1}{\sqrt{2}} (S^-_{+}+S^-_{-})  \nonumber \\ &  S^+_{rel}= \frac{1}{\sqrt{2}} (S^+_{+}-S^+_{-}) ,\;\;\;  S^-_{rel}= \frac{1}{\sqrt{2}} (S^-_{+}-S^-_{-}).  
 \end{eqnarray}
 We first separate the gap function based on the total spin - singlet and triplet components. This will be the same even away from the chiral limit. 
 
 We now define the two-fermion pairing operators
 \beq
 \Pi^\dagger_{\kk,\eta; e_1s_1; e_2, s_2}= \gamma_{\kk, e_1, \eta s_1}^\dagger \gamma^\dagger_{-\kk, e_2, -\eta s_2}
\eneq  and decompose them into singlet and triplet channels, 
 \beq
 \Pi^\dagger_{\kk,\eta, e_1s_1; e_2, s_2}=\sum_{S=0,1; M= -S\ldots S}  C_{s_1 s_2}^{SM} \Pi^\dagger_{\kk, \eta; e_1, e_2, S,M}
 \eneq where $C_{s_1 s_2}^{SM}$ are the Clebsh-Godron Coefficients. The operator form of the $S,M$ pairing channel is then given by
 \begin{eqnarray}
& \Pi^\dagger_{\kk,+;e_1, e_2, 0,0} = \frac{1}{\sqrt{2}} (\gamma_{\kk, e_1, +, \uparrow}^\dagger\gamma_{-\kk, e_2, -, \downarrow}^\dagger - \gamma_{\kk, e_1, +, \downarrow}^\dagger\gamma_{-\kk, e_2, -, \uparrow}^\dagger   ) \nonumber \\ &  \Pi^\dagger_{\kk,+;e_1, e_2, 1,0} = \frac{1}{\sqrt{2}} (\gamma_{\kk, e_1, + \uparrow}^\dagger\gamma_{-\kk, e_2, -\downarrow}^\dagger + \gamma_{\kk, e_1, +, \downarrow}^\dagger\gamma_{-\kk, e_2, -, \uparrow}^\dagger   ) & \nonumber \\ & \Pi^\dagger_{\kk, +; e_1, e_2, 1,1}= \gamma^\dagger_{\kk,e_1,+, \uparrow}\gamma^\dagger_{-\kk,e_2,-, \uparrow}& \nonumber \\ & \Pi^\dagger_{\kk,+;e_1, e_2, 1,-1}= \gamma^\dagger_{\kk,e_1,+, \downarrow}\gamma^\dagger_{-\kk,e_2,-, \downarrow}. \end{eqnarray}
 Besides the usual rotations within spin $1$ by the raising and lowering total spin operators
  \begin{eqnarray}
  S_{tot}^-\Pi^\dagger_{\kk,+;e_1, e_2, 1,1} \ket{0} = \Pi^\dagger_{\kk,+;e_1, e_2, 1,0} \ket{0},\;\;\; (S_{tot}^-)^2\Pi^\dagger_{\kk,+;e_1, e_2, 1,1} \ket{0} = \Pi^\dagger_{\kk,+;e_1, e_2, 1,-1} \ket{0},
 \end{eqnarray} we also have, for $e_1= e_2$, 
 \begin{eqnarray}
 & S_{rel}^-\Pi^\dagger_{\kk,+;e_1, e_1, 1,1 }\ket{0}  = e_1 \Pi^\dagger_{\kk,+;e_1, e_2, 1,0} \ket{0} ;\;\;\;\;  (S_{rel}^-)^2\Pi^\dagger_{\kk,+;e_1, e_1, 1,1 }\ket{0}  =  \Pi^\dagger_{\kk,+;e_1, e_2, 1,-1} \ket{0}
 \end{eqnarray} As such, we can see that the action of $S_{rel}^{-}$ and $S_{tot}^-$ on the pairing states with the same Chern number is identical (albeit some signs).  
 
\emph{However}, for $e_1\ne  e_2$, we have that the singlet and the triplet states are related, due to the relative (rel) $SU(2)$ we find:
\begin{eqnarray}
 & S_{rel}^-\Pi^\dagger_{\kk,+;e_1, -e_1, 1,1 }\ket{0}  = -e_1 \Pi^\dagger_{\kk,+;e_1, -e_1, 0,0} \ket{0} ;\;\;\;\; & S_{rel}^-\Pi^\dagger_{\kk,+;e_1, -e_1, 0,0 }\ket{0}  = e_1 \Pi^\dagger_{\kk,+;e_1, -e_1, 1,-1} \ket{0} ;\;\;\;\;   
 \end{eqnarray} with similar relations for $\Pi^\dagger_{\kk,-;e_1, -e_1, S,M}$ and hence the triplet pairing is actually related to the singlet pairing of the inter-Chern-band pairing, which implies the \emph{inter-Chern-band singlet and triplet pairings to be degenerate}, in the chiral flat-band limit; while the intra-Chern-band triplet and singlet pairings do not need this degeneracy. 

 We now can rewrite the Hamiltonian (\ref{eq:Elel_interaction_ChernBasis}) into the form
\beq
H_{el-el}= - \frac{1}{N_M} \sum_{\kk,\kk'}\sum_{\eta, e_1, e_2, s_1, s_2}V_{\kk,\kk'}^{\eta e_1 e_2}\sum_{S=0,1}\sum_{M=-S\ldots S} \Pi^\dagger_{\kk,\eta;e_1, e_2, S,M}\Pi_{\kk', -\eta;e_1, e_2, S,M}. 
 \eneq
With
\beq
 \Pi^\dagger_{\kk,\eta e_1s_1; e_2, s_2}=- \Pi^\dagger_{-\kk,-\eta; e_2s_2; e_1, s_1}
 \eneq we have
 \beq
 \Pi^\dagger_{\kk, \eta; e_1, e_2,0,0}= \Pi^\dagger_{-\kk, -\eta; e_2, e_1,0,0}; \;\;\; \Pi^\dagger_{\kk, \eta; e_1, e_2,1,M}= -\Pi^\dagger_{-\kk, -\eta; e_2, e_1,1,M}, \;\; M=-1,0,1
 \eneq
Using the reshuffling property, we find that 
  \beq
H_{el-el}= -\frac{2}{N_M} \sum_{\kk,\kk'}\sum_{ e_1, e_2, s_1, s_2}V_{\kk,\kk'}^{+ e_1 e_2}\sum_{S=0,1}\sum_{M=-S\ldots S} 
\Pi^\dagger_{\kk,+;e_1, e_2, S,M}\Pi_{\kk', -;e_1, e_2, S,M}
 \eneq

\subsection{Gap equation and Possible Pairing Channels for the Flat Bands}
\subsubsection{Linearized gap equation}
In this section, we will first drive the linearized gap equation for the TBG Hamiltonian and discuss the properties of the linearized gap equation. 

Now let us consider the general interaction
\beq
H_{el-el}=-\frac{1}{N_M}\sum_{\kk\kk'}\sum_{\alpha \beta \mu \nu} V_{\alpha, \beta, \mu, \nu}(\kk, \kk') \gamma_{\kk,\alpha s_1 }^\dagger \gamma_{-\kk,\beta s_2 }^\dagger \gamma_{-\kk' \mu s_2} \gamma_{\kk' \nu s_1}
\eneq with composite index $\alpha = e_1, \eta; \beta = e_2, -\eta; \mu= e_2, \eta; \nu = e_1 , - \eta $, and the interaction function $V$ is defined as 
\beq
V_{\alpha, \beta, \mu, \nu}(\kk,\kk')= V_{ e_1, \eta; e_2, -\eta;  e_2, \eta; e_1 , - \eta  } (\kk,\kk')= V_{\kk,\kk'}^{\eta e_1 e_2}.
\eneq 

Next we define the gap function as
\beq \label{eq:Decomposition_GapFunctionDef}
 \Delta^{\eta}_{\kk; e_{1} s_1, e_2 s_2}= - \frac{1}{N_M}  \sum_{\kk'} V_{\kk\kk'}^{\eta e_1 e_2} \langle \gamma_{-\kk'  e_2 \eta s_2} \gamma_{\kk' e_1 -\eta s_1}\rangle, 
\eneq
or 
\beq 
 \Delta_{\kk; \alpha s_1, \beta s_2}= - \frac{1}{N_M}  \sum_{\kk'} V_{\alpha\beta\mu\nu}(\kk,\kk') \langle \gamma_{-\kk' \mu s_2} \gamma_{\kk' \nu s_1}\rangle, 
\eneq
in terms of the notation $\alpha,\beta,\mu,\nu$. With the above form of the gap function, we can perform the mean field decomposition of the 
interaction Hamiltonian, 
\beq
H_{el-el}=\frac{1}{N_M}\sum_{\kk}\sum_{\alpha \beta \mu \nu} \left( \Delta_{\kk;\alpha s_1,\beta s_2} \gamma^{\dagger}_{\kk,\alpha s_1} \gamma^{\dagger}_{-\kk,\beta s_2} + \Delta^*_{\kk;\alpha s_1,\beta s_2} \gamma_{-\kk,\beta s_2}\gamma_{\kk,\alpha s_1} - \Delta_{\kk;\alpha s_1,\beta s_2} \langle \gamma^{\dagger}_{\kk,\alpha s_1} \gamma^{\dagger}_{-\kk,\beta s_2} \rangle \right) . 
\eneq
We introduce the anomalous Green's function in the imaginary time $\tau$ as
\beq \mF_{\nu s_1,\mu s_2}(\kk,\tau) = \langle T_{\tau} \gamma_{\kk\nu s_1}(\tau) \gamma_{-\kk \mu s_2}(0)\rangle, 
\eneq
where $\langle ...\rangle$ represents the thermal average, and its Fourier transform 
\beq \mF_{\nu s_1, \mu s_2}(\kk,\tau) = \frac{1}{\beta} \sum_{i\omega_n } \mF_{\nu s_1, \mu s_2}(\kk,i \omega_n) e^{-i \omega_n \tau},  
\eneq
where $\beta=1/k_BT$ and $i\omega_n= (2n+1) \pi/\beta$ is the Matsubara frequency.  

The gap function can be related to the anomalous Green's function by 
\beq \label{eq_Gap_equation_1} \Delta_{\kk; \alpha s_1, \beta s_2}=\sum_{\kk',\mu\nu} V_{\alpha\beta\mu\nu}(\kk \kk') \mF_{\nu s_1, \mu s_2}(\kk',0)
=\frac{1}{\beta} \sum_{\kk',i\omega_n,\mu\nu} V_{\alpha\beta\mu\nu}(\kk \kk') \mF_{\nu s_1, \mu s_2}(\kk',i\omega_n). 
\eneq

With the above definition, we can derive the equation of motion for the anomalous Green's function, given by 
\beq \frac{d\mF_{\nu s_1, \mu s_2}(\kk,\tau)}{d\tau} = -\sum_{\mu'} h_{\kk,\nu\mu'}\mF_{\mu' s_1,\mu s_2}(\kk,\tau) - 2 \sum_{\beta s_1'}
\Delta_{\kk, \nu s_1, \beta s_1'} G_{\mu s_2, \beta s_1'}(-\kk,-\tau), 
\eneq
where $h_{\kk,\nu\mu'}$ is the single-particle Hamiltonian 
\beq H_0=\sum_{\kk,\alpha\beta,s} h_{\kk,\alpha \beta} \gamma^\dagger_{\kk,\alpha,s}\gamma_{\kk,\beta s} \eneq
and the Green's function $G$ is defined as
\beq G_{\alpha s_1, \beta s_2}(\kk,\tau) = - \langle T_{\tau} \gamma_{\kk,\alpha s_1}(\tau) \gamma^{\dagger}_{\kk,\beta s_2}(0)  \rangle 
\eneq
and its Fourier transform is 
\beq G_{\alpha s_1, \beta s_2}(\kk,\tau) = \frac{1}{\beta} \sum_{i\omega_n} G_{\alpha s_1, \beta s_2}(\kk,i\omega_n) e^{-i \omega_n \tau}. 
\eneq
In the frequency space, the equation of motion for the anomalous Green's function then reads
\beq 
\sum_{\mu' } (i\omega_n \delta_{\nu \mu'}-h_{\kk,\nu\mu'})\mF_{\mu' s_1, \mu s_2}(\kk,i\omega_n)= 2 \sum_{\mu' s_1'} \Delta_{\kk, \nu s_1, \mu' s_1'} 
G_{\mu s_2 ,\mu' s_1'}(-\kk,-i\omega_n). 
\eneq

Next we can make the approximation of replacing the full Green's function $G$ by the zero-order Green's function $G_0$, in which
the thermal average is only for the single-particle Hamiltonian $H_0$. From the equation of motion, $G_0$ can be solved analytically
and given by 
\beq G_{0; \alpha s_1, \beta s_2}(\kk, i\omega_n)=\delta_{s_1 s_2} [(i\omega_n - h_{\kk})^{-1}]_{\alpha \beta}
\eneq
Thus, up to the lowest order in the gap function $\Delta$, we find the anomalous Green's function is given by 
\beq \mF_{\nu s_1, \mu s_2}(\kk,i\omega_n) = 2 \sum_{\alpha \beta, s_1' s_2'} G_{0; \nu s_1, \alpha s_1'}(\kk,i\omega_n) \Delta_{\kk, \alpha s_1', \beta' s_2'} 
G_{0; \mu s_2 ,\beta' s_2'}(-\kk,-i\omega_n). 
\eneq

Substituting the above form of $\mF$ into Eq. (\ref{eq_Gap_equation_1}), we obtain the linearized gap equation  
\beq\label{eq:Decomposition_LGE}
\Delta_{\alpha s_1 \beta s_2}(\kk) = \frac{2}{\beta N_M} \sum_{i\omega_n,\kk', \mu, \nu} V_{\alpha, \beta, \mu, \nu}(\kk, \kk') [G_0(\kk', i\omega_n) \Delta(\kk') G_0^T(-\kk', -i\omega_n)]_{\nu s_1, \mu s_2}, \eneq 

It should be noted that in the Chern-band basis, the single-particle Green's function is {\it not} diagonal, and the explicit form of the linearized gap equation
on the Chern-band basis is written as
\beq\label{eq:Decomposition_LGE_Chern}
\Delta_{e_1 s_1, e_2 s_2; \kk}^\eta = \frac{2}{\beta N_M} \sum_{i\omega_n,\kk', e_1',e_2'} V^{\eta,e_1e_2}_{\kk \kk'} G_{0,e_1e_1'}^{-\eta}(\kk',i\omega_n)
\Delta^{-\eta}_{e_1's_1, e_2's_2;\kk'} G^{\eta}_{0,e_2 e_2'}(-\kk', -i\omega_n), \eneq 



It should be noted that the left hand side of the above equation is for gap function $\Delta_{e_1 s_1, e_2 s_2; \kk}^\eta=\Delta_{e_1\eta s_1,e_2 -\eta s_2}$, 
which has the opposite $\eta$ value from $\Delta^{-\eta}_{e_1' s_1, e_2' s_2;\kk'}=\Delta_{e_1' -\eta s_1,e_2' \eta s_2}$ on the right hand side of the equation. 
On the Chern-band basis, the single particle Hamiltonian reads
\beq h_{\eta,\kk}= (d_{0,\eta,\kk}-\mu)\zeta^0+d_{x,\eta,\kk}\zeta^x, \eneq
with
\beq d_{0,\eta,\kk}=(\epsilon_{+,\eta,\kk}+\epsilon_{-,\eta,\kk})/2;  \;\;
d_{x,\eta,\kk}=(\epsilon_{+,\eta,\kk}-\epsilon_{-,\eta,\kk})/2, \eneq
and $\epsilon_{n, \eta,\kk}$ are the single-particle eigen-energies of the BM model $\hat{H}_{0}$ so they should be real. 
Time reversal symmetry requires 
\beq \epsilon_{n, \eta,\kk}=\epsilon_{n, -\eta,-\kk} \implies d_{0,-\eta,-\kk}=d_{0,\eta,\kk};\; d_{x,-\eta,-\kk}=d_{x,\eta,\kk} \implies h_{-\eta,-\kk}=h_{\eta,\kk} .  \eneq 
The corresponding single-particle Green function is given by  
\beq G_{0}^{\eta}(\kk,i\omega_n)=(i\omega_n-h_{\eta,\kk})^{-1}=\sum_m \frac{P_{m,\kk,\eta}}{i\omega_n-\xi_{m,\eta,\kk}}, \eneq 
where $m=\pm$, $\xi_{m,\eta,\kk}=d_{0,\eta,\kk}-\mu+m |d_{x,\eta,\kk}|$ and the projection operator is defined as
\beq \label{eq:Projector_singleHam} P_{m,\kk,\eta}=\frac{1}{2}\left(1+m \frac{d_{x,\eta,\kk}}{|d_{x,\eta,\kk}|} \zeta^x\right). 
\eneq 
With such form of the single-particle Green's function, the linearized gap equation can be written as
\beq \label{eq:LGE_nonflat_Green}
\Delta_{e_1 s_1, e_2 s_2; \kk}^\eta = \frac{2}{N_M} \sum_{\kk', m_1 m_2} V^{\eta,e_1e_2}_{\kk \kk'} \left[P_{m_1,\kk',-\eta}\Delta^{-\eta}_{s_1 s_2,\kk'}P^T_{m_2,-\kk',\eta}\right]_{e_1 e_2} T^{\eta}_{m_1 m_2 \kk'}, \eneq 
where the Mastubara frequency summation can be performed
\beq\label{eq:LGE_Mastubara_summation}
T^{\eta}_{m_1 m_2 \kk'}=\frac{1}{\beta}\sum_{i\omega_n}\frac{1}{i\omega_n-\xi_{m_1,-\eta,\kk'}}\frac{1}{-i\omega_n-\xi_{m_2,\eta,-\kk'}}
=-\frac{n_F(\xi_{m_1, -\eta, \kk'})-n_F(-\xi_{m_2, \eta, -\kk'})}{\xi_{m_1, -\eta, \kk'}+\xi_{m_2, \eta, -\kk'}}. 
\eneq
In the flat band limit, we have $\xi_{\kk,e_1\eta}=\xi_{\kk}\ll k_B T$, and up to the lowest order term, we find 
\beq 
T^{\eta}_{m_1 m_2 \kk'}\approx \frac{\beta}{4}, \eneq
which is independent of $m_1, m_2, \kk', \eta$. In this limit, the gap equation is simplified to 
\beq
\Delta_{e_1 s_1, e_2 s_2; \kk}^\eta = \frac{\beta}{2 N_M} \sum_{\kk', m_1 m_2} V^{\eta,e_1e_2}_{\kk \kk'} \left[P_{m_1,\kk',-\eta}\Delta^{-\eta}_{s_1 s_2,\kk'}P^T_{m_2,-\kk',\eta}\right]_{e_1 e_2}.  \eneq 
Furthermore, one can show that 
\beq 
\sum_{m_1 m_2} P_{m_1,\kk',-\eta}\Delta^{-\eta}_{s_1 s_2,\kk'}P^T_{m_2,-\kk',\eta} = \Delta^{-\eta}_{s_1 s_2,\kk'},
\eneq 
and thus in the flat band limit, the linearized gap equation is simplified as

\beq\label{eq:LGE_gapequation1}
2k_B T\Delta^{\eta}_{\kk;e_1 s_1,e_2 s_2}=\frac{1}{N_M} \sum_{\kk'}V^{\eta e_1 e_2}_{\kk,\kk'} \Delta^{-\eta}_{\kk';e_1 s_1,e_2 s_2}, 
\eneq
where we denote $\Delta^{\eta}_{\kk; e_1 s_1,e_2 s_2}=\Delta_{e_1 \eta s_1,e_2 -\eta s_2}(\kk)$. The above equation can be written in an eigen-problem
form
\begin{eqnarray}\label{eq:LGE_Eigen_GapEquation1}
2k_BT \left(\begin{array}{c}
\Delta^{\eta}_{\kk; e_1 s_1, e_2 s_2}\\
\Delta^{-\eta}_{\kk; e_1 s_1, e_2 s_2}
\end{array}\right)
=\frac{1}{N_M} \sum_{\kk'}\mathcal{V}^{\eta,e_1 e_2}_{\kk \kk'}
\left(\begin{array}{c}
\Delta^{\eta}_{\kk'; e_1 s_1, e_2 s_2}\\
\Delta^{-\eta}_{\kk'; e_1 s_1, e_2 s_2}
\end{array}\right),  
\end{eqnarray}
where 
\begin{eqnarray}
\mathcal{V}^{\eta,e_1 e_2}_{\kk \kk'}=\left(\begin{array}{cc}
0&V^{\eta e_1 e_2}_{\kk \kk'}\\
V^{-\eta e_1 e_2}_{\kk \kk'}&0
\end{array}\right)
\end{eqnarray}
needs to be diagonalized and possesses a off-diagonal form. 

We can decompose the gap function into the orbital part and spin part. Let's denote the spin part as $\mS^{SM}_{s_1 s_2}$, where $S=0,1$ and $M=-S,...,S$. 
Explicitly, $\mS^{00}=i s_y$ for spin singlet and $\mS^{1,\pm 1}=\frac{i}{\sqrt{2}}(s_x \pm i s_y)s_y=\frac{1}{\sqrt{2}}(-s_z\mp s_0), \mS^{1,0}=s_x$ for spin triplet, where $s_0$ is the identity matrix and $s_{x,y,z}$ are three Pauli matrices.  
Then we can decompose
$\Delta^{\eta}_{\kk; e_1 s_1, e_2 s_2}=\sum_{S,M} \Delta^{\eta,SM}_{\kk; e_1 e_2}\mS^{SM}_{s_1 s_2}$. Since $V^{\eta e_1 e_2}_{k k'}$ is independent of spin
and $Tr((\mS^{SM})^\dagger\mS^{S'M'})=2\delta_{SS'}\delta_{MM'}$, the eigen-equation for the gap function is 
\begin{eqnarray}\label{eq:LGE_Eigen_GapEquation2}
2k_BT \left(\begin{array}{c}
\Delta^{\eta,SM}_{\kk; e_1 e_2}\\
\Delta^{-\eta,SM}_{\kk; e_1 e_2}
\end{array}\right)
=\frac{1}{N_M} \sum_{\kk'}\mathcal{V}^{\eta,e_1 e_2}_{\kk \kk'}
\left(\begin{array}{c}
\Delta^{\eta,SM}_{\kk'; e_1 e_2}\\
\Delta^{-\eta,SM}_{\kk'; e_1 e_2}
\end{array}\right) 
\end{eqnarray} for any $S,M$. 

Since we only have two independent $V^{\eta e_1 e_2}_{\kk,\kk'}$ for the fix $\kk,\kk'$, namely one complex $V^{+ + +}_{\kk,\kk'}$ and
one real $V^{+ + -}_{\kk,\kk'}$, we can explicitly write down two independent equations explicitly, namely, 
\begin{eqnarray}\label{eq:LGE_Numericalpp}
2k_BT \left(\begin{array}{c}
\Delta^{+,SM}_{\kk; ++}\\
\Delta^{-,SM}_{\kk; ++}
\end{array}\right)
=\frac{1}{N_M} \sum_{\kk'} \mathcal{V}^{+,++}_{\kk,\kk'}
\left(\begin{array}{c}
\Delta^{+,SM}_{\kk'; ++}\\
\Delta^{-,SM}_{\kk'; ++}
\end{array}\right),  \; \; 
\mathcal{V}^{+,++}_{\kk,\kk'}=\left(\begin{array}{cc}
0&V^{+++}_{\kk \kk'}\\
V^{-++}_{\kk \kk'}&0
\end{array}\right),
\end{eqnarray}
and
\begin{eqnarray}\label{eq:LGE_Numericalpm}
2k_BT \left(\begin{array}{c}
\Delta^{+,SM}_{\kk; +-}\\
\Delta^{-,SM}_{\kk; +-}
\end{array}\right)
=\frac{1}{N_M} \sum_{\kk'}\mathcal{V}^{+,+-}_{\kk,\kk'}
\left(\begin{array}{c}
\Delta^{+,SM}_{\kk'; +-}\\
\Delta^{-,SM}_{\kk'; +-}
\end{array}\right),\;\;
\mathcal{V}^{+,+-}_{\kk,\kk'}=
\left(\begin{array}{cc}
0&V^{++-}_{\kk \kk'}\\
V^{-+-}_{\kk \kk'}&0
\end{array}\right). 
\end{eqnarray}
With the hermitian and reshuffling conditions, one can show $V^{\eta,e_Y,e_Y'}_{\kk,\kk'}=(V^{-\eta,e_Y,e_Y'}_{\kk',\kk})^*$, or more explicitly, 
$V^{-++}_{\kk \kk'}=(V^{+++}_{\kk' \kk})^*$ and $V^{-+-}_{\kk \kk'}=(V^{++-}_{\kk' \kk})^*$, 
so both equations are the eigen-equations for hermitian matrices $\mathcal{V}^{+,e_1 e_2}_{\kk, \kk'}$, 
which can be solved numerically through diagonalization process.

Now let us discuss several properties of the gap function $\Delta^{\eta, SM}_{\kk; e_1 e_2}$. 
Although some results discussed below have been derived in Sec. \ref{Sec:DiscreteSymmetry_GapFunction} from the symmetry view of the
gap function, we will show similar results can be obtained from the linearized gap equation (\ref{eq:LGE_Eigen_GapEquation2}). 
With the reshuffling symmetry Eq. (\ref{eq:GapFun_reshuffling}), $\mS^{00}_{s_1 s_2}=-\mS^{00}_{s_2 s_1}$ and $\mS^{1M}_{s_1 s_2}=\mS^{1M}_{s_2 s_1}$, 
we have 
\beq\label{eq:LGE_antisymmetry_singlet}\Delta^{\eta,00}_{\kk;e_1 e_2}=\Delta^{-\eta,00}_{-\kk;e_2 e_1}\eneq for spin singlet and 
\beq\label{eq:LGE_antisymmetry_triplet}\Delta^{\eta,1M}_{\kk;e_1 e_2}=-\Delta^{-\eta,1M}_{-\kk;e_2 e_1}\eneq for spin triplet. 

As well known for normal superconductors, when there is inversion symmetry, the orbital part of the gap function has definite parity, and 
we only have spin-singlet even-parity or spin-triplet odd-parity gap functions. There are two differences for TBG: (1) 
There is no inversion symmetry for TBG and thus the {\it spinless} $C_{2z}$ symmetry plays the role of inversion; and (2) Due to the
existence of Chern-band index, spin-singlet odd-parity or triplet even-parity states can also exist. Below we will discuss these two points in details. 

Let us first consider the $\hat{C}_{2z}$ symmetry, which requires $V^{\eta e_1 e_2}_{\kk \kk'}=V^{-\eta e_1 e_2}_{-\kk -\kk'}$. Thus, the gap 
equation (\ref{eq:LGE_gapequation1}) can be transformed as 
\begin{eqnarray}
&2k_B T\Delta^{\eta, SM}_{\kk;e_1 e_2}=\sum_{\kk'}V^{\eta e_1 e_2}_{\kk,\kk'} \Delta^{-\eta,SM}_{\kk'; e_1 e_2}
\rightarrow 2k_B T\Delta^{-\eta,SM}_{-\kk;e_1 e_2}=\sum_{\kk'}V^{-\eta e_1 e_2}_{-\kk,-\kk'} \Delta^{\eta,SM}_{-\kk';e_1 e_2}\nonumber\\
&\rightarrow 2k_B T\Delta^{-\eta,SM}_{-\kk;e_1 e_2}=\sum_{\kk'}V^{\eta e_1 e_2}_{\kk,\kk'} \Delta^{\eta,SM}_{-\kk';e_1 e_2}
\end{eqnarray}
In the above derivation, we replace $\kk\rightarrow -\kk, \kk'\rightarrow -\kk', \eta\rightarrow -\eta$ in the first step and use $C_{2z}$ symmetry in the 
second step. We can write the above equation into an eigen-equation form
\begin{eqnarray}
2k_BT \left(\begin{array}{c}
\Delta^{-\eta,SM}_{-\kk; e_1 e_2}\\
\Delta^{\eta, SM}_{-\kk; e_1 e_2}
\end{array}\right)
=\frac{1}{N_M} \sum_{\kk'}
\mathcal{V}^{\eta,e_1 e_2}_{\kk \kk'}
\left(\begin{array}{c}
\Delta^{-\eta,SM}_{-\kk'; e_1 e_2}\\
\Delta^{\eta,SM}_{-\kk'; e_1 e_2}
\end{array}\right),
\end{eqnarray}
which can be compared with Eq. (\ref{eq:LGE_Eigen_GapEquation2}), and one can see that
\begin{eqnarray}\label{eq:C2z_parity}
\Delta^{-\eta,SM}_{-\kk; e_1 e_2}=\lambda_{C_{2z}} \Delta^{\eta,SM}_{\kk; e_1 e_2}
\end{eqnarray}
with $|\lambda_{C_{2z}}|=1$ for any $S,M$. 
If the above procedure is performed twice, we need to go back to the original equation, so $\lambda_{C_{2z}}^2=1$, which means $\lambda_{C_{2z}}=\pm 1$.
This can be understood as the definite parity $\pm 1$ for the gap function with respect to $\hat{C}_{2z}$ symmetry. 

We combine the anti-symmetry condition (Eq. (\ref{eq:LGE_antisymmetry_singlet}) and (\ref{eq:LGE_antisymmetry_triplet})) 
with the definite parity of the orbital part Eq. (\ref{eq:C2z_parity}) and obtain
\beq\label{eq:LGE_constraint_singlet}\Delta^{\eta,00}_{\kk;e_1 e_2}=\lambda_{C_{2z}}\Delta^{\eta,00}_{\kk;e_2 e_1}\eneq for spin singlet and 
\beq\label{eq:LGE_constraint_triplet}\Delta^{\eta,1M}_{\kk;e_1 e_2}=-\lambda_{C_{2z}}\Delta^{\eta,1M}_{\kk;e_2 e_1}\eneq for spin triplet. 

Next let us consider the role of Chern-band index $e_{1,2}$ in the gap function. We decompose the gap function into 
\beq \label{eq:Gapfunction_decomp} \Delta^{\eta}_{\kk;e_1 s_1, e_2 s_2}=\sum_{\mu, SM}\Delta^{\eta,SM}_{\kk;\mu}(\zeta^{\mu})_{e_1 e_2}(\mS^{SM})_{s_1 s_2},
\eneq
where $\mu=0,x,y,z$ and $\zeta^\mu$ are the identity and Pauli matrices for the Chern-band index. The total gap function $\Delta^{\eta}_{\kk;e_1 s_1, e_2 s_2}$ 
needs to be anti-symmetric. For intra-band channels, $\zeta^{\mu}$ should be $\zeta^{0}$ or $\zeta^{z}$, both of which are symmetric matrices. So to make the total gap function being anti-symmetric, the singlet channel $\Delta^{\eta,00}_{\kk;\mu}$ can only be even-parity ($\lambda_{C_{2z}}=+$) and the triplet channel $\Delta^{\eta,1M}_{\kk;\mu}$ can only be odd-parity  ($\lambda_{C_{2z}}=-$) for intra-Chern-band pairings. 
This is consistent with the standard spin-singlet even-parity or spin-triplet odd-parity pairings.

For inter-Chern-band channels, $\zeta^{\mu}$ can be symmetric $\zeta^{x}$ or anti-symmetric $\zeta^{y}$. To make the total gap function anti-symmetric, the singlet channel $\Delta^{\eta,00}_{\kk;\mu}$ can either be even-parity for $\mu=x$ as $\zeta^{x}$ is symmetric, or odd-parity for $\mu=y$ as $\zeta^{y}$ is anti-symmetric. 
Similarly, the triplet channel $\Delta^{\eta,1M}_{\kk;\mu}$ can be either even-parity for $\mu=y$ or odd-parity for $\mu=x$. 
Next we will decompose the linearized gap equation (\ref{eq:LGE_gapequation1}) into the $\zeta^x$ and $\zeta^y$ channels and show that they have the same form. We substitute the decomposition (\ref{eq:Gapfunction_decomp}) into Eq. (\ref{eq:LGE_gapequation1}) and with the orthonormal relation $Tr(\zeta^\mu\zeta^\nu)=2\delta_{\mu\nu}$, we find
\beq 4k_BT\Delta^{\eta,SM}_{\kk;\nu}=\frac{1}{N_M}\sum_{\kk,\mu} 
\left(\sum_{e_1 e_2} V^{\eta e_1 e_2}_{\kk \kk'} (\zeta^\mu)_{e_1 e_2} (\zeta^\nu)_{e_2 e_1} \right) \Delta^{\eta,SM}_{\kk',\mu}. \eneq
For the inter-Chern-band channels, we take $\mu,\nu$ to be $x,y$ and find 
\beq 
\sum_{e_1 e_2} V^{\eta e_1 e_2}_{\kk \kk'} (\zeta^\mu)_{e_1 e_2} (\zeta^\mu)_{e_2 e_1}
=V^{\eta + -}_{\kk \kk'}+ V^{\eta -+}_{\kk \kk'}=2 V^{\eta + -}_{\kk \kk'},\; \mu=x,y
\eneq
\beq 
\sum_{e_1 e_2} V^{\eta e_1 e_2}_{\kk \kk'} (\zeta^x)_{e_1 e_2} (\zeta^y)_{e_2 e_1}
=\sum_{e_1 e_2} V^{\eta e_1 e_2}_{\kk \kk'} \delta_{e_1, -e_2} (-i)\epsilon_{e_2 e_1}
=(-i) (-V^{\eta + -}_{\kk \kk'}+ V^{\eta -+}_{\kk \kk'})=0
\eneq
and
\beq 
\sum_{e_1 e_2} V^{\eta e_1 e_2}_{\kk \kk'} (\zeta^y)_{e_1 e_2} (\zeta^x)_{e_2 e_1}=0, 
\eneq
where we have used $V^{\eta +-}_{kk'}=V^{\eta -+}_{kk'}$ from the constraints of the $V$ function in Eqs. (\ref{eq:Vfun_constraint3}). 
With the above relations, we obtain
\beq 2k_BT\Delta^{\eta,SM}_{\kk;\mu}=\frac{1}{N_M}\sum_{\kk,\mu} V^{\eta + -}_{\kk \kk'} \Delta^{\eta,SM}_{\kk',\mu}, \; \; \mu=x,y.  
\eneq
Thus, $\Delta^{\eta,SM}_{\kk;x}$ and $\Delta^{\eta,SM}_{\kk;y}$ satisfy the same linearized gap equation. 
The largest eigen-value of the above gap equation, which determines the $T_c$ for the inter-Chern-band channel, has a certain $C_{2z}$-parity, 
say even-parity, and then both singlet channel $\Delta^{\eta,00}_{\kk;\mu=x}$ and triplet channel $\Delta^{\eta,1M}_{\kk;\mu=y}$ can exist and are
degenerate with the same $T_c$. This is consistent with the conclusion that singlet and triplet channels are degenerate for the inter-Chern-band pairings in the chiral flat-band limit from the continuous symmetry analysis in Sec. \ref{Sec:constinuous_symmetry_pairing_Channels}.


 \subsubsection{Intra-Chern-band channels from linearized gap equation}\label{sec:LGE_intra}
Now let us discuss the intra-Chern-band channels. Besides Eq. (\ref{eq:LGE_Numericalpp}), we have another linearized gap equation for $V^{\pm,--}_{\kk,\kk'}$, 
\begin{eqnarray}\label{eq:LGE_Numericalmm}
2k_BT \left(\begin{array}{c}
\Delta^{+,SM}_{\kk; --}\\
\Delta^{-,SM}_{\kk; --}
\end{array}\right)
=\frac{1}{N_M} \sum_{\kk'}\left(\begin{array}{cc}
0&V^{+,--}_{\kk \kk'}\\
V^{-,--}_{\kk \kk'}&0
\end{array}\right)
\left(\begin{array}{c}
\Delta^{+,SM}_{\kk'; --}\\
\Delta^{-,SM}_{\kk'; --}.  
\end{array}\right),
\end{eqnarray}
which is directly related to Eq. (\ref{eq:LGE_Numericalpp}) by symmetry.  
Since $C_{2z}T$ requires $V^{\eta,e_Y,e'_Y}_{\kk,\kk'}=(V^{\eta,-e_Y,-e_Y'}_{\kk,\kk'})^*$, the matrix $\mathcal{V}^{+,- - }_{\kk \kk'}$ is just the complex conjugate of that in Eq. (\ref{eq:LGE_Numericalpp}). Therefore, we expect (1) The eigen-equations (\ref{eq:LGE_Numericalpp})
and (\ref{eq:LGE_Numericalmm}) give the same $T_c$, and (2) the corresponding eigen-vectors are related by 
$\left(\begin{array}{c}\Delta^{+,SM}_{\kk; --}\\
\Delta^{-,SM}_{\kk; --}
\end{array}\right)
=\left(\begin{array}{c}
\Delta^{+,SM}_{\kk; ++}\\
\Delta^{-,SM}_{\kk; ++}.  
\end{array}\right)^* $
up to an arbitrary phase factor, which can always be chosen to be $1$. For the convenience of the notation, we define
\begin{eqnarray}\label{eq:Def_PairingFun}
\Psi^{SM}_{\kk, e_Y e'_Y}=\left(\begin{array}{c}\Delta^{+,SM}_{\kk; e_Y e'_Y}\\
\Delta^{-,SM}_{\kk; e_Y e'_Y}
\end{array}\right)
\end{eqnarray}
so the above relation can be simplified as
\begin{eqnarray}\label{eq:LGE_Deltaintra}
\Psi^{SM}_{\kk, - -}=(\Psi^{SM}_{\kk, ++})^*. 
\end{eqnarray}

Before we show the numerical results, we first consider some simplified version for the analytical results. Let us
consider the phonon-mediated el-el interaction with the following form for the intra-Chern-band channel \cite{song2022magic,liu2022future1}
\begin{eqnarray}\label{eq:Interactionform_intra_Chern1}
V^{\eta,e_Y,e_Y}_{\kk,\kk'}=U^*_{0,e_Y,\kk} U_{0,e_Y,\kk'}+U^*_{1,e_Y,\kk} U_{1,e_Y,\kk'};\; U_{0,e_Y,\kk}=\frac{\sqrt{A} b^2}{k^2+b^2}; \; U_{1,e_Y,\kk}=\frac{\sqrt{V_0}}{k^2+b^2} k_{e_Y}^2; \;\;  k_{e_Y}=k_x+i e_Y k_y, 
\end{eqnarray}
where $e_Y=\pm$ and $A$, $V_0$ and $b$ are material dependent parameters with their values $A=0.44 meV$, $V_0=0.4 meV$ and $b=0.185 k_\theta$ obtained from the heavy fermion model for TBG. 
We note that $U_0$ is independent of $e_Y$ index (for s-wave) but the form of $U_1$ critically depends on $e_Y$ index (for d-wave). Both $U_0$ and $U_1$ are independent of the valley $\eta$ index. The corresponding linearized gap equation for intra-Chern-band channel is written as 
\begin{eqnarray}
2k_BT \left(\begin{array}{c}
\Delta^{+,SM}_{\kk,e_Y e_Y}\\
\Delta^{-,SM}_{\kk,e_Y e_Y}
\end{array}\right)
=\frac{1}{N_M}\sum_{\kk',\alpha}U_{\alpha,e_Y,\kk'}U^*_{\alpha,e_Y,\kk}\left(\begin{array}{cc}
0&1\\
1&0
\end{array}\right)
\left(\begin{array}{c}
\Delta^{+,SM}_{\kk',e_Y e_Y}\\
\Delta^{-,SM}_{\kk',e_Y e_Y}
\end{array}\right),
\end{eqnarray}
where $\alpha=0,1$. The above decomposition form of the interaction allows us to define
\beq
\tilde{\Delta}^{\eta, SM}_{\alpha,e_Y}=\frac{1}{N_M}\sum_{\kk'} U_{\alpha,e_Y,\kk'} \Delta^{\eta,SM}_{\kk',e_Y e_Y}
\eneq
for the intra-Chern-band channel and simplify the gap equation as
\begin{eqnarray}\label{eq:LGE_Numericalpp1}
2k_BT \left(\begin{array}{c}
\Delta^{+,SM}_{\kk,e_Y e_Y}\\
\Delta^{-,SM}_{\kk,e_Y e_Y}
\end{array}\right)
=\sum_{\alpha}U^*_{\alpha, e_Y, \kk}\left(\begin{array}{cc}
0&1\\
1&0
\end{array}\right)
\left(\begin{array}{c}
\tilde{\Delta}^{+,SM}_{\alpha,e_Y}\\
\tilde{\Delta}^{-,SM}_{\alpha,e_Y}
\end{array}\right).
\end{eqnarray}
Now we can introduce
\beq \tilde{V}_{\beta\alpha,e_Y}=\frac{1}{N_M}\sum_{\kk}U_{\beta,e_Y,\kk}U^*_{\alpha,e_Y,\kk},
\eneq
and transform the gap equation into an eigen-problem
\begin{eqnarray}
2k_BT \left(\begin{array}{c}
\tilde{\Delta}^{+,SM}_{\beta,e_Y}\\
\tilde{\Delta}^{-,SM}_{\beta,e_Y}
\end{array}\right)
=\sum_{\alpha}\tilde{V}_{\beta\alpha,e_Y}\left(\begin{array}{cc}
0&1\\
1&0
\end{array}\right)
\left(\begin{array}{c}
\tilde{\Delta}^{+,SM}_{\alpha,e_Y}\\
\tilde{\Delta}^{-,SM}_{\alpha,e_Y}
\end{array}\right). 
\end{eqnarray}

Crucially, we find that
\beq 
\tilde{V}_{01,-e_Y}=\tilde{V}_{10,e_Y}=\frac{1}{N_M}\sum_{\kk} \frac{\sqrt{A V_0} b^2 k_{e_Y}^2}{(k^2+b^2)^2}=0
\eneq
after performing the angular integral of the momentum. Thus, two channels (s-wave and d-wave) are decoupled, 
$\tilde{V}_{\beta\alpha,e_Y}=\tilde{V}_{\alpha\alpha,e_Y}\delta_{\alpha\beta}$. Furthermore, we find
\beq \label{eq:LGE_V_intra_1}
\tilde{V}_{00,e_Y}=\frac{1}{N_M}\sum_{\kk} \frac{A b^4}{(k^2+b^2)^2}, \; \; \tilde{V}_{11,e_Y}=\frac{1}{N_M}\sum_{\kk} \frac{V_0 k^4}{(k^2+b^2)^2}, 
\eneq
both of which are independent of the parameter $e_Y$, so we drop the index $e_Y$ for $\tilde{V}_{\alpha\alpha}$ ($\alpha=0,1$) below. 

Now the gap equation becomes
\begin{eqnarray}
2k_BT \left(\begin{array}{c}
\tilde{\Delta}^{+,SM}_{\alpha,e_Y}\\
\tilde{\Delta}^{-,SM}_{\alpha,e_Y}
\end{array}\right)
=\left(\begin{array}{cc}
0&\tilde{V}_{\alpha\alpha}\\
\tilde{V}_{\alpha\alpha}&0
\end{array}\right)
\left(\begin{array}{c}
\tilde{\Delta}^{+,SM}_{\alpha,e_Y}\\
\tilde{\Delta}^{-,SM}_{\alpha,e_Y}
\end{array}\right), 
\end{eqnarray}
with $e_Y=\pm$ and $\alpha=0,1$ for s-wave and d-wave channels, separately. This eigen-problem can be easily solved
with two eigen-solutions $\pm \frac{\tilde{V}_{\alpha\alpha}}{2}$, and we only take the positive value one, namely 
\beq
k_BT_{c,\alpha}=\frac{\tilde{V}_{\alpha\alpha}}{2}.  
\eneq

To get an estimate of $T_c$ in each channel, We can change the momentum summation in Eq. (\ref{eq:LGE_V_intra_1}) to the momentum integral with the cut-off $\sim k_\theta$ and obtain
\beq \tilde{V}_{00}=\frac{S_M}{2\pi} \int_0^{k_\theta} kdk \frac{Ab^4}{(k^2+b^2)^2}, 
\;\; \tilde{V}_{11}=\frac{S_M}{2\pi} \int_0^{k_\theta} kdk \frac{V_0k^4}{(k^2+b^2)^2} \eneq 
where $S_M=\frac{\sqrt{3}}{2} a_M^2$, $a_M$ is the Moire lattice constant

Numerical integral over the momentum gives $\tilde{V}_{00}\approx 0.018 meV$ and $\tilde{V}_{11}\approx 0.387 meV$, which give rise to 
$k_BT_{c,0}\approx0.009 meV$ for s-wave channel and $k_BT_{c,1}\approx 0.193 meV$ for the d-wave channel. Based on this estimate, we conclude that
the d-wave pairing channel will win and give the critical temperature
\beq
k_BT_c^{intra}=\frac{\tilde{V}_{11}}{2}\approx 0.193 meV
\eneq
for the intra-Chern-band channel. 

The corresponding normalized eigen-vector is $(\tilde{\Delta}^{+,SM}_{+}, \tilde{\Delta}^{-,SM}_{+})=\frac{1}{\sqrt{2}}(1,1)$. 
With the normalized eigen-vector and Eq. (\ref{eq:LGE_Numericalpp1}), we have
\begin{eqnarray}\label{eq:gapfun_intra_1}
\Psi^{SM}_{\kk, ++}=\left(\begin{array}{c}
\Delta^{+,SM}_{\kk,++}\\
\Delta^{-,SM}_{\kk,++}
\end{array}\right)
=\frac{\sqrt{V_0}}{\sqrt{2}\tilde{V}_{11} (k^2+b^2)} (k_x-ik_y)^2 \left(\begin{array}{c}
1\\1\end{array}\right).  
\end{eqnarray}
We also consider the gap equation (\ref{eq:LGE_Numericalmm}), which should given the same $T_c$, and the corresponding eigen-state is given by 
\begin{eqnarray}\label{eq:gapfun_intra_2}
\Psi^{SM}_{\kk, - -}=(\Psi^{SM}_{\kk, ++})^*=\frac{\sqrt{V_0}}{\sqrt{2}\tilde{V}_{11} (k^2+b^2)} (k_x+ik_y)^2 \left(\begin{array}{c}
1\\1\end{array}\right).  
\end{eqnarray}
according to Eq. (\ref{eq:LGE_Deltaintra}). It should be noted that both $\Psi^{SM}_{\kk, ++}$ and $\Psi^{SM}_{\kk, --}$ have
even parity, $\lambda_{C_{2z}}=1$, so they can only form spin-singlet pairing ($S,M=0,0$ in the above gap function). 

We emphasize that $\Psi^{SM}_{\kk, ++}$ and $\Psi^{SM}_{\kk, - -}$ are two independent degenerate pairing channels with the same $T_c$,
and they together form a 2D $E_2$ irreducible representation. To see that, we need to consider the operator form of the gap function and define
\begin{eqnarray}
&& H_{\Delta_{e_Y}} = \frac{1}{N_M}  \sum_{\kk, s_1, s_2, \eta } \gamma_{\kk, e_Y, \eta , s_1 }^\dagger \Delta^{\eta, 00}_{\kk; e_Y e_Y} (i s_y)_{s_1 s_2}   
 \gamma_{-\kk, e_Y, -\eta , s_2}^\dagger\nonumber\\
&& =  \frac{\sqrt{2 V_0}}{\tilde{V}_0 }  \sum_{\kk, s_1, s_2 }\frac{k_{-e_Y}^2}{k^2+b^2}
\gamma_{\kk, e_Y, +, s_1 }^\dagger (i s_y)_{s_1 s_2}   \gamma_{-\kk, e_Y, -, s_2}^\dagger
\end{eqnarray}
where $e_Y=\pm$. 
Here $H_{\Delta_{e_Y}}$ only involves the creation operator terms, so the full gap Hamiltonian should be $H_{\Delta_{e_Y}}+H^\dagger_{\Delta_{e_Y}}$ to keep the Hamiltonian hermitian. 
Now we want to check the transformation property of $H_{\Delta_{e_Y}}$ under $C_{3z}$ and $C_{2x}$ (spinless). 
Since we are dealing with the continuous model around $\kk=0$ (excluding Moir\'e BZ boundary), we should be able to choose the phase factors
$\theta(\kk)=\alpha(\kk)=0$ for $C_{3z}$ and $C_{2x}$ in Eq. (\ref{eq:symtrans_Chern1}), so 
$C_{3z} \gamma_{\kk,e_Y\eta s}^\dagger C_{3z}^{-1} = \gamma_{C_3 \kk,e_Y \eta s}^\dagger$ and
$ C_{2x} \gamma_{\kk,e_Y\eta s}^\dagger C_{2x}^{-1} = \gamma_{C_{2x} \kk, -e_Y \eta s}^\dagger $. 
Then, one can show
\beq\label{eq:symmetry_Delta_1}
C_{3z}H_{\Delta_{e_Y}}C_{3z}^{-1} = e^{i e_Y \frac{2\pi}{3}}H_{\Delta_{e_Y}};  C_{2x}H_{\Delta_{e_Y}}C_{2x}^{-1}=H_{\Delta_{-e_Y}}. 
\eneq 
We further define 
\beq\label{eq:transform_Delta}
H_{\Delta_1}=H_{\Delta_+}+H_{\Delta_-};  H_{\Delta_2}=i (H_{\Delta_+}-H_{\Delta_-})
\eneq
and on this basis $(H_{\Delta_1}, H_{\Delta_2})$, we have
\beq
C_{3z}H_{\Delta_{\alpha}}C_{3z}^{-1} = \sum_\beta H_{\Delta_{\beta}} (e^{-i\frac{2\pi}{3}\sigma_y})_{\beta\alpha} ;  C_{2x}H_{\Delta_{\alpha}}C_{2x}^{-1}=\sum_\beta H_{\Delta_{\beta}}(\sigma_z)_{\beta\alpha}, 
\eneq
where $\sigma$ is the Pauli matrix for two components of pairings. 
Since $(H_{\Delta_1}, H_{\Delta_2})$ is even under $C_{2z}$, $\lambda_{C_{2z}}=1$, we conclude $(H_{\Delta_1}, H_{\Delta_2})$ belongs to the 2D $E_2$ irrep
by comparing with the irrep table (Tab. \ref{tab:IrreduciblePairing}). 
It should be emphasized that $H_{\Delta_\alpha}$ ($\alpha=1,2$) {\it cannot} be written in the $k_x^2-k_y^2$ or $k_xk_y$ form because of the Chern-band index. 
For example, the explicit form of $H_{\Delta_1}$ is 
\begin{eqnarray}\label{eq:intra_Chern_pairing_E2}
&&H_{\Delta_1}=H_{\Delta_+}+H_{\Delta_-}\nonumber\\
&&= \frac{\sqrt{2 V_0}}{\tilde{V}_0 N_M} \sum_{\kk, s_1, s_2 }\frac{1}{k^2+b^2} (i s_y)_{s_1 s_2} 
\left(  k_{-}^2 \gamma_{\kk, +, +, s_1 }^\dagger  
 \gamma_{-\kk, +, -, s_2}^\dagger+  k_{+}^2\gamma_{\kk, -, +, s_1 }^\dagger   
 \gamma_{-\kk, -, -, s_2}^\dagger \right),
\end{eqnarray}
and it is clear that we cannot make the summation over $k_-^2$ and $k_+^2$ to get $k_x^2-k_y^2$ form. 

The above analytical solutions fit well with our numerical results as discussed below. The largest three eigen-values of the linearized gap equations (Eqs. \ref{eq:LGE_Numericalpp} and \ref{eq:LGE_Numericalpm}) are shown in Tab. \ref{tab:LGE_numerical}. The largest eigen-value of the intra-Chern-band channel 
correspond to $k_B T_c= 0.16 meV$, which is quite close to the value of $k_B T_c\approx 0.19 meV$ obtained from the effective interaction discussed above. 

\begin{table}[h!]
\centering
\begin{tabular}{|| c | c | c | c || }
 \hline
 Intra-Chern-band channel & 0.16 meV & 0.04 meV & 0.0002 meV \\  
 \hline
 Inter-Chern-band channel & 0.21 meV & 0.008 meV & 0.0079 meV \\
 \hline
\end{tabular}
\caption{ $k_B T$ for the largest three eigen-values of the linearized gap equations (Eqs. \ref{eq:LGE_Numericalpp} and \ref{eq:LGE_Numericalpm}). 
The largest one determines $k_B T_c$. }
\label{tab:LGE_numerical}
\end{table}

We also look at the eigen-vector of the intra-Chern-band channel with the largest eigen-value and show that our numerical result also gives
the 2D $E_2$ irrep for this channel. 
Fig. \ref{fig:gf_intra} (a) shows the gap function $\Delta^{\eta}_{\kk,++}$ as a function of the momentum $\kk$, in which the superconducting gap closes, 
thus revealing a nodal structure at $\Gamma_M$. This is consistent with the 2D $E_2$ irrep, as shown by Eq. (\ref{eq:gapfun_E2_node}) or Eq. (\ref{eq:gapfun_intra_1}). 


To further confirm this conclusion, we test the transformation property of $\Delta^{\eta}_{\kk,++}$ under $C_{3z}$ rotation. From Eq. (\ref{eq:symmetry_Delta_1}), 
we expect
\beq
e^{-i e_Y (\theta(\kk)+\theta(-\kk))} \Delta^{\eta}_{\kk,e_Y e_Y} = e^{i e_Y \frac{2\pi}{3}} \Delta^{\eta}_{C_{3z}\kk, e_Y e_Y}
\eneq
for the 2D $E_2$ irrep channel. We want to test the above expression numerically. We numerically calculate $\theta(\kk)$ from the eigen-state of the flat bands 
in the BM model and solve $\Delta^{\eta}_{\kk,++}$ from the linearized gap equation. With $\theta(\kk)$ and $\Delta^{\eta}_{\kk,++}$, we can compute the phase factor $\Phi_+(\kk)$, defined as
\beq
e^{i \Phi_{+}(\kk)}=\frac{ e^{-i (\theta(\kk)+\theta(-\kk))} \Delta^{\eta}_{\kk,+ +}}{\Delta^{\eta}_{C_{3z}\kk, + +}}, 
\eneq
which is shown as a function of $\kk$ in Fig. \ref{fig:gf_intra} (c). One can see that the phase $\Phi_{+}$ is always
$\frac{2\pi}{3}$ except around $\kk\sim 0$, where $ \Delta^{\eta}_{\kk,++}$ is zero. 
This confirms that our numerical results satisfy the transformation property Eq. (\ref{eq:symmetry_Delta_1})
that corresponds to the 2D $E_2$ irrep pairing channel.


As a comparison, we also consider $\Delta^{\eta}_{\kk,++}$ with the next largest eigen-value, as shown in Fig. \ref{fig:gf_intra}(b). In this case, we
find a node existing at $\KK_M$, which implies that the next highest $T_c$ channel belongs to the 1D irrep. Indeed, Fig. \ref{fig:gf_intra}(d) shows 
that $\Phi_{+}\sim 0$ for the next highest $T_c$ channel (except around $\KK_M$ where $\Delta^{\eta}_{\kk,++}\sim 0$), and thus we should have 
\beq\label{eq:symmetry_Delta_2}
C_{3z}H_{\Delta_{e_Y}}C_{3z}^{-1} = H_{\Delta_{e_Y}};  C_{2x}H_{\Delta_{e_Y}}C_{2x}^{-1}=H_{\Delta_{-e_Y}}
\eneq
in this case. By performing the transformation (\ref{eq:transform_Delta}), we obtain 
\beq
C_{3z}H_{\Delta_{\alpha}}C_{3z}^{-1} =  H_{\Delta_{\alpha}}  ;  C_{2x}H_{\Delta_{\alpha}}C_{2x}^{-1}=\sum_\beta H_{\Delta_{\beta}}(\sigma_z)_{\beta\alpha}. 
\eneq 
Thus,  $H_{\Delta_{1}}$ and  $H_{\Delta_{2}}$ belong to 1D $A_1$ and $A_2$ irreps, respectively, and they are degenerate. 

To understand the above results, we come back to the gap equation. 
In the chiral flat band limit, as we have seen, the linearized gap equation that contains both channels of the intra-Chern-band pairing has the following diagonal form
\begin{eqnarray}\label{eq:LGE_intra_Full_1}
2k_BT \left(\begin{array}{c}
\Psi^{SM}_{++}\\
\Psi^{SM}_{--}
\end{array}\right)
=\left(\begin{array}{cc}
\mathcal{V}^{+,++}_{\kk,\kk'}&0\\
0&\mathcal{V}^{+,--}_{\kk,\kk'}
\end{array}\right)
\left(\begin{array}{c}
\Psi^{SM}_{++}\\
\Psi^{SM}_{--}
\end{array}\right), 
\end{eqnarray}
in which two opposite Chern-band channels do not mix with each other. This form of the el-el interaction can be traced back to the chiral symmetry and flat band approximation used in our derivation. 
With such diagonal form, the linearized gap equation shares a U(1) symmetry created by the generator $\sigma_z$, which acts on the basis of $(\Psi^{SM}_{++},\Psi^{SM}_{--})$. On the other hand, $C_{2x}$ symmetry acts as the Pauli matrix $\sigma_x$ on the above basis (See Eq. \ref{eq:symmetry_Delta_1}).
As $\{\sigma_x,\sigma_z\}=0$ (anti-commutation relation), the eigen-state of the above eigen-problem Eq. (\ref{eq:LGE_intra_Full_1}) must be doubly degenerate.



\begin{figure}[hbt!]
   \centering
    \includegraphics[width=7in]{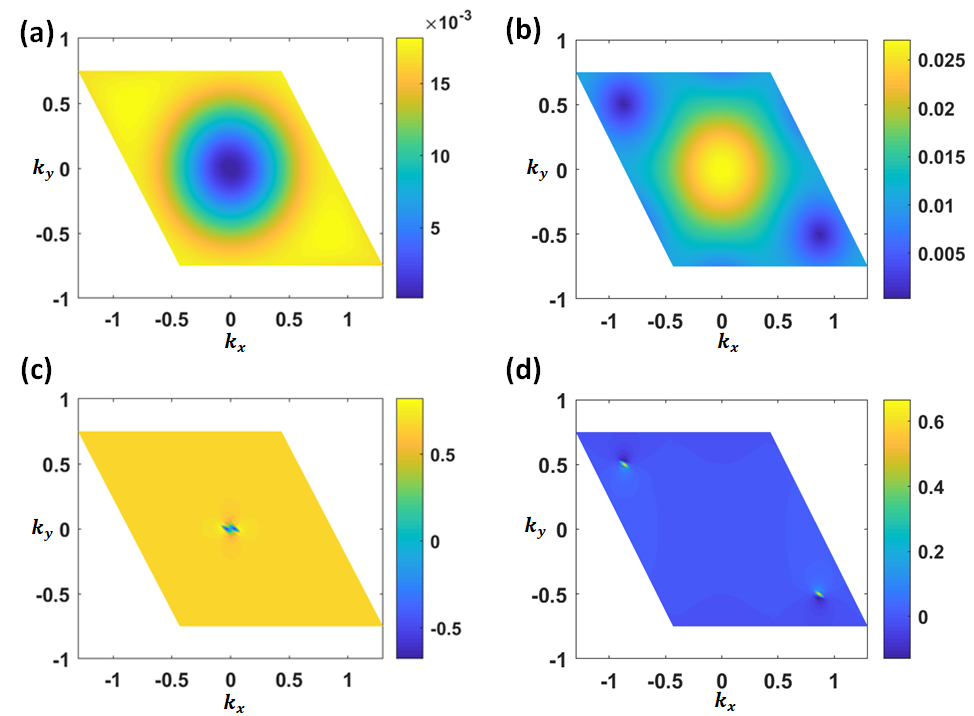} 
    \caption{ (a) and (b) depict the gap function $\Delta^{\eta}_{\kk,++}$ for the largest ($k_BT_c=0.16meV$) and next largest ($k_BT_c=0.04meV$) 
    eigen-values of the linearized gap equation, 
    respectively, for the intra-Chern-band channels. 
    (c) and (d) show the corresponding phase factor $\Phi_+(\kk)/\pi$ due to $C_{3z}$ rotation of the gap function. 
  }
    \label{fig:gf_intra}
\end{figure}


\subsubsection{Inter-Chern-band channels from linearized gap equation} \label{Sec:Inter_LGE}
Next we consider the inter-Chern-band channel, which is determined by the linearized gap equation (\ref{eq:LGE_Numericalpm}). The gap function for
the highest $T_c$ as a function of $\kk$ is shown in Fig. \ref{fig:gf_inter}(a), in which one can easily see that it is almost a constant over the whole Moir\'e BZ, 
so it is s-wave pairing (Fig. \ref{fig:gf_inter}(b)). 
Furthermore, one can show the parity of $C_{2z}$ for the gap function is $\lambda_{Cz}=1$ and the phase of $C_{3z}$ rotation $\Phi_{+-}$ is 0, so we
expect this gap function belongs to the $A_1$ or $A_2$ irreps. As discussed in the early part, we should be able to construct both spin 
singlet and triplet pairings, which are degenerate, for the inter-Chern-band channels. 

To see that more explicitly, we may consider the form of el-el from the heavy fermion model, 
\begin{eqnarray}
V^{\eta,e_Y,-e_Y}_{\kk,\kk'}=U^*_{0,\kk} U_{0,\kk'}+U^*_{1,\kk} U_{1,\kk'}; U_{0,\kk}=\frac{\sqrt{A} b^2}{k^2+b^2}; U_{1,\kk}=\frac{\sqrt{V_0} k^2}{k^2+b^2}. 
\end{eqnarray}
The corresponding linearized gap equation for inter-Chern-band channel is then written as 
\begin{eqnarray}
2k_BT \left(\begin{array}{c}
\Delta^{+,SM}_{\kk,+-}\\
\Delta^{-,SM}_{\kk,+-}
\end{array}\right)
=\frac{1}{N_M}\sum_{\kk',\alpha}U_{\alpha,\kk'}U^*_{\alpha,\kk}\left(\begin{array}{cc}
0&1\\
1&0
\end{array}\right)
\left(\begin{array}{c}
\Delta^{+,SM}_{\kk',+-}\\
\Delta^{-,SM}_{\kk',+-}
\end{array}\right),
\end{eqnarray}
where $\alpha=0,1$. Now we define
\beq
\tilde{\Delta}^{\eta, SM}_{\alpha,+-}=\frac{1}{N_M}\sum_{\kk} U_{\alpha,\kk'} \Delta^{\eta,SM}_{\kk',+-}
\eneq
and rewrite the linearized gap equation as
\begin{eqnarray}
2k_BT \left(\begin{array}{c}
\Delta^{+,SM}_{\kk,+-}\\
\Delta^{-,SM}_{\kk,+-}
\end{array}\right)
=\sum_{\alpha}U^*_{\alpha,\kk}\left(\begin{array}{cc}
0&1\\
1&0
\end{array}\right)
\left(\begin{array}{c}
\tilde{\Delta}^{+,SM}_{\alpha,+-}\\
\tilde{\Delta}^{-,SM}_{\alpha,+-}
\end{array}\right).
\end{eqnarray}
By defining
\beq \tilde{V}_{\beta\alpha}=\frac{1}{N_M}\sum_{\kk}U_{\beta,\kk}U^*_{\alpha,\kk},
\eneq
we obtain an eigen-problem
\begin{eqnarray}
2k_BT \left(\begin{array}{c}
\tilde{\Delta}^{+,SM}_{\beta,+-}\\
\tilde{\Delta}^{-,SM}_{\beta,+-}
\end{array}\right)
=\sum_{\alpha}\tilde{V}_{\beta\alpha}\left(\begin{array}{cc}
0&1\\
1&0
\end{array}\right)
\left(\begin{array}{c}
\tilde{\Delta}^{+,SM}_{\alpha,+-}\\
\tilde{\Delta}^{-,SM}_{\alpha,+-}
\end{array}\right). 
\end{eqnarray}
Explicitly, we have
\begin{eqnarray}
2k_BT \tilde{\Psi}^{SM}_{+-} 
=\left(\begin{array}{cccc}
0&\tilde{V}_{00}&0&\tilde{V}_{01}\\
\tilde{V}_{00}&0&\tilde{V}_{01}&0\\
0&\tilde{V}_{10}&0&\tilde{V}_{11}\\
\tilde{V}_{10}&0&\tilde{V}_{11}&0
\end{array}\right)
\tilde{\Psi}^{SM}_{+-} , \; \;  
\tilde{\Psi}^{SM}_{+-} =\left(\begin{array}{c}
\tilde{\Delta}^{+,SM}_{0,+-}\\
\tilde{\Delta}^{-,SM}_{0,+-}\\
\tilde{\Delta}^{+,SM}_{1,+-}\\
\tilde{\Delta}^{-,SM}_{1,+-}
\end{array}\right)
\end{eqnarray}
with
\beq \tilde{V}_{00}=\frac{1}{N_M}\sum_{\kk} \frac{A b^4}{(k^2+b^2)^2}, \; \tilde{V}_{01}=\tilde{V}_{10}=\frac{1}{N_M}\sum_{\kk} \frac{\sqrt{A V_0} b^2 k^2}{(k^2+b^2)^2}, \; \; \tilde{V}_{11}=\frac{1}{N_M}\sum_{\kk} \frac{V_0 k^4}{(k^2+b^2)^2}. 
\eneq
We can see that $\tilde{V}_{00},\tilde{V}_{11},\tilde{V}_{01}>0$. $\tilde{V}_{00}$ and $\tilde{V}_{11}$ are just the same as the corresponding parameters for the intra-Chern-band pairing. The key difference is that $\tilde{V}_{01}$ is now non-zero, so two channels are coupled with each other since both are s-wave. 
We thus need to solve the whole $4\times 4$ eigen-problem and the largest eigen-value gives $T_c$ as
\beq
k_B T_c^{inter}=\frac{1}{4}\left( \tilde{V}_{00} + \tilde{V}_{11} + \sqrt{(\tilde{V}_{00}-\tilde{V}_{11})^2+4\tilde{V}_{01}^2} \right)
\eneq
Compared to the intra-Chern-band pairing, one can show that 
\beq \label{eq:Tc_inter_1}
k_B T_c^{inter}=\frac{1}{4}\left( \tilde{V}_{00} + \tilde{V}_{11} + \sqrt{(\tilde{V}_{00}-\tilde{V}_{11})^2+4\tilde{V}_{01}^2} \right)>\frac{1}{2}\tilde{V}_{11}=k_B T_c^{intra}. 
\eneq

Numerical calculations give the same values for $\tilde{V}_{00}\approx 0.018 meV$ and $\tilde{V}_{11}\approx 0.387 meV$. In addition, one finds $\tilde{V}_{01}\approx 0.042 meV$. With all these values, the critical temperature for the inter-Chern-band channel 
\beq
k_BT_c^{inter}\approx 0.196 meV
\eneq
is slightly larger than that of the intra-Chern-band channel ($\sim 0.193 meV$). 

Let us compare the $T_c$ calculation from the linearized gap equation from the heavy fermion model with that from the numerical calculations of the full interaction model. We find that the inter-Chern-band s-wave pairing has a higher $T_c$ than the intra-Chern-band d-wave pairing, which is the same for the heavy femrion model and the full interaction model. The heavy fermion model provides a simple explanation of this result. In the above derivation, one can see that  both the $(U^{inter}_{0,\kk})^*U^{inter}_{0,\kk'}$ and $(U^{inter}_{1,\kk})^*U^{inter}_{1,\kk'}$ terms contribute to inter-Chern-band s-wave pairing, but only the $(U^{intra}_{1,\kk})^*U^{intra}_{1,\kk'}$ term contributes to the intra-Chern-band d-wave pairing (Here we add the label $inter$ and $intra$ to distinguish the interactions in different Chern-band channels). Since the interaction parameter in front of $(U^{inter}_{1,\kk})^*U^{inter}_{1,\kk'}$ term is the same as that for  $(U^{intra}_{1,\kk})^*U^{intra}_{1,\kk'}$ term, the $T_c$ of the inter-Chern-band s-wave pairing will always be larger than that of the intra-Chern-band d-wave pairing, as shown in Eq. (\ref{eq:Tc_inter_1}).

The corresponding eigen-vector $\tilde{\Psi}^{SM}_{+-}$ for the inter-Chern-band pairing takes the form
\begin{eqnarray}
\tilde{\Psi}^{SM}_{+-} =\left(\begin{array}{c}
\tilde{\Delta}^{+,SM}_{0,+-}\\
\tilde{\Delta}^{-,SM}_{0,+-}\\
\tilde{\Delta}^{+,SM}_{1,+-}\\
\tilde{\Delta}^{-,SM}_{1,+-}
\end{array}\right)=\frac{1}{\sqrt{2\mathcal{N}}}\left(\begin{array}{c}
2\tilde{V}_{01}\\
2\tilde{V}_{01}\\
\tilde{V}_-+\sqrt{\tilde{V}_-^2+4\tilde{V}_{01}^2}\\
\tilde{V}_-+\sqrt{\tilde{V}_-^2+4\tilde{V}_{01}^2}
\end{array}\right),
\end{eqnarray}
where $\tilde{V}_-=-\tilde{V}_{00}+\tilde{V}_{11}$ and $\mathcal{N}=2\sqrt{\tilde{V}_-^2+4\tilde{V}_{01}^2}\left(\tilde{V}_-+\sqrt{\tilde{V}_-^2+4\tilde{V}_{01}^2}\right)$. 
This gives rise to the gap function
\begin{eqnarray}
 \left(\begin{array}{c}
\Delta^{+,SM}_{\kk,+-}\\
\Delta^{-,SM}_{\kk,+-}
\end{array}\right)
&=&\frac{1}{2k_B T_c^{inter}}\sum_{\alpha}U^*_{\alpha,\kk}
\left(\begin{array}{c}
\tilde{\Delta}^{-,SM}_{\alpha,+-}\\
\tilde{\Delta}^{+,SM}_{\alpha,+-}
\end{array}\right),\nonumber\\
& = & f(|\kk|)
\left(\begin{array}{c}
1\\
1
\end{array}\right), \;\; 
f(|\kk|)=\frac{2\tilde{V}_{01}U_{0,\kk}^*+(\tilde{V}_-+\sqrt{\tilde{V}_-^2+\tilde{V}_{01}^2})U_{1,\kk}^*}{2k_B T_c^{inter}\sqrt{2\mathcal{N}}},
\end{eqnarray}
which is an s-wave pairing.

We can explicitly construct both singlet and triplet pairings as
$\Delta^{\eta}_{\kk;e_1 s_1, e_2 s_2}=f(|\kk|)(\zeta^{x})_{e_1 e_2}(\mS^{00})_{s_1 s_2}$
for singlet pairing and 
$\Delta^{\eta}_{\kk;e_1 s_1, e_2 s_2}=f(|\kk|)(\zeta^{y})_{e_1 e_2}(\mS^{1M})_{s_1 s_2}$
for triplet pairing. Furthermore, $\hat{C}_{2x}$ symmetry reverses the sign of $e_Y$ on the Chern-band basis. 
Since $(\zeta^{x})_{e_1, -e_1}=(\zeta^{x})_{-e_1, e_1}$ and $(\zeta^{y})_{e_1, -e_1}=-(\zeta^{y})_{-e_1, e_1}$, the $\zeta^{x}$
spin singlet channel should belong to $A_1$ irrep while the $\zeta^{y}$ spin triplet channel belongs to $A_2$ irrep.

\begin{figure}[hbt!]
   \centering
    \includegraphics[width=7in]{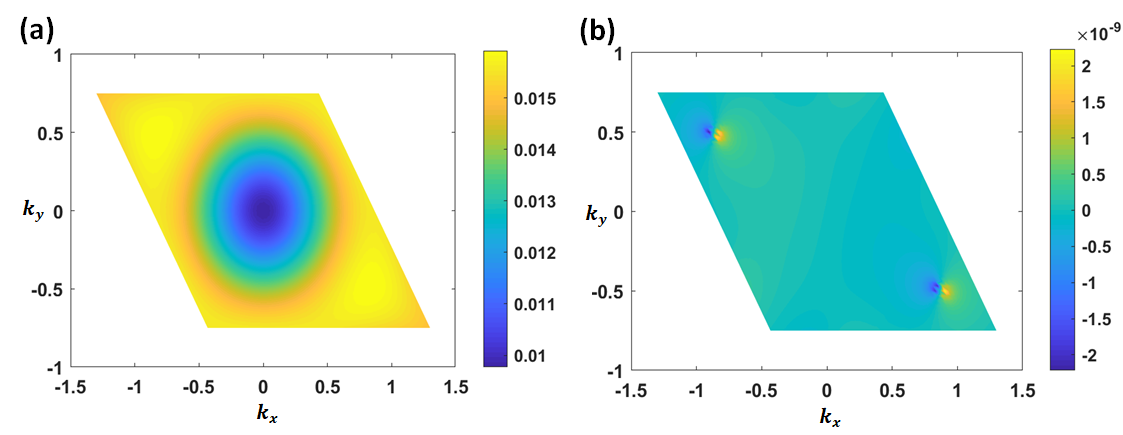} 
    \caption{ (a) depicts the gap function $\Delta^{\eta}_{\kk,+-}$ for the largest eigen-values of the linearized gap equation ($k_BT_c=0.21meV$) for
    the inter-Chern-band channel. (c) shows the corresponding phase factor $\Phi_{+-}(\kk)/\pi$ due to $C_{3z}$ rotation for the gap function. 
  }
    \label{fig:gf_inter}
\end{figure}

It should be emphasized that the degeneracy between singlet and triplet pairings for the inter-Chern-band channels only occurs in the chiral flat band limit. This degeneracy will be removed by a finite kinetic energy term. To see that, we need to go back to the LEG before taking the flat band approximation, namely Eqs. (\ref{eq:LGE_nonflat_Green}) and (\ref{eq:LGE_Mastubara_summation}). Using the decomposition of the gap function (\ref{eq:Gapfunction_decomp}) and $Tr(\zeta^\nu\zeta^\mu)=2\delta_{\mu\nu}$, we have 
\beq 
\Delta_{ \kk;\nu}^{\eta,SM} = \frac{1}{N_M} \sum_{\kk', m_1 m_2, \mu, e_1 e_2} V^{\eta,e_1e_2}_{\kk \kk'} \left[P_{m_1,\kk',-\eta} \zeta^\mu P^T_{m_2,-\kk',\eta}\right]_{e_1 e_2} (\zeta^\nu)_{e_2 e_1} \Delta^{-\eta,SM}_{\kk';\mu} T^{\eta}_{m_1 m_2 \kk'}. \eneq 
With the form of projector operator $P$ in Eq. (\ref{eq:Projector_singleHam}), we find 
\beq P_{m_1,\kk',-\eta} \zeta^\mu P^T_{m_2,-\kk',\eta}=\frac{1}{4} \left( \zeta^\mu + \frac{d_{x,\kk',-\eta}}{|d_{x,\kk',-\eta}|} (m_1 \zeta^x\zeta^\mu + m_2 \zeta^\mu \zeta^x) +m_1 m_2 \zeta^x\zeta^\mu \zeta^x \right). \eneq
Now let us focus on the inter-Chern-band channels, so $\mu, \nu= x, y$, and one can show, in this case, that only the terms $\zeta^\mu$ and $\zeta^x\zeta^\mu \zeta^x $ in $P_{m_1,\kk',-\eta} \zeta^\mu P^T_{m_2,-\kk',\eta}$ can give a non-zero contribution to the LGE. 
We first consider $\mu=x, \nu=y$, so the right-hand side of the LGE is 
\begin{eqnarray} 
&& \frac{1}{4N_M} \sum_{\kk', m_1 m_2, e_1 e_2} V^{\eta,e_1e_2}_{\kk \kk'} \left(\zeta^x+m_1 m_2 \zeta^x\right)_{e_1 e_2} (\zeta^y)_{e_2 e_1} \Delta^{-\eta,SM}_{\kk';x} T^{\eta}_{m_1 m_2 \kk'}\nonumber\\
&& = \frac{1}{4N_M} \sum_{\kk', m_1 m_2, e_1 e_2} V^{\eta,e_1e_2}_{\kk \kk'} \left(1+m_1 m_2\right) \delta_{e_1, -e_2} (-i) e_2 \delta_{e_2,-e_1} \Delta^{-\eta,SM}_{\kk';x} T^{\eta}_{m_1 m_2 \kk'}\nonumber \\
&& = \frac{1}{4N_M} \sum_{\kk', m_1 m_2, e_1 } V^{\eta,e_1,-e_1}_{\kk \kk'} \left(1+m_1 m_2\right) i e_1 \Delta^{-\eta,SM}_{\kk';x} T^{\eta}_{m_1 m_2 \kk'} = 0.
\end{eqnarray}
In the last step, we have used $\sum_{e_1}  V^{\eta,e_1,-e_1}_{\kk \kk'} e_1 =  V^{\eta,+-}_{\kk \kk'}- V^{\eta,-+}_{\kk \kk'}=0$. 
Therefore, $\zeta^x$ and $\zeta^y$ channels can not be mixed with each other even when the kinetic energy term is included.

Next let's consider the LGE for $\mu=\nu=x$ (only for spin singlet), which can be simplified as
\begin{eqnarray}
&&\Delta_{ \kk;x}^{\eta,00} = \frac{1}{4N_M} \sum_{\kk', m_1 m_2, e_1 e_2} V^{\eta,e_1e_2}_{\kk \kk'} (1+m_1 m_2) \zeta^x_{e_1 e_2} (\zeta^x)_{e_2 e_1} \Delta^{-\eta,00}_{\kk';x} T^{\eta}_{m_1 m_2 \kk'}\nonumber\\
&& = \frac{1}{2N_M} \sum_{\kk', m_1 m_2} V^{\eta,+-}_{\kk \kk'} (1+m_1 m_2) \Delta^{-\eta,00}_{\kk';x} T^{\eta}_{m_1 m_2 \kk'}\nonumber\\
&& = \frac{2}{N_M} \sum_{\kk'} V^{\eta,+-}_{\kk \kk'} \Delta^{-\eta,00}_{\kk';x} T^{\eta}_{++ \kk'}
\end{eqnarray}
in the chiral limit, where we have used $T^{\eta}_{-m_1, -m_1, \kk'}=T^{\eta}_{m_1 m_1 \kk'}$. With Eq. (\ref{eq:LGE_Mastubara_summation}), we have
\beq T^{\eta}_{++ \kk'}=\frac{\tanh(\beta \xi_{+,-\eta,\kk'}/2)}{2\xi_{+,-\eta,\kk'}},  \eneq
and the LGE becomes
\beq \label{eq:LGE_xx_nonflat_chiral}
\Delta_{ \kk;x}^{\eta,00} = \frac{1}{N_M} \sum_{\kk'} V^{\eta,+-}_{\kk \kk'} \frac{\tanh(\beta \xi_{+,-\eta,\kk'}/2)}{\xi_{+,-\eta,\kk'}} \Delta^{-\eta,00}_{\kk';x}. 
\eneq

Similar derivation can be applied to $\mu=\nu=y$ for spin triplet channel. One can show 
\beq
\Delta_{ \kk;y}^{\eta,1M} = \frac{1}{N_M} \sum_{\kk', m_1} V^{\eta,+-}_{\kk \kk'} \Delta^{-\eta,1M}_{\kk';y} T^{\eta}_{m_1, -m_1, \kk'}. 
\eneq
To evaluate $T^{\eta}_{m_1, -m_1, \kk'}$, one should first keep a finite $\mu$ and take $\mu\rightarrow 0$ at the end of the calculation, which gives
\beq T^{\eta}_{m_1, -m_1, \kk'}=\frac{\beta e^{\beta \xi_{m_1, -\eta,\kk'}}}{(e^{\beta \xi_{m_1,-\eta, \kk'}}+1)^2}.  \eneq
Thus, 
\beq \label{eq:LGE_yy_nonflat_chiral}
\Delta_{ \kk;y}^{\eta,1M} = \frac{1}{N_M} \sum_{\kk'} V^{\eta,+-}_{\kk \kk'} \frac{\beta}{2\cosh^2(\beta \xi_{+,-\eta,\kk'})} \Delta^{-\eta,1M}_{\kk';y}. 
\eneq

Now one can see that the LGEs for $\Delta_{ \kk;x}^{\eta,00}$ and $\Delta_{ \kk;y}^{\eta,1M}$ are different when the kinetic energy term is taken into account. 
One can easily check that both LGEs (\ref{eq:LGE_xx_nonflat_chiral}) and (\ref{eq:LGE_yy_nonflat_chiral}) reduce to Eq. (\ref{eq:LGE_gapequation1}) in the flat band limit. 
We may regard the band dispersion $\xi_{+,-\eta,\kk'}$ as a perturbation (much smaller than $T_c$) and then we can perform the perturbation expansion of the kinetic energy dependent terms. 
We find 
\beq \frac{\tanh(\beta \xi_{+,-\eta,\kk'}/2)}{\xi_{+,-\eta,\kk'}}\approx \frac{\beta}{2}-\frac{1}{24}\beta^3\xi_{+,-\eta,\kk'}^2 \eneq
and
\beq \frac{\beta}{2\cosh^2(\beta \xi_{+,-\eta,\kk'})}\approx \frac{\beta}{2}-\frac{1}{8}\beta^3\xi_{+,-\eta,\kk'}^2. \eneq
It is clear that the singlet channel $\Delta_{ \kk;x}^{\eta,00}$ has a stronger effective interaction compared to the triplet channel $\Delta_{ \kk;y}^{\eta,1M}$ as
\beq \frac{\tanh(\beta \xi_{+,-\eta,\kk'}/2)}{\xi_{+,-\eta,\kk'}}>\frac{\beta}{2\cosh^2(\beta \xi_{+,-\eta,\kk'})}
\eneq
in the limit $\xi_{+,-\eta,\kk'}\ll k_B T_c$. Thus, we expect the singlet channel $\Delta_{ \kk;x}^{\eta,00}$ has a higher $T_c$ in the chiral non-flat limit for inter-Chern-band pairing channels.

\subsubsection{Bogoliubov-de Gennes Hamiltonian, full self-consistent gap equation and ground state energy}
The above linearized gap equation only gives the behavior of superconductivity near the critical temperature $T_c$. We hope to understand more 
about the superconductivity property at zero temperature. For that purpose, we will derive the full self-consistent gap equation that works at zero temperature, as well as the ground state energy (or condensation energy) for superconductivity, based on the Bogoliubov-de Gennes (BdG) Hamiltonian. 

To do that, let's first write down
the form of the BdG Hamiltonian on the Chern band basis, 
\begin{eqnarray}\label{eq:BdG_Hamiltonian_TBG}
&\mathcal{H}^{\pm}_{BdG} = \sum_{\kk\in MBZ} \psi_{\pm, \kk}^\dagger H_{BdG}^\pm(\kk) \psi_{\pm, \kk},\;\;\; \\&  \psi_{+, \kk}^\dagger = (\gamma^\dagger_{\kk, e_Y = \pm ,+,s =\uparrow \downarrow},\gamma_{-\kk, e_Y=\pm,-,s=\uparrow \downarrow}) ;\;\;\; 
\psi_{-, \kk}^\dagger = (\gamma^\dagger_{\kk, e_Y = \pm ,-,s =\uparrow \downarrow},\gamma_{-\kk, e_Y=\pm,+,s=\uparrow \downarrow}) \nonumber \\ 
& H_{BdG}^+(\kk) =\frac{1}{2}\begin{pmatrix}
h_+(\kk)\otimes s_0 & 2 \Delta_{\kk} \otimes \mS \\
2 \Delta_{\kk}^\dagger\otimes  \mS^\dagger & - h_{-}^\star(-\kk) \otimes s_0
\end{pmatrix} \nonumber \\ &  H_{BdG}^-(\kk) = \frac{1}{2}\begin{pmatrix}
h_-(\kk)\otimes s_0 & -2 \Delta_{-\kk}^T \otimes \mS^T \\
-2 \Delta_{-\kk}^\star \otimes  \mS^\star & - h_{+}^\star(-\kk) \otimes s_0
\end{pmatrix} \nonumber 
\end{eqnarray} 
where $h_\pm(\kk)$ are the non-superconducting effective Hamiltonians for the flat bands at the valley $\eta=\pm$ in TBG, respectively,
while $\Delta_{\kk}=\Delta^{\eta=+}_{\kk, e_1, e_2}$. $H_{BdG}^\pm(\kk)$ are 8-by-8 matrices. 
The superconducting particle-hole symmetry $\hat{\mathcal{P}}$ with $D(\hat{\mathcal{P}})=\rho_x \zeta_0 s_0$, requires
\beq -H_{BdG}^-(\kk) = \rho_x \zeta_0 s_0 H_{BdG}^{+\star}(-\kk) \rho_x \zeta_0 s_0, \eneq
where $s$, $\zeta$ and $\rho$ are the identity and Pauli matrices for the spin, Chern-band basis and superconducting particle-hole basis. 
One should note that the superconducting particle-hole operator $\hat{\mathcal{P}}$ is different from the unitary particle-hole symmetry operator $\hat{P}$
for the non-superconducting TBG Hamiltonian. 

It should be noted that $h_{\eta=\pm}(\kk)$ is a 2-by-2 matrix and does {\it not} possess a diagonal form on the Chern-band basis. 
From the relation (\ref{eq-irrepbasis}) between the eigen-state basis and Chern-band basis, we can show that 
\beq \label{eq:single_particle_Ham1} h_\eta(\kk)= (d_{0,\eta}(\kk)-\mu)\zeta^0+d_{x,\eta}(\kk)\zeta^x, \eneq
where 
\beq d_{0,\eta}(\kk)=(\epsilon_{+,\eta}(\kk)+\epsilon_{-,\eta}(\kk))/2;  \;\;
d_{x,\eta}(\kk)=(\epsilon_{+,\eta}(\kk)-\epsilon_{-,\eta}(\kk))/2, \eneq
and $\epsilon_{n, \eta}(\kk)$ are the single-particle eigen-energies of the BM model $\hat{H}_{0}$ so they should be real. 
Time reversal symmetry requires 
\beq \epsilon_{n, \eta}(\kk)=\epsilon_{n, -\eta}(-\kk) \implies d_{0,-\eta}(-\kk)=d_{0,\eta}(\kk);\; d_{x,-\eta}(-\kk)=d_{x,\eta}(\kk) \implies h_{-\eta}(-\kk)=h_\eta(\kk) .  \eneq 

As discussed above, the spin-singlet and triplet channels are decoupled and below we discuss the singlet and triplet channels, separately. 
We first consider the spin-singlet channels, for which the gap function in $H_{BdG}^+(\kk)$ has the form
\beq \Delta_{\kk} \otimes \mS = \Delta_{\kk} \otimes (i s_y), \text{or} \;\; \Delta^{+}_{\kk;e_1 s_1 e_2 s_2}= \Delta_{\kk,e_1 e_2} (is_y)_{s_1 s_2}, \eneq
where $\Delta_{\kk}$ is a $2\times 2$ matrix on the Chern-band basis. 

Since the single-particle Hamiltonian is diagonal in spin space, the Hamiltonian $H_{BdG}^+(\kk)$ with spin-singlet pairing has a block diagonal form
\begin{eqnarray}\label{eq:HBdG_valleyp1}
H_{BdG}^+(\kk) =\frac{1}{2} \begin{pmatrix}
h_+(\kk)\otimes s_0 & 2 \Delta_{\kk} \otimes i s_y \\
2 \Delta_{\kk}^\dagger\otimes  (-is_y) & - h_{-}^\star(-\kk) \otimes s_0
\end{pmatrix}=\frac{1}{2}\begin{pmatrix}
h_+(\kk) & 0 & 0 & 2 \Delta_{\kk} \\
0 & h_+(\kk) & -2 \Delta_{\kk} & 0 \\
0 & -2 \Delta_{\kk}^\dagger & - h_{-}^\star(-\kk) & 0 \\
2 \Delta_{\kk}^\dagger & 0 &0 & - h_{-}^\star(-\kk) 
\end{pmatrix} 
\end{eqnarray}
on the basis 
$ (\gamma^\dagger_{\kk, e_Y = \pm ,+,\uparrow}, \gamma^\dagger_{\kk, e_Y = \pm ,+,\downarrow}, \gamma_{-\kk, e_Y=\pm,-,\uparrow},\gamma_{-\kk, e_Y=\pm,-,\downarrow})$. 
Thus, we can consider two blocks separately,   
\begin{eqnarray}\label{eq:HBdG_valleyp2}
H_{BdG}^{+,\lambda=\pm}(\kk) = \frac{1}{2}  \begin{pmatrix}
h_+(\kk) & 2 \lambda \Delta_{\kk} \\
2 \lambda \Delta_{\kk}^\dagger & - h_{-}^\star(-\kk)
\end{pmatrix}. 
\end{eqnarray}
Both blocks are 4-by-4 matrices, written on the basis $(\gamma^\dagger_{\kk, e_Y = \pm ,+,s =\uparrow },\gamma_{-\kk, e_Y=\pm,-,s=\downarrow})$
and $(\gamma^\dagger_{\kk, e_Y = \pm ,+,s =\downarrow },\gamma_{-\kk, e_Y=\pm,-,s=\uparrow})$ for the $\lambda=+$ and $-$ blocks, separately. 

To determine the exact pairing form for the temperature well below $T_c$, we need to go back to the full form of the gap equation (\ref{eq:Decomposition_GapFunctionDef}). Since $\Delta^{+}_{\kk; e_{1} s_1, e_2 s_2}=\Delta_{\kk;e_1,e_2} (i s_y)_{s_1 s_2}$ for spin singlet, we have $\Delta^{+}_{\kk; e_{1} \uparrow, e_2 \downarrow}=\Delta_{\kk;e_1,e_2} $ and $\Delta^{+}_{\kk; e_{1} \downarrow, e_2 \uparrow}=-\Delta_{\kk;e_1,e_2}$. 
The gap equation the explicitly reads
\beq \label{eq:gapEqn_general_singlet1} \Delta_{\kk;e_1e_2}= -\Delta^+_{\kk; e_{1} \downarrow, e_2 \uparrow} =  \frac{1}{N_M} \sum_{\kk'} V_{\kk\kk'}^{+ e_1 e_2} \langle \gamma_{-\kk'  e_2 ,+, \uparrow } \gamma_{\kk' e_1, -, \downarrow}\rangle 
\eneq
and
\beq \label{eq:gapEqn_general_singlet2}
 \Delta_{\kk;e_1e_2}=\Delta^{+}_{\kk; e_{1} \uparrow, e_2 \downarrow}= - \frac{1}{N_M}  \sum_{\kk'} V_{\kk\kk'}^{+ e_1 e_2} \langle \gamma_{-\kk'  e_2 + \downarrow} \gamma_{\kk' e_1 - \uparrow}\rangle. 
\eneq
Let's look at the first equation and we need to evaluate $\langle \gamma_{-\kk'  e_1 ,+, \uparrow } \gamma_{\kk' e_1, -, \downarrow}\rangle$ from the BdG Hamiltonian $H_{BdG}^{+,+}(\kk)$ since
\beq
\mathcal{H}_{BdG}^{+,+}(\kk)=\sum_{\kk,e_1e_2} (\gamma^\dagger_{\kk, e_1 ,+,\uparrow },\gamma_{-\kk, e_1, - ,\downarrow}) [H_{BdG}^{+,+}(\kk)]_{e_1,e_2}\left(\begin{array}{c} \gamma_{\kk, e_2 ,+,\uparrow } \\ \gamma^\dagger_{-\kk, e_2, - ,\downarrow} \end{array}\right). 
\eneq
Now let's define 
\beq \label{eq:UMatrix_1}
U_\kk =\begin{pmatrix}
u_\kk & v_\kk \\
w_{\kk} & r_{\kk} 
\end{pmatrix}, \;\; H_{BdG}^{+,+}(\kk)= U_\kk D_\kk U^\dagger_\kk
\eneq
where $u_\kk, v_\kk, w_{\kk}, r_{\kk}$ are two-by-two matrices and $D_\kk$ is a diagonal matrix, $D_\kk=\text{Diag}(E^+_1,E^+_2,E^-_1,E^-_2)$, with $E^+_{n}>0$ and $E^-_{n}<0$ ($n=1,2$). Now let's define Bogoliubov quasi-particle operators
\beq \label{eq:Bogoliubov_transformation_1}
\left(\begin{array}{c} \gamma_{\kk, e_1 ,+,\uparrow } \\ \gamma^\dagger_{-\kk, e_1, - ,\downarrow} \end{array}\right)= \sum_{n} [U_\kk]_{e_1 n} \left(\begin{array}{c} \alpha_{\kk, n} \\ \beta^\dagger_{-\kk, n} \end{array}\right)
\eneq 
so that
\begin{eqnarray}\label{eq:HBdG_valleyp_lambdap}
& \mathcal{H}_{BdG}^{+, +}= \sum_{\kk, e_1 e_2} (\alpha^\dagger_{\kk, e_1 } \; \;\beta_{-\kk, e_1}) [D_\kk]_{e_1,e_2}\left(\begin{array}{c} \alpha_{\kk, e_2  } \\ \beta^\dagger_{-\kk, e_2} \end{array}\right) \nonumber \\
& = \sum_{\kk, n} \left( E^+_{\kk,n} \alpha^\dagger_{\kk, n}  \alpha_{\kk, n }+ (- E^-_{\kk, n}) \beta^\dagger_{-\kk, n} \beta_{-\kk, n}  \right)
+ \sum_{\kk,n } E^-_{\kk,n}. 
\end{eqnarray}
The ground state of superconductors $|GS\rangle$ should be defined by $\alpha_{\kk,n}\ket{GS} = \beta_{\kk,n}\ket{GS} =0$
with $n= 1, 2$, and $\alpha^\dagger_{\kk,n}$ and $\beta^\dagger_{\kk,n}$ create Bogoliubov quasi-particles with the energy $E^+_{\kk,n}$ and $-E^-_{\kk,n}$,
respectively, both of which are positive numbers. 
More explicitly, we have
\beq \gamma_{\kk, e_1 ,+,\uparrow }=\sum_{n} ( u_{\kk,e_1 n} \alpha_{\kk, n } +  v_{\kk,e_1 n} \beta^\dagger_{-\kk, n} );\;\;
\gamma^\dagger_{-\kk, e_1, - ,\downarrow}= \sum_{n} ( w_{\kk,e_1 n} \alpha_{\kk, n } + r_{\kk,e_1 n} \beta^\dagger_{-\kk, n} ). 
\eneq
From the second equation, we can obtain
\beq \gamma_{-\kk, e_1, - ,\downarrow}= \sum_{n} ( w^*_{\kk,e_1 n} \alpha^\dagger_{\kk, n } + r^*_{\kk,e_1 n} \beta_{-\kk, n} ). 
\eneq

Now we can substitute these expressions into the gap equation (\ref{eq:gapEqn_general_singlet1}) and obtain
\begin{eqnarray}\label{eq:gapEqn_singlet_lambdap}
\Delta_{\kk;e_1 e_2} & = & \frac{1}{N_M} \sum_{\kk'} V_{\kk\kk'}^{+ e_1 e_2} \langle \gamma_{-\kk'  e_2 ,+, \uparrow } \gamma_{\kk' e_1, -, \downarrow}\rangle \nonumber \\
& =  & \frac{1}{N_M} \sum_{\kk', n m } V_{\kk\kk'}^{+ e_1 e_2} \langle ( u_{-\kk',e_2 n} \alpha_{-\kk', n } +  v_{-\kk',e_2 n} \beta^\dagger_{\kk', n} ) ( w^*_{-\kk',e_1 m} \alpha^\dagger_{-\kk', m } + r^*_{-\kk',e_1 m} \beta_{\kk', m} )\rangle \nonumber\\
& = & \frac{1}{N_M} \sum_{\kk', n } V_{\kk\kk'}^{+ e_1 e_2} \left( u_{-\kk',e_2 n} w^*_{-\kk',e_1 n} (1 - f(E^+_{-\kk',n})) +
 v_{-\kk',e_2 n} r^*_{-\kk',e_1 n} f(-E^-_{-\kk',n}) \right), 
\end{eqnarray}
where $f(E)=(e^{\beta E}+1)^{-1}$ is the Fermi distribution function. 
At zero temperature, the self-consistent gap equation reads
\beq \label{eq:GapEquation_full_zeroT}
\Delta_{\kk;e_1 e_2}= \frac{1}{N_M} \sum_{\kk', n } V_{\kk\kk'}^{+ e_1 e_2}  u_{-\kk',e_2 n} w^*_{-\kk',e_1 n}. \eneq

This self-consistent gap equation can be applied to both intra-Chern-band and inter-Chern-band channels for spin singlet state. 

Similar procedure can be applied to the gap equation (\ref{eq:gapEqn_general_singlet2}) and the BdG Hamiltonian $H_{BdG}^{+,\lambda=-}(\kk)$. 
From the form of the BdG Hamiltonian in Eq. (\ref{eq:HBdG_valleyp2}), we notice that the $\lambda=\pm$ Hamiltonians are related to each other by a unitary transformation, 
\beq \label{eq:Umatrix_HBdG} H_{BdG}^{+,-}(\kk)=\rho_z H_{BdG}^{+,+}(\kk) \rho_z.  \eneq
Therefore, 
\begin{eqnarray}
&\mathcal{H}_{BdG}^{+,-}(\kk)=\sum_{\kk,e_1e_2} (\gamma^\dagger_{\kk, e_1 ,+,\downarrow },\gamma_{-\kk, e_1, - ,\uparrow}) [H_{BdG}^{+,-}(\kk)]_{e_1,e_2}\left(\begin{array}{c} \gamma_{\kk, e_2 ,+,\downarrow } \\ \gamma^\dagger_{-\kk, e_2, - ,\uparrow} \end{array}\right)\nonumber\\
& =\sum_{\kk,e_1e_2} (\gamma^\dagger_{\kk, e_1 ,+,\downarrow },\gamma_{-\kk, e_1, - ,\uparrow}) [\rho_z  U_\kk D_\kk U^\dagger_\kk \rho_z]_{e_1,e_2}\left(\begin{array}{c} \gamma_{\kk, e_2 ,+,\downarrow } \\ \gamma^\dagger_{-\kk, e_2, - ,\uparrow} \end{array}\right). 
\end{eqnarray}
Thus, the eigen-energies of $H_{BdG}^{+,-}(\kk)$ are the same as those of $H_{BdG}^{+,+}(\kk)$, while the Bogoliubov quasi-particle operators are 
defined by 
\beq \label{eq:Bogoliubov_transformation_2}
\left(\begin{array}{c} \gamma_{\kk, e_1 ,+,\downarrow } \\ -\gamma^\dagger_{-\kk, e_1, - ,\uparrow} \end{array}\right)= \sum_{n} [U_\kk]_{e_1 n} \left(\begin{array}{c} \alpha_{\kk, n} \\ \beta^\dagger_{-\kk, n} \end{array}\right)
\eneq 
with the same $U$-matrix as in Eq. (\ref{eq:UMatrix_1}). With the above transformation, we find 
\beq \langle \gamma_{-\kk'  e_2 + \downarrow} \gamma_{\kk' e_1 - \uparrow}\rangle
=-\sum_n \left( u_{-\kk',e_2 n} w_{-\kk', e_1 n}^* (1-f(E^+_{-\kk',n})) + v_{-\kk',e_2 n} r^*_{-\kk',e_1 n} f(-E^-_{\kk',n}) \right)
\eneq
and thus the gap equation (\ref{eq:gapEqn_general_singlet2}) transforms into 
\begin{eqnarray} 
 \Delta_{\kk;e_1e_2} & = & - \frac{1}{N_M}  \sum_{\kk'} V_{\kk\kk'}^{+ e_1 e_2} \langle \gamma_{-\kk'  e_2 + \downarrow} \gamma_{\kk' e_1 - \uparrow}\rangle\nonumber\\
 & = & \frac{1}{N_M}  \sum_{\kk',n} V_{\kk\kk'}^{+ e_1 e_2} \left( u_{-\kk',e_2 n} w_{-\kk', e_1 n}^* (1-f(E^+_{-\kk',n})) + v_{-\kk',e_2 n} r^*_{-\kk',e_1 n} f(-E^-_{\kk',n}) \right), 
\end{eqnarray}
which is the same as Eq. (\ref{eq:gapEqn_singlet_lambdap}). Thus, both blocks give the same gap equation, which is required for the consistency of the theory.   

We can also calculate the ground state energy or the condensation energy of superconductivity based on the current BdG formalism. 
The BdG Hamiltonian in Eq. (\ref{eq:BdG_Hamiltonian_TBG}) only contains the fermion bilinear terms, and to get the correct form of the ground state energy, one needs to keep all the constant terms for the Hamiltonian, which reads
\beq \label{eq:Htot_SC} \mathcal{H}_{tot}=\sum_{\eta=\pm}\mathcal{H}^{\eta}_{BdG}+\sum_{\kk,\eta,e} [h_{-\eta}(-\kk)]_{ee} - \sum_{\kk \eta e_1 s_1 e_2 s_2} \langle \gamma^\dagger_{\kk e_1 \eta s_1 } \gamma^\dagger_{-\kk e_2 -\eta s_2}\rangle \Delta^{\eta}_{\kk; e_1 s_1 e_2 s_2 }. 
\eneq
Our goal is to calculate the condensation energy, which is defined as 
\beq E_c=\langle \mathcal{H}_{tot} \rangle_s-\langle \mathcal{H}_{tot} \rangle_n, \eneq
where $\langle \mathcal{H}_{tot} \rangle_s=\langle GS|\mathcal{H}_{tot}|GS \rangle$ is the superconductor ground state energy, and
$\langle \mathcal{H}_{tot} \rangle_n$ is the energy of the normal metallic state, which can be obtained from $\langle \mathcal{H}_{tot} \rangle_s$ by 
taking the superconductor gap function $\Delta$ to be zero. As the last two terms in Eq. (\ref{eq:Htot_SC}) are constants, we first focus on calculating
$\langle \mathcal{H}_{BdG}^\eta \rangle_s$. As shown in Eq. (\ref{eq:HBdG_valleyp1}), $H^+_{BdG}$ has a block-digonal form and thus we expect 
$\langle \mathcal{H}_{BdG}^+ \rangle_s=\sum_{\lambda=\pm} \langle \mathcal{H}_{BdG}^{+,\lambda} \rangle_s$. 
In Eq. (\ref{eq:HBdG_valleyp_lambdap}), we rewrite $\mathcal{H}_{BdG}^{+,+}$ in terms of quadratic form of Bogoliubov quasi-particle operators $\alpha_{\kk,n}, \beta_{\kk,n}$. Since the superconductor ground state is defined by $\alpha_{\kk,n}\ket{GS} = \beta_{\kk,n}\ket{GS} =0$, we obtain 
\beq \langle \mathcal{H}_{BdG}^{+,+} \rangle_s = \sum_{\kk,n} E^-_{\kk,n}. 
\eneq
Furthermore, as the $\lambda=-$ block is related to $\lambda=+$ block by a unitary transformation (\ref{eq:Umatrix_HBdG})
and thus the ground state energy of these two blocks are the same, $\langle \mathcal{H}_{BdG}^{+,-} \rangle_s=\langle \mathcal{H}_{BdG}^{+,+} \rangle_s$. 
In addition, one can show that $\mathcal{H}_{BdG}^{\eta=-}$ is equivalent to $\mathcal{H}_{BdG}^{\eta=+}$ in the second quantization form, so  
$\langle\mathcal{H}_{BdG}^{-}\rangle_s=\langle\mathcal{H}_{BdG}^{+}\rangle_s$. 

The second term in Eq. (\ref{eq:Htot_SC}) is related to single-particle Hamiltonian and from Eq. (\ref{eq:single_particle_Ham1}), we obtain 
\beq \sum_{\kk,\eta,e} [h_{-\eta}(-\kk)]_{ee}= 2 \sum_{\kk,\eta} (d_{0,-\eta}(-\kk)-\mu) = -4 \mu N_M +  \sum_{\kk,\eta } (\epsilon_{+,\eta}(\kk)+\epsilon_{-,\eta}(\kk)) = -4 \mu N_M,
\eneq
where we have used the property of unitary particle-hole symmetry that requires $\epsilon_{+,\eta}(\kk)=-\epsilon_{-,\eta}(-\kk)$. 

Now let's look at the last term, for which we need to evaluate $\langle \gamma^\dagger_{\kk e_1 \eta s_1 } \gamma^\dagger_{-\kk e_2 -\eta s_2}\rangle$. 
For spin singlet channel, we have
\begin{eqnarray} 
 & - \sum_{\kk \eta e_1 s_1 e_2 s_2} \langle \gamma^\dagger_{\kk e_1 \eta s_1 } \gamma^\dagger_{-\kk e_2 -\eta s_2}\rangle \Delta^{\eta}_{\kk; e_1 s_1 e_2 s_2 }  = - \sum_{\kk,e_1 s_1 e_2 s_2} ( \langle \gamma^\dagger_{\kk e_1 + s_1 } \gamma^\dagger_{-\kk e_2 - s_2}\rangle \Delta^{+}_{\kk; e_1 s_1 e_2 s_2 } + \langle \gamma^\dagger_{\kk e_1 - s_1 } \gamma^\dagger_{-\kk e_2 + s_2}\rangle \Delta^{-}_{\kk; e_1 s_1 e_2 s_2 } )\nonumber \\
 &= - \sum_{\kk,e_1 e_2} \left( (\langle \gamma^\dagger_{\kk e_1 + \uparrow } \gamma^\dagger_{-\kk e_2 - \downarrow}\rangle- \langle \gamma^\dagger_{\kk e_1 + \downarrow } \gamma^\dagger_{-\kk e_2 - \uparrow }\rangle) \Delta^{+}_{\kk; e_1 e_2 } + (\langle \gamma^\dagger_{\kk e_1 - \uparrow } \gamma^\dagger_{-\kk e_2 + \downarrow }\rangle -\langle \gamma^\dagger_{\kk e_1 - \downarrow } \gamma^\dagger_{-\kk e_2 + \uparrow }\rangle ) \Delta^{-}_{\kk; e_1 e_2 } \right)\nonumber
\end{eqnarray}
where we have used $\Delta^{\eta}_{\kk; e_{1} s_1, e_2 s_2}=\Delta^{\eta}_{\kk;e_1,e_2} (i s_y)_{s_1 s_2}$ for spin singlet, so that $\Delta^{\pm}_{\kk; e_{1} \uparrow, e_2 \downarrow}=\Delta^{\eta}_{\kk;e_1,e_2} $ and $\Delta^{\pm}_{\kk; e_{1} \downarrow, e_2 \uparrow}=-\Delta^{\eta}_{\kk;e_1,e_2}$. 
With the Bogoliubov transformation (\ref{eq:Bogoliubov_transformation_1}) and (\ref{eq:Bogoliubov_transformation_2}), we find 
\begin{eqnarray}
&\langle \gamma^\dagger_{\kk e_1 + \uparrow } \gamma^\dagger_{-\kk e_2 - \downarrow}\rangle = \sum_n v_{\kk,e_1 n}^* r_{\kk, e_2 n}\\
&\langle \gamma^\dagger_{\kk e_1 + \downarrow } \gamma^\dagger_{-\kk e_2 - \uparrow }\rangle = - \sum_n v^*_{\kk,e_1 n} r_{\kk,e_2 n}\\
&\langle \gamma^\dagger_{\kk e_1 - \uparrow } \gamma^\dagger_{-\kk e_2 + \downarrow }\rangle = -\sum_n w_{-\kk,e_1 n} u^*_{-\kk, e_2 n}\\
&\langle \gamma^\dagger_{\kk e_1 - \downarrow } \gamma^\dagger_{-\kk e_2 + \uparrow }\rangle = \sum_n w_{-\kk,e_1n} u^*_{-\kk,e_2 n},
\end{eqnarray}
which leads to 
\begin{eqnarray} 
 & - \sum_{\kk \eta e_1 s_1 e_2 s_2} \langle \gamma^\dagger_{\kk e_1 \eta s_1 } \gamma^\dagger_{-\kk e_2 -\eta s_2}\rangle \Delta^{\eta}_{\kk; e_1 s_1 e_2 s_2 } =
 - 2 \sum_{\kk e_1 e_2 n } \left( v_{\kk,e_1 n}^* r_{\kk, e_2 n}  \Delta^{+}_{\kk; e_1 e_2 } -  w_{-\kk,e_1 n} u^*_{-\kk, e_2 n}  \Delta^{-}_{\kk; e_1 e_2 }
 \right)
\end{eqnarray}
The unitary property $U_{\kk}U^\dagger_{\kk}=I$ ($I$ is an identity matrix) can lead to 
\beq \sum_n v^*_{\kk,e_1n} r_{\kk,e_2 n}= - \sum_n u^*_{\kk,e_1 n} w_{\kk, e_2 n}
\eneq
and together with $\Delta_{\kk; e_1 e_2 }=\Delta^{+}_{\kk; e_1 e_2 } = \Delta^-_{-\kk, e_2 e_1}$ for spin-singlet channel, we eventually obtain 
\begin{eqnarray} 
 & - \sum_{\kk \eta e_1 s_1 e_2 s_2} \langle \gamma^\dagger_{\kk e_1 \eta s_1 } \gamma^\dagger_{-\kk e_2 -\eta s_2}\rangle \Delta^{\eta}_{\kk; e_1 s_1 e_2 s_2 } =
 4 \sum_{\kk e_1 e_2 n }  u_{\kk,e_1 n}^* w_{\kk, e_2 n}  \Delta_{\kk; e_1 e_2 } 
\end{eqnarray}
The superconductor ground state energy is given by  
\beq \label{eq:BdG_Ground_State_Energy} 
\langle\mathcal{H}_{tot}\rangle_s=4\sum_{\kk, n} E^-_{\kk,n} -4 \mu N_M + 4 \sum_{\kk e_1 e_2 n }  u_{\kk,e_1 n}^* w_{\kk, e_2 n}  \Delta_{\kk; e_1 e_2 } . 
\eneq
and the condensation energy is 
\beq E_c=4 \sum_{\kk, n} (E^-_{\kk,n}-\xi^-_{\kk,n})  + 4 \sum_{\kk e_1 e_2 n }  u_{\kk,e_1 n}^* w_{\kk, e_2 n}  \Delta_{\kk; e_1 e_2 },  \eneq
where $\xi^-_{\kk,n}$ is the eigen-energy of the BdG Hamiltonian at zero gap function $\Delta$. 

For spin-triplet channel, we consider the unitary pairing $(\Delta_{\kk} \otimes \mS)^\dagger(\Delta_{\kk} \otimes \mS)\propto I$, and in this case, we can always rotate the spin basis of gap function to $\mS^{10}=s_x$ since the single-particle Hamiltonian part is spin-independent. For the triplet component $\mS^{10}=s_x$, the BdG Hamiltonian for one valley reads 
\begin{eqnarray}\label{eq:HBdG_valleyp1_triplet}
H_{BdG}^+(\kk) =\frac{1}{2} \begin{pmatrix}
h_+(\kk)\otimes s_0 & 2 \Delta_{\kk} \otimes s_x \\
2 \Delta_{\kk}^\dagger\otimes  s_x & - h_{-}^\star(-\kk) \otimes s_0
\end{pmatrix}=\frac{1}{2}\begin{pmatrix}
h_+(\kk) & 0 & 0 & 2 \Delta_{\kk} \\
0 & h_+(\kk) & 2 \Delta_{\kk} & 0 \\
0 & 2 \Delta_{\kk}^\dagger & - h_{-}^\star(-\kk) & 0 \\
2 \Delta_{\kk}^\dagger & 0 &0 & - h_{-}^\star(-\kk) 
\end{pmatrix} 
\end{eqnarray}
on the basis 
$ (\gamma^\dagger_{\kk, e_Y = \pm ,+,\uparrow}, \gamma^\dagger_{\kk, e_Y = \pm ,+,\downarrow}, \gamma_{-\kk, e_Y=\pm,-,\uparrow},\gamma_{-\kk, e_Y=\pm,-,\downarrow})$. This Hamiltonian is also block diagonal and we rewrite each block Hamiltonian as
\begin{eqnarray}\label{eq:HBdG_valleyp2_triplet}
H_{BdG}^{+,\lambda=\pm}(\kk) = \frac{1}{2}  \begin{pmatrix}
h_+(\kk) & 2 \Delta_{\kk} \\
2 \Delta_{\kk}^\dagger & - h_{-}^\star(-\kk)
\end{pmatrix}. 
\end{eqnarray}
The basis functions are $(\gamma^\dagger_{\kk, e_Y = \pm ,+,s =\uparrow },\gamma_{-\kk, e_Y=\pm,-,s=\downarrow})$
and $(\gamma^\dagger_{\kk, e_Y = \pm ,+,s =\downarrow },\gamma_{-\kk, e_Y=\pm,-,s=\uparrow})$ for $\lambda=+$ and $-$, separately. 
Compared to the spin-singlet state, the only difference is that the gap function here is independent of $\lambda$. Thus, all the derivation above
can be directly applied here. 

For spin triplet state, $\Delta^{+}_{\kk; e_{1} s_1, e_2 s_2}=\Delta_{\kk;e_1,e_2} (s_x)_{s_1 s_2}$, so $\Delta_{\kk;e_1,e_2}=\Delta^{+}_{\kk; e_{1} \uparrow, e_2 \downarrow}=\Delta^{+}_{\kk; e_{1} \downarrow, e_2 \uparrow}$. 
The gap equation the explicitly reads
\beq \label{eq:gapEqn_general_triplet1} \Delta_{\kk;e_1e_2}= \Delta^+_{\kk; e_{1} \downarrow, e_2 \uparrow} = - \frac{1}{N_M} \sum_{\kk'} V_{\kk\kk'}^{+ e_1 e_2} \langle \gamma_{-\kk'  e_2 ,+, \uparrow } \gamma_{\kk' e_1, -, \downarrow}\rangle 
\eneq
and
\beq \label{eq:gapEqn_general_triplet2}
 \Delta_{\kk;e_1e_2}=\Delta^{+}_{\kk; e_{1} \uparrow, e_2 \downarrow}= - \frac{1}{N_M}  \sum_{\kk'} V_{\kk\kk'}^{+ e_1 e_2} \langle \gamma_{-\kk'  e_2 + \downarrow} \gamma_{\kk' e_1 - \uparrow}\rangle. 
\eneq

Now one can see that for $\lambda=\pm$, both the BdG Hamiltonian and the gap equation are completely the same. Thus, we only need to look at one of them. Following the procedure above, we obtain
\begin{eqnarray}\label{eq:gapEqn_triplet_lambdap}
\Delta_{\kk;e_1 e_2} & = & - \frac{1}{N_M} \sum_{\kk'} V_{\kk\kk'}^{+ e_1 e_2} \langle \gamma_{-\kk'  e_2 ,+, \uparrow } \gamma_{\kk' e_1, -, \downarrow}\rangle \nonumber \\
& = & - \frac{1}{N_M} \sum_{\kk', n } V_{\kk\kk'}^{+ e_1 e_2} \left( u_{-\kk',e_2 n} w^*_{-\kk',e_1 n} (1 - f(E^+_{-\kk',n})) +
 v_{-\kk',e_2 n} r^*_{-\kk',e_1 n} f(-E^-_{-\kk',n}) \right), 
\end{eqnarray}
At zero temperature, the self-consistent gap equation for spin-triplet channels reads
\beq \label{eq:GapEquation_full_zeroT_triplet}
\Delta_{\kk;e_1 e_2}= - \frac{1}{N_M} \sum_{\kk', n } V_{\kk\kk'}^{+ e_1 e_2}  u_{-\kk',e_2 n} w^*_{-\kk',e_1 n}. \eneq

Although the above formalism looks similar to the spin singlet case, the form of the gap function is different. 
As discussed in the paragraph after the expansion (\ref{eq:Gapfunction_decomp}) of the gap function, we know that for the inter-Chern-band channel, 
the gap function takes the form $\Delta_{\kk;e_1,e_2}=\Delta_{\kk;y}(\zeta^y)_{e_1 e_2}$ for even parity ($\lambda_{C_{2z}}=+1$)
and $\Delta_{\kk;e_1,e_2}=\Delta_{\kk;x}(\zeta^x)_{e_1 e_2}$ for odd parity ($\lambda_{C_{2z}}=-1$) so that the condition  (\ref{eq:LGE_constraint_triplet})
($\Delta_{\kk;e_1,e_2}=-\lambda_{C_{2z}}\Delta_{\kk;e_2,e_1}$ ) is satisfied. 

For spin-triplet pairing, since the spin part is symmetric, $\Delta_{\kk;e_1,e_2}=\sum_\mu \Delta_{\kk;\mu}\zeta^\mu$ needs to satisfy $\Delta_{\kk;e_1,e_2}=-\lambda_{C_{2z}}\Delta_{\kk;e_2,e_1}$ according to Eq. (\ref{eq:LGE_constraint_triplet})
where $\lambda_{C_{2z}}$ is the eigen-value of $C_{2z}$ operator acting on the gap function. 
This requires either anti-symmetric $\zeta^{\mu=y}$ and $\Delta_{\kk;\mu=y}=\lambda_{C_{2z}}\Delta_{-\kk;\mu=y}$ for inter-Chern-band pairing, 
or symmetry $\zeta^{\mu=0,x,z}$ and $\Delta_{\kk;\mu=0,x,z}=-\lambda_{C_{2z}}\Delta_{-\kk;\mu=0,x,z}$ for both intra-Chern-band pairing ($\mu=0,z$) and inter-Chern-band pairing ($\mu=x$). 

For the ground state energy, we expect the first and second terms in Eq. (\ref{eq:Htot_SC}) should still take the same form once the eigen-energy $E^-_{\kk,n}$ is calculated from the BdG Hamiltonian (\ref{eq:HBdG_valleyp2_triplet}) with a different form of $\Delta_{\kk}$. 
For the last term, we have
\begin{eqnarray} 
 & - \sum_{\kk \eta e_1 s_1 e_2 s_2} \langle \gamma^\dagger_{\kk e_1 \eta s_1 } \gamma^\dagger_{-\kk e_2 -\eta s_2}\rangle \Delta^{\eta}_{\kk; e_1 s_1 e_2 s_2 }  = - \sum_{\kk,e_1 s_1 e_2 s_2} ( \langle \gamma^\dagger_{\kk e_1 + s_1 } \gamma^\dagger_{-\kk e_2 - s_2}\rangle \Delta^{+}_{\kk; e_1 s_1 e_2 s_2 } + \langle \gamma^\dagger_{\kk e_1 - s_1 } \gamma^\dagger_{-\kk e_2 + s_2}\rangle \Delta^{-}_{\kk; e_1 s_1 e_2 s_2 } )\nonumber \\
 &= - \sum_{\kk,e_1 e_2} \left( (\langle \gamma^\dagger_{\kk e_1 + \uparrow } \gamma^\dagger_{-\kk e_2 - \downarrow}\rangle + \langle \gamma^\dagger_{\kk e_1 + \downarrow } \gamma^\dagger_{-\kk e_2 - \uparrow }\rangle) \Delta^{+}_{\kk; e_1 e_2 } + (\langle \gamma^\dagger_{\kk e_1 - \uparrow } \gamma^\dagger_{-\kk e_2 + \downarrow }\rangle + \langle \gamma^\dagger_{\kk e_1 - \downarrow } \gamma^\dagger_{-\kk e_2 + \uparrow }\rangle ) \Delta^{-}_{\kk; e_1 e_2 } \right)\nonumber
\end{eqnarray}
where we have used $\Delta^{\eta}_{\kk; e_{1} s_1, e_2 s_2}=\Delta^{\eta}_{\kk;e_1,e_2} (s_x)_{s_1 s_2}$. 
We can again apply the Bogoliubov transformation (\ref{eq:Bogoliubov_transformation_1}) and (\ref{eq:Bogoliubov_transformation_2}) as the Hamiltonian (\ref{eq:HBdG_valleyp2_triplet}) is the same as (\ref{eq:HBdG_valleyp2}) for $\lambda=+$. We find
\begin{eqnarray}
&\langle \gamma^\dagger_{\kk e_1 + \uparrow } \gamma^\dagger_{-\kk e_2 - \downarrow}\rangle = \sum_n v_{\kk,e_1 n}^* r_{\kk, e_2 n}\\
&\langle \gamma^\dagger_{\kk e_1 + \downarrow } \gamma^\dagger_{-\kk e_2 - \uparrow }\rangle = \sum_n v^*_{\kk,e_1 n} r_{\kk,e_2 n}\\
&\langle \gamma^\dagger_{\kk e_1 - \uparrow } \gamma^\dagger_{-\kk e_2 + \downarrow }\rangle = \sum_n w_{-\kk,e_1 n} u^*_{-\kk, e_2 n}\\
&\langle \gamma^\dagger_{\kk e_1 - \downarrow } \gamma^\dagger_{-\kk e_2 + \uparrow }\rangle =  \sum_n w_{-\kk,e_1n} u^*_{-\kk,e_2 n},
\end{eqnarray}
which leads to 
\begin{eqnarray} 
 & - \sum_{\kk \eta e_1 s_1 e_2 s_2} \langle \gamma^\dagger_{\kk e_1 \eta s_1 } \gamma^\dagger_{-\kk e_2 -\eta s_2}\rangle \Delta^{\eta}_{\kk; e_1 s_1 e_2 s_2 } =
 - 2 \sum_{\kk e_1 e_2 n } \left( v_{\kk,e_1 n}^* r_{\kk, e_2 n}  \Delta^{+}_{\kk; e_1 e_2 } +  w_{-\kk,e_1 n} u^*_{-\kk, e_2 n}  \Delta^{-}_{\kk; e_1 e_2 }
 \right)
\end{eqnarray}

From the unitary property 
\beq \sum_n v^*_{\kk,e_1n} r_{\kk,e_2 n}= - \sum_n u^*_{\kk,e_1 n} w_{\kk, e_2 n}
\eneq ($U_{\kk}U^\dagger_{\kk}=I$)
and $\Delta_{\kk; e_1 e_2 }=\Delta^{+}_{\kk; e_1 e_2 } = - \Delta^-_{-\kk, e_2 e_1}$ for spin-triplet channel, we obtain 
\begin{eqnarray} 
 & - \sum_{\kk \eta e_1 s_1 e_2 s_2} \langle \gamma^\dagger_{\kk e_1 \eta s_1 } \gamma^\dagger_{-\kk e_2 -\eta s_2}\rangle \Delta^{\eta}_{\kk; e_1 s_1 e_2 s_2 } =
 4 \sum_{\kk e_1 e_2 n }  u_{\kk,e_1 n}^* w_{\kk, e_2 n}  \Delta_{\kk; e_1 e_2 } 
\end{eqnarray}
The superconductor ground state energy is given by  
\beq  
\langle\mathcal{H}_{tot}\rangle_s=4\sum_{\kk, n} E^-_{\kk,n} -4 \mu N_M + 4 \sum_{\kk e_1 e_2 n }  u_{\kk,e_1 n}^* w_{\kk, e_2 n}  \Delta_{\kk; e_1 e_2 } . 
\eneq
and the condensation energy is 
\beq E_c=4 \sum_{\kk, n} (E^-_{\kk,n}-\xi^-_{\kk,n})  + 4 \sum_{\kk e_1 e_2 n }  u_{\kk,e_1 n}^* w_{\kk, e_2 n}  \Delta_{\kk; e_1 e_2 },  \eneq
where $\xi^-_{\kk,n}$ is the eigen-energy of the BdG Hamiltonian at zero gap function $\Delta$. 
So we find the same form of the ground state energy for spin-triplet pairing as that of spin-singlet pairing.

\subsubsection{Bogoliubov-de Gennes Hamiltonian for the Intra-Chern-band Channel}
The linearized gap equation has shown the form of the spin-singlet intra-Chern-band pairing channels in Eq. (\ref{eq:gapfun_intra_1}) and (\ref{eq:gapfun_intra_2}) around $T_c$, which belong to 2D $E_2$ irrep. There are degenerate pairing channels, and this degeneracy is expected to break when the temperature is below $T_c$. So here we apply the BdG formalism to the intra-Chern-band pairing channels to study the pairing state at zero temperature. We keep the general form 
\beq (\Delta_{\kk})_{e_1 e_2} \otimes \mS = \Delta_{\kk,e_1}\delta_{e_1 e_2} i s_y = \left(\begin{array}{cc} \Delta_{\kk,+}&0\\
0 & \Delta_{\kk,-}\end{array}\right) is_y =  (\Delta_{\kk,0}\zeta^0+\Delta_{\kk,z}\zeta^z)_{e_1,e_2}i s_y . \eneq
For the $\hat{C}_{2z}\hat{T}$-invariant pairing, the intra-Chern-band gap function need to satisfies 
\beq \Delta_{\kk,+}=(\Delta_{\kk,-})^\star  \implies \Delta_{\kk,0}=\Delta_{\kk,0}^\star; \Delta_{\kk,z}=-\Delta_{\kk,z}^\star \eneq 
according to the constraint (\ref{eq:C2zT_Gapfunction}). 

With the single-particle Hamiltonian (\ref{eq:single_particle_Ham1}), the BdG Hamiltonian reads
\begin{eqnarray}\label{eq:HBdGpup}
& H_{BdG}^{+,\lambda}(\kk) = \frac{1}{2}\begin{pmatrix}
(d_{0,\kk}-\mu)\zeta^0+d_{x,\kk}\zeta^x & 2\lambda (\Delta_{0,\kk}\zeta^0+\Delta_{z,\kk}\zeta^z )\\
2\lambda (\Delta^\star_{0,\kk}\zeta^0+\Delta^\star_{z,\kk}\zeta^z) & - (d_{0,\kk}-\mu)\zeta^0-d_{x,\kk}\zeta^x 
\end{pmatrix}\nonumber\\
& = \frac{1}{2} \begin{pmatrix}
d_{0,\kk}-\mu & d_{x,\kk} & 2\lambda \Delta_{\kk,+} & 0 \\
d_{x,\kk} & d_{0,\kk}-\mu & 0 & 2\lambda \Delta_{\kk,-} \\
2\lambda \Delta^\star_{\kk,+} & 0 & - (d_{0,\kk}-\mu) & - d_{x,\kk} \\
0 & 2\lambda \Delta^\star_{\kk,-} & - d_{x,\kk} & - (d_{0,\kk}-\mu)
\end{pmatrix},
\end{eqnarray}
where $\lambda=\pm$, $d_{0(x),\kk}=d_{0(x),+}(\kk)$, and $\Delta_{\kk,e_Y}=\Delta_{0,\kk}+e_Y \Delta_{z,\kk},\; e_Y=\pm$
for two intra-Chern-band channels with the Chern number $e_Y$. 
Below we mainly focus on two types of pairing forms: (1) the Euler pairing with $|\Delta_{\kk,+}|=|\Delta_{\kk,-}|$\cite{yu2022euler2}, which breaks three-fold rotation and thus is a nematic phase, and (2) the chiral d-wave pairing with $\Delta_{\kk,+}\neq 0, \Delta_{\kk,-}=0$ or $\Delta_{\kk,-}\neq 0, \Delta_{\kk,+}=0$, which respect three-fold rotation, but break TR. 

We first want to discuss the general conditions for the existence of point nodes in the BdG Hamiltonian (\ref{eq:HBdGpup}). The eigen-energy $E^\pm_{\kk,n}$ of the BdG Hamiltonian (\ref{eq:HBdGpup}) can be solved explicitly for two independent order parameters $\Delta_{+,\kk}$
and $\Delta_{-,\kk}$,
\begin{eqnarray} \label{eq:BdGEnergy1}
& E^\pm_{\kk,n}= \pm \frac{1}{\sqrt{2}}\sqrt{A+\frac{1}{2}\tilde{d}_{0,\kk}^2+\frac{1}{2}d_{x,\kk}^2 + n \sqrt{A^2-4B^2+ \tilde{d}^2_{0,\kk}d^2_{x,\kk}+d_{x,\kk}^2(A-2B \cos(\Phi_{\kk,-}-\Phi_{\kk,+}))}}\nonumber\\
& = \pm \frac{1}{2}\sqrt{2A+\tilde{d}_{0,\kk}^2+d_{x,\kk}^2 + 2 n \sqrt{A^2-4B^2+ \tilde{d}^2_{0,\kk}d^2_{x,\kk}+d_{x,\kk}^2(A-2B \cos(\Phi_{\kk,-}-\Phi_{\kk,+}))}}
\end{eqnarray}
where $n=\pm$, $\tilde{d}_{0,\kk}=d_{0,\kk}-\mu$, $A=|\Delta_{\kk,+}|^2+|\Delta_{\kk,-}|^2$ and $B=|\Delta_{\kk,+}\Delta_{\kk,-}|$. We also denote 
\beq \label{eq:dwave_ansatz1} \Delta_{\kk,e_Y}=|\Delta_{\kk,e_Y}| e^{i\Phi_{\kk,e_Y}}. \eneq
From the linearized gap equation discussed in Sec. \ref{sec:LGE_intra}, we have shown that the inter-Chern-band channel will favor d-wave $E_2$ pairing with and the gap function ansatz 
\beq \label{eq:dwave_ansatz} \Delta_{\kk;e_Y} = \Delta_{e_Y} \frac{k^2_{-e_Y}}{k^2+b^2}, \eneq
where $e_Y=\pm$ and $\Delta_{e_Y}=|\Delta_{e_Y}|e^{i\varphi_{e_Y}}$. With this gap function ansatz, we find 
\beq \Phi_{\kk,e_Y}=\varphi_{e_Y}-2 e_Y \theta_{\kk} \eneq
with the momentum $\kk=k(\cos\theta_{\kk},\sin\theta_{\kk})$ and 
\beq |\Delta_{\kk,e_Y}|=|\Delta_{e_Y}| \frac{k^2}{k^2+b^2}.  \eneq
For the above eigen-energy, the gapless points at zero energy require
\begin{eqnarray} 
& (2A+\tilde{d}_{0,\kk}^2+d_{x,\kk}^2)^2= 4 (A^2-4B^2+ \tilde{d}^2_{0,\kk}d^2_{x,\kk}+d_{x,\kk}^2(A-2B \cos(\Phi_{\kk,-}-\Phi_{\kk,+}))) \nonumber \\
& \rightarrow 16B^2+4 A \tilde{d}_{0,\kk}^2+(\tilde{d}_{0,\kk}^2-d_{x,\kk}^2)^2+8 B d_{x,\kk}^2 \cos(\Phi_{\kk,-}-\Phi_{\kk,+}) = 0  \nonumber \\
& \rightarrow (4B + \tilde{d}_{0,\kk}^2-d_{x,\kk}^2)^2 + 4 \tilde{d}_{0,\kk}^2 (|\Delta_{\kk,+}|-|\Delta_{\kk,-}|)^2 + 16 B d_{x,\kk}^2 \cos^2\left(\frac{\Phi_{\kk,-}-\Phi_{\kk,+}}{2}\right) = 0. 
\end{eqnarray}
where we have used $A-2B=(|\Delta_{\kk,+}|-|\Delta_{\kk,-}|)^2$. One can see that all three terms in the above equation have to be equal or larger than 0, so the above equation can be satisfied for the following conditions, either (1) 
\begin{eqnarray} 
&4B + \tilde{d}_{0,\kk}^2=d_{x,\kk}^2; \label{eq:nodecondi_11} \\
&|\Delta_{\kk,+}|=|\Delta_{\kk,-}|; \label{eq:nodecondi_12}\\
&\cos\left(\frac{\Phi_{\kk,-}-\Phi_{\kk,+}}{2}\right)=0,  \label{eq:nodecondi_13}
\end{eqnarray}
or (2)
\begin{eqnarray} 
&4B = d_{x,\kk}^2; \label{eq:nodecondi_21} \\
&\tilde{d}_{0,\kk}=0; \label{eq:nodecondi_22} \\ 
&\cos\left(\frac{\Phi_{\kk,-}-\Phi_{\kk,+}}{2}\right)=0. \label{eq:nodecondi_23} 
\end{eqnarray}

For the case (1), we find that Eq. (\ref{eq:nodecondi_12}) is just the condition of Euler pairing. Let us define $|\Delta_{\kk,+}|=|\Delta_{\kk,-}|=\tilde{\Delta}_{0,k}$ and then Eq. (\ref{eq:nodecondi_11}) is simplified to
\beq \label{Eq:Euler_constr1} 4\tilde{\Delta}_{0,k}^2 + \tilde{d}_{0,\kk}^2=d_{x,\kk}^2, \eneq
which determines the the amplitude of the momentum for nodes. In addition, Eq. (\ref{eq:nodecondi_13}) requires
\begin{eqnarray} \label{Eq:Euler_constr2} 
&\cos\left(\frac{\Phi_{\kk,-}-\Phi_{\kk,+}}{2}\right)=0 \rightarrow \Phi_{\kk,-}-\Phi_{\kk,+}=4\theta_{\kk}+\varphi_--\varphi_+= (2n+1) \pi,  \nonumber \\
&\rightarrow \theta_{\kk}=\frac{1}{4} ( (2n+1) \pi + \varphi_+ - \varphi_- ) 
\end{eqnarray}
with $n\in \mathcal{Z}$ (any integer number), which fixes the momentum angle $\theta_{\kk}$ for the location of the nodes in the 2D momentum space. Thus, we generally have point nodes in the 2D momentum space for the Hamiltonian (\ref{eq:HBdGpup}), with their locations fixed by Eqs. (\ref{Eq:Euler_constr1}) and (\ref{Eq:Euler_constr2}). 

For the case (2), both Eqs. (\ref{eq:nodecondi_21}) and (\ref{eq:nodecondi_22}) determine the required momentum amplitude for the nodes, which can not be satisfied in general. However, in the chiral limit $d_{0,\kk}=0$ and with zero chemical potential $\mu=0$, $\tilde{d}_{0,\kk}=0$ for any momentum. In this case, Eq. (\ref{eq:nodecondi_21}) fixes the momentum amplitude and Eq. (\ref{eq:nodecondi_23}) gives the momentum angle of the nodes, so the gapless condition in this case does not require the Euler pairing. For any $(\Delta_{\kk,+}, \Delta_{\kk,-})$, the nodes can exist in the momentum space once the conditions (\ref{eq:nodecondi_21}) and (\ref{eq:nodecondi_23}) are satisfied. 

In Fig. \ref{fig:BdG_nematic_chiral}(a) and (c), we numerically calculate the energy spectrum of BdG Hamiltonian, which shows nodes for the Euler pairing and a full gap for the chiral d-wave pairing. In our numerical calculation, we choose $\Delta_{e_Y}$ to be real ($\varphi_{\pm}=0$), so that $\theta_{\kk}=\pm\frac{\pi}{4}, \pm\frac{3\pi}{4}$, and indeed we find the nodes are located at the diagonal lines in the $k_x$-$k_y$ space in Fig. \ref{fig:BdG_nematic_chiral}b. One can further check the amplitude of the momentum for the nodes and find that the locations of nodes for Euler pairing match the above two conditions (Eqs. (\ref{Eq:Euler_constr1}) and (\ref{Eq:Euler_constr2})). 

\begin{figure}[hbt!]
   \centering
    \includegraphics[width=7in]{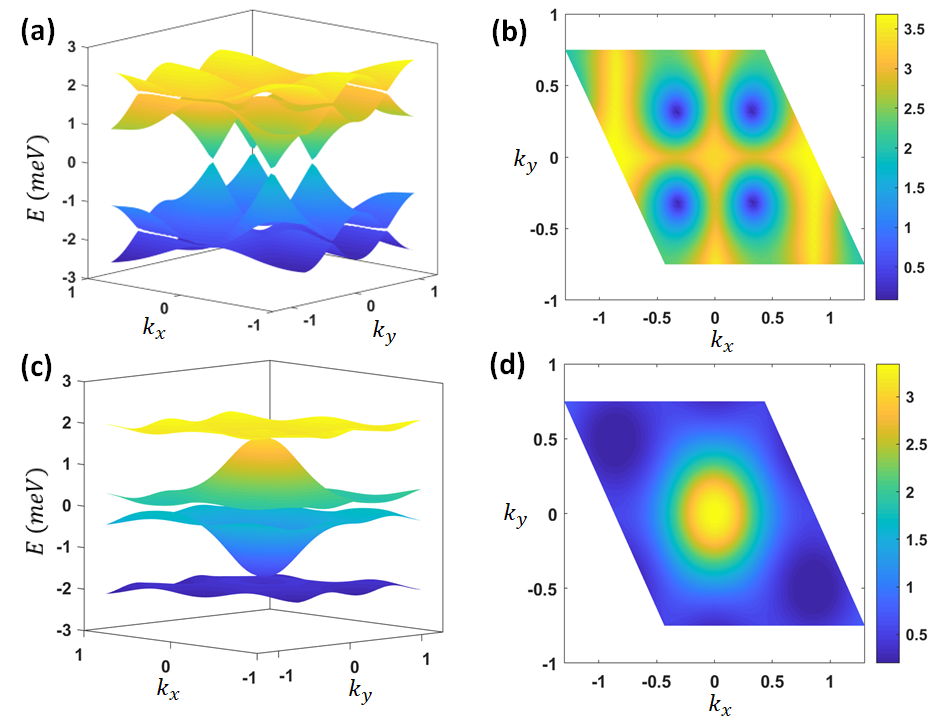} 
    \caption{ (a) and (b) The BdG spectrum and the BdG gap of nematic pairing. (c) and (d) The BdG spectrum and the BdG gap of chiral d-wave pairing. 
    The chemical potential is chosen at $\mu=0.1 meV$. 
  }
    \label{fig:BdG_nematic_chiral}
\end{figure}

The linearized gap equation only tells us the pairing channels around $T_c$ and for the intra-Chern-band channels, different components of the $E_2$ irrep pairings share the same $T_c$. To determine the exact pairing form for the temperature well below $T_c$, we apply the full self-consistent gap equation (\ref{eq:GapEquation_full_zeroT}) to the intra-Chern-band channel, which takes the form 
\beq \Delta_{\kk;e_1}= \frac{1}{N_M} \sum_{\kk', n } V_{\kk\kk'}^{+ e_1 e_1}  u_{-\kk',e_1 n} w^*_{-\kk',e_1 n} \eneq
at zero temperature.
Now let us take the d-wave component ($U^*_{1,e_Y,\kk}U_{1,e_Y,\kk'}$) of the interaction in (\ref{eq:Interactionform_intra_Chern1}) and the gap function ansatz 
Eq. (\ref{eq:dwave_ansatz}). The gap equation can be simplified to 
\beq \label{eq:GapEquation_intra_dwave1} \Delta_{e_1}= \frac{V_0 }{N_M} \sum_{\kk', n } \frac{{k'}^2_{e_1}}{k'^2+b^2}  u_{-\kk',e_1 n} w^*_{-\kk',e_1 n}. \eneq

The above gap function ansatz and gap equation can also help to simplify the ground state energy and condensation energy for superconductivity. Let us consider the last term in the ground state energy of Eq. (\ref{eq:BdG_Ground_State_Energy}) for intra-Chern-band pairing, $e_1=e_2$. With the ansatz (\ref{eq:dwave_ansatz}), we find 
\beq 
 4 \sum_{\kk e_1 n }  u_{\kk,e_1 n}^* w_{\kk, e_1 n}  \Delta_{\kk; e_1} =  4 \sum_{\kk e_1 n }  u_{\kk,e_1 n}^* w_{\kk, e_1 n} \Delta_{e_1} \frac{k^2_{-e_1}}{k^2+b^2} \\
 =  4 \sum_{e_1 } \Delta_{e_1} \left(\sum_{\kk, n} u_{\kk,e_1 n} w_{\kk, e_1 n}^*  \frac{k^2_{e_1}}{k^2+b^2} \right)^* = \frac{4 N_M}{V_0} \sum_{e_1 } |\Delta_{e_1}|^2,
\eneq
where we have used the gap equation (\ref{eq:GapEquation_intra_dwave1}) in the last step. With this simplification, we can write the superconductor ground state energy and condensation energy for intra-Chern-band pairing as
\beq 
\langle\mathcal{H}_{tot}\rangle_s=4 \left(\sum_{\kk, n} E^-_{\kk,n} - \mu N_M + \frac{N_M}{V_0} \sum_{e_1 } |\Delta_{e_1}|^2\right). 
\eneq
and
\beq \label{eq:condensation_energy_intra1} 
E_c=4 \left( \sum_{\kk, n} (E^-_{\kk,n}-\xi^-_{\kk,n})  + \frac{ N_M}{V_0} \sum_{e_1 } |\Delta_{e_1}|^2 \right).  \eneq

The above gap equation can be solved numerically, as discussed in details in the main text. Our numerical calculations of the full self-consistent gap equation was performed in the chiral limit with $w_0=0$, but go beyond the flat-band limit. 
With the parameter set listed in the main text, we find the bandwidth of the nearly flat bands at $\theta=1.06^\circ$ is less than $0.5 meV$. It turns out that the superconductivity is quite sensitive to the band width of the flat bands. 
As shown in Fig. \ref{fig:bands_free}, we find the profile of the band dispersion for the nearly flat bands around the magic angle keeps similar and only the magnitudes of the band width are changing. Thus, for our numerical calculations, we introduce a parameter $\xi$ to manually tune the band width of the single particle Hamiltonian, 
\begin{eqnarray}\label{eq:HBdGpup_1}
& H_{BdG}^{+,\lambda}(\kk) = \frac{1}{2} \begin{pmatrix}
-\mu & \xi d_{x,\kk} & 2\lambda \Delta_{\kk,+} & 0 \\
\xi d_{x,\kk} & -\mu & 0 & 2\lambda \Delta_{\kk,-} \\
2\lambda \Delta^\star_{\kk,+} & 0 & \mu & - \xi d_{x,\kk} \\
0 & 2\lambda \Delta^\star_{\kk,-} & - \xi d_{x,\kk} & \mu
\end{pmatrix},
\end{eqnarray}
where $d_{0,\kk}=0$ in the chiral limit. We can numerically solve the self-consistent solutions of the gap functions from Eq. (\ref{eq:GapEquation_intra_dwave1}) and the corresponding condensation energies from Eq. (\ref{eq:condensation_energy_intra1} ), 
as shown in the Fig. 2 of the main text, where we choose $\xi=0.6$ corresponding to a band width $0.3 meV$. We also show the self-consistent solutions and ground state energy for $\xi=1$ corresponding to a band width $0.5 meV$ in Fig. \ref{fig_gapself_lambda05}. 
In this case, all the superconducting regime is found to have nodes. 

\begin{figure}
	\includegraphics[width=0.9\textwidth,angle=0]{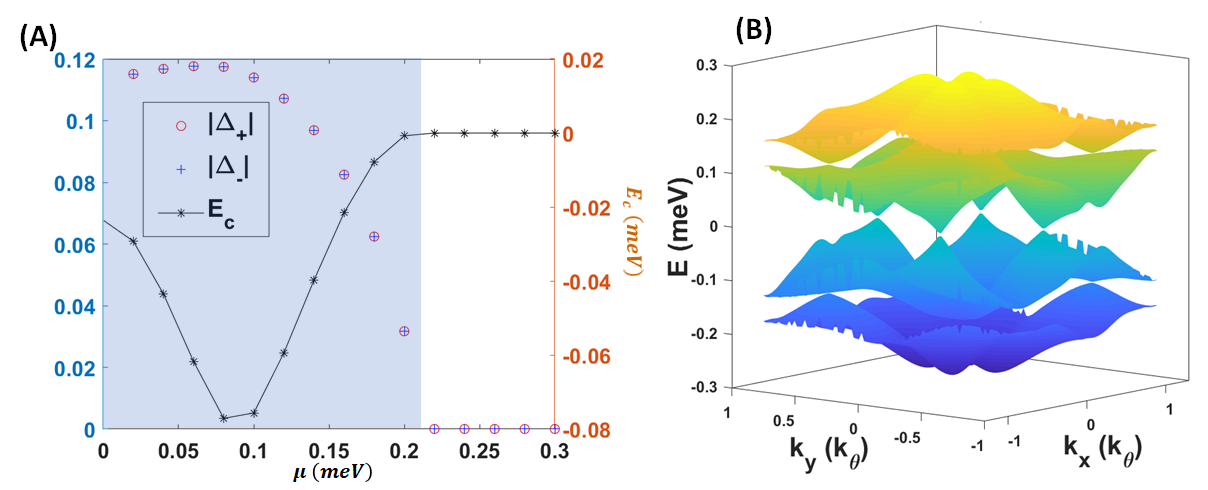}
	\centering
	\caption{ (A) Red circles and blue crosses show the gap functions $\Delta_{+}$ and $\Delta_{-}$ of the self-consistent gap equation as a function of the chemical potential $\mu$ for the intra-Chern-band pairing. Black stars show the condensation energy $E_c$ as a function of $\mu$. (B) show the BdG spectrum for the chemical potential $\mu=0.02 meV$. Here the bandwidth parameter $\xi=1$ corresponding to $0.5 meV$ band width. Here $\theta=1.06^\circ$.  }
	\label{fig_gapself_lambda05}
\end{figure}

One surprising result is that the Euler pairing is always energetically favored, which is in sharp contrast to the chiral d-wave pairing in doped graphene\cite{black2014chiral,nandkishore2012chiral}. Below we will look at the chiral flat band limit $w_0=0$ with zero kinetic energy, in which $H_{BdG}^{+,+}(\kk)$ is block diagonal, 
\begin{eqnarray}
H_{BdG}^{+,+}(\kk) =\begin{pmatrix}
0 & 0 & \Delta_{\kk,+} & 0 \\
0 & 0 & 0 & \Delta_{\kk,-} \\
\Delta_{\kk,+}^\dagger & 0 & 0 & 0 \\
0 & \Delta_{\kk,-}^\dagger & 0 &0  
\end{pmatrix},
\end{eqnarray}
so we can completely solve the problem. One can show that the eigen-energy spectrum and eigen-state of the BdG Hamiltonian $H_{BdG}^{+,+}(\kk)$ are given by 
\beq
E^{\pm}_{e_1}=\pm |\Delta_{e_1}|\frac{k^2}{k^2+b^2}, \;\; U^{(0)}_{\kk,e_1 e_2}=U^{(0)}_{\kk,e_1}\delta_{e_1 e_2}, \; \; 
U^{(0)}_{\kk,e_1}=\frac{1}{\sqrt{2}}\begin{pmatrix} e^{i(\varphi_{e_1}-2 e_1 \theta_\kk)} & - e^{i(\varphi_{e_1}-2 e_1 \theta_\kk)} \\
1 & 1 \end{pmatrix},
\eneq
where $\Delta_{e_1}=|\Delta_{e_1}|e^{i \varphi_{e_1}}$ and $k_{e_1}=k_x+i e_1 k_y = k e^{ie_1 \theta_\kk}$. The gap equation is
then changed to 
\beq \label{eq:Delta_e_chiralLimit} |\Delta_{e_1}| = \frac{V_0}{2 N_M }\sum_{\kk' } \frac{{k'}^2 }{k'^2+b^2} e^{i 2 e_1 \theta_{\kk'}} e^{-i 2 e_1 \theta_{-\kk'} }= \frac{V_0}{2  N_M}\sum_{\kk' } \frac{{k'}^2 }{k'^2+b^2} , \eneq
and the numerical evaluation of this integral gives $|\Delta_{e_1}|\approx 0.21 meV$. We note that the above expression
is independent of the Chern-band index $e_1=\pm$, which means that the amplitudes for the gap functions of the $e_1=\pm$ channels should be equal to each other
$|\Delta_{+}|=|\Delta_{-}|$, while the phases of the gap functions are arbitrary since they do not enter into the gap equation. Thus, the superconductivity here
is always in the nematic phase of Euler pairing.

One can calculate the ground state energy in this limit. With Eq. (\ref{eq:BdG_Ground_State_Energy}), we find the ground state energy at $\mu=0$ is 
\beq \langle\mathcal{H}_{tot}\rangle_s=4 \sum_{\kk, e_1} E^-_{\kk,e_1} + 4 \sum_{\kk e_1 e_2 n }  u_{\kk,e_1 n}^* w_{\kk, e_2 n}  \Delta_{\kk; e_1 e_2 }
=4 \sum_{\kk, e_1} E^-_{\kk,e_1} + 4 N_M \sum_{ e_1 }  \frac{|\Delta_{e_1}|^2}{V_0},  
\eneq
In the last step of the above derivation, we have used the Eqs. (\ref{eq:dwave_ansatz}) and the gap equation (\ref{eq:GapEquation_intra_dwave1}). 
With 
\beq E^-_{e,\kk}=-|\Delta_{\kk,e}|,
\eneq
we obtain
\beq  \langle\mathcal{H}_{tot}\rangle_s = - 4N_M \sum_{ e} \frac{|\Delta_{e}|^2}{V_0},
\eneq
in which we have used Eq. (\ref{eq:Delta_e_chiralLimit}). 
The above expression includes a summation over the Chern-band index $e$, so gapping out each channel will give rise to an energy saving of 
$4N_M\frac{|\Delta_e|^2}{V_0}$. Therefore, in this chiral flat-band limit, due to the decoupling between two channels, we can easily see that the Euler pairing will 
be energetically favored compared to the chiral d-wave pairing.

\subsubsection{Bogoliubov-de Gennes Hamiltonian for the Inter-Chern-band Channel}
In this section, we will apply the BdG formalism to the inter-Chern-band pairing channels. Inter-Chern-band pairing allows for both spin-singlet and spin-triplet pairings, which will be discussed separately below. For spin-singlet pairings, the gap function has the form
\beq \Delta_{\kk} \otimes \mS = \Delta_{\kk} \zeta^x i s_y, \eneq
and thus the BdG Hamiltonian is also block diagonal and we label each block by $\lambda=\pm$, of which the Hamiltonian reads
\begin{eqnarray}\label{eq:HBdGpup_inter_singlet}
& H_{BdG}^{+,\lambda}(\kk) = \frac{1}{2}\begin{pmatrix}
(d_{0,\kk}-\mu)\zeta^0+d_{x,\kk}\zeta^x & 2\lambda \Delta_\kk \zeta^x\\
2\lambda \Delta^*_\kk \zeta^x & - (d_{0,\kk}-\mu)\zeta^0-d_{x,\kk}\zeta^x 
\end{pmatrix}\nonumber\\
& = \frac{1}{2} \begin{pmatrix}
d_{0,\kk}-\mu & d_{x,\kk} & 0 & 2\lambda \Delta_{\kk} \\
d_{x,\kk} & d_{0,\kk}-\mu & 2\lambda \Delta_{\kk} & 0 \\
0 & 2\lambda \Delta^\star_{\kk} & - (d_{0,\kk}-\mu) & - d_{x,\kk} \\
2\lambda \Delta^\star_{\kk} & 0 & - d_{x,\kk} & - (d_{0,\kk}-\mu)
\end{pmatrix}. 
\end{eqnarray}
This Hamiltonian leads to the eigen-energy spectrum
\beq \label{eq:BdG_inter_Eigen_singlet} E^{\pm}_{\kk,n=\pm}=\pm \sqrt{(d_{0,\kk}-\mu+ n d_{x,\kk})^2+|\Delta_{\kk}|^2}.
\eneq
Since the gap function $\Delta_{\kk}$ for the inter-Chern-band channel has no nodes, the above BdG spectrum is thus fully gapped. 

Next we consider the spin-triplet pairing with the gap function
\beq \Delta_{\kk} \otimes \mS = \Delta_{\kk} \zeta^y \mS^{1M}, \eneq
with $M=0,\pm 1$. If we only consider the unitary pairing $(\Delta_{\kk} \otimes \mS)^\dagger(\Delta_{\kk} \otimes \mS)\propto I$, we can always rotate
the spin basis of gap function to $\mS^{10}=s_x$ since the single-particle Hamiltonian part is spin-independent. For the triplet component $\mS^{10}=s_x$, the Hamiltonian is block diagonal and each block (labelled by $\lambda$) reads
\begin{eqnarray}\label{eq:HBdGpup_inter_triplet}
& H_{BdG}^{+,\lambda}(\kk) = \frac{1}{2}\begin{pmatrix}
(d_{0,\kk}-\mu)\zeta^0+d_{x,\kk}\zeta^x & 2 \Delta_\kk \zeta^y\\
2 \Delta^*_\kk \zeta^y & - (d_{0,\kk}-\mu)\zeta^0-d_{x,\kk}\zeta^x 
\end{pmatrix}\nonumber\\
& = \frac{1}{2} \begin{pmatrix}
d_{0,\kk}-\mu & d_{x,\kk} & 0 & -2i \Delta_{\kk} \\
d_{x,\kk} & d_{0,\kk}-\mu & 2i \Delta_{\kk} & 0 \\
0 & -2i \Delta^\star_{\kk} & - (d_{0,\kk}-\mu) & - d_{x,\kk} \\
2i \Delta^\star_{\kk} & 0 & - d_{x,\kk} & - (d_{0,\kk}-\mu)
\end{pmatrix}. 
\end{eqnarray}
The eigen energy spectrum of this Hamiltonian is given by 
\beq \label{eq:BdG_inter_Eigen_triplet} E^{\pm}_{\kk,n}=\pm \left(\sqrt{(d_{0,\kk}-\mu)^2+|\Delta_{\kk}|^2} + n |d_{x,\kk}|\right),
\eneq
where $n=\pm$. The gapless condition of the above energy spectrum is 
\beq (d_{0,\kk}-\mu)^2+|\Delta_{\kk}|^2=d_{x,\kk}^2. 
\eneq
Since there is only one constraint equation in 2D momentum space, we expect a nodal line for the inter-Chern-band triplet pairing channel. 

The above two eigen-energies of the BdG Hamiltonian for spin singlet and triplet states will be the same in the flat-band limit and with zero chemical potential ($d_{0,\kk}=d_{x,\kk}=\mu=0$). Then we expect singlet and triplet states will still be degenerate at zero temperature, and this is consistent with our continuous symmetry argument in Sec. \ref{Sec:constinuous_symmetry_pairing_Channels}. 

Next let's compare the ground state energy between singlet and triplet pairing in the limit of the weak pairing potential (the gap function is much smaller than the band width). In this case, we mainly compare the eigen-energy spectrum of the BdG Hamiltonian. 
From the eigen-energies (\ref{eq:BdG_inter_Eigen_singlet}) and (\ref{eq:BdG_inter_Eigen_triplet}), we find the energy difference is 
\begin{eqnarray} &&\Delta E_\kk= \sum_n (E^{-,singlet}_{\kk,n}-E^{-,triplet}_{\kk,n})=\sum_n \left( - \sqrt{(d_{0,\kk}-\mu+ n d_{x,\kk})^2+|\Delta_{\kk}|^2} + \left(\sqrt{(d_{0,\kk}-\mu)^2+|\Delta_{\kk}|^2}+n |d_{x,\kk}|\right)\right)\nonumber\\
&&= - \sqrt{(d_{0,\kk}-\mu+ d_{x,\kk})^2+|\Delta_{\kk}|^2} -\sqrt{(d_{0,\kk}-\mu- d_{x,\kk})^2+|\Delta_{\kk}|^2} + 2 \sqrt{(d_{0,\kk}-\mu)^2+|\Delta_{\kk}|^2}
\end{eqnarray}

In the limit $|d_{x,\kk}|\gg |\Delta_{\kk}|, |d_{0,\kk}-\mu| $, we have 
\beq \Delta E_\kk\approx -2 |d_{x,\kk}|<0, \eneq
and we would expect the singlet pairing should be energetically favored. This is consistent with our early conclusion that the singlet pairing will have higher $T_c$ in the chiral non-flat band limit in Sec.\ref{Sec:Inter_LGE}.

\subsubsection{Estimate of screened Coulomb Interaction}
In TBG, the Coulomb interaction takes the form 
\beq 
H_I=\frac{1}{2N_{tot}}\sum_{q\in MBZ, G\in\mathcal{Q}_0 } V(q+G) \delta\rho_{q+G}\delta\rho_{-q-G},
\eneq
where $V(q)=\pi \xi^2 U_{\xi} \frac{\tanh{(\xi q/2)}}{(\xi q/2)}$, and
\beq
\delta\rho_{q+G}=\sum_{\eta s k n m}M^{(\eta)}_{nm}(k,q+G) \left(\gamma^{\dagger}_{k+q,n,\eta,s}\gamma_{k,m,\eta,s}-\frac{1}{2}\delta_{q,0}\delta_{nm}\right)
\eneq
with the form factor $M^{(\eta)}_{nm}(k,q+G)=\sum_{\alpha, Q_{l\eta}}(u^n_{k+q,Q_{l\eta}-G,\alpha,\eta})^*u^m_{k,Q_{l\eta},\alpha,\eta}$.
The above expression has included the screening from external gates and we choose the interaction parameter $U_{\xi}=\frac{e^2}{\epsilon \xi}\approx 24 meV$ 
and the screening length $\xi\approx 10 nm$. 

We next consider the bubble diagram for the Coulomb interaction, which describes from the screening of the Fermi surface of the nearly flat bands. 
We define the Green's function as
\beq
\mathcal{G}^{\eta s}_{0,n}(k,\tau)=-\langle T_{\tau} \gamma_{k,n,\eta,s}(\tau)\gamma^{\dagger}_{k,n,\eta,s}(0)\rangle
\eneq
and its Fourier transform 
\beq 
\mathcal{G}^{\eta s}_{0,n}(k,i\omega_n)=\int^{\beta}_{0}d\tau e^{i\omega_n \tau}\mathcal{G}^{\eta s}_{0,n}(k,\tau). 
\eneq
The bubble diagram for electric susceptibility (Fig. \ref{fig:bubble_Diagram}) is then given by 
\beq
\chi_0(q+G,i\nu_m)=\frac{1}{\beta N_0}\sum_{k,i\omega_n,n m \eta s} M^{(\eta)}_{nm}(k,q+G)\mathcal{G}^{\eta s}_{0,m}(k,i\omega_n)
M^{(\eta)}_{mn}(k+q,-q-G)\mathcal{G}^{\eta s}_{0,n}(k+q,i\omega_n+i\nu_m)
\eneq
and the screened Coulomb interaction is given by 
\beq 
V^{RPA}(q+G,i\nu_m)=\frac{V(q+G)}{1-V(q+G)\chi_0(q+G,i\nu_m)}. 
\eneq

\begin{figure}[hbt!]
   \centering
    \includegraphics[width=5in]{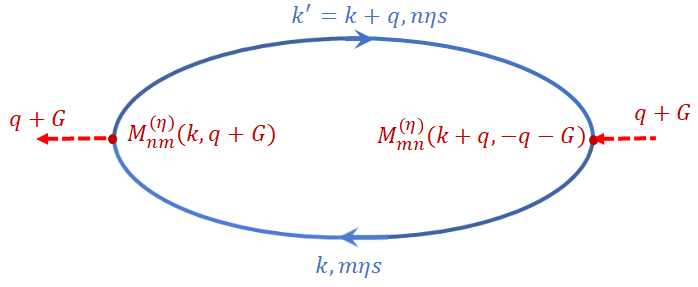} 
    \caption{ Bubble diagram for electric susceptibility. 
  }
    \label{fig:bubble_Diagram}
\end{figure}

We can perform the frequency summation of $i\omega_n$ in $\chi_0$ and the electric susceptibility is transformed as
\beq 
\chi_0(q+G,i\nu_m)=\sum_{n m \eta s} \int_{MBZ}\frac{d^2k}{(2\pi)^2} \frac{n_F(\varepsilon^{\eta s}_{m,k})-n_F(\varepsilon^{\eta s}_{n,k+q})}
{i\nu_m+\varepsilon^{\eta s}_{m,k}-\varepsilon^{\eta s}_{n,k+q}}|M^{(\eta)}_{nm}(k,q+G)|^2, 
\eneq
where $n_F(\varepsilon)$ is the Fermi distribution function and $\varepsilon^{\eta s}_{n,k}$ is the eigen-energy of the TBG Hamiltonian. 
Here we have used the Hermitian condition $M^{(\eta)}_{mn}(k+q,-q-G)=(M^{(\eta)}_{nm}(k,q+G))^*$. 

With $\varepsilon^{\eta s}_{m,k}=\varepsilon^{\eta }_{m,k}=\varepsilon^{-\eta }_{m,-k}$, we can simplify both the summations over $s$ and $\eta$
as $\chi_0=\chi^+_0(q+G)+\chi^-_0(q+G)=\chi^+_0(q+G)+\chi^+_0(-q-G)$, where 
\beq 
\chi^{\eta}_0(q+G,i\nu_m)=2\sum_{n m } \int_{MBZ}\frac{d^2k}{(2\pi)^2} \frac{n_F(\varepsilon^{\eta s}_{m,k})-n_F(\varepsilon^{\eta s}_{n,k+q})}
{i\nu_m+\varepsilon^{\eta s}_{m,k}-\varepsilon^{\eta s}_{n,k+q}}|M^{(\eta)}_{nm}(k,q+G)|^2. 
\eneq
Thus, we only need to consider $\chi^+_0$ part. Furthermore, we only consider static screening ($i\nu_m\rightarrow 0$) and the long 
wavelength limit $G=0, q\rightarrow 0$, and the above expression can be simplified as
\beq 
\chi^{+}_0(G)=2\sum_{m} \int_{MBZ}\frac{d^2k}{(2\pi)^2} \frac{\partial n_F(\varepsilon^{+}_{m,k})}
{\partial \varepsilon^{+}_{m,k}}|M^{(+)}_{mm}(k,G)|^2+2\sum_{n\neq m } 
\int_{MBZ}\frac{d^2k}{(2\pi)^2} \frac{n_F(\varepsilon^{\eta s}_{m,k})-n_F(\varepsilon^{\eta s}_{n,k})}
{i\nu_m+\varepsilon^{\eta s}_{m,k}-\varepsilon^{\eta s}_{n,k}}|M^{(\eta)}_{nm}(k,G)|^2. 
\eneq
Here the first term is the intra-band contribution while the second term is the inter-band contribution. For $G=0$, the form factor
is $M^{(+)}_{mm}(k,G=0)=1$ and $M^{(+)}_{n\neq m}(k,G=0)=0$. Thus, we only need to consider the intra-band contribution for $G=0$, so
\beq \label{Eq:Coulomb_chi0}
\chi_0(E_f)=\chi^+_0+\chi^-_0=4\sum_{m} \int_{MBZ}\frac{d^2k}{(2\pi)^2} \frac{\partial n_F(\varepsilon^{+}_{m,k})}{\partial \varepsilon^{+}_{m,k}}
=-4\sum_{m} \int_{MBZ}\frac{d^2k}{(2\pi)^2} \delta(\varepsilon^+_{m,k}-E_f)=-4D(E_f),
\eneq
where the last equality is for zero temperature and $D(E_f)$ is the density of state (DOS) per valley per spin at the Fermi energy $E_f$
and the number 4 is from spin and valley. 
In the limit $q\rightarrow 0$, the screened Coulomb interaction is then given by 
\beq 
V^{RPA}(q\rightarrow 0)=\frac{V(q\rightarrow 0)}{1-V(q\rightarrow 0)\chi_0(E_f)}\approx \frac{\pi \xi^2 U_{\xi}}{1-\pi \xi^2 U_{\xi}\chi_0(E_f)}
=\frac{\pi \xi^2 U_{\xi}}{1+4\pi \xi^2 U_{\xi}D(E_f)}. 
\eneq
We may give an estimate of the above expression with $U_{\xi}\approx 24 meV$ and $\xi\approx 10 nm$. We may estimate the average value of $D(E_f)$
as $\bar{D}(E_f)\approx \frac{\mathcal{A}_{MBZ}}{(2\pi)^2 w_b}\approx 0.0021 nm^{-2} meV^{-1}$, where the band width $w_b$ of the nearly flat bands is chosen around $3 meV$ 
(Fig. \ref{fig:Coulomb_screening}(a)), and the Moir\'e BZ area is $\mathcal{A}_{MBZ}=\frac{3\sqrt{3}}{2}k_{\theta}^2\approx 0.25 nm^{-2}$. 
Let's define relative dielectric constant as 
$V^{RPA}(q\rightarrow 0)=\frac{\pi \xi^2 U_{\xi}}{\bar{\epsilon}}$, so $\bar{\epsilon}=1+4\pi \xi^2 U_{\xi}D(E_f)\approx 65.4$. 
Thus, With the bubble diagram, we expect the averaged effective interaction strength, defined as $V^{RPA}(q\rightarrow 0)=\pi \xi^2 U^{RPA}_{\xi}$, is estimated as
$U^{RPA}_{\xi}=\frac{U_{\xi}}{\bar{\epsilon}}\approx 0.4 meV$. 

The above estimate is only for an averaged value, but due to the rapid change of the DOS, we expect the screening effect will also be very different for different chemical potential. 
We may further numerically evaluate Eq. (\ref{Eq:Coulomb_chi0}) and the calculated $\chi_0$, $\epsilon$ and $U^{RPA}_{\xi}$ as a function of the Fermi energy
$E_f$ are shown in Fig. \ref{fig:Coulomb_screening}. We find near the peak of DOS, the effective Coulomb interaction strength $U^{RPA}_{\xi}$ is indeed significantly suppressed.

\begin{figure}[hbt!]
   \centering
    \includegraphics[width=6in]{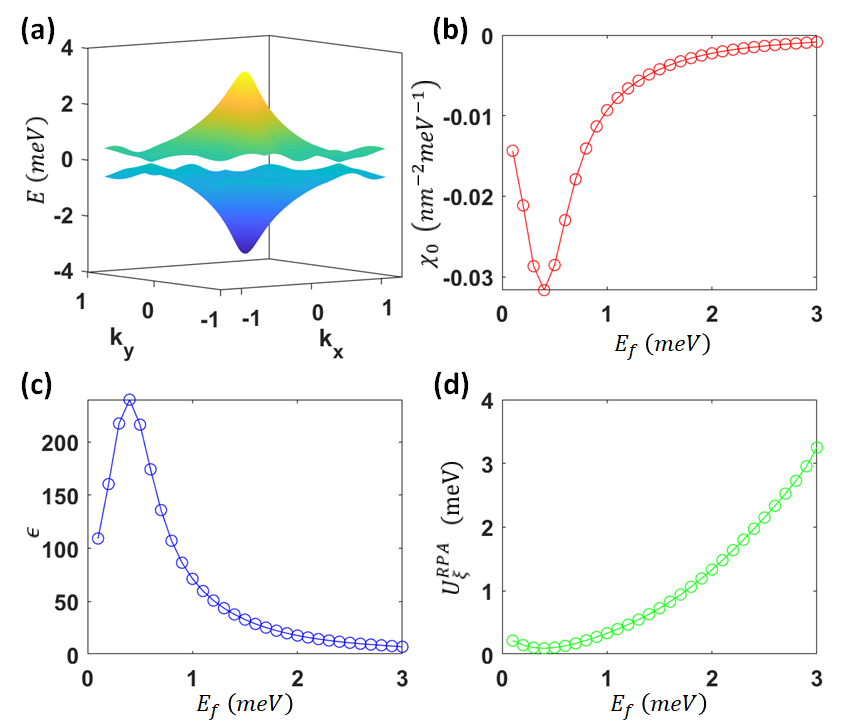} 
    \caption{ (a) Energy dispersion for two nearly flat bands of TBG ($w_0=0.7 w_1$, $\theta=1.05^\circ$). Here $k_x$ and $k_y$ are in the unit of $k_\theta$. (b) Electric susceptibility $\chi_0$ as a function of $E_f$; (c) Relative dielectric constant $\epsilon$ as a function of $E_f$; and (d)    the RPA interaction strength $U^{RPA}_{\xi}$ as a function of $E_f$. 
  }
    \label{fig:Coulomb_screening}
\end{figure}

\end{document}